\documentclass[aps,prd,onecolumn,%
superscriptaddress,preprintnumbers,showpacs,%
floatfix,nofootinbib,longbibliography,noeprint,%
amsmath,amssymb,eqsecnum]{revtex4-2}
\usepackage[T1]{fontenc}
\usepackage[utf8]{inputenc}
\usepackage[english]{babel}
\usepackage{lmodern}
\usepackage[normalem]{ulem}
\usepackage{graphicx,fancybox}
\usepackage[dvipsnames]{xcolor}
\usepackage{pgf}
\usepackage{url}
\usepackage[colorlinks=true, pdfstartview=FitV, linkcolor=red, citecolor=blue, urlcolor=blue]{hyperref}
\usepackage{amsmath,amssymb,amsfonts,mathrsfs,slashed,bbold, bm,braket,xfrac,lineno}
\usepackage{longtable,boldline}
\usepackage{cellspace} % It appears that this is necessary for longtable in revtex
\usepackage{multirow}
\usepackage{colortbl}
\usepackage{hhline}
\usepackage{animate}
\usepackage{placeins}
\usepackage{dcolumn}% Align table columns on decimal point
\usepackage{tikz}
\usetikzlibrary{matrix}
\usepackage{wasysym}
\usepackage{wrapfig}
\usepackage{listings}
\usepackage[absolute,verbose,overlay]{textpos}
\usepackage[title]{appendix}
\usepackage{acronym}
\acrodef{ASQTAD}[Asqtad]{$a^2$ tadpole-improved staggered fermion}
\acrodef{BC}[BC]{Borici-Creutz}
\acrodef{BCF}[BCF]{\acl{BC} fermion}
\acrodef{ChPT}[ChPT]{chiral perturbation theory}
\acrodef{DTO}[DTO]{dropped \acl{TO}}
\acrodef{DTOF}[DTOF]{\acl{DTO} fermion}
\acrodef{DW}[DW]{domain-wall}
\acrodef{DWF}[DWF]{\acl{DW} fermion}
\acrodef{EFT}[EFT]{effective field theory}
\acrodef{GW}[GW]{Ginsparg-Wilson}
\acrodef{GWF}[GWF]{\acl{GW} Fermions}
\acrodef{HISQ}[HISQ]{highly-improved staggered quarks}
\acrodef{LCP}[LCP]{lines of constant physics}
\acrodef{LFT}[LFT]{lattice field theory}
\acrodef{LGT}[LGT]{lattice gauge theory}
\acrodef{LO}[LO]{leading order}
\acrodef{LQCD}[LQCD]{lattice \acs{QCD}}
\acrodef{LW}[LW]{L{\"u}scher-Weisz}
\acrodef{KS}[KS]{Kogut-Susskind}
\acrodef{KSF}[KSF]{\acl{KS} fermion}
\acrodef{KW}[KW]{Karsten-Wilczek}
\acrodef{KWF}[KWF]{\acl{KW} fermion}
\acrodef{MDF}[MDF]{minimally doubled fermion}
\acrodef{MITP}[MITP]{Mainz Institute for Theoretical Physics}
\acrodef{NG}[NG]{Nambu-Goldstone}
\acrodef{NLO}[NLO]{next-to-leading order}
\acrodef{NTM}[NTM]{nontopological modes}
\acrodef{QCD}[QCD]{Quantum Chromodynamics}
\acrodef{QFT}[QFT]{quantum field theory}
\acrodef{stag}[stag.]{staggered}
\acrodef{SF}[SF]{staggered fermion}
\acrodef{TO}[TO]{twisted-ordering}
\acrodef{TOF}[TOF]{\acl{TO} fermion}
\acrodef{TD}[TD]{taste diagonal}
\acrodef{T-D}[TD]{taste-diagonal}
\acrodef{TI}[TI]{taste-isovector}
\acrodef{TN}[TN]{taste-nondiagonal}
\acrodef{TS}[TS]{taste singlet}
\acrodef{T-S}[TS]{taste-singlet}
\acrodef{TSB}[TSB]{taste-symmetry breaking}
\acrodef{TT}[TT]{taste-nondiagonally twisted}
\acrodef{ST}[ST]{spin-taste}
\acrodef{WBZ}[WBZ]{would-be zero}
\definecolor{hu-berlin-blue}{RGB}{0,65,137} % HEX 004189
\definecolor{hu-berlin-green}{RGB}{150,190,20} % HEX 93C11A 
\definecolor{hu-berlin-grey}{RGB}{169,169,169}
\definecolor{hu-berlin-brown}{RGB}{82,79,60}
\definecolor{hu-berlin-red}{RGB}{180,0,0}
\definecolor{wsu-green}{RGB}{0,89,76}
\definecolor{mc-gill-red}{RGB}{237, 27, 47}
\definecolor{tum-blue}{RGB}{0,101,189}
\definecolor{clustergray}{RGB}{117,128,145}
\definecolor{clusterorange}{RGB}{255,158,37}
\newcommand{\ee}{\textrm{e}} % for e
\DeclareMathOperator{\Tr}{Tr}
\DeclareMathOperator{\Imag}{\mathrm{Im}}
\DeclareMathOperator{\Real}{\mathrm{Re}}

\newcommand{\bmat}{\left(\begin{array}}
\newcommand{\emat}{\end{array}\right)}
\newcommand{\bvec}{\left(\begin{array}{c}}
\newcommand{\evec}{\end{array}\right)}

\newcommand{\markit}{}

\newcommand{\txtw}{\textwidth}

\newcommand{\pad}{\partial}

\newcommand{\inv}{^{-1}}
\newcommand{\pri}{^\prime}
\newcommand{\crc}{^\circ}
\renewcommand{\dag}{^\dagger}
\newcommand{\daginv}{^{\dagger-1}}
\newcommand{\bare}{^{\text{bare}}}
\newcommand{\ren}{^{\text{ren}}}

\newcommand{\gaf}{\gamma_{5}}
\newcommand{\Gaf}{\Gamma_{5}}
\newcommand{\mMf}{\mM_{5}}
\newcommand{\gaD}{\gamma_{D}}
\newcommand{\GaD}{\Gamma_{D}}
\newcommand{\mMD}{\mM_{D}}
\newcommand{\gaDf}{\gamma_{D5}}
\newcommand{\GaDf}{\Gamma_{D5}}
\newcommand{\mMDf}{\mM_{D5}}
\newcommand{\gap}{\gamma^{\prime}}
\newcommand{\Gap}{\Gamma^{\prime}}

\newcommand{\nab}{\nabla}
\newcommand{\lap}{\triangle}
\newcommand{\hDj}{h_{Dj}}

\newcommand{\al}{\alpha}

\newcommand{\ga}{\gamma}
\newcommand{\Ga}{\Gamma}
\newcommand{\de}{\delta}

\newcommand{\ep}{\epsilon}

\newcommand{\ze}{\zeta}
\newcommand{\et}{\eta}
\newcommand{\etD}{\tilde{\eta}_{D}}
\newcommand{\etf}{\tilde{\eta}_{5}}
\newcommand{\etDf}{\tilde{\eta}_{D5}}
\newcommand{\etj}{\tilde{\eta}_{j_{0}}}
\newcommand{\etjf}{\tilde{\eta}_{j_{0}5}}

\newcommand{\ka}{\kappa}
\newcommand{\ska}{\sin(\kappa)}
\newcommand{\sqka}{\sin^2(\kappa)}
\newcommand{\la}{\lambda}
\newcommand{\La}{\Lambda}

\renewcommand{\vr}{\varrho}
\newcommand{\si}{\sigma}
\newcommand{\Si}{\Sigma}

\newcommand{\vp}{\varphi}

\newcommand{\ps}{\psi}
\newcommand{\om}{\omega}

\newcommand{\bdm}{\begin{displaymath}}
\newcommand{\edm}{\end{displaymath}}
\newcommand{\bea}{\begin{eqnarray}}
\newcommand{\eea}{\end{eqnarray}}
\newcommand{\beq}{\begin{equation}}
\newcommand{\eeq}{\end{equation}}

\newcommand{\mr}{\mathrm}

\newcommand{\ri}{\mr{i}}
\newcommand{\rr}{\mr{r}}
\newcommand{\Nf}{N_{\!f}}%{{N_{\!f}}}
\newcommand{\aD}{a_{D}}%{{n_{ D }}}
\newcommand{\nD}{n_{D}}%{{n_{ D }}}
\newcommand{\ND}{N_{D}}%{{N_{ D }}}
\newcommand{\DN}{D_\text{N}}
\newcommand{\DW}{D_\text{W}}
\newcommand{\DKW}{D_\text{KW}}

\newcommand{\mDKW}{\mD_\text{KW}}

\newcommand{\mB}{\mathcal{B}}
\newcommand{\mC}{\mathcal{C}}
\newcommand{\mD}{\mathcal{D}}

\newcommand{\mK}{\mathcal{K}}

\newcommand{\mM}{\mathcal{M}}

\newcommand{\mO}{\mathcal{O}}
\newcommand{\mP}{\mathcal{P}}

\newcommand{\mR}{\mathcal{R}}

\newcommand{\mT}{\mathcal{T}}

\newcommand{\Id}{\text{Id}}

\newcommand{\nequn}[1]{{\normalsize\begin{equation}#1\end{equation}}}

\newcommand{\nalign}[1]{\vskip-0ex{\normalsize\begin{align}#1\end{align}}}

\newcommand{\tm}[1]{{${#1}$}}
\begin{document}

\title{Spin-taste representation of minimally doubled fermions from first principles: Karsten-Wilczek fermions}
% \title{Taste structure of \aclp{MDF}: \acl{KW} fermions}
\author{Johannes~Heinrich~Weber} 
\affiliation{Institut f{\"u}r Physik, Humboldt-Universit{\"a}t zu Berlin \& IRIS Adlershof, Zum gro{\ss}en Windkanal 2, D-12489 Berlin, Germany}
\affiliation{Institut f{\"u}r Kernphysik, Technische Universit{\"a}t Darmstadt, Schlossgartenstra{\ss}e 2, D-64289 Darmstadt, Germany}

\preprint{TUM-EFT 104/17; HU-EP-25/09-RTG}
\date{\today}

\begin{abstract}
Minimally doubled fermions realize one pair of Dirac fermions on the lattice. 
Similarities to staggered fermions exist, namely, spin and taste degrees of freedom become intertwined, and a peculiar nonsinglet chiral symmetry and ultralocality are maintained. 
However, charge conjugation, some space-time reflection symmetries and isotropy are broken by the cutoff. 
We address the most simple variant, Karsten-Wilczek (KW) fermions, in its parallel and in its perpendicular version. 
We derive the correct spin-taste representation from first principles. 
% For two variants, i.e., \acf{KW} or \acf{BC} fermions, a tasted charge conjugation symmetry can be identified, and the respective representations of the spin-taste algebra can be constructed explicitly. 
% In the case of \acs{BC} fermions, the tasted symmetry indicates that amendments to the published counterterms are necessary.
The spin-taste representation on the quark level permits construction of local or extended hadron interpolating operators for any spin-taste combination, albeit with contamination by parity partners and taste-symmetry breaking. 
We classify all interpolating operators for mesons and diquarks, and give examples for baryons. 
We also discuss the appropriate discretizations for taste-singlet or taste-isovector mass or chemical potential terms. 
We explain the counterterms in the spin-taste framework, and derive generic constraints on the parametric form and cutoff effects from the KW determinant and hadronic correlation functions. 
We derive how and why nonperturbative tuning schemes for the counterterms work, and obtain analytic, assumption-free, nonperturbative predictions for taste-symmetry breaking and other hadronic properties from first principles. 
In particular, we identify the origin and nature of two different types of taste-symmetry breaking cutoff effects. 
The few available numerical results for KW fermions validate these predictions. 
% For a third variant called \aclp{TOF} a similar tasted charge conjugation symmetry cannot be identified. 
\end{abstract}

\maketitle
\tableofcontents
%%%%%%%%%%%%%%%%%%%%%%%%%%%%%%%%%%%%%%%%%%%%%%%%%%%%%%%%%%%%%%%%%%%%%%%%%%%%%%%%

%%%%%%%%%%%%%%%%%%%%%%%%%%%%%%%%%%%%%%%%%%%%%%%%%%%%%%%%%%%%%%%%%%%%%%%%%%%%%%%%
% \input{\INCDIR/intro}
%% intro.tex

\section{Introduction}
\label{sec:intro}

Any lattice operator describing relativistic fermions fails to realize some of the favorable properties of the corresponding continuum operator. 
The question of an \emph{optimal choice} for a lattice fermion operator has been around since the inception of \acl{LGT}~\cite{Wilson:1974sk}. 
The answer has always been that \emph{optimal} depends on the details of the problem one wants to study, and it has always come out as a compromise between ideals of theoretical cleanliness and practical requirements of a computation. 
The Nielsen-Ninomiya No-Go theorem~\cite{Nielsen:1980rz, Nielsen:1981xu} is the ultimate arbiter which favorable properties can be realized together. 
\vskip1ex

A naive approach to putting Dirac fermions on the lattice leads to the infamous \emph{doubling problem}, as there are \tm{2^D} species of Dirac fermions that become degenerate in the continuum limit. \tm{D} is the dimension of the underlying space-time grid (\tm{D=4} for \ac{QCD}). 
In reminiscence of the \emph{flavor} quantum number these fermion species are referred to as \emph{tastes}. 
In the free theory an internal \emph{taste group} structure of \tm{\text{U}(2^{\sfrac{D}{2}}) \times \text{U}(2^{\sfrac{D}{2}})} arises among these~\cite{Kawamoto:1981hw, Kluberg-Stern:1981zya}, similar to the \emph{flavor group} structure \tm{\text{U}(\Nf)} in the continuum. 
One of these two internal taste groups is realized as the \emph{doubling symmetry} group~\cite{Karsten:1980wd} and can be diagonalized exactly for arbitrary gauge fields~\cite{Kawamoto:1981hw, Sharatchandra:1981si}. 
Thinning out the degrees of freedom of the naive discretization by keeping one component at each site (after \emph{spin diagonalization}~\cite{Kawamoto:1981hw}) in the \emph{\acl{SF}} approach the number of species is reduced from \tm{2^D} to \tm{2^{\sfrac{D}{2}}}. 
The other internal taste group mixes with the spin group, and gives rise to a nontrivial \emph{\ac{ST} structure} specific to each fermion discretization. 
While a taste multiplet structure is not realized for nonsmooth gauge fields, it could be recovered to a high degree by smoothing any given gauge field configuration~\cite{Durr:2004as}. 
However, since the \emph{\ac{TSB}} fluctuates configuration by configuration, neither can the taste overcounting be divided out for an inexact degeneracy nor can an exact degeneracy be recovered. 
The approximate or explicit \ac{TSB} usually comprises the dominant discretization errors.
\vskip1ex

In \ac{QCD} thermodynamics, the \emph{\acl{stag}} or \emph{\acl{KS}} approach~\cite{Susskind:1976jm} is commonly used in some of its improved variants. 
One full hypercube, i.e. \tm{2^{D}} sites, are necessary to accommodate the full field content of \tm{2^{\sfrac{D}{2}}} Dirac fermions. 
Unless one can afford to deal with multiples of \tm{2^{\sfrac{D}{2}}} Dirac fermions, either degenerate or nondegenerate with an extended \emph{\acl{TI} mass term}~\cite{Golterman:1984cy, Chreim:2024duv}, the \acl{stag} approach must be amended. 
This is usually dealt with by taking a root of the \acl{stag} determinant obtained after the Grassmann variables are formally integrated out, although it needs to be pointed out that this \emph{\acl{stag} rooting} approach has some severe drawbacks. 
First of all, while the \ac{TSB} between the eigenvalues becomes exponentially small for smooth gauge fields~\cite{Ammer:2024yro, Durr:2024ttb}, there is no formal proof of correctness for \emph{\acl{stag} rooting} in any nonperturbative circumstances due to the inexact degeneracy within eigenvalue multiplets at finite lattice spacing \tm{a}. 
Second, there is no local operator realizing rooted \aclp{SF}, and therefore, simulations with \aclp{SF} suffer from partial quenching, e.g. an inequivalence between valence and sea quarks that leads to unitarity violations at finite lattice spacing \tm{a}. 
At least for vacuum physics, or for thermodynamics at zero density this does not appear to be a practical problem. 
Formally, the approximate degeneracy results in soft \ac{TSB} cutoff effects in the (rooted) determinant at \tm{\mO(a^2)}, and all reported, continuum extrapolated results with \aclp{SF} are in excellent agreement with calculations using other fermion formulations. 
Many improved variants of \aclp{SF} have been developed that diminish \ac{TSB} cutoff effects at the price of reducing locality in a well-controlled fashion~\cite{Orginos:1999cr, Follana:2006rc}. 
However, at real chemical potential this approach fails completely~\cite{Golterman:2006rw}, once the flavor-singlet or light-quark chemical potential becomes similar to half the pion mass~\cite{Borsanyi:2023tdp}. 
At present it is unclear whether the drawbacks of \emph{staggered rooting} could eventually be overcome with rigorously controlled systematic uncertainties. 
Given the various issues with any other known formulation of lattice fermions, the \acl{LQCD} thermodynamics community has started to ponder the viability of the \emph{\ac{MDF}} approach. 
\vskip1ex

\ac{MDF} are yet another particularly interesting family of strictly local discretizations, which maintain a specific \emph{\acl{TI} chiral symmetry} between two surviving tastes of opposite chirality in any dimension \tm{D}, and thereby stay within the No-Go theorem's requirements. 
This is achieved by adding a symmetric (restricted) Laplacian term with a nontrivial gamma structure that preserves this \emph{chiral symmetry}, while it violates hypercubic symmetry in order to achieve the necessary \ac{TSB}. 
Due to the product of a gamma matrix times a Laplacian, each \ac{MDF} breaks \emph{charge conjugation symmetry} and some \emph{reflection symmetry} while maintaining the CPT symmetry.
Thus, one has to choose whether one deliberately breaks parity, time reflection invariance, or both.
\ac{MDF} have the spinor structure of Dirac fermions with \tm{2^{\sfrac{D}{2}}} components at each site, like Wilson-type fermions. 
Due to the minimal doubling, at least two sites (of opposite lattice parity% , i.e. p(n)=(-1)^{\sum_\mu n_\mu} for a lattice site labeled by the vector n% 
) are necessary to accommodate the full field content. 
Together the family of such modified operators is subsumed under the term \ac{MDF}. 
\vskip1ex

These ideas have been introduced originally by Karsten~\cite{Karsten:1981gd} and Wilczek~\cite{Wilczek:1987kw} with an extra term breaking time-reflection invariance (\tm{\mT}). 
The surviving \acl{WBZ} modes are located at the two naive propagator poles on the temporal axis. 
This \emph{\ac{KWF}} proposal has a dimensionless \tm{r}~parameter~\cite{Wilczek:1987kw} similar to its analog for Wilson fermions. 
With \tm{r^2>\sfrac{1}{4}} the \emph{\ac{KW}} operator is minimally doubled (in arbitrary \tm{D}), see~\cite{Durr:2020yqa}. 
Questions concerning \ac{KWF} and taste were addressed for the first time by Pernici in an analytic study of a parity~(\tm{\mP})-breaking variant with the surviving tastes distributed along one spatial axis~\cite{Pernici:1994yj}. 
For reasons we spell out later and to make the distinction explicit in line with conventions introduced a long time ago~\cite{Weber:2015oqf}, we refer to Pernici's variant as perpendicular \ac{KWF}, while naming the usual variant (parallel) \ac{KWF}. 
\ac{MDF} had been practically forgotten for two decades until the \ac{MDF} revival brought about in the wake of the experimental discovery of graphene~\cite{Novoselov:2005kj}, where carbon atoms form a hexagonal lattice in a two-dimensional layer, within which a pair of electrons propagates with the relativistic dispersion relation of massless degrees of freedom~\cite{CastroNeto:2007fxn}. 
After Creutz generalized the concept to the \tm{D=4} case~\cite{Creutz:2007af}, new ideas proliferated quickly, and Borici reformulated these on the usual hypercubic lattices~\cite{Borici:2007kz}. 
Thanks to Creutz's physical insight~\cite{Creutz:2008sr}, this discretization has become decoupled from the original notion of a generalized graphene, and is usually referred to as the \emph{\ac{BCF}} proposal. 
With some care an \tm{r} parameter can be introduced in the \emph{\ac{BC}} operator to control the doubling. 
For \tm{r^2>\sfrac{1}{2}} (resp. \tm{\sfrac{1}{3}}) \ac{BCF} are minimally doubled (at least in \tm{D=4} resp. \tm{2})~\cite{Durr:2020yqa}.
It has been realized that \ac{BCF} with \tm{r^2=1} have an additional symmetry (for both \tm{D=4} or \tm{2}) that is lost for general \tm{r}~\cite{Weber:2023kth}, where Creutz's original proposal~\cite{Creutz:2008sr} corresponds to \tm{r=-1}. 
The surviving \acl{WBZ} modes of this class of operators cannot be located at any pair of the naive propagator poles, but are located instead on a hypercube diagonal axis that is selected in the operator construction. 
The lifting patterns of \ac{BCF} or \ac{KWF} clearly show that these are two distinct variants~\cite{Durr:2020yqa}. 
\ac{BCF} simultaneously break parity (\tm{\mP}) and time-reflection (\tm{\mT}) invariance, and the forward or backward orientations on the hypercube diagonal axis are inequivalent, such that no variant of \ac{BCF} could have as much symmetry as \ac{KWF} that preserve at least one of the two discrete reflection symmetries. 
\vskip1ex

During an enthusiastic initial period of the \ac{MDF} revival, a large number of researchers studied many different aspects of \ac{MDF}~\cite{Cichy:2008gk, Bedaque:2008xs, Bedaque:2008jm, Kimura:2009di, Kimura:2009qe, Chakrabarti:2009sa, Capitani:2009yn, Tiburzi:2010bm, Capitani:2010nn, Creutz:2010cz, Creutz:2010qm, Misumi:2012uu, Misumi:2012ky, Capitani:2013zta, Capitani:2013iha, Weber:2013tfa, Weber:2015hib, Weber:2015oqf, Basak:2017oup, Hyka:2018oza, Durr:2020yqa, Osmanaj:2022mzs}. 
The generalizations of the underlying concepts led to a classification of the known variants of \ac{MDF} on a hypercubic lattice within a common framework, where new types of \emph{\acl{TO}} or \emph{\acl{DTO}} fermions were proposed~\cite{Creutz:2010cz} and generalized further~\cite{Seibt:2024a}, which are all related to \ac{BCF} or \ac{KWF}, respectively, and are identical to these in \tm{D=2}~\cite{Durr:2020yqa}. 
Therefore, the family of \ac{MDF} has for \tm{D>2} at least four distinct branches that cannot be deformed into one another by any discrete transforms at finite lattice spacing. 
Improvement in terms of the Symanzik \ac{EFT}~\cite{Symanzik:1983dc, Symanzik:1983gh} and Brillouin~\cite{Durr:2010ch} or Naik~\cite{Naik:1986bn} improvement schemes have been explored~\cite{Weber:2015hib, Vig:2024umj, Bakenecker:2024a}. 
% Whether these novel branches of \ac{MDF} are of any practical utility remains to be seen, but at least the Naik improvement of the \ac{KW} operator appears to be a valuable extension. 
Exploratory \tm{D=4} or \tm{D=2} (partially) quenched or dynamical simulations with \ac{KWF} or \ac{BCF} have been performed, too~\cite{Weber:2013tfa, Weber:2015hib, Weber:2015oqf, Durr:2022mnz, Godzieba:2024uki, Vig:2024umj, Ammer:2024yro, Durr:2024ttb, Kishore:2025fxt}. 
More interesting results based on fermions with minimal number of doublers have been obtained for naive or \acl{stag} operators complemented with extended nonsinglet mass terms~\cite{Golterman:1984cy, Adams:2009eb, Chreim:2024duv, Hoelbling:2010jw, Creutz:2011cd, deForcrand:2012bm}, both directly, or as kernels of overlap or \ac{DW} operators~\cite{Adams:2010gx, Creutz:2010bm, Hoelbling:2016dug, Chreim:2022wlm}. 
Ongoing interest in \ac{MDF} led to new considerations of central branch Wilson fermions~\cite{Misumi:2020eyx} that of course break chiral symmetry, which can be generalized to forms that actually realize minimal doubling for arbitrary dimension \tm{D}~\cite{Durr:2022mnz}. 
While any of these operators are or could be made minimally doubled, they are structurally different from the family of \ac{MDF}. 
\vskip1ex

\ac{KWF} are optimally suited for setups where \emph{anisotropic lattices} are desirable, since \ac{KWF}---on the contrary to any other lattice fermion approach---do not run into extra complications. 
This is a very attractive proposition for \ac{QCD} spectroscopy at finite temperature, since a good resolution of temporal correlation functions---providing enough information to resolve spectral features related to real-time in-medium dynamics---requires a large number of points at a fixed physical length of the Euclidean time direction that sets the inverse temperature. 
Similar to the earlier mentioned case of thermodynamics of \emph{isospin-symmetric two-flavor \ac{QCD} at real baryon chemical potential}, where importance sampling with \acl{SF} sea and staggered rooting runs into difficulties, \ac{KWF} present themselves as an almost natural solution, both for the quark sea and for the valence sector. 
While including strange quarks in the sea in such a setup with \ac{KW} light quarks may be conceptually more challenging, it is much less problematic in a physical sense due to the much smaller impact of the strange quark on \ac{QCD}'s critical behavior. 
However, we have to first understand the contribution of \ac{KWF} to the \tm{\Nf=2} vacuum and with an extra strange quark, before we may understand the subtle changes at finite temperature or finite density. 
\vskip1ex

It has to be pointed out that the symmetries claimed for \ac{MDF} in various publications are incompatible, often incomplete, and in many cases even incorrect. 
The temporal link-reflection symmetry correctly reported for perpendicular \ac{KWF}~\cite{Pernici:1994yj} was incorrectly claimed for parallel \ac{KWF} as well~\cite{Bedaque:2008xs}. 
Moreover, the peculiar site-reflection symmetry claimed for parallel \ac{KWF}~\cite{Tiburzi:2010bm} and reiterated in the construction of chiral Lagrangians for \ac{KWF}~\cite{Shukre:2024tkw} does not exist. 
The \ac{ST} decomposition through a time-smearing procedure claimed for \ac{KWF} would give rise to a \acl{TD} structure of the \ac{KW} operator, i.e. as the sum of two independent operators for the two time-smeared tastes~\cite{Tiburzi:2010bm}. 
These claims are inconsistent with the results derived from first principles, which clearly show that the two extra terms in the \ac{KW} operator have two different \ac{ST} structures, a \acl{TD} and a \acl{TN} isovector one~\cite{Kimura:2011ik, Weber:2023kth}. 
Since a first attempt to construct a chiral Lagrangian for \ac{KWF}~\cite{Shukre:2024tkw} was built on the incorrect \ac{ST} structure~\cite{Tiburzi:2010bm}, it failed to realize the nontrivial intertwining of the two tastes and advertised---incorrectly---a two-taste chiral Lagrangian consisting of two independent copies of single-taste Lagrangians. 
Moreover, three out of the seven \tm{\mO(a)} operators in the Symanzik \ac{EFT} of~\cite{Shukre:2024tkw} are prohibited by the underlying chiral symmetry~\cite{Weber:2015hib}, and the chiral Lagrangian contains inexistent terms prohibited by symmetry, but missed other terms that should be among the actual \ac{LO} corrections. 
A similar \ac{ST} decomposition for \ac{BCF} would give rise to a \acl{TD} structure of the \ac{BC} operator~\cite{Bedaque:2008xs}, too, but is inconsistent with the results derived from first principles, which clearly show that the two extra terms in the \ac{BC} operator have different \ac{ST} structures~\cite{Kimura:2011ik, Weber:2023kth}. 
As a consequence, the tasted charge-conjugation symmetry claimed for \ac{BCF}~\cite{Bedaque:2008xs} is inconsistent with the one derived from first principles~\cite{Weber:2023kth}. 
There is a major problem with the \ac{ST} identification~\cite{Kimura:2011ik} as well. 
It includes fake \ac{TSB} terms that appear at one power higher or lower in \tm{a} versus the corresponding genuine hopping terms, i.e of the derivative or Laplacian operators. 
These are analogous to the corresponding ones that plague the widely used \ac{ST} identification of \aclp{SF}~\cite{Kluberg-Stern:1983lmr}, and arise due to choices of extended stencils to extract tastes from the spinors~\cite{Lepage:2011vr}.
While these accompany each hopping term, they are not caused by the underlying operator, and they occlude its \ac{ST} structure and its cutoff effects. 
The \emph{ad hoc} point-splitting procedures for \ac{KWF}~\cite{Creutz:2010cz, Creutz:2010qm} or \ac{BCF}~\cite{Basak:2017oup, Hyka:2018oza, Osmanaj:2022mzs} also fail to separate the tastes correctly, since they do not start from the underlying symmetries.  
Both \ac{BCF} and \ac{KWF} have a \emph{natural \acl{ST} representation} that can be derived from a \emph{tasted charge-conjugation symmetry}. 
The proper Hermitian representation of the taste algebra for \ac{BCF} is a very recent result~\cite{Weber:2023kth}, while the corresponding representation for \ac{KWF} is known much longer~\cite{Weber:2016dgo}.
The discrepancies, fake \ac{TSB} cutoff effects, or inabilities to achieve the desired taste projections---in particular, due to misinterpreting in principle known properties or reiterating already debunked falsehoods---indicate that the ongoing confusion about \ac{MDF} must be overcome through a systematic discussion based on first principles and the correct underlying symmetries. 
\vskip1ex

This paper contains this much needed discussion for \ac{KWF} and is organized as follows. 
Firstly we recapitulate well-known features of naive and Wilson fermions and introduce most of our notation in Sec.~\ref{sec:NW fermions}. 
Sec.~\ref{sec:KW_quark} begins the main content of this treatise, where we derive the \ac{KW} representation of the taste generators and the \ac{ST} structure of the spinor field from first principles. 
Next, in Sec.~\ref{sec:KW_bilinears}, we study fermion bilinear operators, and derive the \ac{ST} structures of mesons and diquarks and discuss some interpolating operators for baryons. 
There we also derive appropriate \acl{T-S} or \acl{TI} mass terms and chemical potential terms. 
At last, in Sec.~\ref{sec:KW_gauge}, we study the implications of coupling \ac{KWF} to gauge fields, which involves issues of counterterms and renormalization, the fermion determinant and cutoff effects. 
We continue with tuning of the relevant counterterm and signatures of its mistuning, and finally analyze the \ac{LO} \ac{TSB} cutoff effects in the pseudoscalar sector. 
Some lengthier calculations have been deferred to the Appendix~\ref{app:KW_other_pNGB_details}. 
We close the paper with a summary in Sec.~\ref{sec:summary}. 
A similar treatise focusing on \ac{BCF} is expected to require extensive effort, but should be straightforward based on~\cite{Weber:2023kth} and this paper. 
Whether or not there is sufficient physical interest in and relevance of a such work for \ac{BCF} is an open question. 
\vskip1ex

A few days before finishing this treatise a preprint of a numerical study of \ac{TSB} with \ac{KWF} was posted to the arXiv~\cite{Borsanyi:2025big}, which continues previous studies published in conference proceedings~\cite{Godzieba:2024uki, Vig:2024umj} 
It is the first study to actually implement our previously incomplete description of \ac{ST} structure of \ac{KWF} that was published in our previous conference proceedings~\cite{Weber:2023kth}.
We comment on its results in the context of our predictions in the summary. 
% \clearpage
%%%%%%%%%%%%%%%%%%%%%%%%%%%%%%%%%%%%%%%%%%%%%%%%%%%%%%%%%%%%%%%%%%%%%%%%%%%%%%%%

%%%%%%%%%%%%%%%%%%%%%%%%%%%%%%%%%%%%%%%%%%%%%%%%%%%%%%%%%%%%%%%%%%%%%%%%%%%%%%%%
% \input{\INCDIR/naive}
%% naive.tex
\acresetall

\section{Naive or Wilson-type fermions}
\label{sec:NW fermions}

A discretized fermion action with a lattice Dirac operator \tm{D(x,y)} is given by 
\nequn{
S = \big(\prod_{\mu} a_\mu \big)\sum\limits_{x,y} \bar\ps(x)~D(x,y)~\ps(y)
~.
}%nequn
The naive or Wilson Dirac operators with vanishing bare mass are defined~\cite{Wilson:1974sk} as 
\nalign{
\DN(x,y)
&=\sum_\mu \ga_\mu \xi_{f0,\mu} \nab_\mu(x,y)
,~\label{eq:D_naiv}\\
\DW(x,y)
&=\sum_\mu \ga_\mu \xi_{f0,\mu} \nab_\mu(x,y)
-\frac{r a_{D}}{2} \sum_\mu \xi_{f0,\mu}^2 \lap_\mu(x,y)
~.\label{eq:D_wils}
}%nalign
where we have introduced bare fermion anisotropies in the time direction 
\nequn{
\xi_{f0,\mu} 
\equiv \frac{a_{\mu}}{a_{D}}
~.~\label{eq:xif0mu}
}%nequn
We will have to keep track of anisotropies in the later discussion of \aclp{MDF}, since these are always anisotropic and generate further anisotropic contributions that require renormalization. 
The coordinates run over the ranges \tm{x_{\mu}\in \{0,\ldots,L_{\mu}-a_{\mu}\}}. 
We denote the one-step hopping or shift operator as \tm{h_{\mu}^{\pm}}, which acts on all of the fields \tm{\ps(x)} to its right
as \tm{(h_{\mu}^{\pm}\ps)(x)=\ps(x\pm\hat{\mu})} and is rendered gauge covariant in the obvious manner  
\nequn{
(h_{\mu}^{\pm}\ps)(x)
=
\sum\limits_{y} h_{\mu}^{\pm}(x,y)\ps(y)
=U_{\pm\mu}(x)\ps(x\pm\hat\mu)
,\label{eq:shift}
}%nequn
where \tm{U_{+\mu}(x) \equiv U_{\mu}(x)} is the parallel transporter from \tm{x+\hat\mu} to \tm{x}, 
and \tm{U_{-\mu}(x) \equiv (U_{\mu}^{-1})\dag(x-\hat\mu)} is the inverse parallel transporter from \tm{x-\hat\mu} to \tm{x}.
\tm{\hat\mu} denotes \tm{a_{\mu}} times the unit-vector \tm{e_{\mu}}. 
The corresponding adjoint shift \tm{\bar{h}_{\mu}^{\pm}} acts on all of the fields to its left as \tm{(\bar\ps \bar{h}_{\mu}^{\pm})(x)=\bar\ps(x\pm\hat{\mu})U_{\pm\mu}\dag(x\pm\hat{\mu})}. 
The shifts and adjoint shifts are related to the inverse shifts under summation of the \tm{x_\mu} direction with periodic boundary conditions, 
\tm{\sum_{x_\mu} (\bar\ps \bar{h}_{\mu}^{\pm})(x)\ps(x) = \sum_{x_\mu} \bar\ps(x) (h_{\mu}^{\mp}\ps)(x)}.
We use greek letters for Euclidean directions ranging from \tm{1} to \tm{D} and roman letters for the spatial directions ranging from \tm{1} to \tm{(D-1)}. 
Unless the forward or backward orientation is of specific relevance, we omit the sign and simply write \tm{h_{\mu}} for a shift operator. 
Later on, in our discussion of symmetries and \acl{ST} structure, we preferably use instead of the dimensionful coordinate \tm{x_{\mu}} the site index \tm{n_{\mu}=\sfrac{x_{\mu}}{a_{\mu}}} running over the ranges \tm{n_{\mu}\in \{0,\ldots,N_{\mu}-1\}}, and understand that the shift acts as \tm{h_{\mu}^{\pm}n=n\pm e_\mu}. 
We use \tm{\pad_\mu} or \tm{\pad_\mu^\ast} to denote the discrete forward or backward derivatives, respectively.
Their symmetric counterpart is 
\nequn{
\nab_\mu
= \frac{(\pad_\mu+\pad_\mu^*)}{2}
= \frac{(h_{\mu}^{+}-h_{\mu}^{-})}{2a_{\mu}}
~.~\label{eq:deriv}
}
Similarly, the Laplacian 
\tm{\lap_\mu=\pad_\mu^\ast\pad_\mu=\pad_\mu\pad_\mu^\ast=\sfrac{(h_{\mu}^{+}+h_{\mu}^{-}-2)}{a_{\mu}^2}} denotes the second discrete derivative,
and we refer to its symmetrized hopping or shift part as 
\nequn{
h_{\mu}^{c}=\frac{(h_{\mu}^{+}+h_{\mu}^{-})}{2}
,~\label{eq:cos_shift}
}%nequn
i.e. \tm{\lap_\mu=\sfrac{2(h_{\mu}^{c}-1)}{a_{\mu}^2}}, and may write
\nequn{
\DW(x,y)
=\sum_\mu \ga_\mu \xi_{f0,\mu} \nab_\mu(x,y)
+\frac{r}{a_{D}} \Big(D\de(x,y)-\sum_\mu h_{\mu}^{c}(x,y) \Big)
~.\label{eq:D_wils_taste}
}%nequn 

The actions with either operator are obviously invariant under discrete translations (for periodic boundary conditions) and all discrete rotation reflections constituting the hypercubic group \tm{W_4}. In particular, the reflections \tm{\mR_{\mu}} act as 
\nequn{
\begin{aligned}
\mR_{\mu} \ps(n_\nu,n_\mu) 
&= \ga_{\mu} \ga_5 \ps(n_\nu,-n_\mu) 
,\\~
\bar{\ps}(n_\nu,n_\mu) \mR_{\mu}\dag  
&= \bar{\ps}(n_\nu,-n_{\mu})  \ga_5 \ga_{\mu}
,\\~
\mR_{\mu} U_{\pm\la}(n_\nu,n_\mu) \mR_{\mu}\dag  
&= U_{\pm(-1)^{\de_{\mu\la}}\la}(n_\nu,-n_\mu) 
,
\end{aligned}\label{eq:Rmu_act}~
}%nequn
and include parity (\tm{\mP}) or (Euclidean) time reflection (\tm{\mT}). 
Lastly, charge conjugation (\tm{\mC}) is a separate standard symmetry. 
These transformations leave naive of Wilson fermion actions invariant, see e.g.~\cite{Gattringer:2010zz}. 
We denote the charge conjugated fields as \tm{\ps^{c}}. 
Furthermore, they are \tm{\gaf} Hermitian, e.g.~\tm{\gaf \DN \gaf=\DN\dag} or \tm{\gaf \DW \gaf=\DW\dag}. 
\tm{\DN} also satisfies a chiral symmetry with (unmodified) \tm{\gaf}. 
This chiral symmetry is broken for Wilson fermions, i.e.~\tm{[\gaf,\DW]_+=(\DW+\DW\dag)\gaf=-ra_{D}\gaf\sum_\mu \xi_{f0,\mu}^2 \lap_{\mu}}. 
In momentum space the two operators read 
\nalign{
\DN(p)
&=\ri \sum_\mu \ga_\mu \frac{\xi_{f0,\mu}}{a_\mu}\sin(a_\mu p_\mu)
= \ri \sum_\mu \ga_\mu \xi_{f0,\mu} \bar{p}_\mu
,\label{eq:D_naiv_mom}\\
\DW(p)
&=\ri \sum_\mu \ga_\mu \frac{\xi_{f0,\mu}}{a_\mu}\sin(a_\mu p_\mu)
+\frac{r}{a_{D}}\Big(D-\sum_\mu \cos(ap_\mu) \Big)
= \ri \sum_\mu \ga_\mu \xi_{f0,\mu} \bar{p}_\mu
+\frac{r a_{D}}{2} \sum_\mu \xi_{f0,\mu}^2 \hat{p}_\mu^2
,\label{eq:D_wils_mom}
}%nalign
where \tm{a_{\mu}\bar{p}_\mu=\sin(a_{\mu} p_{\mu})} and \tm{a_{\mu} \hat{p}_\mu=2\sin(\sfrac{a_{\mu} p_{\mu}}{2})}. 
The Wilson term enforces the periodicity of a boson dispersion relation with periodicity \tm{2\pi}, and the Wilson operator describes for \tm{r^2>\sfrac{1}{4}} only one remaining Dirac fermion in the continuum limit, which corresponds to the zero mode at \tm{a_{\mu}\hat{p}_\mu=0~\forall\mu} in the free theory. 
Therefore, all \tm{2^{\sfrac{D}{2}}} spinor components on each site can be attributed to the one surviving taste. 
The absence of the Wilson term in \tm{\DN(p)} implies a fermion dispersion relation with periodicity \tm{\pi}, which permits a nontrivial taste symmetry. 
Therefore, a total of \tm{2^{\sfrac{3D}{2}}} spinor components distributed over \tm{2^{{D}}} sites must be attributed to the \tm{2^{{D}}} surviving tastes. 
We define the \ac{ST} rotations that give rise to the \emph{doubling symmetries}~\cite{Karsten:1980wd, Kawamoto:1981hw} through involutions \tm{\tau_{\mu}}, 
\nequn{
\tau_{\mu}\equiv\tau_{\mu}(n)=\tau_{\mu}\dag(n)= \ga_{\mu5} (-1)^{n_\mu}
,\label{eq:taumu}
}%nequn 
with matrices \tm{\ga_{\mu\nu} \equiv \ri \ga_{\mu} \ga_{\nu}} for \tm{1 \le \mu < \nu \le 5}. 
In the chiral representation of the matrices \tm{\ga_{\mu}} (see e.g.~\cite{Gattringer:2010zz}, 
which we will use unless otherwise stated, the charge conjugation matrix is \tm{C\equiv\ga_{24}}. 
These involutions \tm{\tau_{\mu}} act as
\nequn{
\begin{aligned}
\tau_{\mu}(n) \ps(n_\nu,n_\mu) 
&= \ga_{\mu5} (-1)^{n_\mu} \ps(n_\nu,n_\mu) 
,\\~
\bar{\ps}(n_\nu,n_\mu) \tau_{\mu}\dag(n)  
&= \bar{\ps}(n_\nu,n_\mu) (-1)^{n_\mu} \ga_{\mu5}
,\\~
\tau_{\mu}(n_\nu,n_\mu) U_{\pm\mu}(n_\nu,n_\mu) \tau_{\mu}\dag(n_\nu,n_\mu+e_{\mu}) 
&= U_{\pm\mu}(n_\nu,+n_\mu)
~.  
\end{aligned}\label{eq:taumu_act}~
}%nequn
It is important to stress that these transformation rules, Eq.~\eqref{eq:taumu_act}, are imposed and do not transform \tm{\ps(n)} or \tm{\ps\dag(n)} coherently, i.e. with adjoint factors. 
However, on the level of the taste components the relevant factors are adjoint to one another, as we shall demonstrate later in Sec.~\ref{sec:KW_spinors}. 
The \tm{\tau_{\mu}} leave \tm{\DN} invariant, and thus play the role of the \tm{\mathfrak{su}(2^{\sfrac{D}{2}})} generators for one \tm{\text{U}(2^{\sfrac{D}{2}})} taste group. 
This group is not broken by gauge interactions, and its fundamental components get decoupled in the exact diagonalization~\cite{Kawamoto:1981hw, Sharatchandra:1981si}. 
The naive action has \emph{mirror fermion symmetries}~\cite{Pernici:1994yj}, too, i.e. it is invariant under the product 
\nequn{
\mM_{\mu}=\ri\mR_{\mu}\tau_{\mu}
\label{eq:mirror_mu}
}%nequn 
for any \tm{\mu}. 
These are obviously not independent of the separate symmetries under \tm{\mR_{\mu}} or \tm{\tau_{\mu}}, but just a rephrased version of the doubling symmetries. 
All \tm{\mM_{\mu}} commute with one another. 
Thus, we also define products as \tm{\mM_{\mu\nu} \equiv \mM_{\mu}\mM_{\nu}} and even \tm{\mMf \equiv \prod_{\mu} \mM_{\mu}}. 
In the free theory the mirror fermion symmetries are just the combination of a geometric inversion and a phase, i.e. 
\nequn{
\begin{aligned}
\mM_{\mu} \ps(n_\nu,n_\mu) 
&= (-1)^{n_{\mu}} \ps(n_\nu,-n_\mu) 
,\\~
\bar{\ps}(n_\nu,n_\mu) \mM_{\mu}\dag  
&= \bar{\ps}(n_\nu,-n_{\mu})  (-1)^{n_{\mu}}
,\\~
\mM_{\mu} U_{\pm\la}(n_\nu,n_\mu) \mM_{\mu}\dag  
&= U_{\pm(-1)^{\de_{\mu\la}}\la}(n_\nu,-n_\mu) 
~.
\end{aligned}
~\label{eq:mirror_mu_act}
}%nequn 
Thus, the mirror fermions not only have support at different Fermi points in the Brillouin zone, but have local support in different areas of the lattice, and thus see different local gauge fields. 
We see how the Wilson term lifts this naive taste symmetry, as the \tm{\tau_{\mu}} do not map the Wilson operator onto itself, 
\nequn{
\tau_{\nu}(x) \DW(x,y) \tau_{\nu}\dag(y) 
= \sum_\mu \ga_\mu \xi_{f0,\mu} \nab_\mu(x,y)
+\frac{r}{a_{D}}\Big(D\de(x,y)-\sum_\mu (-1)^{\de_{\mu,\nu}} h_{\mu}^{c}(x,y) \Big)
~.
}%nequn
Since naive fermions describe \tm{2^{D}} (in the free theory) degenerate Dirac fermions, one taste algebra \tm{\mathfrak{su}(2^{\sfrac{D}{2}})} cannot contain all of the taste-symmetry generators. 
In fact, the missing taste-symmetry generators are products of the \acl{stag} phases 
\nequn{
\ze_{\mu}(n) = \prod_{\nu > \mu} (-1)^{n_{\nu}}
~\label{eq:KS_zeta}
}%nequn
and shifts. 
If these shifts are defined as internal within a hypercube, i.e.
\nequn{
(h_{\mu}^{I}\ps)(m)
=
\sum\limits_{m} 
\left\{\begin{array}{lr} 
h_{\mu}^{+}(m,n)\ps(n)
& \text{for}~n_\mu~\text{mod}~2=0
,~\\
h_{\mu}^{-}(m,n)\ps(n)
& \text{for}~n_\mu~\text{mod}~2=1
,\label{eq:int_shift}
\end{array}\right.
}%nequn
they happen to be involutions.\footnote{The shift \tm{h_{\mu}^{I}} and adjoint shift \tm{\bar{h}_{\mu}^{I}} coincide under summation of the \tm{n_\mu} direction.} 
Then we combine those with the \acl{stag} phases \tm{\ze_\mu} as \tm{t_{\mu} = \ze_{\mu}h_{\mu}^{I}}.
The set \tm{\{t\}} with multiplication constitutes in the free theory a closed representation of the \acl{stag} taste Clifford algebra \tm{[t_{\mu},t_{\nu}]_+=2 \de_{\mu,\nu}}. 
This shift algebra is more familiar from its role as the \acl{stag} shift-symmetry of \aclp{SF}~\cite{Golterman:1984cy}. 
The other set of \acl{stag} phases 
\nequn{
\et_{\mu}(n) = \prod_{\nu < \mu} (-1)^{n_{\nu}}
,\label{eq:KS_eta}
}%nequn 
which are the remnants of the gamma matrices after the exact diagonalization~\cite{Kawamoto:1981hw, Sharatchandra:1981si}, are combined with the shifts to \tm{s_{\mu} = \et_{\mu}h_{\mu}^{I}}. 
The set \tm{\{s\}} with multiplication constitutes in the free theory a closed representation of the \acl{stag} spin Clifford algebra \tm{[s_{\mu},s_{\nu}]_+=2 \de_{\mu,\nu}}.
Each of the two \acl{stag} Clifford algebras has its \tm{D}-shift chiral element \tm{\{s_{5}\}} or \tm{\{t_{5}\}}, respectively, each corresponding to \tm{\gaf} in the respective space. 
The product of both is a local, i.e. single-site operator, and coincides with the site parity \tm{\et_{5}(n)} defined via 
\nequn{
\et_{5}(n) = \prod_{\mu} (-1)^{n_{\mu}}
~\label{eq:site_parity}
}%nequn
\tm{\DN} has besides its \tm{\gaf} Hermiticity an \tm{\et_{5}} Hermiticity, too, i.e. \tm{\et_{5} \DN \et_{5}=\DN\dag}. 
Yet the Wilson operator does not have \tm{\et_{5}} Hermiticity, since \tm{\et_{5} \DW \et_{5}= -\sum_\mu \ga_\mu \xi_{f0,\mu} \nab_\mu(x,y)
+r a_{D}\sum_\mu [ \de(x,y) + h_\mu^{c}(x,y) ]}. 
% \clearpage
%%%%%%%%%%%%%%%%%%%%%%%%%%%%%%%%%%%%%%%%%%%%%%%%%%%%%%%%%%%%%%%%%%%%%%%%%%%%%%%%
% \input{\INCDIR/kawi_quark}
%% kawi_quark.tex
\acresetall

\section{Karsten-Wilczek fermions on the quark level}
\label{sec:KW_quark}

The oldest \acl{MDF} approach is the \ac{KW} proposal~\cite{Karsten:1981gd,Wilczek:1987kw} that restricts the Laplacian in (\ref{eq:D_wils}) to its spatial components
\nequn{
\DKW(x,y)=\sum\limits_{\mu=1}^D \ga_\mu \xi_{f0,\mu} \nab_\mu(x,y)
-\ri\frac{r \aD}{2}\gaD \sum_{j=1}^{D-1} \xi_{f0,j}^2 \lap_j(x,y)
~.\label{eq:D_KW}
}%\nequn
For real \tm{r}, which we assume hereafter, the extra factor \tm{\ri\gaD} makes the operator anti-Hermitian and anti-commuting with \tm{\gaf}, but at the price of breaking symmetries of the \ac{KW} action. 
With the operator described in Eq.~\eqref{eq:D_KW} symmetries under \emph{charge conjugation} (\tm{\mC}) or (Euclidean) \emph{time reflection} (\tm{\mT \equiv \mR_{D}}) are broken. 
The \ac{KW} action with the operator described in Eq.~\eqref{eq:D_KW} is obviously invariant under discrete translations (for periodic boundary conditions) or discrete spatial rotation-reflections constituting the cubic group \tm{W_3}. 
In particular, while \emph{parity} (\tm{\mP \equiv \ri \mR_{1}\mR_{2}\mR_{3}}) is a symmetry of this \ac{KW} action on its own, we shall see later that some caution is required with its \acl{ST} interpretation. 
The product of (Euclidean) time reflection \tm{\mT} and charge conjugation \tm{\mC} is yet another symmetry of this \ac{KW} action. 
\emph{Anisotropy} between temporal \tm{D} and spatial \tm{1 \le j \le (D-1)} directions is a consequence of, among other less important aspects, the individually broken symmetry under \tm{\mT} that maps the \tm{r}-parameter as \tm{r \to -r}. 
\tm{\DKW} is \tm{\gaf} Hermitian,~\tm{\gaf \DKW \gaf=\DKW\dag}, 
and chiral,~\tm{\gaf \DKW \gaf=-\DKW}; the appropriate \acl{ST} interpretation of the corresponding chirality matrix \tm{\gaf} shall be derived later on. 
Finally, the \ac{KW} action inherits from the naive action one \emph{mirror fermion symmetry}, i.e. \tm{\mMD \equiv \ri\mT \tau_{D}}, see Eq.~\eqref{eq:mirror_mu_act}. 
While all other mirror fermion symmetries (generated by any \tm{\mM_{j} \equiv \ri\mR_{j} \tau_{j}}) are broken by the extra term, they transform Eq.~\eqref{eq:D_KW} into inequivalent operators that still share the same spectrum in the free theory; but their spectra differ in the interacting theory. 
We shall derive the \acl{ST} interpretation of these mirror fermion symmetries later. 
Contrary to claims in the literature, the \ac{KW} action with the operator described in Eq.~\eqref{eq:D_KW} does neither satisfy time link-reflection invariance~\cite{Bedaque:2008xs} or site-reflection invariance~\cite{Tiburzi:2010bm}.
All of these properties carry over to improved variants of \ac{KW} fermions~\cite{Vig:2024umj, Bakenecker:2024a}.
\vskip1ex

For \tm{r^2 > \sfrac{1}{4}}, the \ac{KW} operator is minimally doubled with the survivors located at \tm{\aD p_{D}=0,\pi} and \tm{a_j p_j=0} in the Brillouin zone. 
This is apparent once the \ac{KW} operator is expressed in momentum space, 
\nequn{
\DKW(p)
=\ri \sum\limits_{\mu=1}^D \ga_\mu \frac{\xi_{f0,\mu}
}{a_\mu}\sin(a_\mu p_\mu)
+\ri\frac{r}{\aD}\gaD \sum_{j=1}^{D-1} \big[ 1-\cos(a_j p_j) \big]
= \ri \sum\limits_{\mu=1}^D \ga_\mu \xi_{f0,\mu} \bar{p}_\mu
+\ri \frac{r \aD}{2}\gaD \sum_{j=1}^{D-1} \xi_{f0,j}^2  \hat{p}_j^2
~.\label{eq:D_KW_mom}
}%nequn
The free \ac{KW} propagator is obtained by inverting Eq.~\eqref{eq:D_KW_mom}, 
\nequn{
\DKW\inv(p)
= \frac{-\ri \sum\limits_{\mu=1}^D \ga_\mu \xi_{f0,\mu} \bar{p}_\mu
-\ri \frac{r \aD}{2}\gaD \sum\limits_{j=1}^{D-1} \xi_{f0,j}^2  \hat{p}_j^2}
{\sum\limits_{j=1}^{D-1} \big( \xi_{f0,j} \bar{p}_j \big)^2
+\big( \xi_{f0,D} \bar{p}_D + \frac{r \aD}{2} \sum\limits_{j=1}^{D-1} \xi_{f0,j}^2  \hat{p}_j^2 \big)^2}
~.\label{eq:D_KW_prop}
}%nequn
While the notation via \tm{\hat{p}^2} nicely highlights the approach to the naive \tm{\aD \to 0} limit, it occludes more than it reveals about the nontrivial \acf{ST} structure encoded in the Laplacian. 
This is indeed quite similar to the case of Wilson fermions discussed in Sec.~\ref{sec:NW fermions}.   
Thus, calling back to mind Eq.~\eqref{eq:D_wils_taste}, we take the Laplacian apart and rewrite Eq.~\eqref{eq:D_KW} in terms of the naive operator \tm{\DN} and two extra ones that are both multiplied by \tm{r},
\nequn{
\DKW(x,y)
=\sum_\mu \ga_\mu \xi_{f0,\mu} \nab_\mu(x,y)
+\ri\frac{r}{\aD}\gaD \Big( (D-1)\de(x,y) -\sum_{j=1}^{D-1} h_{j}^{c}(x,y) \Big)
~.\label{eq:D_KW_taste}
}%nequn 

An alternative version of the \ac{KW} operator has been proposed by Pernici~\cite{Pernici:1994yj}. 
It lifts the doublers through an extra term \tm{-\ri\tfrac{r\aD}{2}\ga_{1} \sum_{\nu=2}^{D} \xi_{f0,\nu}^2\lap_\nu(x,y)}. 
The differences to the symmetries of the standard version described in Eq.\eqref{eq:D_KW} include the obvious ones including time link-reflection invariance. 
Yet it is worthwhile to highlight that Pernici's proposal satisfies \emph{link-reflection positivity} in the Euclidean time direction at the expense of breaking parity. 
We promote Pernici's proposal to a slightly more generic form by replacing the \tm{1}-direction by the generic spatial \tm{j_{0}}-direction, i.e. we define 
\nequn{
\DKW^{(j_{0})}(x,y)=\sum_\mu \ga_\mu \xi_{f0,\mu} \nab_\mu(x,y)
-\ri\frac{r \aD}{2}\ga_{j_{0}} \sum_{\nu\neq j_{0}} \xi_{f0,\nu}^2  \lap_\nu(x,y)
~.\label{eq:D_KW_per}
}%\nequn 
Given that lattice correlation functions are usually studied along the time direction, we refer to Pernici's version as \emph{perpendicular \ac{KW} fermions}. 
We refer to the standard version simply as \emph{(parallel) \ac{KW} fermions} following convention introduced a long time ago~\cite{Weber:2015oqf}. 
Hereafter, we always assume \tm{\nu \neq j_{0}} when discussing perpendicular \ac{KW} fermions.

\subsection{Hints of the Karsten-Wilczek spin-taste structure}\label{sec:KW_hints}

It seems natural to associate the two poles of the free theory propagator with the two tastes of the free theory. 
As the two poles of the propagator must correspond to opposite chiralities (via the standard \tm{\gaf}) due to the Nielsen-Ninomiya No-Go theorem~\cite{Nielsen:1980rz, Nielsen:1981xu}, one may define---in analogy to \aclp{BCF}~\cite{Borici:2007kz}---a \emph{dual representation of gamma matrices} as 
\nequn{
\gap_{\mu} = \gaD \ga_{\mu} \gaD = -(-1)^{\de_{\mu,D}} \ga_{\mu}
,\qquad [\gap_{\mu}, \gap_{\nu}]_+ =  2{\de_{\mu,\nu}}
,\qquad [\gap_{\mu}, \ga_{\nu}]_+ =  -(-1)^{\de_{\mu,D}}2{\de_{\mu,\nu}}
,~\label{eq:KW_dual}
}%nequn
with dual \tm{\gap_{5}=-\gaf}. 
We thus introduce the dual spinor \tm{\ps^\prime=\gaD\ps} and the dual \ac{KW} operator \tm{\DKW^\prime=\gaD\DKW\gaD}, which will turn out convenient when interpreting the \acl{ST} structure. \tm{\DKW} and  \tm{\DKW^\prime} have the same eigenvalues,
\nequn{
\begin{split}
\DKW \ps_i &\equiv \la_{i} \ps_i \\
~\Rightarrow~\DKW^\prime \ps^\prime_i &= \gaD\DKW\gaD \gaD\ps_i = \gaD\la_{i} \ps_i = \la_{i} \ps^\prime_i \\
~\Rightarrow~\DKW^\prime \ps^\prime_i &= \la_{i} \ps^\prime_i
~.~
\end{split}
\label{eq:KW_dual_eigen}
}%nequn
This dual representation corresponds to flipping the sign of all spatial momenta, and is thereby related to parity.  
\vskip1ex

A total of \tm{2^{\sfrac{D}{2}+1}} spinor components distributed over two sites must be attributed to the two tastes of Dirac fermions. 
It is suggestive to define a \ac{KW} \emph{fermion hypersite} \tm{n^{H}} as two boson sites one lattice step apart in the \tm{D}-direction, 
i.e. \tm{n_{j}^{H}=n_{j}}, \tm{\nD=2\nD^{H}+\nD^{I}} with \tm{\nD^{I}\in\{0,1\}} and \tm{\nD^{H}\in\{0,(\sfrac{\ND}{2}-1)\}}. 
Yet any two boson sites of opposite site-parity will do, once appropriate site-dependent phase factors are taken into account. 
Thus, we define a fermion hypersite consisting of two boson sites separated one step in a fixed \tm{\mu_{H}} direction as 
\nequn{
\left\{
\begin{aligned}
n_{\mu_{H}}
&=2n_{\mu_{H}}^{H}+n_{\mu_{H}}^{I}
,~
n_{\mu_{H}}^{H}=\sfrac{n_{\mu_{H}}}{2}
,~
n_{\mu_{H}}^{I}=\{0,1\}
,&~\mu=\mu_{H}
~\\
n_{\mu}
&=\phantom{2}n_{\mu}^{H}~
,&~\mu\neq\mu_{H}
~
\end{aligned}\right\}
,~\mu_{H}~\text{fixed}
~.\label{eq:KW_hypersite}
}%nequn
In particular, as we shall see later, a spatially split hypersite with \tm{\mu_{H}\neq D} could be particularly attractive for interpolating operators on a single time slice. 
Correlation functions will require a two-step transfer matrix anyway and a time-slab formalism~\cite{Pernici:1994yj}. 
As each site provides \tm{2^{\sfrac{D}{2}}} spinor components as degrees of freedom, one may ask if each spinor component on a particular site can be unambiguously associated with the low-lying modes of either of the two tastes, or whether each component contributes to all tastes as in the case of \acl{stag} fermions~\cite{Kluberg-Stern:1983lmr}. 
\vskip1ex

The \ac{KW} operator inherits from the naive operator \tm{\DN} given in Eq.~\eqref{eq:D_naiv} exactly one \emph{mirror fermion symmetry} \tm{\mMD} given in Eq.~\eqref{eq:mirror_mu}, which was first noted for the perpendicular \ac{KW} operator~\cite{Pernici:1994yj}. 
This symmetry is obvious from Eq.~\eqref{eq:D_KW_mom}, and it can be understood as the consecutive transform under \tm{\mR_{D}=\mT} defined in Eq.~\eqref{eq:Rmu_act} and \tm{\tau_D} defined in Eq.~\eqref{eq:taumu_act}. 
Since \tm{\mC\mT} is the generator of a nontrivial discrete symmetry of the \ac{KW} action and \tm{\ri\mT\tau_{D}} is another, this action has a \emph{peculiar charge conjugation symmetry} under \tm{\tau_{D}\mC} that acts nontrivially in some internal group space. 
We stress this version of the symmetry, since a corresponding peculiar charge conjugation symmetry that acts nontrivially in some internal group space also exists for a specific version of \aclp{BCF}~\cite{Weber:2023kth}. %; see Sec.~\ref{sec:BC}. 
This internal group space corresponds to the \ac{KW} taste group \tm{\text{U}(2)}. 
\vskip1ex

\tm{\tau_{D}} is the operator that exchanges the two poles of the \ac{KW} propagator and leaves \tm{\DN} invariant, but flips the \tm{r} parameter in front of the Laplacian in Eq.~\eqref{eq:D_KW}.
Hence, we are led to interpret \tm{\tau_D} as one generator of the \ac{KW} taste group. 
All of the generators must leave the \tm{\aD \to 0} limit invariant, since it is a taste singlet, and must lead to well-defined behavior for any extra terms in Eq.~\eqref{eq:D_KW_taste}. 
We introduce a taste-flipped representation of gamma matrices,
\nequn{
\ga_{\mu}^{\tau} \equiv \tau_{D} \ga_{\mu} \tau_{D} = (-1)^{\de_{\mu,D}} \ga_{\mu}
,\qquad [\ga_{\mu}^{\tau}, \ga_{\nu}^{\tau}]_+ =  2{\de_{\mu,\nu}}
,\qquad [\ga_{\mu}^{\tau}, \ga_{\nu}]_+ =  (-1)^{\de_{\mu,D}}2{\de_{\mu,\nu}}
,~\label{eq:KW_tflip}
}%nequn
with taste-flipped \tm{\ga_{5}^{\tau}=-\gaf}, and introduce a taste-flipped spinor \tm{\ps^{\tau}=\tau_{D}\ps} and a taste-flipped \ac{KW} operator \tm{\DKW^{\tau}=\tau_{D}\DKW\tau_{D}\dag}, which has the same eigenvalues as \tm{\DKW}, to be shown along similar lines as in Eq.~\eqref{eq:KW_dual_eigen},
\nequn{
\DKW \ps_i = \la_{i} \ps_i
~\Rightarrow~\DKW^{\tau} \ps^{\tau}_i = \la_{i} \ps^{\tau}_i
~.
}%nequn 
Due to the symmetries under \tm{\mC\tau_{D}} or \tm{\mT\tau_{D}}, both \tm{\mC\DKW\mC\dag} and \tm{\mT\DKW\mT\dag} share this same spectrum, too. 
Note that since \tm{\DKW^{\tau}} differs from \tm{\DKW} only by the sign of \tm{r}, the two tastes perceive opposite signs of \tm{r}. 
Any operators of odd engineering dimension in Eq.~\eqref{eq:D_KW_taste} are accompanied by an odd power of \tm{r}, and therefore must be of explicit \acl{TI} nature. 
This observation does not rule out contributions of implicit \acl{TI} nature accompanied by even powers of \tm{r}. 
All odd powers in \tm{r}, and thus in \tm{\aD}, must cancel exactly in an average of the two tastes---or equivalently of \tm{\DKW^{\tau}} and \tm{\DKW}---such as any \acl{TS} observable, if the underlying distribution of field configurations is itself an even function of \tm{r}. 
This is indeed the case in \acs{QCD} with \ac{KW} fermions~\cite{Weber:2016dgo}. 
We return to this topic in Sec.~\ref{sec:KW_gauge}. 
\vskip1ex

\subsection{Karsten-Wilczek spin-taste decomposition of operators}\label{sec:KW_generators}

Different and in part inconsistent patterns of \acl{ST} structure have been claimed in the literature, either~\cite{Tiburzi:2010bm, Kimura:2011ik, Weber:2023kth}.
While the incorrect \acl{ST} structure of Refs.~\cite{Tiburzi:2010bm} has not been ruled out thus far, the following first-principles derivation exposes its fallacy and the consistency of~\cite{Kimura:2011ik, Weber:2023kth}.

\subsubsection{Generators of spin-taste structure}\label{sec:KW_gen_par}

In the following we specialize to \tm{D=4}. 
We construct the entire \emph{\ac{KW} representation of taste generators} \tm{\{\vr\}} that lead to well-defined behavior for the extra terms in Eq.~\eqref{eq:D_KW_taste} and satisfy an \tm{\mathfrak{su}(2)} algebra \tm{[\vr_{i},\vr_{j}]_-=2\ri\ep_{ijk}\vr_{k}} as  
\nequn{
\begin{aligned}
\vr_{1} 
&\equiv \gaD \etD 
\phantom{{}_{5}}\phantom{{}_{5}}\hspace{-1pt} = \si_1 \otimes \Id_{2\times2}~ \etD
,\\
\vr_{2} 
&\equiv \gaDf \etDf = \si_2 \otimes \Id_{2\times2}~ \etDf
,\\
\vr_{3} 
&\equiv \gaf \etf 
\phantom{{}_{D}}\phantom{{}_{D}}\hspace{-1pt} = \si_3 \otimes \Id_{2\times2}~ \etf
\end{aligned}
~\label{eq:KW_vr}
}%nequn
where we have invoked in the last equalities the chiral representation of the set of gamma matrices \tm{\{\ga\}}. 
The notation regarding the phases is inspired by the \acl{stag} phases defined in Eq.~\eqref{eq:KS_eta}, i.e. \tm{\etD=\et_{D}} and \tm{\etf=\et_{5}}. 
The set of phases closes under multiplication, since \tm{\etDf(n)\equiv(\etD\etf)(n)=(-1)^{\nD}}. 
These phases and the identity factors \tm{\Id_{2\times2}} trivially multiply the \tm{\{\si\}}, whose \tm{\mathfrak{su}(2)} algebra is underlying the one of the \tm{\{\vr\}}. 
We stress that the taste generators \tm{\{\vr\}} are obtained as products of the \tm{\tau_{\mu}} defined in Eq.~\eqref{eq:taumu}, and thus inherit their property of trivially commuting with gauge fields, thus being unaffected by gauge interactions, see Eq.\eqref{eq:taumu_act}. 
Careful comparison of \tm{\vr_{1}} defined in Eq.~\eqref{eq:KW_vr} reveals that it combines with parity to a mirror fermion operator, \tm{\mMDf = \mP\vr_{1}}. 
We may define from \tm{\{\vr\}} genuine \emph{taste-projection operators} \tm{P^{\vr_{i}}_{\pm}} that remain exact in the interacting theory. 
We shall only make use of a specific one, 
\nequn{
P^{\vr}_{\pm}(n) \equiv P^{\vr_{3}}_{\pm}(n)
,\qquad
P^{\vr_{i}}_{\pm}(n) 
\equiv \frac{\Id_{\vr}\pm \vr_{i}(n)}{2}
,\qquad
\vr_{i} P^{\vr_{j}}_{\pm} = 
\left\{\begin{aligned}
P^{\vr_{j}}_{\mp} \vr_{i}
,~ i \neq j
,~
\\
P^{\vr_{j}}_{\pm} \vr_{i}
,~ i = j
~.~
\end{aligned}\right.
\label{eq:KW_Pvr}
}%nequn
The \ac{KW} taste generators and projectors and their extremely suggestive interpretation in the chiral representation have been known for a couple of years~\cite{Weber:2016dgo}. 
Yet we cannot claim at this point, whether the Pauli matrices \tm{\{\si\}} act in the taste or spin space or in both, or whether the phase factors of the \tm{\{\vr\}} belong to the taste or spin space or neither. 
Furthermore, we cannot exclude so far that some generators might act nontrivially both in taste and spin space, or claim definite knowledge of site-dependent phases, as the \tm{\{\vr\}} act on each site separately. 
Nevertheless, Eq.~\eqref{eq:KW_vr} already provides hints that both spatial (site) parity, indicated by \tm{\etD(n)}, and (Euclidean) site parity, indicated by \tm{\etf(n)}, play important roles. 
We shall prove in Sec.~\ref{sec:KW_spinors} that the \tm{\{\vr\}} act exclusively in taste space, and find the correct, i.e. spatial site-dependent phases from first principles in the following new derivation. 
\vskip1ex

\begin{table}
\setlength{\extrarowheight}{2pt}
\begin{tabular}{c|c|c|c}
Generator & \tm{\{\ga\}} & \tm{\{h\}} & Symmetry \\
\hline
\tm{\vr_{1}}   & \tm{\gaD,\ga_{j5}}      & \tm{h_D}    & \tm{\mP,\mR_{j}} \\
\tm{\vr_{2}}   & \tm{\ga_{j},\gaDf}      & \tm{h_j}    & \tm{\mT,\mR_{D}} \\
\tm{\vr_{3}}   & \tm{\gaf,\ga_{jD}}      & \tm{\hDj} & \tm{\mC,\mP\mT}\\
\tm{\Id_{\vr}} & \tm{\Id_{\ga},\ga_{kl}} & \tm{\Id_{h}} \\
\end{tabular}
\caption{Transformation behavior of standard gamma %\tm{\text{GL}(4,\mathbb{C})} 
matrices \tm{\{\ga\}} and the shifts \tm{\{h\}}, when sandwiched between generators as \tm{\vr_{i} X \vr_{i}\dag} for parallel \ac{KW} fermions. 
The operators \tm{X \in \{\ga, h\}} in each row transform like the generator (first column). 
The three spatial directions \tm{j,~k,~l} are mutually orthogonal. 
Each mentioned discrete symmetry operator leaves only one \tm{\vr_{i}} invariant. 
\label{tab:KW_vr_sim}
} 
\end{table}

Any fermion bilinear consists of a product of a standard gamma matrix, whose set we shall label \tm{\{\ga\}}, a shift whose set we shall label \tm{\{h\}}, and some site-dependent phases, all sandwiched between two spinors, for mesons \tm{\bar\ps} and \tm{\ps}, or for diquark operators \tm{(\ps^c)^T} and \tm{\ps}.  
The \acl{ST} structure of each fermion bilinear could be understood as a direct product of \tm{\mathfrak{su}(2)} generators corresponding to \tm{(1+\sfrac{D}{2})} different \tm{\mathfrak{su}(2)} algebras, i.e. for \tm{D=4} any interpolating fermion bilinear has a \acl{ST} interpretation
\nequn{
\{ \ga \} \otimes \{ h \} \sim \{\rho\} \otimes \{\chi\} \otimes \{\omega\} \sim \{\rho\} \otimes \{\Ga\}
,\label{eq:KW_ST_int}
}%nequn
where the \tm{\{\rho\}} act in taste space, the \tm{\{\chi\}} act on the two Weyl spinors of each taste, and the \tm{\{\omega\}} act on each Weyl spinor's two components. 
We denote the set of the spin gamma matrices as \tm{ \{\Ga\} \equiv \{\chi\} \otimes \{\omega\}}, which act separately on each taste, and label its generators as \tm{\Ga_{\mu}}, etc. 
We label the identity in each set as \tm{\Id}, i.e. \tm{\Id_{\vr} = \Id_{\ga} = \mathbb{1}_{4\times4}},~\tm{\Id_{h} = \de(x,y)},~\tm{\Id_{\rho},~\Id_{\chi},~\Id_{\omega}} or \tm{\Id_{\Ga} = \Id_{\chi} \otimes \Id_{\omega}}. 
For \tm{D=2} there is obviously no corresponding set \tm{\{\om\}}. 
The sets of the standard gamma matrices \tm{\{\ga\}} or of the shifts \tm{\{h\}} satisfy nontrivial commutation relations with the set of taste generators \tm{\{\vr\}}. 
Therefore, we can group their properties as in Table~\ref{tab:KW_vr_sim}. 
While the appropriate products of a gamma matrix and a shift with the same Euclidean index contribute---by construction---as \acl{TS} spin-vector operators in the naive \tm{\aD \to 0} limit of Eq.~\eqref{eq:D_KW_taste}, 
\nequn{
\begin{aligned}
\phantom{{}_{j}}\phantom{{}_{j}} \gaD h_{D} 
= \Id_{\rho} \otimes \GaD\phantom{{}_{j}} 
&= \Id_{\rho} \otimes \chi_{1} \otimes \Id_{\om}
,~\\  
\phantom{{}_{D}}\phantom{{}_{D}}  \ga_{j} h_{j} 
= \Id_{\rho} \otimes \Ga_{j}\phantom{{}_{D}} 
&= \Id_{\rho} \otimes \chi_{2} \otimes \om_{j}
,~\\
\Id_{\ga} \Id_{h} 
= \Id_{\rho} \otimes \Id_{\Ga} ~
&= \Id_{\rho} \otimes \Id_{\chi} \otimes \Id_{\om}
,
\end{aligned}
~\label{eq:KW_singlet_constraint}
}%nequn 
the extra operators in Eq.~\eqref{eq:D_KW_taste} clearly exhibit different \acl{TI} structures, i.e. they transform under \tm{\{\vr\}} as 
\nequn{
\gaD \Id_{h} \sim \vr_{1}
,~
\gaD h_{j} \sim \vr_{1}\vr_{2} \sim \vr_{3} 
~.
~\label{eq:KW_nonsinglet_operators}
}%nequn 
According to Tab.~\ref{tab:KW_vr_sim} the chirality operator \tm{\gaf} induces a \acl{TD} isovector chiral symmetry. 
Moreover, the standard discrete symmetry operators, i.e. \tm{\mC}, and \tm{\mP,~\mT} or \tm{\mR_{\mu}}, act nontrivially on the set of taste generators \tm{\{\vr\}} and vice versa. 
We include those discrete symmetry operators in Table~\ref{tab:KW_vr_sim}, too. 
We see that \acl{TS} symmetry operators can be defined by including suitable orthogonal translations, e.g. \tm{h_{D}} with any \tm{\mR_{j}} (or \tm{\mP}) or any (odd power or product of) \tm{h_{j}} with \tm{\mR_{D}}. 
Thus, we may define \acl{T-S} parity operators \tm{\mP_{\pm D}} that act as
\nequn{
\begin{aligned}
\mP_{\pm D} \ps(\nD,\bm{n})  
&= \gaD (h_{D}^{\pm} \ps)(\nD,-\bm{n}) 
 = \gaD U_{\pm D}(\nD,-\bm{n}) \ps(\nD \pm 1,-\bm{n}) 
,\\~
\bar{\ps}(\nD,\bm{n})  \mP_{\pm D} \dag  
&= (\bar{\ps} \bar{h}_{D}^{\pm})(\nD,-\bm{n}) \gaD
 = \bar{\ps}(\nD \pm 1,-\bm{n}) U_{\pm D}\dag(\nD,-\bm{n}) \gaD
,\\~
\mP_{\pm D} U_{\pm\la}(\nD,\bm{n}) \mP_{\pm D}\dag  
&= U_{\mp(-1)^{\de_{\la D}}\la}(\nD \pm 1,-\bm{n}) 
,~
\end{aligned}\label{eq:KW_Ppm_act}~
}%nequn
where the \tm{\pm} sign of the Euclidean index \tm{\pm\la} of the link is independent of the \tm{\pm} sign associated with the operator \tm{\mP_{\pm D}}. 
\tm{\mP_{\pm D}} are Hermitian, since the orthogonal translations and reflections commute, but they are nonunitary. 
They are inverse to each other:\footnote{If we use instead the hypercube-internal shifts defined in Eq.~\eqref{eq:int_shift}, i.e. 
\tm{\mP_{D}^{I} \ps(\nD,\bm{n}) = \gaD (h_{D}^{I} \ps)(\nD,-\bm{n})}, 
then the \acl{T-S} parity operator is indeed an involution: \tm{(\mP_{D}^{I})\inv=\mP^{I}_{D}}~.} 
{\markit\tm{\mP\inv_{\pm D}=\mP_{\mp D}}}. 
Both operators \tm{\mP_{\pm D}} are symmetry operators for the \ac{KW} action defined with the operator in Eq.~\eqref{eq:D_KW}. 
We return to the \acl{TS} parity operators \tm{\mP_{\pm D}} in Sec.~\ref{sec:KW_baryon}. 
A \acl{T-S} time reflection operator \tm{\mT_{\pm j}} can be constructed in a similar way,  
\nequn{
\begin{aligned}
\mT_{\pm j} \ps(\nD,\bm{n})  
&= \gaDf (h_{j}^{\pm} \ps)(-\nD,\bm{n})
 = \gaDf U_{\pm j}(-\nD,\bm{n}) \ps(-\nD,\bm{n} \pm e_{j}) 
,\\~
\bar{\ps}(\nD,\bm{n})  \mT_{\pm j} \dag  
&= (\bar{\ps} \bar{h}_{j}^{\pm})(-\nD,\bm{n})\gaDf
 = \bar{\ps}(-\nD,\bm{n} \pm e_{j}) U_{\pm j}\dag(-\nD,\bm{n}) \gaDf 
,\\~
\mT_{\pm j} U_{\pm\la}(\nD,\bm{n}) \mT_{\pm j}\dag  
&= U_{\pm(-1)^{\de_{\la D}}\la}(-\nD,\bm{n} e_{j}) 
,~
\end{aligned}\label{eq:KW_Tpm_act}~
}%nequn
and each term of the \ac{KW} operator in Eq.~\eqref{eq:D_KW_taste} transforms analogously as for standard \tm{\mT} with the same change \tm{r \to -r} in the action (after appropriately relabeling the site index). 
We shall return to the \acl{TS} time reflection operator \tm{\mT_{\pm j}} when discussing chemical potential in Sec.~\ref{sec:KW_chempot}. 
\vskip1ex

\subsubsection{Generators of spin-taste structure for perpendicular Karsten-Wilczek fermions}\label{sec:KW_gen_per}

\begin{table}
\setlength{\extrarowheight}{2pt}
\begin{tabular}{c|c|c|c}
Generator & \tm{\{\ga\}} & \tm{\{h\}} & Symmetry \\
\hline
\tm{\vr_{1}\pri \sim \vr_{2}\crc}   & \tm{\ga_{j_{0}},\ga_{\nu  5}}      & \tm{h_{j_{0}}} & \tm{\mR_{\nu}}\\
\tm{\vr_{2}\pri \sim \vr_{1}\crc}   & \tm{\ga_{\nu},\ga_{j_{0} 5}}   & \tm{h_\nu} & \tm{\mR_{j_0}}\\
\tm{\vr_{3}}                        & \tm{\gaf,\ga_{\nu j_{0} }}      & \tm{h_{j_{0}\nu}} & \tm{\mC,\mP\mT} \\
\tm{\Id_{\vr}}                      & \tm{\Id_{\ga},\ga_{\mu\nu}} & \tm{\Id_{h}}\\
\end{tabular}
\caption{Transformation behavior of standard \tm{\text{GL}(4,\mathbb{C})} matrices \tm{\{\ga\}} and the shifts \tm{\{h\}}, when sandwiched between generators as \tm{\vr_{i}\pri X (\vr_{i}\pri)\dag} or  \tm{\vr_{i}\crc X (\vr_{i}\crc)\dag} for perpendicular \ac{KW} fermions. 
The operators \tm{X \in \{\ga, h\}} in each row transform like the generator in the first column. 
The three directions \tm{j_{0},~\mu,~\nu} are mutually orthogonal. 
Each mentioned discrete symmetry operator leaves only one \tm{\vr_{i}\pri} (resp. \tm{\vr_{i}\crc}) invariant. 
\label{tab:KW_vr_sim_per}
} 
\end{table}

An analogous derivation of the \acl{ST} decomposition of operators can be repeated for perpendicular \ac{KW} fermions.  
Namely, we swap generators as \tm{\gaD \etD \to \ga_{j_{0}}\etj}, \tm{\gaDf\etDf \to \ga_{j_{0} 5}\etjf}, while \tm{\gaf\etf} remains as is. 
The phase factors here are not the corresponding \acl{stag} phases of Eq.~\eqref{eq:KS_eta}, but must be understood as a generalization of \tm{\etD} and \tm{\etDf}, i.e. \tm{\etj(n)\equiv\prod_{\nu\neq j_{0}}(-1)^{n_\nu}} and \tm{\etjf(n)\equiv(-1)^{n_{j_{0}}}}. 
Naturally the shift \tm{h_{j_{0}}} plays a special role in this case. 
Inspired by the analogy to parallel \ac{KW} fermions we could identify these operators as a representation \tm{\{\vr\pri\}} of the taste algebra via 
\nequn{
\begin{aligned}
\vr_{1}\pri 
&\equiv \ga_{j_{0}}\etj
\phantom{{}_{5}}\phantom{{}_{5}}\hspace{-1pt} 
= \phantom{-}\si_2 \otimes \si_{j_{0}}\phantom{_{\times2}}~ \etj
~~ \equiv \phantom{-}\vr_{2}\crc
,\\
\vr_{2}\pri  
&\equiv \ga_{j_{0} 5}\etjf
= -\si_1 \otimes \si_{j_{0}}\phantom{_{\times2}}~ \etjf
~ \equiv -\vr_{1}\crc
,\\
\vr_{3}\pri  
&\equiv \gaf \etf 
\phantom{{}_{j_{0}}}\phantom{{}_{j_{0}}}\hspace{-1pt} 
= \phantom{-}\si_3 \otimes \Id_{2\times2}~ \etf
~~~\, \equiv \phantom{-}\vr_{3}\crc
~.~
\end{aligned}
~\label{eq:KW_vr_per}
}%nequn
We have invoked in the second equalities the chiral representation of the gamma matrices \tm{\{\ga\}}, and in the third equality a different map of the operators to yet another representation \tm{\{\vr\crc\}} of the taste algebra, that is more aligned with the chiral representation of the gamma matrices \tm{\{\ga\}} (instead of the construction analogous to parallel \ac{KW} fermions). 
We can group the transformation properties of the sets of gamma matrices \tm{\{\ga\}} and shifts \tm{\{h\}} as in Table~\ref{tab:KW_vr_sim_per}. 
The extra operators in Eq.~\eqref{eq:D_KW_per} clearly exhibit different \acl{TI} structures, i.e. they transform under \tm{\{\vr\pri\}} or \tm{\{\vr\crc\}} as 
\nequn{
\ga_{j_{0}} \Id_{h} \sim \vr_{1}\pri \sim \vr_{2}\crc
,~
\ga_{j_{0}} h_{\nu} \sim \vr_{1}\pri\vr_{2}\pri \sim \vr_{1}\crc\vr_{2}\crc \sim \vr_{1}\vr_{2} \sim \vr_{3} 
,~\label{eq:KW_nonsinglet_operators_per}
}%nequn 
where the \acl{ST} structure of the former depends on the chosen representation \tm{\{\vr\pri\}} or \tm{\{\vr\crc\}}, while the latter does not. 
Choosing either \tm{\{\vr\pri\}} or \tm{\{\vr\crc\}} implies different phase factors for the spinors, see Sec.~\ref{sec:KW_spinors_per}, and some differences in the \acl{TN} sector, see Sec.~\ref{sec:KW_bilinears_per}. 
Again, certain discrete symmetry operators, i.e. \tm{\mC}, and \tm{\mR_{j_{0}},~\mR_{\nu}}, act nontrivially on the \tm{\{\vr\pri\}} (resp. \tm{\{\vr\crc\}}) and vice versa. 
We include those discrete symmetry operators in Table~\ref{tab:KW_vr_sim_per}, too. 
The same \acl{TS} parity \tm{\mP_{\pm D}} or time reflection \tm{\mT_{\pm j}} defined in Eqs.~\eqref{eq:KW_Ppm_act} or~\eqref{eq:KW_Tpm_act} can be used, too.
\vskip1ex

\subsection{Karsten-Wilczek spin-taste decomposition of spinors}\label{sec:KW_spinors}

\subsubsection{Weyl spinors and taste}\label{sec:KW_Weyl}

The blockwise structure of both the generators in Eq.~\eqref{eq:KW_vr} and the operators in the naive \tm{\aD \to 0} limit of the action, Eq.~\eqref{eq:KW_singlet_constraint}, suggest a blockwise taste decomposition of spinors in the chiral representation,
\nequn{
\setlength{\extrarowheight}{2pt}
\left.\begin{array}{l|cc}
(\etD, \etf)(n) & \bar\ps(n) & \ps(n) 
\\[2pt]
\hline
(+,+) & (\bar\ps_{\uparrow L},~\bar\ps_{\downarrow B}^{\rr} \vp_{++}^\ast) & \bmat{c} \ps_{\uparrow R}\\ \vp_{++} \ps_{\downarrow A}^{\rr} \emat 
\\[2pt]
(-,+) & (\bar\ps_{\uparrow L},~\bar\ps_{\downarrow B}^{\rr} \vp_{-+}^\ast) & \bmat{c} \ps_{\uparrow R}\\ \vp_{-+} \ps_{\downarrow A}^{\rr} \emat 
\\[2pt]
(-,-) & (\bar\ps_{\downarrow A}^{\rr} \vp_{--}^\ast,~\bar\ps_{\uparrow R} ) & \bmat{c} \vp_{--} \ps_{\downarrow B}^{\rr} \\ \ps_{\uparrow L} \emat 
\\[2pt]
(+,-) & (\bar\ps_{\downarrow A}^{\rr} \vp_{+-}^\ast,~\bar\ps_{\uparrow R}) & \bmat{c} \vp_{+-} \ps_{\downarrow B}^{\rr} \\ \ps_{\uparrow L} \emat 
\\[2pt]
% \hline
\end{array}\right.
,~\label{eq:KW_decomp_trial}
}%nequn 
where we have named the two tastes suggestively \tm{\ps_{\uparrow}} or \tm{\ps_{\downarrow}}, having isospin in mind.
We have chosen the overall phase of the spinors at one even-even site \tm{n}, i.e. with \tm{\etD(n)=\etf(n)=+1}, such that the arbitrarily chosen first taste \tm{\ps_{\uparrow }} does not have any extra phase and contributes with positive chirality in the upper Weyl spinor, namely with the standard chirality one would anticipate for a single-taste spinor. 
As a consequence of Eq.~\eqref{eq:KW_singlet_constraint}, \tm{\ps_{\uparrow }} contributes in the upper Weyl spinor for even-even or odd-odd sites, i.e. with \tm{\etf(n)=+1}, and in the lower Weyl spinor for even-odd or odd-even sites, i.e. with \tm{\etf(n)=-1}, and is not accompanied by any phase factors for any sites \tm{n}.  
Hence, 
\tm{P^{\vr}_{+}\ps(n)=\ps_{\uparrow }(n)}, such that chirality is intertwined with site-parity \tm{\etf(n)}. 
The parity operator \tm{\mP} anticommutes with \tm{\vr_{3}} and thus exchanges the projection operators 
\tm{\mP P_{\pm}^{\vr} = P_{\mp}^{\vr} \mP}, reallocating \tm{\ps_{\uparrow }(n)} to the other Weyl spinor of \tm{\ps(n)}. 
Namely, this shows that parity \tm{\mP} or time reflection \tm{\mT} are \acl{TN} operators for \ac{KW} fermions. 
Their \acl{TS} counterparts have been introduced in Eqs.~\eqref{eq:KW_Ppm_act} or~\eqref{eq:KW_Tpm_act}, respectively. 
\vskip1ex

The other taste projector, \tm{P^{\vr}_{-}\ps(n)}, returns exclusively the second taste component \tm{\ps_{\downarrow }^{\rr}(n)}, however, up to a phase factor \tm{\vp_{\etD,\etf}}. 
We have not specified yet which chirality of \tm{\ps_{\downarrow }^{\rr}(n)} is associated with the other Weyl spinor for each site-parity \tm{\etf(n)}, i.e. the Weyl spinors \tm{(A,B)} could correspond to chiral components \tm{(R,L)} or to \tm{(L,R)}. 
Then we have indexed the spinor of the other taste component \tm{\ps_{\downarrow }^{\rr}(n)} with a representation index \tm{\rr}, as it may also be in yet another representation of the 
%\tm{\text{GL}(4,\mathbb{C})} 
spin gamma set \tm{\{\Ga\}}. 
Thus, since these specifics of phase factors \tm{\vp_{\etD,\etf}}, association of Weyl spinors \tm{(A,B)} to chirality, and representation \tm{\rr} can only be meaningfully answered together, we derive these answers step by step. 
Readers not interested in this derivation may skip directly to Sec.~\ref{sec:KW_doublet}. 
\vskip1ex

\subsubsection{Representation, chirality, and phase factors}\label{sec:KW_rep}

Using the decomposition of Eq.~\eqref{eq:KW_decomp_trial} we analyze which phase factors \tm{\vp_{\etD,\etf}}, chiralities \tm{(A,B)}, and representations \tm{\rr} are permitted. 
We determine the phase factors \tm{\vp_{\etD,\etf}} by looking at the \acl{TN} generator bilinears with \tm{\vr_{1,2}},
\nequn{
\setlength{\extrarowheight}{2pt}
\left.\begin{array}{l|l|l}
(\etD, \etf)(n) & (\bar\ps \vr_{1} \ps)(n)  & (\bar\ps \vr_{2} \ps)(n) 
\\[2pt]
\hline
(+,+) 
& \bar\ps_{\uparrow L}~(+\vp_{++})~\ps_{\downarrow A}^{\rr} + \bar\ps_{\downarrow B}^{\rr}~(+\vp_{++}^\ast)~\ps_{\uparrow R} 
& \bar\ps_{\uparrow L}~(-\ri\vp_{++})~\ps_{\downarrow A}^{\rr} + \bar\ps_{\downarrow B}^{\rr}~(+\ri\vp_{++}^\ast)~\ps_{\uparrow R} 
\\[2pt]
(-,+) 
& \bar\ps_{\uparrow L}~(-\vp_{-+})~\ps_{\downarrow A}^{\rr} + \bar\ps_{\downarrow B}^{\rr}~(-\vp_{-+}^\ast)~\ps_{\uparrow R} 
& \bar\ps_{\uparrow L}~(+\ri\vp_{-+})~\ps_{\downarrow A}^{\rr} + \bar\ps_{\downarrow B}^{\rr}~(-\ri\vp_{-+}^\ast)~\ps_{\uparrow R} 
\\[2pt]
(-,-) 
& \bar\ps_{\uparrow R}~(-\vp_{--})~\ps_{\downarrow B}^{\rr} + \bar\ps_{\downarrow A}^{\rr}~(-\vp_{--}^\ast)~\ps_{\uparrow L} 
& \bar\ps_{\uparrow R}~(+\ri\vp_{--})~\ps_{\downarrow B}^{\rr} + \bar\ps_{\downarrow A}^{\rr}~(-\ri\vp_{--}^\ast)~\ps_{\uparrow L} 
\\[2pt]
(+,-) 
& \bar\ps_{\uparrow R}~(+\vp_{+-})~\ps_{\downarrow B}^{\rr} + \bar\ps_{\downarrow A}^{\rr}~(+\vp_{+-}^\ast)~\ps_{\uparrow L} 
& \bar\ps_{\uparrow R}~(-\ri\vp_{+-})~\ps_{\downarrow B}^{\rr} + \bar\ps_{\downarrow A}^{\rr}~(+\ri\vp_{+-}^\ast)~\ps_{\uparrow L} 
\\[2pt]
% \hline
\end{array}\right.
,~\label{eq:KW_vr_decomp}
}%nequn
where the row-by-row alternating signs are due to phase factors included in the \tm{\vr_{i}}, \tm{\etD(n)} or \tm{\etDf(n)}, respectively. 
We see that we require real \tm{\vp(n)=\pm\etD(n)} in order to have \tm{\vr_{i}} act as \tm{\pm\rho_{i}}. 
Without reason for the negative sign we choose the positive one. 
Thus, we find that the \tm{\vr_{i}} act as \tm{\rho_{i}\otimes\Id_{\Ga}^{\rr}} for \tm{(A,B)=(R,L)} or as \tm{\rho_{i}\otimes\GaD^{\rr}} for \tm{(A,B)=(L,R)}. 
\vskip1ex

We relate the chiralities \tm{(A,B)} and representation \tm{\rr} by inspecting the bilinears corresponding to the naive operator,
\nequn{
\setlength{\extrarowheight}{2pt}
\left.\begin{array}{l|l|l}
(\etD, \etf)(n) & (\bar\ps\gaD h_{D}^{\pm} \ps)(n)  & (\bar\ps \ga_{j} h_{j}^{\pm} \ps)(n) 
\\[2pt]
\hline
(+,+) 
& \bar\ps_{\uparrow L}~\ps_{\uparrow L} + \bar\ps_{\downarrow B}^{\rr}~\vp_{++}^\ast \vp_{+-}~\ps_{\downarrow B}^{\rr} 
& \bar\ps_{\uparrow L}~(-\ri\si_{j})~\ps_{\uparrow L} + \bar\ps_{\downarrow B}^{\rr}~(+\ri\si_{j}\vp_{++}^\ast \vp_{--})~\ps_{\downarrow B}^{\rr} 
\\[2pt]
(-,+) 
& \bar\ps_{\uparrow L}~\ps_{\uparrow L} + \bar\ps_{\downarrow B}^{\rr}~\vp_{-+}^\ast \vp_{--}~\ps_{\downarrow B}^{\rr} 
& \bar\ps_{\uparrow L}~(-\ri\si_{j})~\ps_{\uparrow L} + \bar\ps_{\downarrow B}^{\rr}~(+\ri\si_{j}\vp_{-+}^\ast \vp_{+-})~\ps_{\downarrow B}^{\rr} 
\\[2pt]
(-,-) 
& \bar\ps_{\uparrow R}~\ps_{\uparrow R} + \bar\ps_{\downarrow A}^{\rr}~\vp_{--}^\ast \vp_{-+}~\ps_{\downarrow A}^{\rr}
& \bar\ps_{\uparrow R}~(+\ri\si_{j})~\ps_{\uparrow R} + \bar\ps_{\downarrow A}^{\rr}~(-\ri\si_{j}\vp_{--}^\ast \vp_{++})~\ps_{\downarrow A}^{\rr}
\\[2pt]
(+,-) 
& \bar\ps_{\uparrow R}~\ps_{\uparrow R} + \bar\ps_{\downarrow A}^{\rr}~\vp_{+-}^\ast \vp_{++}~\ps_{\downarrow A}^{\rr}
& \bar\ps_{\uparrow R}~(+\ri\si_{j})~\ps_{\uparrow R} + \bar\ps_{\downarrow A}^{\rr}~(-\ri\si_{j}\vp_{+-}^\ast \vp_{-+})~\ps_{\downarrow A}^{\rr}
\\[2pt]
% \hline
\end{array}\right.
~.~\label{eq:KW_deriv_decomp}
}%nequn 
The \acl{TS} nature of the \tm{\bar\ps(n) \gaD h_{D}^{\pm} \ps(n\pm e_{D})} bilinears requires \tm{\vp_{++}^\ast\vp_{+-}=\vp_{-+}^\ast\vp_{--}=+1}, 
% Thus, the \tm{\vp(n)} may not depend on \tm{\etf(n)}, but on \tm{\etD(n)}. 
but we learn nothing new from that. 
Because \tm{\vp_{++}^\ast\vp_{--}=\vp_{-+}^\ast\vp_{+-}=-1}, the \acl{TS} nature of the \tm{\bar\ps(n) \ga_{j} h_{j}^{\pm} \ps(n\pm e_{j})} bilinears requires either \tm{(A,B)=(R,L)}, if one were to assume that the second taste uses the \emph{chiral representaion} of the gamma matrices \tm{\{\Ga\}}, or alternatively \tm{(A,B)=(L,R)}, if one were to assume that the second taste uses the \emph{dual representation} of the gamma matrices \tm{\{\Gap\}} that we defined in Eq.~\eqref{eq:KW_dual}, i.e. where the spatial gamma matrices (and the chirality matrix) have acquired an extra minus sign. 
Since the \emph{dual representation} implies that chirality is flipped w.r.t. the chiral representation, both lead to the same association between chirality and site parity \tm{\etf(n)}, after the second taste \tm{\ps_{\downarrow }^{\rr}(n)} is converted back to the chiral representation. 
Therefore, we understand that the association \tm{(A,B)=(R,L)} is the one where both tastes use the same representation of gamma matrices \tm{\{\Ga\}}, and therefore, where the direct product \acl{ST} interpretation of Eq.~\eqref{eq:KW_ST_int} can be applied. 
\vskip1ex

We may decipher the \acl{ST} structure of the two extra terms in the \ac{KW} operator, Eq.~\eqref{eq:D_KW_taste}, 
\nequn{
\setlength{\extrarowheight}{2pt}
\left.\begin{array}{l|l|l}
(\etD, \etf)(n) & (\bar\ps \gaD \ps)(n)  & (\bar\ps \gaD h_{j}^{\pm} \ps)(n) 
\\[2pt]
\hline
(+,+) 
& \bar\ps_{\uparrow L}~(\vp_{++})~\ps_{\downarrow A}^{\rr} + \bar\ps_{\downarrow B}^{\rr}~(\vp_{++}^\ast)~\ps_{\uparrow R} 
& \bar\ps_{\uparrow L}~\ps_{\uparrow L} + \bar\ps_{\downarrow B}^{\rr}~(\vp_{++}^\ast \vp_{--})~\ps_{\downarrow B}^{\rr} 
\\[2pt]
(-,+) 
& \bar\ps_{\uparrow L}~(\vp_{-+})~\ps_{\downarrow A}^{\rr} + \bar\ps_{\downarrow B}^{\rr}~(\vp_{-+}^\ast)~\ps_{\uparrow R} 
& \bar\ps_{\uparrow L}~\ps_{\uparrow L} + \bar\ps_{\downarrow B}^{\rr}~(\vp_{-+}^\ast \vp_{+-})~\ps_{\downarrow B}^{\rr} 
\\[2pt]
(-,-) 
& \bar\ps_{\uparrow R}~(\vp_{--})~\ps_{\downarrow B}^{\rr} + \bar\ps_{\downarrow A}^{\rr}~(\vp_{--}^\ast)~\ps_{\uparrow L} 
& \bar\ps_{\uparrow R}~\ps_{\uparrow R} + \bar\ps_{\downarrow A}^{\rr}~(\vp_{--}^\ast \vp_{++})~\ps_{\downarrow A}^{\rr}
\\[2pt]
(+,-) 
& \bar\ps_{\uparrow R}~(\vp_{+-})~\ps_{\downarrow B}^{\rr} + \bar\ps_{\downarrow A}^{\rr}~(\vp_{+-}^\ast)~\ps_{\uparrow L} 
& \bar\ps_{\uparrow R}~\ps_{\uparrow R} + \bar\ps_{\downarrow A}^{\rr}~(\vp_{+-}^\ast \vp_{-+})~\ps_{\downarrow A}^{\rr}
\\[2pt]
% \hline
\end{array}\right.
,~\label{eq:KW_lap_decomp}
}%nequn 
where either \tm{(A,B)=(R,L)} implies that \tm{(\bar\ps\gaD\ps)(n)} acts a \tm{\rho_{1}\otimes\Id_{\Ga}^{\rr} \etD}, or alternatively \tm{(A,B)=(L,R)} implies that \tm{(\bar\ps \gaD \ps)(n)} acts as \tm{\rho_{1}\otimes\GaD^{\rr} \etD}. 
We note that the remnant phase factor \tm{\etD(n)} requires further careful discussion that will follow in Sec.~\ref{sec:KW_doublet}. 
On the contrary, \tm{\bar\ps(n) \GaD h_{j}^{\pm} \ps(n\pm e_{j})} acts as \tm{\rho_{3}\otimes\GaD^{\rr}} anyways. 
\vskip1ex

We may construct another representation of the taste \tm{\mathfrak{su}(2)} algebra by combining shifts with phases, too. 
The algebra closes if we use the hypercube-internal shifts defined in Eq.~\eqref{eq:int_shift}. 
Judging from Table~\ref{tab:KW_vr_sim}, in order to have a nontrivial Lie product, we have to include either a phase factor \tm{\etD(n)} or \tm{(\etDf)(n)} with each one-step shift, or we have to include a factor \tm{\etf(n)} with one arbitrarily chosen one-step shift and with the two-step shift \tm{\hDj^{I}=h_{D}^{I}h_{j}^{I}}. 
As an involution the one-step shift with the phase factor in the same direction needs to include \tm{\pm\ri}. 
In line with the taste decomposition of Eq.~\eqref{eq:KW_decomp_trial}, we choose \tm{\etD} for the shift operators, and have these satisfy the \tm{\mathfrak{su}(2)} algebra 
\nequn{
[\etD h_{D}^{I}, \ri\etD h_{j}^{I}]_- = +2\ri \hDj^{I}
~\label{eq:KW_shift_alg}
}%nequn 
in the free theory, which is broken by interactions. 
The \acl{ST} structure of the one-step shifts is given by 
\nequn{
\setlength{\extrarowheight}{2pt}
\left.\begin{array}{l|l|l}
(\etD, \etf)(n) & (\bar\ps \etD h_{D}^{\pm} \ps)(n) & (\bar\ps \etD\ri h_{j}^{\pm} \ps)(n)
\\[2pt]
\hline
(+,+) 
& \bar\ps_{\uparrow L}~(+\vp_{+-})~\ps_{\downarrow B}^{\rr} + \bar\ps_{\downarrow B}^{\rr}~(+\vp_{++}^\ast)~\ps_{\uparrow L} 
& \bar\ps_{\uparrow L}~(+\ri\vp_{--})~\ps_{\downarrow B}^{\rr} + \bar\ps_{\downarrow B}^{\rr}~(+\ri\vp_{++}^\ast)~\ps_{\uparrow L} 
\\[2pt]
(-,+) 
& \bar\ps_{\uparrow L}~(-\vp_{--})~\ps_{\downarrow B}^{\rr} + \bar\ps_{\downarrow B}^{\rr}~(-\vp_{-+}^\ast)~\ps_{\uparrow L} 
& \bar\ps_{\uparrow L}~(-\ri\vp_{+-})~\ps_{\downarrow B}^{\rr} + \bar\ps_{\downarrow B}^{\rr}~(-\ri\vp_{-+}^\ast)~\ps_{\uparrow L} 
\\[2pt]
(-,-) 
& \bar\ps_{\uparrow R}~(-\vp_{--})~\ps_{\downarrow A}^{\rr} + \bar\ps_{\downarrow A}^{\rr}~(-\vp_{--}^\ast)~\ps_{\uparrow R} 
& \bar\ps_{\uparrow R}~(-\ri\vp_{++})~\ps_{\downarrow A}^{\rr} + \bar\ps_{\downarrow A}^{\rr}~(-\ri\vp_{--}^\ast)~\ps_{\uparrow R}
\\[2pt]
(+,-) 
& \bar\ps_{\uparrow R}~(+\vp_{++})~\ps_{\downarrow A}^{\rr} + \bar\ps_{\downarrow A}^{\rr}~(+\vp_{+-}^\ast)~\ps_{\uparrow R} 
& \bar\ps_{\uparrow R}~(+\ri\vp_{-+})~\ps_{\downarrow A}^{\rr} + \bar\ps_{\downarrow A}^{\rr}~(+\ri\vp_{+-}^\ast)~\ps_{\uparrow R}
\\[2pt]
% \hline
\end{array}\right.
,~\label{eq:KW_onestep_decomp}
}%nequn 
where either \tm{(A,B)=(R,L)} implies that \tm{(\bar\ps \etD)(n)h_{D}^{\pm} \ps(n\pm e_{D})} or \tm{(\bar\ps \etD)(n)\ri h_{j}^{\pm} \ps(n\pm e_{j})}
act as \tm{\rho_{1}\otimes\GaD^{\rr}} or \tm{\rho_{2}\otimes\GaD^{\rr}}, respectively, or alternatively \tm{(A,B)=(L,R)} implies that \tm{(\bar\ps \etD)(n)h_{D}^{\pm} \ps(n\pm e_{D})} or \tm{(\bar\ps \etD)(n)\ri h_{j}^{\pm} \ps(n\pm e_{j})} act as \tm{\rho_{1}\otimes\Id_{\Ga}^{\rr}} or \tm{\rho_{2}\otimes\Id_{\Ga}^{\rr}}, respectively. 
This is exactly the reverse of the behavior of the local generators \tm{\vr_{1,2}},  
% \vskip1ex
%
%
The bilinears with \acl{TD} generators, \tm{(\bar\ps\vr_{3}\ps)(n)} or \tm{(\bar\ps \etD)(n)h_{D}^{\pm}~\etD(n\pm e_{D})h_{j}^{\pm} \ps(n\pm e_{D} \pm e_{j})},
% }%nequn 
both %\tm{(\bar\ps\vr_{3}\ps)(n)} or \tm{(\bar\ps)(n)h_{D}^{\pm}h_{j}^{\pm} \ps(n\pm e_{D} \pm e_{j})} 
act as \tm{\rho_{3}\otimes\Id_{\Ga}} independent of the association of chirality \tm{(A,B)} and representation \tm{\rr}. 
\vskip1ex

\subsubsection{Taste-doublet structure}\label{sec:KW_doublet}

We conclude the preceding derivation and summarize:
\begin{enumerate}
  \item 
  Each taste contributes at each site, and only one taste is accompanied by a phase factor \tm{\varphi_{\etD,\etf} = \etD}.
  \item 
  Associating each Weyl spinor with one chiral component of one taste does not lead to inconsistencies. 
  \item 
  Site-parity \tm{\etf(n)} determines the correspondence between Weyl spinor and chiral components of the tastes.
  \item 
  Whether a specific taste is the upper or lower Weyl spinor at one particular site is not determined by the \ac{KW} operator or action, but is a matter of choice; both choices are related by parity \tm{\mP}. 
\end{enumerate}
We juxtapose two options: 
either we have \tm{(A,B)=(R,L)}, both tastes use the same---namely chiral---representation of spin matrices \tm{\{\Ga\}}, the \acl{TN} generators \tm{\{\vr\}} act exclusively on the taste group, while hypercube-internal translations \tm{h^{I}_{\mu_H}} act both on spin and taste, such that we can indeed write Eq.~\eqref{eq:KW_ST_int}. 
Or we have \tm{(A,B)=(L,R)}, the tastes use different representations of spin matrices (the second taste using dual representation), the \acl{TN} generators \tm{\{\vr\}} act on the spin group, too, while hypercube-internal translations \tm{h^{I}_{\mu_H}} act exclusively on taste, such that we cannot simply rephrase Eq.~\eqref{eq:KW_ST_int} as \tm{\{\rho\} \otimes \{\Ga\}}.  
We choose the former such that the \ac{KW} \acl{ST} decomposition simplifies to 
\nequn{
\setlength{\extrarowheight}{2pt}
\left.\begin{array}{l|cc}
\etf(n) & \bar\ps(n) & \ps(n) 
\\[2pt]
\hline
+ & (\bar\ps_{\uparrow L},~\bar\ps_{\downarrow L} \etD) & \bmat{c} \ps_{\uparrow R}\\ \etD \ps_{\downarrow R} \emat 
\\[2pt]
- & (\bar\ps_{\downarrow R} \etD,~\bar\ps_{\uparrow R} ) & \bmat{c} \etD \ps_{\downarrow L} \\ \ps_{\uparrow L} \emat 
\\[2pt]
% \hline
\end{array}\right.
,~\label{eq:KW_decomp}
}%nequn 
or equivalently, by expressing it with the taste-projection operators defined in Eq.~\eqref{eq:KW_Pvr}:
\nequn{
\begin{aligned}
\bar\ps(n) 
&= \Big(\bar\ps_{\uparrow }P^{\vr}_{+} + \bar\ps_{\downarrow }\vr_{1}P^{\vr}_{-}\Big)(n)
= \Big(\bar\ps_{\uparrow }P^{\vr}_{+} + \bar\ps_{\downarrow }P^{\vr}_{+}\vr_{1}\Big)(n)
,~\\
\ps(n) 
&= \Big( P^{\vr}_{+}\ps_{\uparrow } + P^{\vr}_{-}\vr_{1}\ps_{\downarrow }\Big)(n)
= \Big(P^{\vr}_{+}\ps_{\uparrow } + \vr_{1}P^{\vr}_{+}\ps_{\downarrow }\Big)(n)
~.~
\end{aligned}\label{eq:KW_proj}
}%nequn
The phase factor \tm{\etD} are introduced by \tm{\vr_{1}}, which flips the chirality of the second taste's Weyl spinors such that both tastes encode the same chirality at each site. 
It is the \tm{P^{\vr}_{\pm}} that reduce the degrees of freedom of each taste by a factor two, from Dirac to Weyl, such that each taste component contributes one Weyl spinor at each site, i.e. \tm{\big(P^{\vr}_{+} \ps_{f}\big)(n)=\ps_{f}(n)}, \tm{\big(P^{\vr}_{-} \ps_{f}\big)(n)=0}. 
Moreover, we can convert factors of \tm{\etf} and \tm{\gaf} into one another, when they act on the projectors \tm{P^{\vr}_{\pm}} or taste components \tm{\ps_{f}}, 
\nequn{
\big(\etf P^{\vr}_{\pm}\big)(n) = \gaf \big(\vr_{3} P^{\vr}_{\pm}\big)(n) = \pm \gaf P^{\vr}_{\pm}(n)
~.\label{eq:KW_parity_mixing}
}%nequn
Equation~\eqref{eq:KW_parity_mixing} has profound consequences, since the parity operator \tm{\mP} commutes with \tm{\etf}, but not with \tm{\gaf}. 
Therefore any interpolating operator built using \ac{KW} fermions necessarily excites states of both parities. 
\vskip1ex

\begin{figure}%
\begin{tikzpicture}
  \matrix (m) [matrix of math nodes,row sep=3em,column sep=3em,minimum width=12em]
  {
    \ps_{\uparrow }(a_j p_j,\aD p_D) & \gaf\ps_{\uparrow }(a_j p_j+\pi,ap_D+\pi) \\
    \gaD\gaf\ps_{\downarrow }(a_j p_j,\aD p_D+\pi) & \gaD\ps_{\downarrow }(a_j p_j+\pi,\aD p_D) \\};
  \path[<->]
    (m-1-1) 
            edge [thick] node [left] {\tm{\vr_{1}\hskip1em}} (m-2-2)
            edge [thick] node [left] {\tm{\vr_{2}}} (m-2-1)
            edge [thick] node [above] {\tm{\vr_{3}}} (m-1-2)
    (m-2-1) 
            edge [thick] node [right] {\tm{\hskip1em\vr_{1}}} (m-1-2)
            edge [thick] node [below] {\tm{\vr_{3}}} (m-2-2)
    (m-2-2) 
            edge [thick] node [right] {\tm{\vr_{2}}} (m-1-2)
%     (m-2-1.east|-m-2-2) edge node [below] {$\mathcal{B}_T$}
%             node [above] {$\exists$} (m-2-2)
%     (m-1-2) edge node [right] {$\mathcal{B}_T$} (m-2-2)
%             edge [dashed,-] (m-2-1)
  ;
\end{tikzpicture}
\caption{%
Action of the taste generators \tm{\{\vr\}} in momentum space.% 
\label{fig:KW_vr_mom}%
} 
\end{figure}
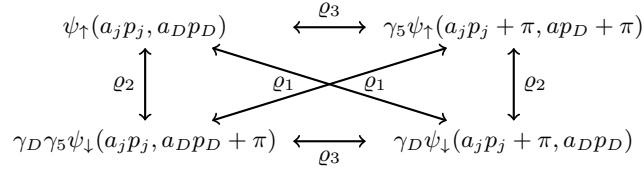

We can invert the relation between the \ac{KW} spinor \tm{\ps} and the tastes and project out the two tastes \tm{\ps_{f}} as function of \tm{\ps} in coordinate space. 
However, in order to permit both chiralities we have to use \tm{\ps} on a full \ac{KW} fermion hypersite \tm{n^{H}}, i.e. \emph{ne need both \tm{\ps(n^{H})} and \tm{\ps(n^{H}+e_{\mu_{H}})}} with appropriate gauge links included (in the \tm{h^{I}_{\mu_{H}}}), 
\nequn{
\begin{aligned}
\bar\ps_{\uparrow }(n^{H}) 
&= \big(\bar\ps P^{\vr}_{+}\big)(n^{H})  
+~ \big(\bar\ps P^{\vr}_{+}U_{-\mu_{H}}\big)(n+e_{\mu_{H}})
&=~&
\Big(\bar\ps P^{\vr}_{+}\big(1+h^{I}_{\mu_{H}}\big)\Big)(n^{H}) 
,~\\
\bar\ps_{\downarrow }(n^{H})  
&= \big(\bar\ps P^{\vr}_{-}\vr_{1}U_{-\mu_{H}}\big)(n^{H}+e_{\mu_{H}}) 
+~ \big(\bar\ps P^{\vr}_{-}\vr_{1}\big)(n^{H}) 
&=~&
\Big(\bar\ps P^{\vr}_{-}\vr_{1}\big(h^{I}_{\mu_{H}}+1\big)\Big)(n^{H})  
,~
\\
\ps_{\uparrow }(n^{H})  
&= \big(P^{\vr}_{+}\ps\big)(n^{H})  
+~ U_{\mu_{H}}(n)\big(P^{\vr}_{+}\ps\big)(n^{H}+e_{\mu_{H}})
&=~&
\Big(\big(1+h^{I}_{\mu_{H}}\big)P^{\vr}_{+}\ps\Big)(n^{H})  
,~\\
\ps_{\downarrow }(n^{H})  
&= U_{\mu_{H}}(n)\big(\vr_{1}P^{\vr}_{-}\ps\big)(n^{H}+e_{\mu_{H}}) 
+~ \big(\vr_{1}P^{\vr}_{-}\ps\big)(n^{H}) 
&=~&
\Big(\big(h^{I}_{\mu_{H}}+1\big)\vr_{1}P^{\vr}_{-}\ps\Big)(n^{H}) 
~.~
\end{aligned}
~\label{eq:KW_taste_comp}
}%nequn
The first or second terms correspond to the right-handed or left-handed components of \tm{\ps_{f}}, respectively, or vice versa of \tm{\bar\ps_{f}}.  
These two tastes can be contracted with one another or with spinors of any other single-taste fermion discretization using the standard gamma matrices (chiral representation). 
Note that \tm{\big(1+h^{I}_{\mu_{H}}\big)} is proportional to 
\nequn{
P_{+}^{H}(n_{H})
\equiv
\frac{1+h^{I}_{\mu_{H}}}{2} 
= \left(\frac{1+h^{I}_{\mu_{H}}}{2}\right)^2
=\left(P_{+}^{H}(n_{H})\right)^2
,~\label{eq:KW_hypersite_avg}
}%nequn
which is a gauge covariant projector that replaces the spinor fields on the boson sites by their average over a fermion hypersite (while accounting for the phase factors \tm{\etD} via \tm{\vr_{1}}) even in the interacting theory. 
For bilinears mixing \ac{KW} fermions and single-taste spinors it seems natural---albeit not necessary---to average the single-taste fields via \tm{P_{+}^{H}(n_{H})} over the \ac{KW} fermion hypersite. 
We rearrange the Weyl spinors of \tm{\ps} across the fermion hypersite or the chiral components of the tastes \tm{\ps_{f}(n^{H})} into a taste-doublet notation consistent with the interpretation in Eq.~\eqref{eq:KW_ST_int}, 
\nequn{
\bar\ps(n^{H}) =
\bmat{c}
\bar\ps_{\uparrow L}~
\bar\ps_{\uparrow R}~
\bar\ps_{\downarrow L}~
\bar\ps_{\downarrow R}
\emat(n^{H})
,\quad
\ps(n^{H}) =
\bmat{c}
\ps_{\uparrow R}  \\
\ps_{\uparrow L}  \\
\ps_{\downarrow R} \\
\ps_{\downarrow L}  
\emat(n^{H})
~.~
\label{eq:KW_taste_doublet}
}%nequn
This notation is underlying the \acl{ST} analysis of \ac{KW} fermion bilinear operators in Sec.~\ref{sec:KW_bilinears}. 
It is particularly remarkable that the outer \tm{\mathfrak{su}(2)} matrix of the set \tm{\{\ga\}} corresponds to the \tm{\mathfrak{su}(2)} algebra \tm{\{\rho\}}, while the inner \tm{\mathfrak{su}(2)} matrix of the set \tm{\{\ga\}} corresponds to the inner \tm{\mathfrak{su}(2)} matrix of the set \tm{\{\Ga\}},  whereas the shifts correspond to the outer \tm{\mathfrak{su}(2)} matrix of the set \tm{\{\Ga\}}. 
% \vskip1ex
%
%
We may Fourier transform Eq.~\eqref{eq:KW_proj} to obtain the momentum space representation in the free theory, 
\nequn{
\begin{aligned}
\bar\ps(p) 
&= 
\bar\ps_{\uparrow }\Big(p\Big) + \bar\ps_{\downarrow }\Big(p+\frac{\pi}{a_j}e_{j}\Big)\gaD
 = 
\bar\ps_{\uparrow }\Big(p\Big) + \bar\ps_{\downarrow }\Big(p+\frac{\pi}{\aD}e_{D}\Big)(+\ri\gaDf)
,~\\
\ps(p) 
&= 
\ps_{\uparrow }\Big(p\Big) + \gaD \ps_{\downarrow }\Big(p+\frac{\pi}{a_j}e_{j}\Big)
 = 
\ps_{\uparrow }\Big(p\Big) +(-\ri\gaDf) \ps_{\downarrow }\Big(p+\frac{\pi}{\aD}e_{D}\Big)
,~
\end{aligned}\label{eq:KW_proj_mom}
}%nequn
where each taste component has support from two maximally separated Fermi points arising from the projectors \tm{P^{\vr}_{+}},
\nequn{
% \begin{aligned}
\bar\ps_{f}\Big(p\Big) 
%&
= 
\frac{\bar\ps_{f}\Big({\markit{p}}\Big) + \bar\ps_{f}\Big(p+\frac{\pi}{a_{\mu}}e_{\mu}\Big) \gaf}{2}
,%~\\
\quad
\ps_{f}\Big(p\Big) 
%&
= 
\frac{\ps_{f}\Big(p\Big) + \gaf \ps_{f}\Big(p+\frac{\pi}{a_{\mu}}e_{\mu}\Big)}{2}
~.~
% \end{aligned}
\label{eq:KW_comp_mom}
}%nequn
We see in Fig.~\ref{fig:KW_vr_mom} how the generators \tm{\{\vr\}} swap these contributions in momentum space. 
Therefore, the \acl{ST} structure cannot be understood from looking only at the two survivor modes. 
Neither of the earlier ad hoc attempts~\cite{Creutz:2010cz, Creutz:2010qm} to separate the tastes of \ac{KW} fermions has taken this key feature into account. 
\vskip1ex

We are now ready to spell out the \ac{KW} operator in this taste-doublet notation as
\nequn{
\DKW(x,y) \equiv 
\sum_\mu \left( \Id_{\rho} \otimes \Ga_\mu \xi_{f0,\mu} \right) \nab_\mu(x,y) 
+\left( {\rho}_{1} \otimes \Id_{\Ga} \etD \right) \ri\frac{r}{\aD} (D-1) \de(x,y)
-\left( {\rho}_{3} \otimes \GaD \right) \ri\frac{r}{\aD} \sum_{j=1}^{D-1} h_{j}^{c}(x,y)
~.~
\label{eq:D_KW_comp}
}%nequn
The rightmost (hopping) operator's \acl{ST} structure \tm{\big( {\rho}_{3} \otimes \GaD \big)} suggests its interpretation as a \acl{TD} isovector quark number density \tm{\tfrac{1}{D-1}\bar\ps \left( {\rho}_{3} \otimes \GaD \right) \ps} multiplying an imaginary \acl{TD} isovector chemical potential \tm{\propto-\frac{\ri r}{a}(D-1)} that diverges in the continuum limit. 
Odd contributions of \tm{\left( {\rho}_{3} \otimes \GaD \right)} cancel in taste-symmetric quantities. 
The second-to-right (single-site) operator's \acl{ST} structure \tm{\big( {\rho}_{1} \otimes \Id_{\Ga} \etD \big)} with remnant phase factor \tm{\etD} is indistinguishable from \tm{\big( {\rho}_{1} \otimes \Gaf \etDf \big)} via Eq.~\eqref{eq:KW_parity_mixing}. 
In either way it multiplies an imaginary \acl{TN} isovector density or source term \tm{\propto\tfrac{\ri r}{a}(D-1)} that diverges in the continuum limit.
\tm{\bar\ps \big( {\rho}_{1} \otimes \Gaf \etDf \big) \ps } has to be interpreted as an imaginary \acl{TN} yet symmetric isovector pseudoscalar source operator alternating between consecutive time slices such that its odd powers cancel in time-averaged quantities. 
Finite results in the continuum limit can be obtained if and only if the contributions from both divergent densities (or condensates) cancel. 
This is far from trivial, since both operators encode different \acl{ST} structures. 
Both of these individual cancellations suggest that the \ac{KW} fermion determinant, which is a four-volume averaged \acl{TS} observable, perceives only even contributions from such operators. 
We will show in Sec.~\ref{sec:KW_determinant} that this is indeed the case. 
Nonetheless, such cancellations may fail for some correlation functions, see Sec.~\ref{sec:KW_splittings}. 
\vskip1ex

Equation~\eqref{eq:D_KW_comp} supersedes the \acl{ST} structure claimed in a previous publication~\cite{Tiburzi:2010bm},\footnote{%
Ref.~\cite{Tiburzi:2010bm} identified the isovector nature of the extra terms w.r.t. \tm{\vr_{2}}, but failed to realize that this does not imply that they share the same isovector structure. 
Moreover, the attribution of maximally separated opposing corners of the Brillouin zone to the same taste given in Eq.~\eqref{eq:KW_comp_mom} had not been realized, as the Brillouin zone had been incorrectly cut into two disjoint \emph{balls} separated along the temporal direction to represent the tastes in momentum space~\cite{Tiburzi:2010bm}. 
The \emph{balls} fail to separate the two tastes and the \acl{TN} single-site term is misidentified.} 
whose inconsistent \acl{ST} analysis does not affect the correctness of the claims made about the axial anomaly~\cite{Tiburzi:2010bm}, since the misidentified \acl{ST} structure of the divergent single-site extra term does not contribute to the axial anomaly at the one-loop level. 
Equation~\eqref{eq:D_KW_comp} is consistent with the physical terms of the \acl{ST} identification~\cite{Kimura:2011ik} inspired by similar analyses for \aclp{SF}~\cite{Kluberg-Stern:1983lmr}. 
The two conserved currents are the usual ones~\cite{Capitani:2010nn}. 
Each conserved current can be generated by a transformation rule without site-dependent phase factors and is coherent across all components of \tm{\bar{\ps}} and \tm{\ps} at the same boson site,
\nalign{
% \begin{split}
\bar{\ps}(n) \to \bar\ps(n) \ee^{-\ri \al_{V}^{0}(n) }
= \bar\ps(n^{H}) \ee^{-\ri \al_{V}^{0}(n^{H}) \big(\Id_{\rho} \otimes \Id_{\Ga} \big)}
,&\quad
\ps(n) \to \ee^{+\ri \al_{V}^{0}(n) } \ps(n)
= \ee^{+\ri \al_{V}^{0}(n^{H}) \big(\Id_{\rho} \otimes \Id_{\Ga} \big)} \ps(n^{H})
~,~\label{eq:KW_conserved_v0}\\
\bar{\ps}(n) \to \bar\ps(n) \ee^{-\ri \al_{A}^{3}(n) \gaf}
= \bar\ps(n^{H}) \ee^{+\ri \al_{A}^{3}(n^{H}) \big(\rho_{3} \otimes {\Gaf} \big)} 
,&\quad
\ps(n) \to \ee^{+\ri \al_{A}^{3}(n) \gaf} \ps(n)
= \ee^{+\ri \al_{A}^{3}(n^{H}) \big(\rho_{3} \otimes {\Gaf} \big)} \ps(n^{H})
~.~\label{eq:KW_conserved_a3}
% \end{split}~
}% 
Yet only the respective temporal components can be cleanly identified as either \acl{TS} vector or \acl{TI} axial densities. 
The spatial components include according Tab.~\ref{tab:KW_vr_sim} an admixture of a \acl{TI} vector or a \acl{TS} axial current, respectively.
The nonconserved currents are only partially consistent with the \acl{ST} identification~\cite{Kimura:2011ik}, since their association to vector or axial transformations is the opposite. 
Each can be generated by a transformation rule with a site-dependent phase factor \tm{\etf}. 
Thus, they are incoherent across all components of \tm{\bar{\ps}} and \tm{\ps} at the same boson site. 
In this formalism these are generated by the transforms 
\nalign{
% \begin{split}
\bar{\ps}(n) \to \bar\ps(n) \ee^{-\ri \al_{V}^{3}(n) \vr_{3}(n)}
= \bar\ps(n^{H}) \ee^{-\ri \al_{V}^{3}(n^{H}) \big(\rho_{3} \otimes \Id_{\Ga} \big)}
,&\quad
\ps(n) \to \ee^{+\ri \al_{V}^{3}(n) \vr_{3}(n)}\ps(n)
= \ee^{+\ri \al_{V}^{3}(n^{H}) \big(\rho_{3} \otimes \Id_{\Ga} \big)} \ps(n^{H})
,~\label{eq:KW_non_conserved_v3}\\
\bar{\ps}(n) \to \bar\ps(n) \ee^{+\ri \al_{A}^{0}(n) \etf(n)}
= \bar\ps(n^{H}) \ee^{+\ri \al_{A}^{0}(n^{H}) \big(\Id_{\rho} \otimes {\Gaf} \big)}
,&\quad
\ps(n) \to \ee^{+\ri \al_{A}^{0}(n) \etf(n)}\ps(n)
= \ee^{+\ri \al_{A}^{0}(n^{H}) \big(\Id_{\rho} \otimes {\Gaf} \big)} \ps(n^{H}) 
~.~\label{eq:KW_non_conserved_a0}
% \end{split}
}%
As we have seen in the discussion of naive fermions, the transformation rules of Eq.~\eqref{eq:taumu_act} for taste-rotations do not transform \tm{\ps(n)} or \tm{\ps\dag(n)} coherently, i.e. with adjoint factors and accounting for \tm{\bar\ps = \ps\dag\gaD}, but do so only on the level of the taste components. 
This coherence of the transformation rules on the level of taste components is apparent upon inspection of Eq.~\eqref{eq:KW_decomp}. 
We see now which operators or condensates violate the conservation of which current. 
The vector current in Eq.~\eqref{eq:KW_conserved_v0} is always conserved as it has to be due to baryon number conservation.
The axial current in Eq.~\eqref{eq:KW_conserved_a3} is broken by a \acl{TS} or \acl{TI} scalar (or pseudoscalar) operator, i.e. by any \acl{TD} mass term or any \acl{TD} scalar condensate.
The vector current in Eq.~\eqref{eq:KW_non_conserved_v3} is broken by a \acl{TN} scalar (or pseudoscalar) operator, i.e. by the second to rightmost operator structure in Eq.~\eqref{eq:D_KW_comp} and the corresponding condensate. 
Thus, although baryon number is conserved, each taste's quark number is not. 
Finally, the axial current in Eq.~\eqref{eq:KW_non_conserved_a0} is broken by any of these terms. 
Hence, although taste-isovector axial charge is conserved in the chiral limit, topology can change due to the nonconservation of each taste's axial charge. 
\vskip1ex

\subsubsection{Decomposition of spinors for perpendicular Karsten-Wilczek fermions}\label{sec:KW_spinors_per}

The entire derivation of the \acl{ST} structure carries over to perpendicular \ac{KW} fermions with a few straightforward changes. 
First, due to Eq.~\eqref{eq:KW_vr_per} we need to include a Pauli matrix \tm{\om_{j_{0}}}.
Second, we have to replace the phase factors \tm{\etD} by \tm{\ri\etjf} or \tm{\etj}, depending on the generator's representation. 
We underline the parts varying due to choosing either \tm{\{\vr\pri\}} or \tm{\{\vr\crc\}} representations. 
We only summarize the results for the explicit decomposition in Eq.~\eqref{eq:KW_decomp_per}, 
\nequn{
\setlength{\extrarowheight}{4pt}
\left.\begin{array}{l|cc}
\{\vr\pri\} \\[2pt]
\hline
\etf(n) & \bar\ps(n) & \ps(n) 
\\[2pt]
\hline
+
 & (\bar\ps_{\uparrow L},~\underline{-\ri\etj} \bar\ps_{\downarrow L} \om_{j_{0}} ) 
 & \bmat{c} \ps_{\uparrow R}\\ \underline{+\ri\etj} \om_{j_{0}} \ps_{\downarrow R} \emat 
\\[2pt]
-
 & (\underline{-\ri\etj} \bar\ps_{\downarrow R} \om_{j_{0}} ,~\bar\ps_{\uparrow R} ) 
 & \bmat{c} \underline{+\ri\etj} \om_{j_{0}} \ps_{\downarrow L} \\ \ps_{\uparrow L} \emat 
\\[2pt]
% \hline
\end{array}\right.
\quad,\quad
\left.\begin{array}{l|cc}
\{\vr\crc\} \\[2pt]
\hline
\etf(n) & \bar\ps(n) & \ps(n) 
\\[2pt]
\hline
+
 & (\bar\ps_{\uparrow L},~\underline{\etjf}\bar\ps_{\downarrow L} \om_{j_{0}} ) 
 & \bmat{c} \ps_{\uparrow R}\\ \underline{\etjf} \om_{j_{0}} \ps_{\downarrow R} \emat 
\\[2pt]
-
 & ( \underline{\etjf} \bar\ps_{\downarrow R} \om_{j_{0}},~\bar\ps_{\uparrow R} ) 
 & \bmat{c} \underline{\etjf} \om_{j_{0}} \ps_{\downarrow L} \\ \ps_{\uparrow L} \emat 
\\[2pt]
% \hline
\end{array}\right.
,~\label{eq:KW_decomp_per}
}%nequn 
which takes exactly the same form with either representation \tm{\{\vr\pri\}} or \tm{\{\vr\crc\}} in projector notation, 
\nequn{
\begin{aligned}
\bar\ps(n) 
&= \Big(\bar\ps_{\uparrow }P^{\vr}_{+} - \bar\ps_{\downarrow } P^{\vr}_{+} \underline{\vr_{1}\pri}\Big)(n)
 = \Big(\bar\ps_{\uparrow }P^{\vr}_{+} - \bar\ps_{\downarrow } P^{\vr}_{+} \underline{\vr_{1}\crc}\Big)(n)
,~\\
\ps(n) 
&= \Big( P^{\vr}_{+}\ps_{\uparrow } -\underline{\vr_{1}\pri} P^{\vr}_{+} \ps_{\downarrow }\Big)(n)
 = \Big(P^{\vr}_{+}\ps_{\uparrow } - \underline{\vr_{1}\crc} P^{\vr}_{+} \ps_{\downarrow }\Big)(n)
~.
\end{aligned}\label{eq:KW_proj_per}
}%nequn
In line with this taste decomposition of Eq.~\eqref{eq:KW_decomp_per}, we define the corresponding shift representations of generators that satsify the \tm{\mathfrak{su}(2)} algebra \tm{[\etj h_{j_{0}}^{I}, \ri\etj h_{\nu}^{I}]_- = +2\ri h_{j_{0}\nu}^{I}} or \tm{[\ri h_{j_{0}}^{I}\etjf , h_{\nu}^{I}\etjf ]_- = +2\ri h_{j_{0}\nu}^{I}} in the free theory, which is broken by interactions. 
The choice between representations \tm{\{\vr\pri\}} or \tm{\{\vr\crc\}} of the taste algebra becomes irrelevant in the momentum space representation (stressing that \tm{\nu \neq j_{0}}),
\nalign{
\begin{aligned}
\bar\ps(p) 
&= 
\bar\ps_{\uparrow }\Big(p\Big) +\bar\ps_{\downarrow }\Big(p+\underline{\frac{\pi}{a_{j_{0}}}e_{j_{0}}}\Big)\underline{\ga_{j_{0} 5}}
 = 
\bar\ps_{\uparrow }\Big(p\Big) +\bar\ps_{\downarrow }\Big(p+\underline{\frac{\pi}{a_{\nu}}e_{\nu}}\Big)\underline{(-\ri\ga_{j_{0}})}
,
~\\
\ps(p) 
&= \ps_{\uparrow }\Big(p\Big) +\underline{\ga_{j_{0} 5}} \ps_{\downarrow }\Big(p+\underline{\frac{\pi}{a_{j_{0}}}e_{j_{0}}}\Big)
 = \ps_{\uparrow }\Big(p\Big) +\underline{(+\ri\ga_{j_{0}})} \ps_{\downarrow }\Big(p+\underline{\frac{\pi}{a_{\nu}}e_{\nu}}\Big)
,~
\end{aligned}\label{eq:KW_proj_mom_per}
}%nalign
where the two taste components are defined as in Eq.~\eqref{eq:KW_comp_mom} and have support from two maximally separated Fermi points arising from the projectors \tm{P^{\vr}_{+}}. 
The perpendicular \ac{KW} operator in taste-doublet notation follows from Eq.~\eqref{eq:D_KW_comp} with the obvious replacements. 
Similarly, the discussion of conserved or nonconserved symmetry currents follows from Eqs.~\eqref{eq:KW_conserved_v0}~-~\eqref{eq:KW_non_conserved_a0} with obvious replacements. 
\vskip1ex
% \clearpage
%%%%%%%%%%%%%%%%%%%%%%%%%%%%%%%%%%%%%%%%%%%%%%%%%%%%%%%%%%%%%%%%%%%%%%%%%%%%%%%%
% \input{\INCDIR/kawi_bilinear}
%% kawi_bilinear.tex
%\acresetall

\section{Karsten-Wilczek bilinears and hadron interpolating operators}\label{sec:KW_bilinears}

Now we are in a position to understand the \acl{ST} structure of any \ac{KW} fermion bilinears. 
If we contract two spinors of \ac{KW} fermions together, we do not have to explicitly project out the taste components as in Eq.~\eqref{eq:KW_taste_comp}. 
Instead we may construct interpolating operators from bilinears of \ac{KW} spinors \tm{\bar\ps,~\ps} for mesons or from \tm{\ps^{T},~\ps} for diquarks and use Eq.~\eqref{eq:KW_proj} to understand the involved \acl{ST} components. 
Each interpolating operator involves a product of gamma matrices (in chiral representation), shifts and appropriate phase factors \tm{\etD},~\tm{\etf},~or \tm{\etDf}.
We note that each combination of a gamma matrix and a shift excites low-lying states of both parities, where the interpolating operators exciting the parity partners differ by a factor \tm{\etf} due to Eq.~\eqref{eq:KW_parity_mixing}. 
Thus, one of the parity partners has a large momentum at the lattice cutoff and oscillates in the temporal direction. 
While it is still necessary to include the two boson sites that make up the fermion hypersite to include both chiralities, we may directly construct \ac{KW} spinor bilinears of any given \acl{ST} structure without explicit hypersite average via Eq.~\eqref{eq:KW_hypersite_avg} at each end of each propagator.   
\vskip1ex

\subsection{Karsten-Wilczek spin-taste structure of mesons}\label{sec:KW_meson}

\begin{table}
\setlength{\extrarowheight}{2pt}
\begin{tabular}{l||l|l|c||l|l|c||l|l|c||l|l|c}
& \multicolumn{3}{c||}{\tm{\Id_{h}}} & \multicolumn{3}{c||}{\tm{\etD h_{D}}} & \multicolumn{3}{c||}{\tm{\ri\etD h_{j}}} & \multicolumn{3}{c}{\tm{\hDj}} 
\\[2pt]\hline\hline
%%%%%%%%%%%%%%%%%%%%%%%%%%%%%%%%%%%%%%%%%%%%%%%%%%%%%%%%%%%
%%%%%%%%%%%%%%%%%%%%%%%%%%%%%%%%%%%%%%%%%%%%%%%%%%%%%%%%%%%
\tm{\Id_{\ga}}   
& %\tm{\Id_{\rho}\otimes\Id_{\chi}\otimes\Id_{\om}}
\tm{~~\Id_{\rho}\otimes\Id_{\Ga}} & \tm{0^{++}} & \multirow{2}{*}{0}
& %\tm{\rho_{1}\otimes\chi_{1}\otimes\Id_{\om}}
\tm{~~~\rho_{1}\otimes\GaD} & \tm{0^{+}} & \multirow{2}{*}{1(S)}
& %\tm{\rho_{2}\otimes\chi_{1}\otimes\Id_{\om}}
\tm{~~~\rho_{2}\otimes\GaD} & \tm{0^{+}} & \multirow{2}{*}{1(A)}
& %\tm{\rho_{3}\otimes\Id_{\chi}\otimes\Id_{\om}}
\tm{~~~\rho_{3}\otimes\Id_{\Ga}} & \tm{0^{++}} & \multirow{2}{*}{1(0)}
\\[2pt]
\tm{\Id_{\ga}\etf}   
& %\tm{\Id_{\rho}\otimes\chi_{3}\otimes\Id_{\om}}
\tm{~~\Id_{\rho}\otimes{\Gaf}} & \tm{0^{-+}} & %0
& %\tm{\ri\rho_{1}\otimes\chi_{2}\otimes\Id_{\om}}
\tm{~~\ri\rho_{1}\otimes\GaDf} & \tm{0^{-}} & %1(S) 
& %\tm{\ri\rho_{2}\otimes\chi_{2}\otimes\Id_{\om}}
\tm{~~\ri\rho_{2}\otimes\GaDf} & \tm{0^{-}} & %1(A)
& %\tm{\rho_{3}\otimes\chi_{3}\otimes\Id_{\om}}
\tm{~~~\rho_{3}\otimes{\Gaf}} & \tm{0^{-+}} & %1(0) 
\\[2pt]\hline
%%%%%%%%%%%%%%%%%%%%%%%%%%%%%%%%%%%%%%%%%%%%%%%%%%%%%%%%%%%
\tm{\gaD\etD}   
& %\tm{\rho_{1}\otimes\Id_{\chi}\otimes\Id_{\om}}
\tm{~~~\rho_{1}\otimes\Id_{\Ga}} & \tm{0^{+}} & \multirow{2}{*}{1(S)}
& %\tm{\Id_{\rho}\otimes\chi_{1}\otimes\Id_{\om}}
\tm{~~\Id_{\rho}\otimes\GaD} & \tm{0^{+-}} & \multirow{2}{*}{0}
& %\tm{+\ri\rho_{3}\otimes\chi_{1}\otimes\Id_{\om}}
\tm{+\ri\rho_{3}\otimes\GaD} & \tm{0^{+-}} & \multirow{2}{*}{1(0)}
& %\tm{-\ri\rho_{2}\otimes\Id_{\chi}\otimes\Id_{\om}}
\tm{-\ri\rho_{2}\otimes\Id_{\Ga}} & \tm{0^{+}} & \multirow{2}{*}{1(A)}
\\[2pt]
\tm{\gaD\etDf}   
& %\tm{\rho_{1}\otimes\chi_{3}\otimes\Id_{\om}}
\tm{~~~\rho_{1}\otimes\Gaf} & \tm{0^{-}} & %1(S) 
& %\tm{~\ri\Id_{\rho}\otimes\chi_{2}\otimes\Id_{\om}}
\tm{~\ri\Id_{\rho}\otimes\GaDf} & \tm{0^{-+}} & %0 
& %\tm{-\rho_{3}\otimes\chi_{2}\otimes\Id_{\om}}
\tm{~-\rho_{3}\otimes\GaDf} & \tm{0^{-+}} & %1(0)
& %\tm{-\ri\rho_{2}\otimes\chi_{3}\otimes\Id_{\om}}
\tm{-\ri\rho_{2}\otimes\Gaf} & \tm{0^{-}} & %1(A) 
\\[2pt]\hline
%%%%%%%%%%%%%%%%%%%%%%%%%%%%%%%%%%%%%%%%%%%%%%%%%%%%%%%%%%%
\tm{\gaDf\etD}   
& %\tm{\rho_{2}\otimes\chi_{3}\otimes\Id_{\om}}
\tm{~~~\rho_{2}\otimes\Gaf} & \tm{0^{-}} & \multirow{2}{*}{1(A)}
& %\tm{-\rho_{3}\otimes\chi_{2}\otimes\Id_{\om}}
\tm{~~~\rho_{3}\otimes\GaDf} & \tm{0^{-+}} & \multirow{2}{*}{1(0)}
& %\tm{\ri\Id_{\rho}\otimes\chi_{2}\otimes\Id_{\om}}
\tm{~\ri\Id_{\rho}\otimes\GaDf} & \tm{0^{-+}} & \multirow{2}{*}{0}
& %\tm{\rho_{3}\otimes\chi_{3}\otimes\Id_{\om}}
\tm{~~\ri\rho_{1}\otimes\Gaf} & \tm{0^{-}} & \multirow{2}{*}{1(S)}
\\[2pt]
\tm{\gaDf\etDf}   
& %\tm{~\rho_{2}\otimes\Id_{\chi}\otimes\Id_{\om}}
\tm{~~~\rho_{2}\otimes\Id_{\Ga}} & \tm{0^{+}} & %1(A) 
& %\tm{-\ri\rho_{3}\otimes\chi_{1}\otimes\Id_{\om}}
\tm{-\ri\rho_{3}\otimes\GaD} & \tm{0^{+-}} & %1(0) 
& %\tm{\Id_{\rho}\otimes\chi_{1}\otimes\Id_{\om}}
\tm{~~\Id_{\rho}\otimes\GaD} & \tm{0^{+-}} & %0 
& %\tm{\rho_{1}\otimes\Id_{\chi}\otimes\Id_{\om}}
\tm{~~\ri\rho_{1}\otimes\Id_{\Ga}} & \tm{0^{+}} & %1(S) 
\\[2pt]\hline
%%%%%%%%%%%%%%%%%%%%%%%%%%%%%%%%%%%%%%%%%%%%%%%%%%%%%%%%%%%
\tm{\gaf}   
& %\tm{\rho_{3}\otimes\chi_{3}\otimes\Id_{\om}}
\tm{~~~\rho_{3}\otimes\Gaf} & \tm{0^{-+}} & \multirow{2}{*}{1(0)}
& %\tm{-\rho_{2}\otimes\chi_{2}\otimes\Id_{\om}}
\tm{~-\rho_{2}\otimes\GaDf} & \tm{0^{-}} & \multirow{2}{*}{1(A)}
& %\tm{+\rho_{1}\otimes\chi_{2}\otimes\Id_{\om}}
\tm{~+\rho_{1}\otimes\GaDf} & \tm{0^{-}} & \multirow{2}{*}{1(S)}
& %\tm{\rho_{3}\otimes\chi_{3}\otimes\Id_{\om}}
\tm{~~\Id_{\rho}\otimes\Gaf} & \tm{0^{-+}} & \multirow{2}{*}{0}
\\[2pt]
\tm{\gaf\etf}   
& %\tm{\rho_{3}\otimes\Id_{\chi}\otimes\Id_{\om}}
\tm{~~~\rho_{3}\otimes\Id_{\Ga}} & \tm{0^{++}} & %1(0) 
& %\tm{\ri\rho_{2}\otimes\chi_{1}\otimes\Id_{\om}}
\tm{~~~\ri\rho_{2}\otimes\GaD} & \tm{0^{+}} & %1(A) 
& %\tm{-\ri\rho_{1}\otimes\chi_{1}\otimes\Id_{\om}}
\tm{-\ri\rho_{1}\otimes\GaD} & \tm{0^{+}} & %1(S) 
& %\tm{\Id_{\rho}\otimes\Id_{\chi}\otimes\Id_{\om}}
\tm{~~\Id_{\rho}\otimes\Id_{\Ga}} & \tm{0^{++}} & %0 
\\[2pt]\hline\hline
%%%%%%%%%%%%%%%%%%%%%%%%%%%%%%%%%%%%%%%%%%%%%%%%%%%%%%%%%%%
%%%%%%%%%%%%%%%%%%%%%%%%%%%%%%%%%%%%%%%%%%%%%%%%%%%%%%%%%%%
\tm{
%\ep_{jkl}
\ga_{kl}}   
& %\tm{\Id_{\rho}\otimes\Id_{\chi}\otimes\om_{j}}
\tm{~~\Id_{\rho}\otimes
%\ep_{jkl}
\Ga_{kl}} & \tm{1^{+-}} & \multirow{2}{*}{0}
& %\tm{\rho_{1}\otimes\chi_{1}\otimes\om_{j}}
\tm{~~~~\rho_{1}\otimes\Ga_{j5}} & \tm{1^{+}} & \multirow{2}{*}{1(S)}
& %\tm{-\rho_{2}\otimes\chi_{1}\otimes\om_{j}}
\tm{~~~\rho_{2}\otimes\Ga_{j5}} & \tm{1^{+}} & \multirow{2}{*}{1(A)}
& %\tm{-\rho_{3}\otimes\Id_{\chi}\otimes\om_{j}}
\tm{~~~\rho_{3}\otimes
%\ep_{jkl}
\Ga_{kl}} & \tm{1^{+-}} & \multirow{2}{*}{1(0)}
\\[2pt]
\tm{
%\ep_{jkl}
\ga_{kl}\etf}   
& %\tm{-\Id_{\rho}\otimes \otimes\om_{j}}
\tm{-\Id_{\rho}\otimes
%\ep_{jkl}
\Ga_{jD}} & \tm{1^{ -- }} & %0
& %\tm{-\ri\rho_{1}\otimes \otimes\om_{j}}
\tm{~-\ri\rho_{1}\otimes\Ga_{j}} & \tm{1^{-}} & %1(S)
& %\tm{-\rho_{2}\otimes \otimes\om_{j}}
\tm{-\ri\rho_{2}\otimes\Ga_{j}} & \tm{1^{-}} & %1(A)
& %\tm{-\rho_{3}\otimes \otimes\om_{j}}
\tm{~~~\rho_{3}\otimes
%\ep_{jkl}
\Ga_{jD}} & \tm{1^{ -- }} & %1(0)
\\[2pt]\hline
%%%%%%%%%%%%%%%%%%%%%%%%%%%%%%%%%%%%%%%%%%%%%%%%%%%%%%%%%%%
\tm{\ga_{j5}\etD}   
& %\tm{\rho_{1}\otimes\Id_{\chi}\otimes\om_{j}}
\tm{~~~\rho_{1}\otimes
%\ep_{jkl}
\Ga_{kl}} & \tm{1^{+}} & \multirow{2}{*}{1(S)} 
& %\tm{-\Id_{\rho}\otimes\chi_{1}\otimes\om_{j}}
\tm{~~~\Id_{\rho}\otimes\Ga_{j5}} & \tm{1^{+-}} & \multirow{2}{*}{0} 
& %\tm{-\ri\rho_{3}\otimes\chi_{1}\otimes\om_{j}}
\tm{~~~\ri\rho_{3}\otimes\Ga_{j5}} & \tm{1^{+-}} & \multirow{2}{*}{1(0)}
& %\tm{\ri\rho_{2}\otimes\Id_{\chi}\otimes\om_{j}}
\tm{-\ri\rho_{2}\otimes
% \ep_{jkl}
\Ga_{kl}} & \tm{1^{+}} & \multirow{2}{*}{1(A)} 
\\[2pt]
\tm{\ga_{j5}\etDf}   
& %\tm{-\rho_{1}\otimes \otimes\om_{j}}
\tm{~-\rho_{1}\otimes
%\ep_{jkl}
\Ga_{iD}} & \tm{1^{-}} & %1(S) 
& %\tm{-\Id_{\rho}\otimes \otimes\om_{j}}
\tm{+\ri\Id_{\rho}\otimes\Ga_{j}} & \tm{1^{ -- }} & %0 
& %\tm{-\ri\rho_{3}\otimes \otimes\om_{j}}
\tm{~~~~\rho_{3}\otimes\Ga_{j}} & \tm{1^{ -- }} & %1(0)
& %\tm{\ri\rho_{2}\otimes \otimes\om_{j}}
\tm{~~~\ri\rho_{2}\otimes
% \ep_{jkl}
\Ga_{jD}} & \tm{1^{-}} & %1(A) 
\\[2pt]\hline
%%%%%%%%%%%%%%%%%%%%%%%%%%%%%%%%%%%%%%%%%%%%%%%%%%%%%%%%%%%
\tm{\ga_{j}\etD}   
& %\tm{\rho_{2}\otimes\chi_{3}\otimes\om_{j}}
\tm{~~~\rho_{2}\otimes\Ga_{jD}} & \tm{1^{-}} & \multirow{2}{*}{1(A)} 
& %\tm{\rho_{3}\otimes\chi_{2}\otimes\om_{j}}
\tm{~~~~\rho_{3}\otimes\Ga_{j}} & \tm{1^{ -- }} & \multirow{2}{*}{1(0)} 
& %\tm{\ri\Id_{\rho}\otimes\chi_{2}\otimes\om_{j}}
\tm{~\ri\Id_{\rho}\otimes\Ga_{j}} & \tm{1^{ -- }} & \multirow{2}{*}{0}
& %\tm{\ri\rho_{1}\otimes\chi_{3}\otimes\om_{j}}
\tm{~~~\ri\rho_{1}\otimes\Ga_{jD}} & \tm{1^{-}} & \multirow{2}{*}{1(S)} 
\\[2pt]
\tm{\ga_{j}\etDf}   
& %\tm{\rho_{2}\otimes\chi_{3}\otimes\om_{j}}
\tm{~-\rho_{2}\otimes\Ga_{kl}} & \tm{1^{+}} & %1(A) 
& %\tm{\rho_{3}\otimes\chi_{2}\otimes\om_{j}}
\tm{~~~\ri\rho_{3}\otimes\Ga_{j5}} & \tm{1^{+-}} & %1(0) 
& %\tm{\ri\Id_{\rho}\otimes\chi_{2}\otimes\om_{j}}
\tm{-\Id_{\rho}\otimes\Ga_{j5}} & \tm{1^{+-}} & %0
& %\tm{\ri\rho_{1}\otimes\chi_{3}\otimes\om_{j}}
\tm{~~~\ri\rho_{1}\otimes\Ga_{kl}} & \tm{1^{+}} & %1(S) 
\\[2pt]\hline
%%%%%%%%%%%%%%%%%%%%%%%%%%%%%%%%%%%%%%%%%%%%%%%%%%%%%%%%%%%
\tm{\ga_{jD}}   
& %\tm{\rho_{3}\otimes\chi_{3}\otimes\om_{j}}
\tm{~~~\rho_{3}\otimes\Ga_{jD}} & \tm{1^{ -- }} & \multirow{2}{*}{1(0)} 
& %\tm{-\rho_{2}\otimes\chi_{2}\otimes\om_{j}}
\tm{~~-\rho_{2}\otimes\Ga_{j}} & \tm{1^{-}} & \multirow{2}{*}{1(A)} 
& %\tm{\rho_{1}\otimes\chi_{2}\otimes\om_{j}}
\tm{~~~~\rho_{1}\otimes\Ga_{j}} & \tm{1^{-}} & \multirow{2}{*}{1(S)} 
& %\tm{\Id_{\rho}\otimes\chi_{3}\otimes\om_{j}}
\tm{~~~\Id_{\rho}\otimes\Ga_{jD}} & \tm{1^{ -- }} & \multirow{2}{*}{0}
\\[2pt]
\tm{\ga_{jD}\etf}   
& %\tm{\rho_{3}\otimes\chi_{3}\otimes\om_{j}}
\tm{~-\rho_{3}\otimes\Ga_{kl}} & \tm{1^{+-}} & %1(0) 
& %\tm{-\rho_{2}\otimes\chi_{2}\otimes\om_{j}}
\tm{~~~~\rho_{2}\otimes\Ga_{j5}} & \tm{1^{+}} & %1(A) 
& %\tm{\rho_{1}\otimes\chi_{2}\otimes\om_{j}}
\tm{~~~\ri\rho_{1}\otimes\Ga_{j5}} & \tm{1^{+}} & %1(S) 
& %\tm{\Id_{\rho}\otimes\chi_{3}\otimes\om_{j}}
\tm{~~~\Id_{\rho}\otimes\Ga_{kl}} & \tm{1^{+-}} & %0
\\[2pt]\hline\hline
%%%%%%%%%%%%%%%%%%%%%%%%%%%%%%%%%%%%%%%%%%%%%%%%%%%%%%%%%%%
%%%%%%%%%%%%%%%%%%%%%%%%%%%%%%%%%%%%%%%%%%%%%%%%%%%%%%%%%%%
& \tm{~~\{\rho\}\otimes\{\Ga\}} & \tm{J^{PC}} & \tm{I(m_I)} 
& \tm{~~\{\rho\}\otimes\{\Ga\}} & \tm{J^{PC}} & \tm{I(m_I)} 
& \tm{~~\{\rho\}\otimes\{\Ga\}} & \tm{J^{PC}} & \tm{I(m_I)} 
& \tm{~~\{\rho\}\otimes\{\Ga\}} & \tm{J^{PC}} & \tm{I(m_I)} 
\\[2pt]
\hline
\end{tabular}
\caption{
\Acl{ST} structure of meson interpolating operators constructed from two \ac{KW} spinor bilinears as \tm{\bar\ps X \ps}. 
The rows or columns indicate the gamma matrices (in chiral representation) or shifts used in the interpolating operators, with phase factors \tm{\etD}, \tm{\etf}, or \tm{\etDf}, where they are appropriate. 
For each combination we show the resulting \acl{ST} structure according to Eq.~\eqref{eq:KW_ST_int}, the quantum numbers \tm{J^{PC}} and the isospin quantum numbers \tm{I(m_I)}, where \tm{(S,A)} indicate the symmetric or antisymmetric combination of \tm{m_I=\pm 1}. 
The upper or lower lines in each block contain the states expected for a single-taste theory or due to the parity partners. 
Different spatial indices in a row, i.e. \tm{j,k,l} are always assumed to be antisymmetrized. 
\label{tab:KW_meson}
} 
\end{table}

\begin{figure}%
% \begin{minipage}{1.\txtw}
% \begin{minipage}{.48\txtw}
\includegraphics[width=0.38\txtw]{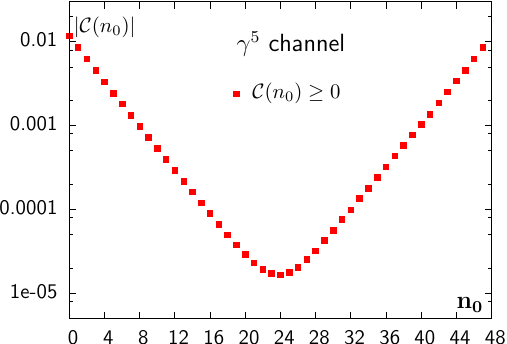}
\hskip1ex
\includegraphics[width=0.38\txtw]{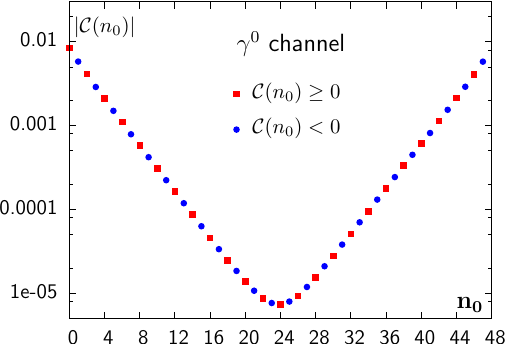}
% \end{minipage}
\vskip1ex
% \begin{minipage}{.48\txtw}
\includegraphics[width=0.38\txtw]{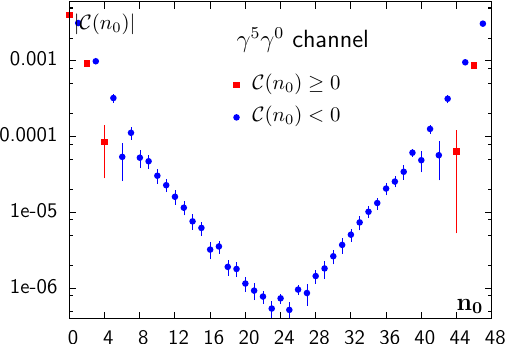}
\hskip1ex
\includegraphics[width=0.38\txtw]{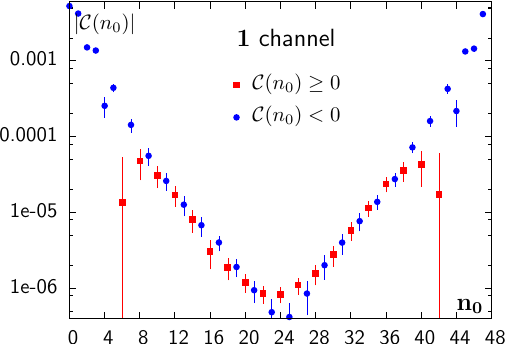}
% \end{minipage}
% \end{minipage}
\caption{%
Spin \tm{0} correlation functions of local bilinear operators with parallel \ac{KW} fermions on a pure gauge background with Jacobi smearing at the source. 
Figure from~\cite{Weber:2015oqf}. For details of the ensemble, see~\cite{Weber:2015oqf}, for a discussion see the text.% 
\label{fig:KW_corr_mes}%
} 
\end{figure}%

We collect the \ac{KW} fermion bilinears for meson interpolating operators in Table~\ref{tab:KW_meson}. 
The first lesson that we take out of Table~\ref{tab:KW_meson} is that we can access the same set of hadrons through the parity partners, in perfect analogy to \aclp{SF}. 
Namely, the same \acl{ST} structure (up to an irrelevant, constant phase) can be  accessed by changing the gamma matrix by \tm{\vr_{3}} and the number of suitable shifts by two, i.e. for \ac{KW} fermions by a spatial and a temporal one, see Table~\ref{tab:KW_meson}.  
In particular, such behavior was observed a long time ago~\cite{Weber:2015oqf}, where leaving out factors \tm{\etDf} or \tm{\etf} from the interpolating operators resulted in oscillating correlation functions. 
Since these interpolating operators used gauge-invariant Jacobi smearing~\cite{Gusken:1989ad} at the source and did not account for appropriate phase factors \tm{\etD} or \tm{\etf}, the taste identification may be slightly ambiguous. 
Nevertheless, the point-sink is expected to rule out contributions from odd numbers of shifts in the interpolating operators. 
We reproduce the corresponding plots in Fig.~\ref{fig:KW_corr_mes}. 
The plots in the upper row correspond to \tm{\big(\rho_{3}\otimes \Gaf\big)} (upper left) or to an oscillating contribution of 
\tm{\big(\rho_{1}\otimes \Gaf\big)} with a missing phase factor \tm{\etDf} (upper right). 
No further low-lying (excited) states could be resolved in either case. 
The plots in the lower row are less clear due to smearing and missing phase factors. 
The one on the lower left may correspond to \tm{\big(\rho_{2}\otimes \Gaf\big)} with a missing phase factor \tm{\etD} 
and includes a global minus sign. 
Some remnants of an oscillating contribution are clearly visible, as it may contain oscillating opposite parity contributions corresponding to  \tm{\big(\rho_{2}\otimes \Id_{\Ga}\big)} with a missing phase factor \tm{\etDf}.
The one on the lower right may correspond to oscillating contributions of \tm{\big(\Id_{\rho}\otimes \Gaf\big)} with a missing phase factor \tm{\etf}.
The presence of some nonoscillating contribution is undeniable, as it may contain opposite parity contributions corresponding to  \tm{\big(\Id_{\rho}\otimes \Id_{\Ga}\big)}.
\vskip1ex

\tm{\big(\rho_{3}\otimes \Gaf\big)} (upper left) contains the lightest state in line with a pseudo-\ac{NG} boson related to a spontaneously broken chiral symmetry unaffected by the cutoff effects. 
These plots demonstrate that \ac{KW} fermions have two different sources of \ac{TSB}, since the corresponding plots are within each row degenerate for naive (or \acl{stag}) fermions up to an oscillating phase factor \tm{\etDf(n)}~\cite{Weber:2015oqf}. 
While the nondegeneracy between the two rows increases if the extra operators are switched off in the naive \tm{r\to 0} limit, this observation should not be overinterpreted; first, since the changes are of a similar order as the change between the pseudo-\ac{NG} boson masses, and second, since the two discretizations at the same bare quark mass do not correspond to the same \acl{LCP}. 
Equation~\eqref{eq:KW_proj} delivers the first-principles explanation for this observed \acl{ST} structure and supersedes the previous suggestions for the \acl{ST} decomposition or identification~\cite{Tiburzi:2010bm, Creutz:2010cz, Creutz:2010qm, Kimura:2011ik, Weber:2015oqf}. 
We later return to these four pseudoscalar correlation functions in Sec.~\ref{sec:KW_splittings} and derive some of these features from the underlying symmetries. 
\vskip1ex

One has to expect that the mass splittings between different taste hadrons vanish in the continuum limit. 
Until now the approach to the continuum limit has been studied only for the mass splitting \tm{\big(m_{\rho_{1}\otimes \Gaf}^2-m_{\rho_{3}\otimes \Gaf}^2\big)} at various quark masses.  
Results are so far inconclusive, i.e. the most simple \tm{a^2} scaling could neither be verified with thin links on pure gauge~\cite{Weber:2015oqf} nor with stout-smeared~\cite{Morningstar:2003gk} links on staggered backgrounds~\cite{Godzieba:2024uki}. 
These continuum extrapolations end up slightly below zero, albeit inclusion of an \tm{\mO(a^4)} term seems to solve this issue. 
Given the two different causes of \ac{TSB} understanding the continuum limit of these splittings requires a systematic study and new data. 
\vskip1ex

\subsection{Mass terms for Karsten-Wilczek fermions}\label{sec:KW_mass}

The second lesson we take out of Table~\ref{tab:KW_meson} is that we understand how and why we can write down \acl{T-S} or \acl{T-D} isovector mass terms via \tm{\bar\ps \Id_{\ga} \ps} or \tm{\bar\ps (\Id_{\ga}\hDj) \ps}, respectively. 
The \acl{T-S} mass term is just the usual local one and requires no links, 
\nequn{
M_0(x,y) = m_0 \de(x,y)
,~\label{eq:KW_M0}
}%nequn
while the \acl{T-D} isovector one requires an even number of links, which must include an odd number in the time direction and an odd number in any spatial one. 
In order to retain discrete spatial isotropy, the \tm{(D-1)} spatial directions ought to be treated symmetrically, which may be done via a two-shift~\cite{Weber:2016dgo} or a \tm{D}-shift term,
\nalign{
M_{3}^{2\text{s}}(x,y)
&= \frac{m_3}{2(D-1)} \sum\limits_{j=1}^{D-1} \big[ h_{D},h_{j} \big]_+ (x,y)
,~\label{eq:KW_M3_2s}\\
M_{3}^{\text{Ds}}(x,y)
&= \frac{m_3}{D!} \prod\limits_{\mu=1}^{D}h_{\mu}(x,y)
,~\label{eq:KW_M3_Ds}
}%nalign
where either \tm{M_{3}^{2\text{s}}} or \tm{M_{3}^{D\text{s}}} may be constructed using the symmetrized shift of Eq.~\eqref{eq:cos_shift} or the hypercube-internal shift of Eq.~\eqref{eq:int_shift}. 
It is obvious from Table~\ref{tab:KW_vr_sim} that both \tm{M_{3}^{2\text{s}}} or \tm{M_{3}^{D\text{s}}} transform as \tm{\vr_{3}}, since any odd number of (appropriately symmetrized or hypercube-internal) shifts in the temporal direction transforms as \tm{\vr_{1}} or in the spatial directions as \tm{\vr_{2}}. 
In \tm{D=2}, where two-shift or \tm{D}-shift terms coincide, the mass term with symmetrized shift has been shown to be sensititve to the spectral flow and permits access to topology when combined with \tm{\gaf}~\cite{Durr:2022mnz, Kishore:2025fxt}. 
A construction of a \acl{TI} mass term involving shifts in only one direction~\cite{Creutz:2010bm} cannot be made \acl{TD} or diagonal in chirality, see Table~\ref{tab:KW_meson}. 
The massive \ac{KW} action acquires only an overall sign under the chiral symmetry of the massless case, if all mass parameters are simultaneously transformed as \tm{m_{0,3} \to - m_{0,3}}. 
A Symanzik \ac{EFT}~\cite{Weber:2015hib} and the construction of chiral Lagrangians have to take this into account. 
Whether a computationally convenient zero-shift, \acl{T-D} isovector scalar operator 
\nequn{
M_{3}^{0\text{s}}(x,y) = m_\vr \vr_{3} \de(x,y)
,~\label{eq:KW_M3_0s}
}%nequn 
could be a suitable option for a mass term is not obvious. 
While it breaks parity \tm{\mP} or time reflection \tm{\mT} symmetries, the action with it is invariant under \acl{T-S} parity \tm{\mP_{\pm D}} or \acl{T-S} time reflection \tm{\mT_{\pm j}}. 
Moreover, the free \ac{KW} operator would not be diagonal in momentum space anymore. 
Thus, Eq.~\eqref{eq:KW_M3_0s} should not be adopted without careful studies. 
Equations~\eqref{eq:KW_M0},~\eqref{eq:KW_M3_Ds}, or~\eqref{eq:KW_M3_0s} work for perpendicular \ac{KW} fermions, too, while Eq.~\eqref{eq:KW_M3_2s} has to be adapted in the obvious manner.
\vskip1ex

\subsection{Chemical potential terms for Karsten-Wilczek fermions}\label{sec:KW_chempot}

Chemical potential breaks charge conjugation \tm{\mC} and time reflection \tm{\mT} symmetries by introducing an asymmetric weight between forward or backward propagation in time. 
Implementing it through a constant, additive term gives rise to a quadratic power-law divergence~\cite{Hasenfratz:1983ba}. 
It can be implemented by modifying the time derivative \tm{\gaD\nab_D} as  
\nequn{
D^{[\mu_{0}]}
\equiv
  \gaD\nab_D^{[\mu_{0}]}
= \gaD\frac{h_{D}^{+} f(\aD\mu_{0}) - h_{D}^{-} f\inv(\aD\mu_{0})}{2\aD}
,~\label{eq:KW_deriv_chem}
}%nequn
where we treat \tm{\mu_{0}} as a physical parameter of mass dimension one. 
The known solutions~\cite{Gavai:1985ie} 
\nequn{
% \begin{aligned}
\bar{f}(x) = \ee^{x}
,\quad
% \\ 
\tilde{f}(x) = \frac{(1+x)}{\sqrt{1-x^2}}
% \end{aligned}
\label{eq:f_chempot}
}%nequn 
yield the correct small \tm{\aD\mu_{0}} behavior \tm{D^{[\mu_{0}]} = \gaD (\nab_D+\mu_0) +\mO(a^2,\mu_{0}^2)} corresponding to \tm{\tfrac{\ri}{\aD}\gaD \sin(\aD (p_D+\ri \mu_0))} and avoid a quadratic power-law divergence in the \tm{\aD \to 0} limit. 
The chemical potential acts like the temporal component of a constant, imaginary Abelian field. 
Thus, the factors \tm{f^{\pm1}(\aD\mu_{0})} could be absorbed into effective link variables \tm{U_{\pm D}^{[\mu_{0}]}} seen by the fermions. 
These are not elements of a special unitary group anymore, i.e. 
\nequn{
% \begin{aligned}
U_{+D}^{[\mu_{0}]}(x) = U_{+D}(x) f(\aD\mu_{0}) 
,\quad
% ~\\
U_{-D}^{[\mu_{0}]}(x) = U_{-D}(x) f\inv(\aD\mu_{0}) 
~.~
% \end{aligned}
\label{eq:KW_Vlink_mu0}
}%nequn
Equation~\eqref{eq:KW_deriv_chem} is \acl{TS} due to \tm{\gaD h_{D}^{\pm}}, see Table~\ref{tab:KW_meson}, and transforms as \tm{D^{[\mu_{0}]} \to D^{[-\mu_{0}^\ast]}} under  
\tm{\mC}, \tm{\mT} or \tm{\gaf}. 
The complex conjugation of \tm{\mu_{0}} under \tm{\mC} or \tm{\mT} results from transforming the effective link variables \tm{U_{\pm D}^{[\mu_{0}]}} instead of the original gauge links \tm{U_{\pm D}}. 
Thus, \tm{\mu_{0}} is a \acl{T-S} chemical potential. 
As usual, \acl{T-S} real chemical potential yields a complex determinant and a sign problem, while \acl{T-S} imaginary chemical potential is unproblematic in that regard. 
Moreover, since \acl{T-S} real chemical potential can be reduced to a mere temporal boundary condition through a field redefinition, it does not require renormalization.
\vskip1ex

Finding a suitable discretization for isovector chemical potential is less obvious for \ac{KW} fermions. 
While we might naively want to replace \tm{f(\aD\mu_{0})} by \tm{f(\aD\mu_{3}\rho_{3})}, this requires more diligence due to the intertwining of the \acl{ST} structure and the lattice. 
We begin by explaining why an existing proposal~\cite{Misumi:2012uu, Misumi:2012ky} cannot work, and finish with two proposals that could or should work.
% \vskip1ex
%
%
It has been argued for implementing a specific form of a taste-nonsinglet chemical potential~\cite{Misumi:2012uu, Misumi:2012ky} through the extra terms in Eq.~\eqref{eq:D_KW_taste}, i.e. 
\nequn{
O_{m}^{[\mu_{m}]} = \mu_{m} \gaD 
,~\label{eq:KW_mis_isochem}
}%nequn
where we treat \tm{\mu_{m}} as a physical parameter of mass dimension one. 
According to Table~\ref{tab:KW_meson}, this term implements the \acl{ST} structure \tm{\rho_{1} \otimes \Id_{\Ga}} with a phase \tm{\etD} or \tm{\rho_{1} \otimes \Gaf} with a phase \tm{\etDf}, but not the appropriate \acl{ST} structure \tm{\rho_{3} \otimes \GaD}. 
Indeed, \tm{\vr_{1}} leaves Eq.~\eqref{eq:KW_mis_isochem} invariant, yet \tm{\vr_{2}}, \tm{\vr_{3}}, \tm{\mT}, or \tm{\mC} transform it as \tm{O^{[\mu_{m}]} \to O^{[-\mu_{m}]}}, while \tm{\gaf} transforms it as \tm{O^{[\mu_{m}]} \to O^{[-\mu_{m}^\ast]})}. 
Thus, \tm{\mu_{m}} corresponds to chemical potential \tm{\sim \vr_{1}}. 
Equation~\eqref{eq:KW_mis_isochem} is affected by the power-law divergence in the \tm{\aD \to 0} limit~\cite{Hasenfratz:1983ba} and mixes with the first extra term in Eq.~\eqref{eq:D_KW_taste}. 
Thus, \tm{\mu_{m}\bare} would have a nonzero imaginary part and require fine tuning due to additive, complex renormalization as \tm{\aD\mu_{m}\ren(r)=Z_{\mu_{m}}(r)\aD\mu_{m}\bare+\mu_{m}^{\text{crit}}(r)} with both \tm{Z_{\mu_{m}}(r)} and \tm{\mu_{m}^{\text{crit}}(r)} being complex. 
Only for purely imaginary \tm{\mu_{m}\bare} one would have real \tm{Z_{\mu_{m}}(r^2)} and purely imaginary \tm{\mu_{m}^{\text{crit}}(r)} and \tm{\mu_{m}\ren(r)}. 
Although this is somewhat analogous to the additive renormalization of the bare quark mass for Wilson fermions, there is no tuning option similarly convenient as the pseudoscalar masses. 
\vskip1ex

Instead \tm{\rho_{3} \otimes \GaD} could be implemented directly through \tm{\gaD h_{j}} or \tm{\ri\gaDf\etf h_{D}} according to Table~\ref{tab:KW_meson}. 
Taking up the first suggestion, we might explore 
\nequn{
O_{h}^{[\mu_{h}]} = \frac{\mu_{h}}{(D-1)} \gaD \sum\limits_{j=1}^{D-1} h_{j}^{c}
,~\label{eq:KW_h_isochem}
}%nequn
where we treat the coefficient \tm{\mu_{h}} as a physical parameter of mass dimension one. 
The term is invariant under \tm{\vr_{3}}, and transforms under \tm{\vr_{1}}, \tm{\vr_{2}}, \tm{\mC}, \tm{\mT} as \tm{O_{h}^{[\mu_{h}]} \to O_{h}^{[-\mu_{h}]}}. 
Moreover, it transforms as \tm{O_{h}^{[\mu_{h}]} \to O_{h}^{[-\mu_{h}^\ast]}} under \tm{\gaf}. 
Thus, \tm{\mu_{h}} corresponds to chemical potential \tm{\sim \vr_{3}}. 
Equation~\eqref{eq:KW_h_isochem} is affected by the power-law divergence in the \tm{\aD \to 0} limit~\cite{Hasenfratz:1983ba} and mixes with the second extra term in Eq.~\eqref{eq:D_KW_taste}. 
The issues of additive, complex renormalization of \tm{\mu_{h}\bare} are analogous to \tm{\mu_{m}\bare}.
% \vskip1ex
%
%
Both suggestions, i.e. Eqs.~\eqref{eq:KW_mis_isochem} and \eqref{eq:KW_h_isochem}, effectively modify the \tm{r} parameter in front of either extra term in Eq.~\eqref{eq:D_KW_taste}. 
Thus, a misalignment between the two modified \tm{r} parameters in front of the two extra terms with different \acl{ST} structure could realize a \acl{T-D} isovector chemical potential \tm{\mu_{h}\ren+\mu_{m}\ren} at some parameter \tm{r+\aD\mu_{m}\ren}, or a \acl{TN} isovector chemical potential \tm{\mu_{h}\ren+\mu_{m}\ren} at different \tm{r-\aD\mu_{h}\ren}. 
Thus, at fixed lattice spacing theories with \tm{\mu_{h}\ren} or \tm{\mu_{m}\ren} differ only through their respective \tm{r} parameters, but supposedly describe different physics.  
Thus, on physical grounds they cannot have a continuum limit.
\vskip1ex

Neither among Eqs.~\eqref{eq:KW_mis_isochem} nor~\eqref{eq:KW_h_isochem} is a viable approach for phenomenology of hot-dense \ac{QCD}, and isovector chemical potential for \ac{KW} fermions requires fresh ideas that we shall discuss in the following. 
This leaves us with pursuing the second suggestion, i.e. implementing \tm{\ri\gaDf\etf h_{D} = \gaD h_{D}\vr_{3}}.
This might be simply incorporated in the time derivatives through \tm{f(a\mu_{\vr}\vr_{3})}. 
Namely, we could implement a \acl{T-D} isovector chemical potential through a one-shift derivative as 
\nequn{
D^{[\mu_{\vr}\vr_{3}]}
\equiv
  \gaD\nab_D^{[\mu_{\vr}\vr_{3}]}
= \gaD\frac{h_{D}^{+} f(\aD\mu_{\vr}\vr_{3}) - h_{D}^{-} f\inv(\aD\mu_{\vr}\vr_{3})}{2\aD}
,~\label{eq:KW_vrderiv_isochem}
}%nequn
where we treat the coefficient \tm{\mu_{\vr}} as a physical parameter of mass dimension one. 
As required \tm{D^{[\mu_{\vr}\vr_{3}]}} is invariant under \tm{\vr_{3}} and transforms as \tm{D^{[\mu_{\vr}\vr_{3}]} \to D^{[-\mu_{\vr}\vr_{3}]}} for \tm{\vr_{1}}, \tm{\vr_{2}}. 
It also transforms as \tm{D^{[\mu_{\vr}\vr_{3}]} \to D^{[-\mu_{\vr}^\ast\vr_{3}]}} under \tm{\mC} or \tm{\gaf}. 
However, we have \tm{D^{[\mu_{\vr}\vr_{3}]} \to D^{[-\mu_{\vr}\vr_{3}]}} under \tm{\mP} and \tm{D^{[\mu_{\vr}\vr_{3}]} \to D^{[\mu_{\vr}^\ast\vr_{3}]}} under \tm{\mT} due to the nontrivial taste structure of both reflection operators, see Sec.~\ref{sec:KW_generators}. 
While this is clearly not the desired behavior, it is amended if \acl{T-S} parity \tm{\mP_{\pm D}} of Eq.~\eqref{eq:KW_Ppm_act} or \acl{T-S} time reflection \tm{\mT_{\pm j}} of Eq.~\eqref{eq:KW_Tpm_act} are used instead. 
This \acl{T-D} isovector chemical potential \tm{\mu_{\vr}} cannot be absorbed into a temporal boundary condition, and it may be subject to renormalization, since there is no conserved charge density associated with it, see Eq.~\eqref{eq:KW_non_conserved_v3}. 
Due to its different symmetries, Eq.~\eqref{eq:KW_vrderiv_isochem} cannot mix with the extra terms in Eq.~\eqref{eq:D_KW_taste}, and thus \tm{\mu_{\vr}\bare} avoids additive renormalization, while its multiplicative renormalization factor is a real function of \tm{r^2}, i.e. \tm{a\mu_{\vr}\ren=Z_{\mu_{\vr}}(r^2)\aD\mu_{\vr}\bare}. 
Combining Eqs.~\eqref{eq:KW_deriv_chem} and~\eqref{eq:KW_vrderiv_isochem} should be straightforward and should work out of the box as 
\nequn{
D^{[\mu_{0}+\mu_{\vr}\vr_{3}]}
= \gaD \nab_D^{[\mu_{0}+\mu_{\vr}\vr_{3}]}
=
\gaD (\nab_D+\mu_0+\mu_{\vr}\vr_{3}) +\mO(\aD^2, \mu_{0}^2, \mu_{\vr}^2, \mu_{0}\mu_{\vr}\vr_{3})
~.~\label{eq:KW_fullchem_vrderiv}
}%nequn
Yet since Eq.~\eqref{eq:KW_vrderiv_isochem} faces similar difficulties as the zero-shift \acl{T-D} isovector scalar operator suggested in Eq.~\eqref{eq:KW_M3_0s}, it is not obvious whether or not this is a viable option for realizing isovector chemical potential. 
\vskip1ex

We identify yet another discretization of isovector chemical potential by implementing the \acl{T-D} isovector operator through the shift \tm{\hDj}. 
We remember that any meromorphic function \tm{f(xI)} breaks apart as 
\nequn{
f(xI) = f_e(x) +f_o(x)I
}%nequn 
for any \tm{I} that is an involution, where \tm{f_e(x)=f_e(-x)} and \tm{f_o(x)=-f_o(-x)}. 
This applies to \tm{f(a\mu_{3}\rho_{3})}. 
Then the rest of the operator (besides the parts of \tm{f(a\mu_{3}\rho_{3})}) transforms as \tm{\Id_{\rho} \otimes \GaD}. 
The even part could be straightforwardly realized according to Table~\ref{tab:KW_meson} as an average of \tm{\pm\mu_{3}} through 
\nequn{
\gaD\nab_D^{[\mu_{3}]_e}
= \gaD\frac{\nab_D^{[+\mu_{3}]}+\nab_D^{[-\mu_{3}]}}{2}
= \gaD\frac{h_{D}^{+} f^{+1}_e - h_{D}^{-} f\inv_e}{2\aD}
= \gaD\nab_D f_e
,~\label{eq:KW_even_isochem_onelink}
}%nequn
where we have omitted the argument of \tm{f^{\pm1}_e(\aD\mu_{3})}. 
It is crucial to take the inverse first and then the average of \tm{\pm\mu_{3}}. 
For the two suggested solutions in Eq.~\eqref{eq:f_chempot} we have 
\nequn{
% \begin{aligned}
\bar{f}_e(x) = \bar{f}\inv_e(x)=\cosh(x)
,\quad
%~\\
\tilde{f}_e(x) = \tilde{f}\inv_e(x) = \frac{1}{\sqrt{1-x^2}}
~.~
% \end{aligned}
}%nequn 
As required \tm{\gaD\nab_D^{[\mu_{3}]_e}} is invariant under \tm{\{\vr\}} and transforms as \tm{\gaD\nab_D^{[\mu_{3}]_e} \to \gaD\nab_D^{[\mu_{3}^\ast]_e}} under \tm{\mC}, \tm{\mT}, or \tm{\gaf}. 
The odd part is more subtle, as we need, according to Table~\ref{tab:KW_meson}, an odd number of spatial shifts and an even number of temporal shifts to get the \acl{ST} structure right, and a nonzero net number of temporal shifts to permit that the chemical potential acts like the temporal component of a constant, imaginary Abelian field. 
This implies at least two temporal shifts, and thus, either \tm{f_o(2\aD\mu_3)} or \tm{f_o^2(\aD\mu_3)}, which are the same for \tm{\bar{f}_o} and differ at \tm{\mO((\aD\mu_3)^2)} for \tm{\tilde{f}_o}. 
The most simple symmetrized discretization suited to these requirements is 
\nalign{
\gaD\nab_D^{[\mu_{3}]_o}
&= -\gaD\sum\limits_{j=1}^{D-1}
\frac{2\alpha \left[h_{D}^{+}h_{j}^{c}h_{D}^{+}f^{+2}_o - h_{D}^{-}h_{j}^{c}h_{D}^{-}f^{-2}_o\right]+(1-\alpha)\left[[h_{D}^{2+},h_{j}^{c}]_+ f^{+2}_o - [h_{D}^{2-},h_{j}^{c}]_+ f^{-2}_o\right]}{4(D-1)\aD}
\\
&= -\gaD\sum\limits_{j=1}^{D-1}\frac{2\alpha \left[h_{D}^{+}h_{j}^{c}h_{D}^{+} + h_{D}^{-}h_{j}^{c}h_{D}^{-}\right]+(1-\alpha)\left[[h_{D}^{2+},h_{j}^{c}]_+ + [h_{D}^{2-},h_{j}^{c}]_+ \right]}{4(D-1)\aD} f^{+2}_o
,~\label{eq:KW_odd_isochem}
}%nalign
where we have omitted the argument of \tm{f^{\pm 2}_o(\aD\mu_{3})}. 
While \tm{\alpha} is an \emph{a priori} arbitrary real parameter, optimized choices might diminish computational cost, cutoff effects, or both. 
We repeat that it is crucial to take the inverse first and then the average of \tm{\pm\mu_{3}}. 
For the two suggested solutions in Eq.~\eqref{eq:f_chempot} we have \nequn{
\bar{f}^{+2}_o(x) = -\bar{f}^{-2}_o(x)=\sinh(2x)
,\quad%~\\
\tilde{f}^{+2}_o(x) = -\tilde{f}^{-2}_o(x) = \tfrac{2x}{1-x^2}
~.~
}%nequn
As required \tm{\gaD\nab_D^{[\mu_{3}]_o}} is invariant under \tm{\vr_{3}},
and transforms as \tm{\gaD\nab_D^{[\mu_{3}]_o} \to \gaD\nab_D^{[-\mu_{3}]_o} = -\gaD\nab_D^{[\mu_{3}]_o}} for \tm{\vr_{1}}, \tm{\vr_{2}}.
It also transforms as \tm{\gaD\nab_D^{[\mu_{3}]_o} \to \gaD\nab_D^{[\mu_{3}^\ast]_o}} under \tm{\mC}, \tm{\mT}, or \tm{\gaf}.
We see that the sum of Eqs.~\eqref{eq:KW_even_isochem_onelink} and~\eqref{eq:KW_odd_isochem} indeed yields the correct small \tm{\aD\mu_{3}} behavior, namely,
\nequn{
D^{[\mu_{3}\rho_{3}]}
\equiv
\gaD\nab_D^{[\mu_{3}]_e} + \gaD\nab_D^{[\mu_{3}]_o}
= \gaD (\nab_D+\mu_{3}\rho_3) +\mO(a^2, \mu_{3}^2)
,~
\label{eq:KW_eoderiv_isochem}
}%nequn
which is straightforwardly demonstrated in momentum space, e.g. for \tm{f = \bar{f}} we get 
\nalign{
D^{[\mu_{3}\rho_{3}]}(p_D,\mu_{3})
&=
\ri\gaD \left[\frac{\sin(\aD p_D)}{\aD} \cosh(\aD \mu_{3})
+\cos(2\aD p_D)\frac{\ri\sinh(2\aD \mu_{3})}{2\aD}\sum\limits_{j=1}^{D-1}\frac{\cos(a_j p_j)}{D-1}  \right]
\\
&=
\ri\gaD \frac{\sin(\aD p_D) \cos(\ri\aD \mu_{3})
+\cos(\aD p_D)\sin(\ri\aD \mu_{3})\rho_{3}  +\mO(a^3)}{\aD}
\\
&=
\ri\gaD \frac{\sin\left[\aD (p_D+\ri\mu_{3}\rho_{3}) \right]}{\aD}+\mO(\aD^2)
~.~
}%nalign
If we treat the coefficient \tm{\mu_{3}} as a physical parameter of mass dimension one, Eq.~\eqref{eq:KW_eoderiv_isochem} realizes \acl{T-D} isovector chemical potential. 
As required \tm{D^{[\mu_{3}\rho_{3}]}} is invariant under \tm{\vr_{3}} or parity \tm{\mP} and transforms as \tm{D^{[\mu_{3}\rho_{3}]} \to D^{[-\mu_{3}\rho_{3}]}} for \tm{\vr_{1}}, \tm{\vr_{2}}. 
It also transforms as \tm{D^{[\mu_{3}\rho_{3}]} \to D^{[-\mu_{3}^\ast\rho_{3}]}} under \tm{\mC}, \tm{\mT}, or \tm{\gaf}. 
This \acl{T-D} isovector chemical potential \tm{\mu_{3}} cannot be absorbed into a temporal boundary condition, and it may be subject to renormalization, since there is no conserved charge density associated with it, see Eq.~\eqref{eq:KW_non_conserved_v3}. 
Due to its different symmetries, Eq.~\eqref{eq:KW_eoderiv_isochem} cannot mix with the extra terms in Eq.~\eqref{eq:D_KW_taste}, and thus \tm{\mu_{3}\bare} avoids additive renormalization, while its multiplicative renormalization factor is a real function of \tm{r^2}, i.e. \tm{\aD\mu_{3}\ren=Z_{\mu_{3}}(r^2)\aD\mu_{3}\bare}. 
Yet there is a more subtle issue, namely that Eq.~\eqref{eq:KW_odd_isochem} has three-link terms (with one spatial link), while  Eq.~\eqref{eq:KW_even_isochem_onelink} has a one-link term. 
One could reformulate Eq.~\eqref{eq:KW_even_isochem_onelink} and implement it as a three-link term, but would have either zero or two spatial links instead. 
Thus, any implementations of even or odd powers of the \acl{T-D} isovector chemical potential are expected to renormalize differently and require fine-tuning.  
When the \acl{T-S} and \acl{TI} chemical potential terms of Eqs.~\eqref{eq:KW_deriv_chem} and \eqref{eq:KW_eoderiv_isochem} are combined, every temporal shift \tm{h_D^{\pm}} has to be assigned one factor of \tm{f^{\pm1}(\aD\mu_{0})}, respectively. 
This correct combination yields 
\nequn{
D^{[\mu_{0}+\mu_{3}\rho_{3}]}
\equiv
\gaD \left[\nab_D^{[\mu_0,\mu_{3}]_e} + \nab_D^{[\mu_0,\mu_{3}]_o}
\right] 
= \gaD (\nab_D+\mu_0+\mu_{3}\rho_3) +\mO(\aD^2, \mu_{0}^2, \mu_{3}^2, \mu_{0}\mu_{3}\rho_{3})
~.~
\label{eq:KW_fullchem}
}%nequn
Which one of the two implementations of the \acl{TI} chemical potentials, Eqs.~\eqref{eq:KW_vrderiv_isochem} or~\eqref{eq:KW_eoderiv_isochem}, fares better or worse in practice is unclear at present. 
One might guess that the problems of either do not matter too much in practice, since the determinant itself is only sensitive to even powers of the chemical potential, so the problematic odd powers contribute only on the level of valence quarks. 
Reaching a robust conclusion requires a detailed numerical study. 
\vskip1ex

Equations~\eqref{eq:KW_deriv_chem},~\eqref{eq:KW_vrderiv_isochem}, and~\eqref{eq:KW_even_isochem_onelink} immediately work for perpendicular \ac{KW} fermions, too, while Eqs.~\eqref{eq:KW_odd_isochem}, and~\eqref{eq:KW_eoderiv_isochem} have to be adapted in the obvious manner, namely by replacing the average over spatial directions \tm{j} by only the special \tm{j_{0}} direction.
\vskip1ex

\subsection{Karsten-Wilczek spin-taste structure of baryons}\label{sec:KW_baryon}

\begin{table}
\setlength{\extrarowheight}{2pt}
\begin{tabular}{c||l|l|c||l|l|c||l|l|c||l|l|c}
& \multicolumn{3}{c||}{\tm{\Id_{h}}} & \multicolumn{3}{c||}{\tm{\etD h_{D}}} & \multicolumn{3}{c||}{\tm{\ri\etD h_{j}}} & \multicolumn{3}{c}{\tm{\hDj}} 
\\[2pt]\hline\hline
%%%%%%%%%%%%%%%%%%%%%%%%%%%%%%%%%%%%%%%%%%%%%%%%%%%%%%%%%%%
%%%%%%%%%%%%%%%%%%%%%%%%%%%%%%%%%%%%%%%%%%%%%%%%%%%%%%%%%%%
\tm{C\gaf}   
& \textbf{\tm{\bm{ ~~\Id_{\rho}\otimes C\Gaf}}} & \textbf{\tm{\bm{ 0^{+}}}} & \multirow{2}{*}{\textbf{ 0}} 
& \textbf{\tm{\bm{ ~~~\ri\rho_{1}\otimes C\GaDf}}} & \textbf{\tm{\bm{ 0^{+}}}} & \multirow{2}{*}{1(S)} 
& \tm{\,~~~\ri\rho_{2}\otimes C\GaDf} & \tm{0^{+}} & \multirow{2}{*}{1(A)} 
& \textbf{\tm{\bm{ ~~\rho_{3}\otimes C\Gaf}}} & \textbf{\tm{\bm{ 0^{+}}}} & \multirow{2}{*}{\textbf{ 1(0)}} 
\\[2pt]
\tm{C\gaf\etf}   
& \textbf{\tm{\bm{ ~~\Id_{\rho}\otimes C}}} & \textbf{\tm{\bm{ 0^{-}}}} & %\textbf 0
& \tm{ ~~~~~\rho_{1}\otimes C\GaD} & \tm{ 0^{-}} & %\textbf{ 1(S)}
& \textbf{\tm{\bm{ ~~~\rho_{2}\otimes C\GaD}}} & \textbf{\tm{\bm{ 0^{-}}}} & %1(A)
& \textbf{\tm{\bm{ ~~\rho_{3}\otimes C}}} & \textbf{\tm{\bm{ 0^{-}}}} & %\textbf{ 1(0)}
\\[2pt]\hline
%%%%%%%%%%%%%%%%%%%%%%%%%%%%%%%%%%%%%%%%%%%%%%%%%%%%%%%%%%%
\tm{C\gaDf\etD}  
& \textbf{\tm{\bm{-\ri\rho_{1}\otimes C\Gaf}}} & \textbf{\tm{\bm{ 0^{+}}}} & \multirow{2}{*}{\textbf{ 1(S)}} 
& \textbf{\tm{\bm{ ~~~ \Id_{\rho}\otimes C\GaDf}}} & \textbf{\tm{\bm{ 0^{+}}}} & \multirow{2}{*}{\textbf{ 0}} 
& \textbf{\tm{\bm{ ~~~ \rho_{3}\otimes C\GaDf}}} & \textbf{\tm{\bm{ 0^{+}}}} & \multirow{2}{*}{\textbf{ 1(0)}} 
& \tm{~-\rho_{2}\otimes C\Gaf} & \tm{0^{+}} & \multirow{2}{*}{1(A)} 
\\[2pt]
\tm{C\gaDf\etDf}  
& \textbf{\tm{\bm{-\ri\rho_{1}\otimes C}}} & \textbf{\tm{\bm{ 0^{-}}}} & %\textbf{ 1(S)}
& \textbf{\tm{\bm{ -\ri\Id_{\rho}\otimes C\GaD}}} & \textbf{\tm{\bm{ 0^{-}}}} & %\textbf{ 0}
& \textbf{\tm{\bm{ ~~~ \rho_{3}\otimes C\GaD}}} & \textbf{\tm{\bm{ 0^{-}}}} & %\textbf{ 1(0)}
& \tm{~-\rho_{2}\otimes C} & \tm{0^{-}} & %1(A)
\\[2pt]\hline
%%%%%%%%%%%%%%%%%%%%%%%%%%%%%%%%%%%%%%%%%%%%%%%%%%%%%%%%%%%
\tm{C\gaD\etD}   
& \tm{~~~\ri\rho_{2}\otimes C} & \tm{0^{-}} & \multirow{2}{*}{1(A)} 
& \tm{~~~~~\rho_{3}\otimes C\GaD} & \tm{0^{-}} & \multirow{2}{*}{1(0)} 
& \tm{\,~~\ri\Id_{\rho}\otimes C\GaD} & \tm{0^{-}} & \multirow{2}{*}{0} 
& \textbf{\tm{\bm{ -\rho_{1}\otimes C}}} & \textbf{\tm{\bm{ 0^{-}}}} & \multirow{2}{*}{\textbf{ 1(S)}} 
\\[2pt]
\tm{C\gaD\etDf}   
& \tm{~~~\ri\rho_{2}\otimes C\Gaf} & \tm{0^{+}} & %1(A)
& \textbf{\tm{\bm{ ~~~\ri\rho_{3}\otimes C\GaDf}}} & \textbf{\tm{\bm{ 0^{+}}}} & %1(0)
& \textbf{\tm{\bm{ -\Id_{\rho}\otimes C\GaDf}}} & \textbf{\tm{\bm{0 ^{+}}}} & %0
& \textbf{\tm{\bm{ -\rho_{1}\otimes C\Gaf}}} & \textbf{\tm{\bm{ 0^{+}}}} & %\textbf{ 1(S)}
\\[2pt]\hline
%%%%%%%%%%%%%%%%%%%%%%%%%%%%%%%%%%%%%%%%%%%%%%%%%%%%%%%%%%%
\tm{C}   
& \textbf{\tm{\bm{ ~~~\rho_{3}\otimes C}}} & \textbf{\tm{\bm{ 0^{-}}}} & \multirow{2}{*}{\textbf{ 1(0)}} 
& \textbf{\tm{\bm{ ~~~ \ri\rho_{2}\otimes C\GaD}}} & \textbf{\tm{\bm{ 0^{-}}}} & \multirow{2}{*}{1(A)} 
& \tm{\,~-\ri\rho_{1}\otimes C\GaD} & \tm{0^{-}} & \multirow{2}{*}{1(S)} 
& \textbf{\tm{\bm{ ~\Id_{\rho}\otimes C}}} & \textbf{\tm{\bm{ 0^{-}}}} & \multirow{2}{*}{\textbf{ 0}} 
\\[2pt]
\tm{C\etf}   
& \textbf{\tm{\bm{ ~~~\rho_{3}\otimes C\Gaf}}} & \textbf{\tm{\bm{ 0^{+}}}} & %\textbf{ 1(0)}
& \tm{\,~~-\rho_{2}\otimes C\GaDf} & \tm{ 0^{+}} & %\textbf 1(A)
& \textbf{\tm{\bm{ ~~~\rho_{1}\otimes C\GaDf}}} & \textbf{\tm{\bm{ 0^{+}}}} & %1(S)
& \textbf{\tm{\bm{ ~\Id_{\rho}\otimes C\Gaf}}} & \textbf{\tm{\bm{ 0^{+}}}} & %\textbf{ 0}
\\[2pt]\hline\hline
%%%%%%%%%%%%%%%%%%%%%%%%%%%%%%%%%%%%%%%%%%%%%%%%%%%%%%%%%%%
%%%%%%%%%%%%%%%%%%%%%%%%%%%%%%%%%%%%%%%%%%%%%%%%%%%%%%%%%%%
\tm{C\ga_{jD}}   
& \tm{~~\Id_{\rho}\otimes C\Ga_{jD}} & \tm{1^{-}} & \multirow{2}{*}{0} 
& \tm{ ~~~\rho_{1}\otimes C\Ga_{j}} & \tm{ 1^{-}} & \multirow{2}{*}{1(S)} 
& \textbf{\tm{\bm{ ~~\ri\rho_{2}\otimes C\Ga_{j}}}} & \textbf{\tm{\bm{ 1^{-}}}} & \multirow{2}{*}{1(A)} 
& \tm{~~~\rho_{3}\otimes C\Ga_{jD}} & \tm{1^{-}} & \multirow{2}{*}{1(0)} 
\\[2pt]
\tm{C\ga_{jD}\etf}   
& \tm{-\Id_{\rho}\otimes C\Ga_{kl}} & \tm{1^{+}} & %0
& \textbf{\tm{\bm{ ~~\rho_{1}\otimes C\Ga_{j5}}}} & \textbf{\tm{\bm{ 1^{+}}}} & %\textbf{ 1(S)} 
& \tm{\,~-\rho_{2}\otimes C\Ga_{j5}} & \tm{1^{+}} & %\hugrey 1(A) 
& \tm{~-\rho_{3}\otimes C\Ga_{kl}} & \tm{1^{+}} & %1(0) 
\\[2pt]\hline
%%%%%%%%%%%%%%%%%%%%%%%%%%%%%%%%%%%%%%%%%%%%%%%%%%%%%%%%%%%
\tm{C\ga_{j}\etD}   
& \tm{\,-\ri\rho_{1}\otimes C\Ga_{jD}} & \tm{1^{-}} & \multirow{2}{*}{1(S)} 
& \tm{ ~~\Id_{\rho}\otimes C\Ga_{j}} & \tm{1^{-}} & \multirow{2}{*}{0} 
& \tm{ ~~~\ri\rho_{3}\otimes C\Ga_{j}} & \tm{1^{-}} & \multirow{2}{*}{1(0)} 
& \textbf{\tm{\bm{ -\rho_{2}\otimes C\Ga_{jD}}}} & \textbf{\tm{\bm{ 1^{-}}}} & \multirow{2}{*}{\textbf{ 1(A)}} 
\\[2pt]
\tm{C\ga_{j}\etDf}   
& \tm{~~~\ri\rho_{1}\otimes C\Ga_{kl}} & \tm{1^{+}} & %1(S) 
& \textbf{\tm{\bm{ \ri\Id_{\rho}\otimes C\Ga_{j5}}}} & \textbf{\tm{\bm{ 1^{+}}}} & %0
& \textbf{\tm{\bm{ -\ri\rho_{3}\otimes C\Ga_{j5}}}} & \textbf{\tm{\bm{ 1^{+}}}} & %1(0) 
& \textbf{\tm{\bm{ \,~~\rho_{2}\otimes C\Ga_{kl}}}} & \textbf{\tm{\bm{ 1^{+}}}} & %\textbf{ 1(A)}
\\[2pt]\hline
%%%%%%%%%%%%%%%%%%%%%%%%%%%%%%%%%%%%%%%%%%%%%%%%%%%%%%%%%%%
\tm{C \ga_{j5}\etD}   
& \textbf{\tm{\bm{ ~~\ri\rho_{2}\otimes C \Ga_{kl}}}} & \textbf{\tm{\bm{ 1^{+}}}} & \multirow{2}{*}{\textbf{ 1(A)}} 
& \textbf{\tm{\bm{ ~~{\rho}_{3}\otimes C\Ga_{j5}}}} & \textbf{\tm{\bm{ 1^{+}}}} & \multirow{2}{*}{1(0)} 
& \textbf{\tm{\bm{ ~\ri\Id_{\rho}\otimes C\Ga_{j5}}}} & \textbf{\tm{\bm{ 1^{+}}}} & \multirow{2}{*}{0} 
& \tm{~-\rho_{1}\otimes C
\Ga_{kl}} & \tm{1^{+}} & \multirow{2}{*}{1(S)} 
\\[2pt]
\tm{C \ga_{j5}\etDf}   
& \textbf{\tm{\bm{ -\ri\rho_{2}\otimes C \Ga_{jD}}}} & \textbf{\tm{\bm{ 1^{-}}}} & %\textbf{ 1(A)}
& \tm{ -\ri{\rho}_{3}\otimes C\Ga_{j}} & \tm{1^{-}} & %\hugrey 1(0) 
& \tm{~~~\Id_{\rho}\otimes C\Ga_{j}} & \tm{1^{-}} & %\hugrey 1(0)
& \tm{~~~\rho_{1}\otimes C \Ga_{jD}} & \tm{1^{-}} & %1(S) 
\\[2pt]\hline
%%%%%%%%%%%%%%%%%%%%%%%%%%%%%%%%%%%%%%%%%%%%%%%%%%%%%%%%%%%
\tm{C
\ga_{kl}}   
& \tm{~~~~{\rho}_{3}\otimes C \Ga_{kl}} & \tm{1^{+}} & \multirow{2}{*}{1(0)} 
& \tm{\,~~\ri\rho_{2}\otimes C\Ga_{j5}} & \tm{1^{+}} & \multirow{2}{*}{1(A)} 
& \textbf{\tm{\bm{ -\ri\rho_{1}\otimes C\Ga_{j5}}}} & \textbf{\tm{\bm{ 1^{+}}}} & \multirow{2}{*}{1(S)} 
& \tm{~~\Id_{\rho}\otimes C \Ga_{kl}} & \tm{1^{+}} & \multirow{2}{*}{0} 
\\[2pt]
\tm{C \ga_{kl}\etf}   
& \tm{~~-{\rho}_{3}\otimes C \Ga_{jD}} & \tm{1^{-}} & %1(0)
& \textbf{\tm{\bm{ ~~\rho_{2}\otimes C\Ga_{j}}}} & \tm{1^{-}} & %1(A)
& \tm{~~-\rho_{1}\otimes C\Ga_{j}} & \tm{1^{-}} & % 1(S)
& \tm{-\Id_{\rho}\otimes C \Ga_{jD}} & \tm{1^{-}} & %0
\\[2pt]\hline\hline
%%%%%%%%%%%%%%%%%%%%%%%%%%%%%%%%%%%%%%%%%%%%%%%%%%%%%%%%%%%
%%%%%%%%%%%%%%%%%%%%%%%%%%%%%%%%%%%%%%%%%%%%%%%%%%%%%%%%%%%
& \tm{\{\rho\}\otimes\{\Ga\}} & \tm{J^{PC}} & \tm{I(m_I)} 
& \tm{\{\rho\}\otimes\{\Ga\}} & \tm{J^{PC}} & \tm{I(m_I)} 
& \tm{\{\rho\}\otimes\{\Ga\}} & \tm{J^{PC}} & \tm{I(m_I)} 
& \tm{\{\rho\}\otimes\{\Ga\}} & \tm{J^{PC}} & \tm{I(m_I)} 
\\[2pt]
\hline
\end{tabular}
\caption{
\Acl{ST} structure of diquark interpolating operators constructed from \ac{KW} spinor bilinears as \tm{\ps^{bT} C X \ps^c}. 
The rows or columns indicate the gamma matrices (in chiral representation) or shifts used in the interpolating operators, with phase factors \tm{\etD}, \tm{\etf}, or \tm{\etDf}, where they are appropriate. 
For each combination we show the resulting \acl{ST} structure according to Eq.~\eqref{eq:KW_ST_int}, the quantum numbers \tm{J^{PC}} and the isospin quantum numbers \tm{I(m_I)}, where \tm{(S,A)} indicate the symmetric or antisymmetric combination of \tm{m_I=\pm 1}. 
The upper or lower lines in each block contain the states expected for a single-taste theory or due to the parity partners. 
Different spatial indices in a row, i.e. \tm{j,k,l} are always assumed to be antisymmetrized {\markit{and cyclic}}. 
Entries for diquarks that vanish trivially upon color contraction with an antisymmetric tensor \tm{\ep_{abc}} are in bold font. 
\label{tab:KW_diquark}
} 
\end{table}

We can put together baryon interpolating operators, by using Eq.~\eqref{eq:KW_taste_comp} for the lone quark field that carries the spin index and a diquark for the two remaining quarks. 
We collect the \ac{KW} fermion bilinears for diquark interpolating operators in Table~\ref{tab:KW_diquark}.
A parity projection \tm{P^{D}_{\pm}} should act only on the lone quark field. 
However, its chiral components are spread over different boson sites. Projection operators \tm{P^{\mu}_{\pm}} defined as
\nequn{
P^{\mu}_{\pm} 
= \frac{\Id_{\ga}\pm \ga_{\mu}}{2}
,~\mu=1,\ldots,5
,~\label{eq:KW_Pmu}
}%nequn
could be applied as usual. 
They would mix the two tastes if applied to \tm{\ps}, but they work out of the box for the taste components defined in Eq.~\eqref{eq:KW_taste_comp}, where one chiral component has already been parallel transported. 
We see in this the \acl{TS} parity of Eq.~\eqref{eq:KW_Ppm_act} in action, since \tm{P^{D}_{\pm}} simply averages the two chiral components of one taste that have been parallel transported to the same boson site. 
Thus, parity-projected \ac{KW} fermions can be used just like any parity-projected single-taste formulation, with the exception that there is only one Dirac field for each \ac{KW} taste component for each parity instead of the average of two Dirac fields for a single-taste formulation.  
This lone quark is contracted with a diquark. 
Since a single boson site provides only one chiral component of each taste component, we have to include the missing \acl{ST} components\footnote{%
It might still be possible to determine baryon energy levels or matrix elements without including these missing \acl{ST} components.} of the diquark by including yet another parallel transported diquark operator on a neighboring boson site. 
Thus, the gauge covariant interpolating operator for a proton %or neutron 
can be written directly in terms of the \ac{KW} spinors as  
\nequn{
p^{+(2L)}_{\alpha}(n^{H})
=
\ep_{abc}~\Big[\Big( P^{D}_{+}\big(1+h^{I}_{\mu_{H}}\big)P^{\vr}_{+}\ps \Big)_{\alpha}^{a}
\Big( \big(1+h^{I}_{\mu_{H}}\big) \sum\limits_{j=1}^{D-1} \frac{ \ps^{T} (-\ri C\gaDf\etD \hDj^{I}) \ps}{D-1} \Big)^{bc}\Big](n^{H})
,~
% \\
% &\begin{aligned}
% &=\ep_{abc}~\Big(
% \Big[ P^{D}_{+}P^{5}_{+}\ps(n^{H}) \Big]^{a}_{\alpha} + 
% \Big[ P^{D}_{+}P^{5}_{-} \Big( U_{\mu_{H}}(n^{H})(\etD\ps)(n^{H}+e_{\mu_{H}}) \Big) \Big]^{a}_{\alpha} \Big)
% ~\\
% &\times\sum\limits_{j=1}^{D-1}
% \Big( (\etD\ps^{bT})(n^{H}) (+\ga_{25})
% \Big( \frac{ U_{D}(n^{H}) U_{j}(n^{H}+e_{D}) + U_{j}(n^{H}) U_{D}(n^{H}+e_{j}) }{2(D-1)} \ps(n^{H}+e_{D}+e_{j}) \Big)^{c}
% ~\\
% &\phantom{\times\sum\limits_{j=1}^{D-1}}+\frac{
% \Big( U_{D}(n^{H}) (\etD\ps)(n^{H}+e_{D}) \Big)^{bT} (+\ga_{25})
% \Big( U_{j}(n^{H}) \ps(n^{H}+e_{j}) \Big)^{c} + j \leftrightarrow D }{2(D-1)}
% \Big)
% ,~
% \end{aligned}
~\label{eq:KW_proton_2L}
% \\
% %
% n^{+}_{\alpha}(n^{H})
% &=
% \ep_{abc}~\Big[\Big( P^{D}_{+}\big(h^{I}_{\mu_{H}}+1\big)\vr_{1}P^{\vr}_{-}\ps \Big)_{\alpha}^{a}
% \Big( \big(1+h^{I}_{\mu_{H}}\big) \sum\limits_{j=1}^{D-1} \frac{ \ps^{T} (\ri C\gaDf\etD h_{D}h_{j}) \ps}{D-1} \Big) ^{bc}\Big](n^{H})
% ~\\
% &\begin{aligned}
% &=\ep_{abc}~\Big(
% \Big[ P^{D}_{+}P^{5}_{+} \Big( U_{\mu_{H}}(n^{H})\ps(n^{H}+e_{\mu_{H}}) \Big) \Big]^{a}_{\alpha} + 
% \Big[ P^{D}_{+}P^{5}_{-} (\etD\ps)(n^{H}) \Big) \Big]^{a}_{\alpha} \Big)
% ~\\
% &\times\sum\limits_{j=1}^{D-1}
% \Big( (\ps^{bT}\etD)(n^{H}) (-\ga_{25})
% \Big( \frac{ U_{D}(n^{H}) U_{j}(n^{H}+e_{D}) + U_{j}(n^{H}) U_{D}(n^{H}+e_{j}) }{2(D-1)} \ps(n^{H}+e_{D}+e_{j}) \Big)^{c}
% ~\\
% &\phantom{\times\sum\limits_{j=1}^{D-1}}+\frac{
% \Big( U_{D}(n^{H}) (\ps\etD)(n^{H}+e_{D}) \Big)^{bT} (-\ga_{25})
% \Big( U_{j}(n^{H}) \ps(n^{H}+e_{j}) \Big)^{c} + j \leftrightarrow D }{2(D-1)}
% \Big)
% ,~
% \end{aligned}
% ~\label{eq:KW_neutron}
}%nequn
where we restricted to hypercube-internal shifts that realize the \acl{ST} structure. 
Alternatively, we could just rely on the parity partners and construct a proton operator with a zero-link diquark, 
\nequn{
p^{+(0L)}_{\alpha}(n^{H})
=
\ep_{abc}~\Big[\Big( P^{D}_{+}\big(1+h^{I}_{\mu_{H}}\big)P^{\vr}_{+}\ps \Big)_{\alpha}^{a}
\Big( \big(1+h^{I}_{\mu_{H}}\big) { \ps^{T} (C\gaD\etDf) \ps} \Big)^{bc}\Big](n^{H})
,~
% \end{aligned}
~\label{eq:KW_proton_0L}
}%nequn
where the only necessity for a shift is now due to having to include both chiral components for the tastes. 
If we take the lessons learned from the interpolating operators for some pseudo-\ac{NG} bosons into account, cf. Appendix~\ref{app:KW_other_pNGB_details}, which lead to a vanishing continuum limit of some taste correlators, it is likely that  Eq.~\eqref{eq:KW_proton_0L} faces similar issues, but yields the same energy levels as Eq.~\eqref{eq:KW_proton_2L}. 
We also note that the choice of the direction \tm{\mu_{H}}, in which the \ac{KW} fermion hypersite extends, does not matter for the diquark as long as all spatial directions \tm{j} are averaged in the two-link diquark bilinear. 
For one practical example we spell out the details of the pointlike proton annihilation operator \tm{p^{+(2L)}_{\alpha}(n^{H})}, namely 
\nalign{
p^{+(2L)}_{\alpha}(n^{H})
&=\ep_{abc}~\Big(
\Big[ P^{D}_{+}P^{5}_{+}\ps(n^{H}) \Big]^{a}_{\alpha} + 
\Big[ P^{D}_{+}P^{5}_{-} \Big( U_{\mu_{H}}(n^{H})(\etD\ps)(n^{H}+e_{\mu_{H}}) \Big) \Big]^{a}_{\alpha} \Big)
~\nonumber\\
&\times\sum\limits_{j=1}^{D-1}
\Big( (\etD\ps^{bT})(n^{H}) (+\ga_{25})
\Big( \frac{ U_{D}(n^{H}) U_{j}(n^{H}+e_{D}) + U_{j}(n^{H}) U_{D}(n^{H}+e_{j}) }{2(D-1)} \ps(n^{H}+e_{D}+e_{j}) \Big)^{c}
~\nonumber\\
&\phantom{\times\sum\limits_{j=1}^{D-1}}+\frac{
\Big( U_{D}(n^{H}) (\etD\ps)(n^{H}+e_{D}) \Big)^{bT} (+\ga_{25})
\Big( U_{j}(n^{H}) \ps(n^{H}+e_{j}) \Big)^{c} + j \leftrightarrow D }{2(D-1)}
\Big)
,~
}%nalign
but see no benefit in discussing further ones. 
While the \ac{KW} baryon operators are more involved than their counterparts for single-taste fermions, other baryon interpolating operators can be constructed analogously. 
\vskip1ex

We stress that---unless an explicit isovector mass term along the lines of Eqs.~\eqref{eq:KW_M3_2s} or \eqref{eq:KW_M3_Ds} (or \eqref{eq:KW_M3_0s}) or explicit isovector chemical potential along the lines of Eqs.~\eqref{eq:KW_vrderiv_isochem} or~\eqref{eq:KW_eoderiv_isochem} is included---any distinction between proton and neutron plays a role only at finite lattice spacing, since only higher order terms encode a nontrivial \acl{TI} structure that breaks isospin symmetry. 
\vskip1ex

\subsection{Spin-taste structure of mesons for perpendicular Karsten-Wilczek fermions}\label{sec:KW_bilinears_per}

\begin{table}
\setlength{\extrarowheight}{2pt}
\begin{tabular}{l||l|l|c||l|l|c||l|l|c||l|l|c}
& \multicolumn{3}{c||}{\tm{\Id_{h}}} & \multicolumn{3}{c||}{\tm{\etj h_{j_{0}}}} & \multicolumn{3}{c||}{\tm{\ri\etj h_{\nu}}} & \multicolumn{3}{c}{\tm{h_{j_0} h_{\nu}}} 
\\[2pt]\hline\hline
%%%%%%%%%%%%%%%%%%%%%%%%%%%%%%%%%%%%%%%%%%%%%%%%%%%%%%%%%%%
%%%%%%%%%%%%%%%%%%%%%%%%%%%%%%%%%%%%%%%%%%%%%%%%%%%%%%%%%%%
\tm{\Id_{\ga}}   
& \tm{~~\Id_{\rho}\otimes\Id_{\Ga}} & \tm{0^{++}} & \multirow{2}{*}{0}
& \tm{~-\rho_{2}\otimes\Ga_{j_{0}}} & \tm{1^{-}} & \multirow{2}{*}{1(A)}
& \tm{-\ri\rho_{1}\otimes\Ga_{j_{0}}} & \tm{1^{-}} & \multirow{2}{*}{1(S)}
& \tm{~~~\rho_{3}\otimes\Id_{\Ga}} & \tm{0^{++}} & \multirow{2}{*}{1(0)}
\\[2pt]
\tm{\Id_{\ga}\etf}   
& \tm{~~\Id_{\rho}\otimes{\Gaf}} & \tm{0^{-+}} & %0
& \tm{-\ri\rho_{2}\otimes\Ga_{j_{0}5}} & \tm{1^{+}} & %1(A) 
& \tm{~~~\rho_{1}\otimes\Ga_{j_{0}5}} & \tm{1^{+}} & %1(S)
& \tm{~~~\rho_{3}\otimes{\Gaf}} & \tm{0^{-+}} & %1(0) 
\\[2pt]\hline
%%%%%%%%%%%%%%%%%%%%%%%%%%%%%%%%%%%%%%%%%%%%%%%%%%%%%%%%%%%
\tm{\ga_{j_{0}}\etj}   
& \tm{~-\rho_{2}\otimes\Id_{\Ga}} & \tm{0^{+}} & \multirow{2}{*}{1(A)}
& \tm{~~\Id_{\rho}\otimes\Ga_{j_{0}}} & \tm{1^{ -- }} & \multirow{2}{*}{0}
& \tm{~~~\rho_{3}\otimes\Ga_{j_{0}}} & \tm{1^{ -- }} & \multirow{2}{*}{1(0)}
& \tm{-\ri\rho_{1}\otimes\Id_{\Ga}} & \tm{0^{+}} & \multirow{2}{*}{1(S)}
\\[2pt]
\tm{\ga_{j_{0}}\etjf}   
& \tm{~-\rho_{2}\otimes{\Gaf}} & \tm{0^{-}} & %1(A) 
& \tm{~\ri\Id_{\rho}\otimes\Ga_{j_{0} 5}} & \tm{1^{+-}} & %0 
& \tm{~~\ri\rho_{3}\otimes\Ga_{j_{0} 5}} & \tm{1^{+-}} & %1(0)
& \tm{-\ri\rho_{1}\otimes\Gaf} & \tm{0^{-}} & %1(S) 
\\[2pt]\hline
%%%%%%%%%%%%%%%%%%%%%%%%%%%%%%%%%%%%%%%%%%%%%%%%%%%%%%%%%%%
\tm{\ga_{j_{0}5}\etj}   
& \tm{~~~\rho_{1}\otimes
\Gaf} & \tm{0^{-}} & \multirow{2}{*}{1(S)} 
& \tm{~~~\rho_{3}\otimes\Ga_{j5}} & \tm{1^{+-}} & \multirow{2}{*}{1(0)} 
& \tm{~~~\Id_{\rho}\otimes\Ga_{j5}} & \tm{1^{+-}} & \multirow{2}{*}{0}
& \tm{-\ri\rho_{2}\otimes
\Gaf} & \tm{0^{-}} & \multirow{2}{*}{1(A)} 
\\[2pt]
\tm{\ga_{j_{0}5}\etjf}   
& \tm{~-\rho_{1}\otimes\Id_{\Ga}} & \tm{0^{+}} & %1(S) 
& \tm{+\ri\rho_{3}\otimes\Ga_{j}} & \tm{1^{ -- }} & %1(0) 
& \tm{-\ri\Id_{\rho}\otimes\Ga_{j}} & \tm{1^{ -- }} & %0
& \tm{~~\ri\rho_{2}\otimes\Id_{\Ga}} & \tm{0^{+}} & %1(A) 
\\[2pt]\hline
%%%%%%%%%%%%%%%%%%%%%%%%%%%%%%%%%%%%%%%%%%%%%%%%%%%%%%%%%%%
\tm{\gaf}   
& \tm{~~~\rho_{3}\otimes\Gaf} & \tm{0^{-+}} & \multirow{2}{*}{1(0)}
& \tm{~-\rho_{1}\otimes\Ga_{j_{0}5}} & \tm{1^{+}} & \multirow{2}{*}{1(S)}
& \tm{~~\ri\rho_{2}\otimes\Ga_{j_{0}5}} & \tm{1^{+}} & \multirow{2}{*}{1(A)}
& \tm{~~\Id_{\rho}\otimes\Gaf} & \tm{0^{-+}} & \multirow{2}{*}{0}
\\[2pt]
\tm{\gaf\etf}   
& \tm{~~~\rho_{3}\otimes\Id_{\Ga}} & \tm{0^{++}} & %1(0) 
& \tm{~~~\ri\rho_{1}\otimes\Ga_{j_{0}}} & \tm{1^{-}} & %1(S) 
& \tm{~~~\rho_{2}\otimes\Ga_{j_{0}}} & \tm{1^{-}} & %1(A) 
& \tm{~~\Id_{\rho}\otimes\Id_{\Ga}} & \tm{0^{++}} & %0 
\\[2pt]\hline\hline
%%%%%%%%%%%%%%%%%%%%%%%%%%%%%%%%%%%%%%%%%%%%%%%%%%%%%%%%%%%
%%%%%%%%%%%%%%%%%%%%%%%%%%%%%%%%%%%%%%%%%%%%%%%%%%%%%%%%%%%
\tm{\ga_{kl}}   
& \tm{~~\Id_{\rho}\otimes\Ga_{kl}} & \tm{1^{ -- }} & \multirow{2}{*}{0} 
& \tm{~~~~\rho_{2}\otimes\GaDf} & \tm{0^{-}} & \multirow{2}{*}{1(A)} 
& \tm{~~~\ri\rho_{1}\otimes\GaDf} & \tm{0^{-}} & \multirow{2}{*}{1(S)} 
& \tm{~~~\rho_{3}\otimes\Ga_{kl}} & \tm{1^{ -- }} & \multirow{2}{*}{1(0)}
\\[2pt]
\tm{\ga_{kl}\etf}   
& \tm{-\Id_{\rho}\otimes\Ga_{j_{0}D}} & \tm{1^{+-}} & %0 
& \tm{~-\ri\rho_{2}\otimes\GaD} & \tm{0^{+}} & %1(A) 
& \tm{~~~~\rho_{1}\otimes\GaD} & \tm{0^{+}} & %1(S) 
& \tm{~-\rho_{3}\otimes\Ga_{j_{0}D}} & \tm{1^{+-}} & %1(0)
\\[2pt]\hline
%%%%%%%%%%%%%%%%%%%%%%%%%%%%%%%%%%%%%%%%%%%%%%%%%%%%%%%%%%%
\tm{\gaDf\etj}   
& \tm{~~~\rho_{2}\otimes\Ga_{kl}} & \tm{1^{+}} & \multirow{2}{*}{1(A)} 
& \tm{~~~\Id_{\rho}\otimes\GaDf} & \tm{0^{-+}} & \multirow{2}{*}{0} 
& \tm{~~~~\rho_{3}\otimes\GaDf} & \tm{0^{-+}} & \multirow{2}{*}{1(0)}
& \tm{~~~\ri\rho_{1}\otimes\Ga_{kl}} & \tm{1^{+}} & \multirow{2}{*}{1(S)} 
\\[2pt]
\tm{\gaDf\etjf}   
& \tm{~-\rho_{2}\otimes\Ga_{j_{0}D}} & \tm{1^{-}} & %1(A) 
& \tm{-\ri\Id_{\rho}\otimes\GaD} & \tm{0^{+-}} & %0 
& \tm{~-\ri\rho_{3}\otimes\GaD} & \tm{0^{+-}} & %1(0)
& \tm{~-\ri\rho_{1}\otimes \Ga_{j_{0}D}} & \tm{1^{-}} & %1(S) 
\\[2pt]\hline
%%%%%%%%%%%%%%%%%%%%%%%%%%%%%%%%%%%%%%%%%%%%%%%%%%%%%%%%%%%
\tm{\gaD\etj}   
& \tm{~-\rho_{1}\otimes\Ga_{j_{0}D}} & \tm{1^{-}} & \multirow{2}{*}{1(S)}
& \tm{~~~~\rho_{3}\otimes\GaD} & \tm{0^{+-}} & \multirow{2}{*}{1(0)}
& \tm{~~~\Id_{\rho}\otimes\GaD} & \tm{0^{+-}} & \multirow{2}{*}{0}
& \tm{~~~\ri\rho_{2}\otimes\Ga_{j_{0}D}} & \tm{1^{-}} & \multirow{2}{*}{1(A)}
\\[2pt]
\tm{\gaD\etjf}   
& \tm{~+\rho_{1}\otimes\Ga_{kl}} & \tm{1^{+}} & %1(S) 
& \tm{~~~\ri\rho_{3}\otimes\GaDf} & \tm{0^{-+}} & %0 
& \tm{~~\ri\Id_{\rho}\otimes\GaDf} & \tm{0^{-+}} & %1(0)
& \tm{~-\ri\rho_{2}\otimes\Ga_{kl}} & \tm{1^{+}} & %1(A) 
\\[2pt]\hline
%%%%%%%%%%%%%%%%%%%%%%%%%%%%%%%%%%%%%%%%%%%%%%%%%%%%%%%%%%%
\tm{\ga_{j_{0}D}}   
& \tm{~~~\rho_{3}\otimes\Ga_{j_{0}D}} & \tm{1^{ -- }} & \multirow{2}{*}{1(0)} 
& \tm{~~~~\rho_{1}\otimes\GaD} & \tm{0^{-}} & \multirow{2}{*}{1(A)} 
& \tm{~-\ri\rho_{2}\otimes\GaD} & \tm{0^{-}} & \multirow{2}{*}{1(S)} 
& \tm{~~~\Id_{\rho}\otimes\Ga_{j_{0}D}} & \tm{1^{ -- }} & \multirow{2}{*}{0}
\\[2pt]
\tm{\ga_{j_{0}D}\etf}   
& \tm{~-\rho_{3}\otimes\Ga_{kl}} & \tm{1^{+-}} & %1(0) 
& \tm{~~~\ri\rho_{1}\otimes\GaDf} & \tm{0^{+}} & %1(A) 
& \tm{~~~~\rho_{2}\otimes\GaDf} & \tm{0^{+}} & %1(S) 
& \tm{~-\Id_{\rho}\otimes\Ga_{kl}} & \tm{1^{+-}} & %0
\\[2pt]\hline\hline
%%%%%%%%%%%%%%%%%%%%%%%%%%%%%%%%%%%%%%%%%%%%%%%%%%%%%%%%%%%
%%%%%%%%%%%%%%%%%%%%%%%%%%%%%%%%%%%%%%%%%%%%%%%%%%%%%%%%%%%
\tm{\ga_{lD}}   
& \tm{~~\Id_{\rho}\otimes\Ga_{lD}} & \tm{1^{ -- }} & \multirow{2}{*}{0} 
& \tm{~~~~\rho_{2}\otimes\Ga_{k5}} & \tm{1^{+}} & \multirow{2}{*}{1(A)} 
& \tm{~~~\ri\rho_{1}\otimes\Ga_{k5}} & \tm{1^{+}} & \multirow{2}{*}{1(S)} 
& \tm{~~~\rho_{3}\otimes\Ga_{lD}} & \tm{1^{ -- }} & \multirow{2}{*}{1(0)}
\\[2pt]
\tm{\ga_{lD}\etf}   
& \tm{-\Id_{\rho}\otimes\Ga_{j_{0}k}} & \tm{1^{+-}} & %0 
& \tm{~-\ri\rho_{2}\otimes\Ga_{k}} & \tm{1^{-}} & %1(A) 
& \tm{~~~~\rho_{1}\otimes\Ga_{k}} & \tm{1^{-}} & %1(S) 
& \tm{~-\rho_{3}\otimes\Ga_{j_{0}k}} & \tm{1^{+-}} & %1(0)
\\[2pt]\hline
%%%%%%%%%%%%%%%%%%%%%%%%%%%%%%%%%%%%%%%%%%%%%%%%%%%%%%%%%%%
\tm{\ga_{k 5}\etj}   
& \tm{~~~\rho_{2}\otimes\Ga_{lD}} & \tm{1^{-}} & \multirow{2}{*}{1(A)} 
& \tm{~~~\Id_{\rho}\otimes\Ga_{k5}} & \tm{1^{+-}} & \multirow{2}{*}{0} 
& \tm{~~~~\rho_{3}\otimes\Ga_{k5}} & \tm{1^{+-}} & \multirow{2}{*}{1(0)}
& \tm{~~~\ri\rho_{1}\otimes\Ga_{lD}} & \tm{1^{-}} & \multirow{2}{*}{1(S)} 
\\[2pt]
\tm{\ga_{k 5}\etjf}   
& \tm{~-\rho_{2}\otimes\Ga_{j_{0}k}} & \tm{1^{+}} & %1(A) 
& \tm{-\ri\Id_{\rho}\otimes\Ga_{k}} & \tm{1^{ -- }} & %0 
& \tm{~-\ri\rho_{3}\otimes\Ga_{k}} & \tm{1^{ -- }} & %1(0)
& \tm{~-\ri\rho_{1}\otimes \Ga_{j_{0}k}} & \tm{1^{+}} & %1(S) 
\\[2pt]\hline
%%%%%%%%%%%%%%%%%%%%%%%%%%%%%%%%%%%%%%%%%%%%%%%%%%%%%%%%%%%
\tm{\ga_{k}\etj}   
& \tm{~-\rho_{1}\otimes\Ga_{j_{0}k}} & \tm{1^{+}} & \multirow{2}{*}{1(S)}
& \tm{~~~\rho_{3}\otimes\Ga_{k}} & \tm{1^{ -- }} & \multirow{2}{*}{1(0)}
& \tm{~~~\Id_{\rho}\otimes\Ga_{k}} & \tm{1^{ -- }} & \multirow{2}{*}{0}
& \tm{~~~\ri\rho_{2}\otimes\Ga_{j_{0}k}} & \tm{1^{+}} & \multirow{2}{*}{1(A)}
\\[2pt]
\tm{\ga_{k}\etjf}   
& \tm{~+\rho_{1}\otimes\Ga_{lD}} & \tm{1^{-}} & %1(S) 
& \tm{~~\ri\rho_{3}\otimes\Ga_{k 5}} & \tm{1^{++}} & %0 
& \tm{~~\ri\Id_{\rho}\otimes\Ga_{k 5}} & \tm{1^{++}} & %1(0)
& \tm{~-\ri\rho_{2}\otimes\Ga_{lD}} & \tm{1^{-}} & %1(A) 
\\[2pt]\hline
%%%%%%%%%%%%%%%%%%%%%%%%%%%%%%%%%%%%%%%%%%%%%%%%%%%%%%%%%%%
\tm{\ga_{j_{0}k}}   
& \tm{~~~\rho_{3}\otimes\Ga_{j_{0}k}} & \tm{1^{+-}} & \multirow{2}{*}{1(0)} 
& \tm{~~~~\rho_{1}\otimes\Ga_{k}} & \tm{1^{-}} & \multirow{2}{*}{1(S)} 
& \tm{~-\ri\rho_{2}\otimes\Ga_{k}} & \tm{1^{-}} & \multirow{2}{*}{1(A)} 
& \tm{~~~\Id_{\rho}\otimes\Ga_{j_{0}k}} & \tm{1^{+-}} & \multirow{2}{*}{0}
\\[2pt]
\tm{\ga_{j_{0}k}\etf}   
& \tm{~-\rho_{3}\otimes\Ga_{lD}} & \tm{1^{ -- }} & %1(0) 
& \tm{~~~\ri\rho_{1}\otimes\Ga_{k5}} & \tm{1^{+}} & %1(S) 
& \tm{~~~~\rho_{2}\otimes\Ga_{k5}} & \tm{1^{+}} & %1(A) 
& \tm{~-\Id_{\rho}\otimes\Ga_{lD}} & \tm{1^{ -- }} & %0
\\[2pt]\hline\hline
%%%%%%%%%%%%%%%%%%%%%%%%%%%%%%%%%%%%%%%%%%%%%%%%%%%%%%%%%%%
%%%%%%%%%%%%%%%%%%%%%%%%%%%%%%%%%%%%%%%%%%%%%%%%%%%%%%%%%%%
& \tm{~~\{\rho\}\otimes\{\Ga\}} & \tm{J^{PC}} & \tm{I(m_I)} 
& \tm{~~\{\rho\}\otimes\{\Ga\}} & \tm{J^{PC}} & \tm{I(m_I)} 
& \tm{~~~\{\rho\}\otimes\{\Ga\}} & \tm{J^{PC}} & \tm{I(m_I)} 
& \tm{~~~\{\rho\}\otimes\{\Ga\}} & \tm{J^{PC}} & \tm{I(m_I)} 
\\[2pt]
\hline
\end{tabular}
\caption{
\Acl{ST} structure of meson interpolating operators constructed from two perpendicular \ac{KW} spinor bilinears as \tm{\bar\ps X \ps}. 
The rows or columns indicate the gamma matrices (in chiral representation) or shifts used in the interpolating operators, with phase factors \tm{\etj},~\tm{\etf},~or \tm{\etjf}, where they are appropriate. 
For each combination we show the resulting \acl{ST} structure according to Eq.~\eqref{eq:KW_ST_int}, the quantum numbers \tm{J^{PC}} and the isospin quantum numbers \tm{I(m_I)}, where \tm{(S,A)} indicate the symmetric or antisymmetric combination of \tm{m_I=\pm 1}. 
Different spatial indices in a row, i.e. \tm{j_{0},k,l} are always assumed to be antisymmetrized.
\label{tab:KW_meson_per}
} 
\end{table}

We proceed and analyze the \acl{ST} structure of perpendicular \ac{KW} fermion bilinears using Eq.~\eqref{eq:KW_proj_per}. 
In this case it is necessary to distinguish between the directions \tm{j_{0}}, \tm{k}, \tm{l}, or \tm{D}, which are all taken to be mutually orthogonal. 
Each interpolating operator is a product of gamma matrices (in chiral representation), shifts and appropriate phase factors \tm{\etj},~\tm{\etf},~or \tm{\etjf}. 
We note that any leftover, nontrivial phase factors \tm{\etj},~\tm{\etf},~or \tm{\etjf} entail some spatial momentum components at the cutoff; these have to be accounted for in any spatial sum of the interpolating operators in order to excite low-lying states. 
Otherwise, the parity partners decay rapidly and become numerically irrelevant at large Euclidean times. 
This has to be contrasted with parallel \ac{KW} fermions, see Table~\ref{tab:KW_meson}, where not accounting for overall factors \tm{\etDf} simply yields oscillating contributions from low-lying states in the correlation functions. 
We collect the perpendicular \ac{KW} fermion bilinears for meson interpolating operators in Table~\ref{tab:KW_meson_per}, but skip the discussion of diquark interpolating operators or baryons. 
We note that each combination of a gamma matrix and a shift excites low-lying states of both parities, where the interpolating operators exciting the parity partners differ by a factor \tm{\etf} due to Eq.~\eqref{eq:KW_parity_mixing}. 
On the one hand, all interpolating operators with an even total power of \tm{\ga_k} and \tm{\ga_l} factors excite spin \tm{0} states for an even number of shifts or excite spin \tm{1} states for an odd number of shifts. 
On the other hand, all interpolating operators with an odd total power of \tm{\ga_k} and \tm{\ga_l} factors excite only spin \tm{1} states for any number of shifts. 
It is still necessary to include the two boson sites that make up the fermion hypersite to include both chiralities, and we may directly construct \ac{KW} spinor bilinears of any given \acl{ST} structure without relying on the hypersite average of Eq.~\eqref{eq:KW_hypersite_avg} at each end of each propagator. 
\vskip1ex

\begin{figure}%
\includegraphics[width=0.38\txtw]{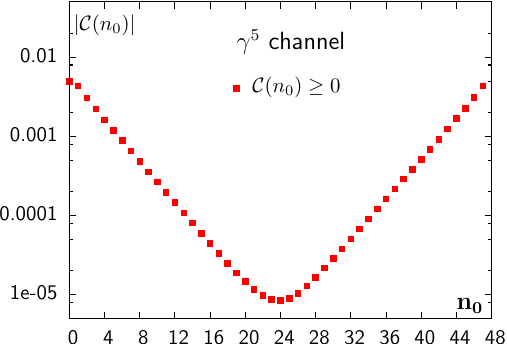}
\hskip1ex
\includegraphics[width=0.38\txtw]{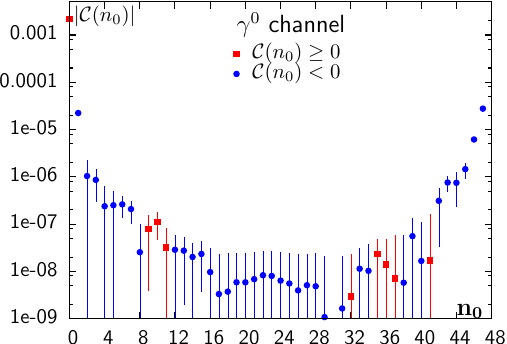}
\vskip1ex
\includegraphics[width=0.38\txtw]{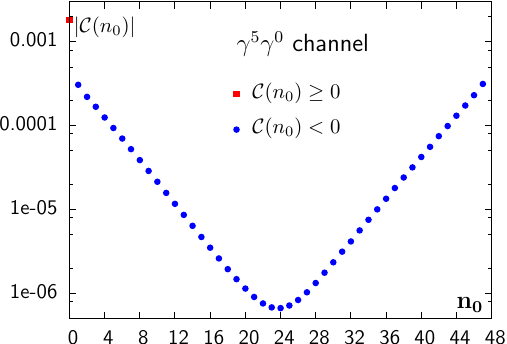}
\hskip1ex
\includegraphics[width=0.38\txtw]{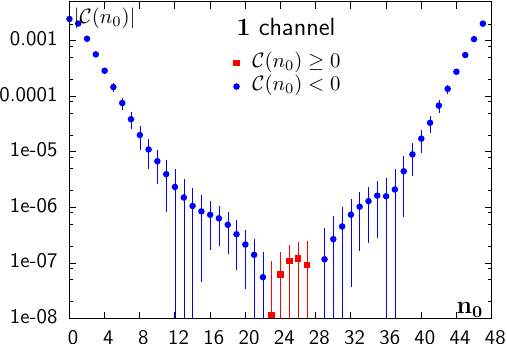}
\caption{%
Spin \tm{0} correlation functions of local bilinear operators with perpendicular \ac{KW} fermions on a pure gauge background with Jacobi smearing at the source. 
Figure from~\cite{Weber:2015oqf}. For details of the ensemble, see~\cite{Weber:2015oqf}, for a discussion see the text.% 
\label{fig:KW_corr_mes_per}%
} 
\end{figure}%

We see in Table~\ref{tab:KW_meson_per} that we can access the same set of hadrons through the parity partners, in perfect analogy to parallel \ac{KW} or \acl{stag} fermions. 
Namely, the same \acl{ST} structure (up to an irrelevant, constant phase) can be  accessed by changing the gamma matrix by \tm{\vr_{3}} and the number of suitable shifts by two, i.e. for \ac{KW} fermions one in the special \tm{j_{0}} direction and one in any other (labeled \tm{\nu}), see Table~\ref{tab:KW_meson_per}.  
This is fully analogous to parallel \ac{KW} fermions, cf. Table~\ref{tab:KW_meson}. 
Such correlation functions were studied a long time ago~\cite{Weber:2015oqf}. 
Since these interpolating operators used gauge-invariant Jacobi smearing~\cite{Gusken:1989ad} at the source and did not account for appropriate phase factors \tm{\etj} or \tm{\etf}, the taste identification may be slightly ambiguous. 
Nevertheless, the point-sink is expected to rule out contributions from odd numbers of shifts in the interpolating operators. 
We reproduce the corresponding plots in Fig.~\ref{fig:KW_corr_mes_per}. 
%%%%%%%%%%%%%%%%%%%%%%%%%%%%%%%%%%%%%%%%%%%%%%%%%%%%%%%%%%%
The plots in the left column correspond to spin \tm{0} via  \tm{\big(\rho_{3}\otimes \Gaf\big)} (upper left) and \tm{\big(\rho_{3}\otimes \GaDf\big)} with an extra minus sign (lower left), whose ground states are degenerate within errors, and degenerate with \tm{\big(\rho_{3}\otimes \Gaf\big)} for parallel \ac{KW} fermions. 
Yet these correlation functions would yield different spin \tm{1} partners for odd number of shifts,
namely 
spin \tm{1} via \tm{\big(\rho_{2}\otimes \Ga_{j_{0}}\big)} with a minus sign (upper panel) or 
spin \tm{1} via \tm{\big(\rho_{2}\otimes \Ga_{j_{0}D}\big)} (lower panel), both missing a phase \tm{\etjf}. 
Thus, these nonoscillating terms would contribute with opposite signs, which qualitatively matches the observed short time behavior in Fig.~\ref{fig:KW_corr_mes_per}. 
%%%%%%%%%%%%%%%%%%%%%%%%%%%%%%%%%%%%%%%%%%%%%%%%%%%%%%%%%%%
The oscillating contributions would be, on the one hand, due to 
spin \tm{0} via \tm{\big(\rho_{3}\otimes \Id_{\Ga}\big)} missing a factor \tm{\etf} or due to
spin \tm{1} via \tm{\big(\rho_{2}\otimes \Ga_{j_{0}5}\big)} missing a factor \tm{\etj} (upper panel), or, on the other hand, due to 
spin \tm{0} via \tm{\big(\rho_{3}\otimes \GaD\big)} missing a phase \tm{\etf} or due to 
spin \tm{1} via \tm{\big(\rho_{2}\otimes \Ga_{kl}\big)} missing a phase \tm{\etj} (lower panel). 
There are no indications that any of these are numerically relevant beyond the observed short time behavior in Fig.~\ref{fig:KW_corr_mes_per}. 
%%%%%%%%%%%%%%%%%%%%%%%%%%%%%%%%%%%%%%%%%%%%%%%%%%%%%%%%%%%
The plots in the right column are less clear. 
The one on the upper right may correspond to 
spin \tm{0} via \tm{\big(\Id_{\rho}\otimes \GaD\big)} with a minus sign (or would correspond to 
spin \tm{1} via  \tm{\big(\rho_{1}\otimes \Ga_{kl}\big)} with a missing phase factor \tm{\etjf} for odd number of shifts). 
The one on the lower right may correspond to 
spin \tm{0} through \tm{\big(\Id_{\rho}\otimes \Id_{\Ga}\big)} (or would correspond to 
spin \tm{1} via \tm{\big(\rho_{1}\otimes \Ga_{j_{0}5}\big)} for odd number of shifts). 
It seems indeed that the upper or lower panels are dominated by rather different states. 
On the one hand, the oscillating contributions correspond to 
spin \tm{0} via \tm{\big(\Id_{\rho}\otimes \GaDf\big)} with a missing phase factor \tm{\etf} (or would correspond to 
spin \tm{1} via \tm{\big(\rho_{1}\otimes \Ga_{j_{0}D}\big)} with a missing phase factor \tm{\etj} for odd number of shifts), 
or, on the other hand, correspond to 
spin \tm{0} via \tm{\big(\Id_{\rho}\otimes \Gaf\big)} with a missing factor \tm{\etf}, (or would correspond to 
spin \tm{1} via \tm{\big(\rho_{1}\otimes \Ga_{j_{0}}\big)} with a missing factor \tm{\etj} for odd number of shifts). 
%%%%%%%%%%%%%%%%%%%%%%%%%%%%%%%%%%%%%%%%%%%%%%%%%%%%%%%%%%%
Thus, Eq.~\eqref{eq:KW_proj} delivers the first-principles explanation of the \acl{ST} structure and supersedes the previous \emph{ad hoc} picture~\cite{Weber:2015oqf}. 
\vskip1ex
% \clearpage
%%%%%%%%%%%%%%%%%%%%%%%%%%%%%%%%%%%%%%%%%%%%%%%%%%%%%%%%%%%%%%%%%%%%%%%%%%%%%%%%
% \input{\INCDIR/kawi_gauge}
%% kawi_gauge.tex
% \acresetall

\section{Gauge theories with Karsten-Wilczek fermions}\label{sec:KW_gauge}

In this section we address questions concerning the \acl{ST} structure of \ac{KW} fermions in an interacting setup. 
It has been known for a long time that \ac{KW} fermions require anisotropic counterterms, and how they feed their anisotropy back to the gauge fields. 
It has been observed that some correlation functions with \ac{KW} fermions show visible oscillations and yield different rest energies depending on the direction. 
Both effects have been used to tune the anisotropic counterterms. 
Lastly, it has been seen that the pseudoscalar rest energies show a distinctive lattice spacing dependence, while they do not seem to scale as naively expected; see Sec.~\ref{sec:KW_meson}. 
However, it was not clear at all how these different phenomena could be understood as effects of the underlying \acl{ST} structure. 
We address these aspects in the following. 
In Sec.~\ref{sec:KW_renormalization} we discuss the \acl{ST} structure of the counterterms. 
In Sec.~\ref{sec:KW_determinant} we discuss how the \acl{ST} structure controls the parametric form of the \ac{KW} determinant. 
In Sec.~\ref{sec:KW_mistuning} we derive the correct interpretation of mistuned \ac{KW} fermions from the \acl{ST} structure. 
In Sec.~\ref{sec:KW_oscillations} we derive the origin of nonstandard oscillations in hadronic correlation functions with mistuned \ac{KW} fermions from the \acl{ST} structure, and show how this can be used for tuning the counterterms. 
In Sec.~\ref{sec:KW_dispersion} we derive the origin of nonstandard quark dispersion relations (for mistuned) \ac{KW} fermions and their parametric form, which we juxtapose to the parameter dependence of the pseudo-\ac{NG} boson's rest energy.
In Sec.~\ref{sec:KW_splittings} we study taste splittings and \ac{LO} cutoff effects of pseudoscalar correlators, while leaving some technical calculations to Appendix~\ref{app:KW_other_pNGB_details}. 
\vskip1ex

\subsection{Renormalization of Karsten-Wilczek fermions}\label{sec:KW_renormalization}

Interacting \ac{KW} fermions require anisotropic renormalization. 
Renormalization factors and counterterms for standard \ac{KW} fermions with \tm{r=+1} coupled to \tm{\text{SU}(3)} gauge fields with Wilson plaquette action are known at \ac{NLO}~\cite{Capitani:2010nn}. 
Furthermore, renormalization factors and mixing of the local currents (S,~P,~V,~A,~T) are known at \ac{NLO} with exact chiral symmetry being confirmed, and the conservation of the vector and axial Noether currents has been verified~\cite{Capitani:2010nn}.
The three anisotropic counterterms read in Symanzik \ac{EFT} 
\nalign{
S_{\text{KW}}^{(3)}
&=c_{3}(r^2, g^2)~\big(\prod_{\mu} a_\mu \big)\sum\limits_{x,y} \bar\ps(x)~\Big(\ri \frac{r}{\aD} \gaD \de(x,y)\Big)~\ps(y)
,~\label{eq:S_KW_3}\\
S_{\text{KW}}^{(4)}
&=c_{4}(r^2, g^2)~\big(\prod_{\mu} a_\mu \big)\sum\limits_{x,y} \bar\ps(x)~\Big(\gaD \xi_{f0,D} D_{D}(x,y)\Big)~\ps(y)
,~\label{eq:S_KW_4}\\
S_{\text{KW}}^{(G)}
&=c_{4g}(r^2, g^2)~\sum\limits_{x} \sum_{j=1}^{D-1}a_j^2\aD^2 \Tr{ \Big(F_{jD}^2}(x) \Big)
~.~\label{eq:S_KW_4g}
}%nalign 
We break with previous convention and include an extra factor of the \tm{r} parameter in Eq.~\eqref{eq:S_KW_3}.\footnote{ 
The overall \ac{NLO} coefficients of the operators are odd or even functions in \tm{r} for the relevant or marginal counterterms, respectively~\cite{Weber:2015oqf}.} 
Thus, the three coefficients \tm{c_{3}(r^2, g^2)},~\tm{c_{4}(r^2, g^2)}, and \tm{c_{4g}(r^2, g^2)} are even functions of \tm{r} in our conventions. 
This is true nonperturbatively as a consequence of the \acl{ST} structure of the operators, see Table~\ref{tab:KW_vr_sim}. 
In the following, we reorganize the coefficients as 
\nequn{
r_0 \equiv r(D-1+{c_{3}})
,\quad 
r_1 \equiv r
,~\label{eq:KW_r01}
}%nequn
since this is convenient for exposing the symmetries of hadronic observables through analytical considerations. 
The two marginally relevant (dimension-four) counterterms in Eqs.~\eqref{eq:S_KW_4} or \eqref{eq:S_KW_4g} simply account for renormalization of the speed of light for fermions or gauge fields, respectively, where the former is transmitted to the latter through fermion loops, i.e. the fermion determinant. 
Both are \acl{TS} operators that survive in the continuum limit. 
For Eq.~\eqref{eq:S_KW_4} this implies that the derivative has to be discretized with an odd number of steps\footnote{
This discretization has been assumed~\cite{Capitani:2010nn} and used in all numerical studies without ever discussing the underlying first-principle reason.} 
according to Eq.~\eqref{eq:KW_singlet_constraint}. 
Their nonperturbative tuning has been established in the first dynamical simulations with \ac{KW} fermions~\cite{Vig:2024umj}. 
The marginal coefficient \tm{c_{4}(r^2, g^2)} can be absorbed into the renormalization factor of the fermion anisotropy, i.e. 
\nequn{
\xi_{f1,D}(r^2, g^2) = Z_{\xi_{f,D}}(r^2,g^2)\xi_{f0,D}(r^2, g^2) = \big(1+c_{4}(r^2, g^2)\big)Z_{\xi_{f0,D}}(r^2,g^2)\xi_{f0,D}(r^2, g^2)
~.~\label{eq:KW_xif1D}
}%nequn 
These arguments trivially carry over to perpendicular \ac{KW} fermions after the obvious exchanges of directions. 
\vskip1ex

The relevant (dimension-three) counterterm is at the same time more important and less trivial. 
It is evident without calculation that the two extra terms cannot have a joint renormalization factor. 
According to Eq.~\eqref{eq:KW_nonsinglet_operators} they transform differently under the taste generators \tm{\{\vr\}}, so they generally incur different higher order corrections. 
One may reach a similar conclusion by contemplating the number of links in these, namely, zero or one. 
These are two separate physical arguments, i.e. the former argument is independent of the strength of gauge interactions. 
This argument would apply in any statistical lattice model %without gauge fields, but another form of local interactions that give rise to 
that requires renormalization. 
The latter argument suggests that their nonperturbative renormalization factors must approach \tm{1} according to the weak-coupling expansion of a gauge theory, if the links are smoothed or tadpole improved. 
This has been verified in terms of the \ac{NLO} coefficients with tadpole-improved boosted coupling~\cite{Weber:2015oqf}, and it has been shown that stout smearing~\cite{Morningstar:2003gk} largely ameliorates the need for renormalization~\cite{Godzieba:2024uki}. 
One has to readjust the coefficients for one of the two extra terms\footnote{%
Adjusting the coefficient in front of the local (zero-link) operator, namely through Eq.~\eqref{eq:S_KW_3}, is better for algorithmic reasons. 
} to retain the cancellation between the two extra terms at the spatial momentum \tm{ap_j=0} corresponding to the two tree-level poles with \tm{\bar{p}_{D}=0}. 
We discuss this in detail in Sec.~\ref{sec:KW_mistuning}, and then derive from first principles why and how the various tuning schemes function in Secs.~\ref{sec:KW_oscillations} and \ref{sec:KW_dispersion}.  
\vskip1ex

\subsection{Karsten-Wilczek fermion determinant}\label{sec:KW_determinant}

We employ the line of reasoning established in~\cite{Weber:2016dgo} and begin with the structure of the full \ac{KW} operator 
\nalign{
\mDKW 
&= \Big( \DKW + \DKW^{(3)} + \DKW^{(4)} + M_{0} + M_{3}^{D\text{L}} \Big)
~\label{eq:KW_full}\\
&=\Big( 
\sum_{j=1}^{D-1} \xi_{f0,j} \ga_j \nab_j(x,y)
+ \xi_{f1,D} \gaD \nab_D^{[\mu_{0}+\mu_{3}\rho_{3}]}(x,y) 
~\\
&\phantom{=\det \Big(}
+\frac{\ri}{\aD}\gaD\sum_{j=1}^{D-1} \big[r_0\de(x,y)-r_1 h_{j}^{c}(x,y) \big]
+m_0 \de(x,y) 
+m_3 \frac{1}{D!} \prod\limits_{\mu=1}^{D}h_{\mu} 
\Big)
,~
}%nalign
where we have included the mass terms of Eqs.~\eqref{eq:KW_M0} and~\eqref{eq:KW_M3_Ds}, the chemical potential terms of Eqs.~\eqref{eq:KW_deriv_chem} and~\eqref{eq:KW_eoderiv_isochem}, and the counterterms of Eqs.~\eqref{eq:S_KW_3} and~\eqref{eq:S_KW_4}. 
The following discussion is practically unchanged if we used instead of \tm{M_{3}^{D\text{s}}} another \acl{TD} isovector mass term, i.e. \tm{M_{3}^{2\text{s}}} of Eq.~\eqref{eq:KW_M3_2s} or \tm{M_{3}^{0\text{s}}} of Eq.~\eqref{eq:KW_M3_0s}, or instead of \tm{\nab_D^{[\mu_{0}+\mu_{3}\rho_{3}]}} the other \acl{TD} isovector chemical potential term, i.e. \tm{\gaD \nab_D^{[\mu_{0}+\mu_{\vr}\vr_{3}]}} of Eq.~\eqref{eq:KW_fullchem_vrderiv}. 
We highlight the different structures via Eqs.~\eqref{eq:KW_r01} and \eqref{eq:KW_xif1D}. 
Chemical potentials are split as \tm{\mu_{t} = \mu_{tR}+\ri \mu_{tI}} for \tm{t\in\{0,3\}}.
The determinant \tm{|\mDKW|} is a function of the parameters \tm{g^2,~m_0,~m_3,~\mu_{0},~\mu_{3},~r_0,~r_1,~\xi_{f0,j}}, and \tm{\xi_{f1,D}}. 
\tm{|\mDKW|} is a \acl{TS} observable with vacuum quantum numbers, \tm{0^{++}}, despite the \ac{KW} operator's explicit \acl{TI} contributions.
Since we know that odd powers in \tm{\aD} are exclusively accompanied by odd powers in \tm{r} and are always associated with \acl{TI} contributions that break charge conjugation and time reflection symmetries, these have to conspire to contribute only in even powers. 
\vskip1ex

We shall prove that indeed \tm{|\mDKW| = |\mDKW|(r^2)}. 
We use involutions that each change the sign of some parameters multiplying subsets of operators contributing in Eq.~\eqref{eq:KW_full} to expose symmetries of \tm{|\mDKW|}. 
\tm{\vr_{3}} delivers  
\nequn{
\vr_{3} ~\Rightarrow~ |\mDKW| 
~=~ |\mDKW|\big(~g^2,
+m_{0},+m_{3},+\mu_{0R},+\mu_{3R},+\mu_{0I},+\mu_{3I},
-r_0,+r_1,
+\xi_{f0,j},+\xi_{f1,D}
\big)
~.~\label{eq:KW_det_vr3}
}%nequn
Due to Eq.~\eqref{eq:KW_det_vr3} \tm{|\mDKW|} is independent of the sign of \tm{r_0}.
This permits us to conclude a nontrivial property from the existence of the renormalized \ac{KW} operator's finite, nonzero continuum limit. 
Existence of this continuum limit implies cancellations between all divergent contributions, namely between the power-law divergent terms multiplying the \tm{r} parameter. 
As one of these relevant coefficients is restricted to even powers, i.e. \tm{r_0}, the other relevant coefficient, i.e. \tm{r_1}, has to be restricted to even powers, too. 
Thus, we may strengthen the constraint of Eq.~\eqref{eq:KW_det_vr3} to 
\nequn{
\vr_{3} ~\Rightarrow~ |\mDKW|\
~=~ |\mDKW|\big(~g^2,
+m_{0},+m_{3},+\mu_{0},+\mu_{3},
~r_0^2,~r_1^2,
+\xi_{f0,j},+\xi_{f1,D}
\big)
~.~\label{eq:KW_det_vr3_upgrade}
}%nequn
Thus, the determinant \tm{|\mDKW|} is an even function of \tm{r_0} and \tm{r_1}, and thereby of \tm{r} and \tm{\aD}. 
Thus, \tm{|\mDKW|} is automatically \tm{\mO(\aD)} improved, and in fact any odd powers in \tm{\aD} or \tm{r} are generally prohibited in the gauge sector. 
The gauge fields suffer from no further symmetry breaking other than the anisotropy of the \ac{KW} fermion determinant that has to be canceled by the gauge counterterm given in Eq.~\eqref{eq:S_KW_4g}, and discretization effects scale parametrically as \tm{\mO(\aD^2)}. 
This is analogous to the case of twisted-mass Wilson fermions at maximal twist~\cite{Aoki:2006nv}, where the automatic \tm{\mO(a)} improvement occurs on the level of \tm{\mO(a)} fluctuations compensating one another in the ensemble averages. 
Nonetheless, the statistical errors due to such \tm{\mO(a)} fluctuations can be ameliorated by explicitly including the corresponding irrelevant operators from the Symanzik improvement program~\cite{Becirevic:2006ii}. 
For discussions about improving the \ac{KW} operator, see~\cite{Weber:2015hib, Vig:2024umj, Bakenecker:2024a}.
Averaging calculations using opposite signs of \tm{r} in the valence sector removes odd powers in \tm{\aD} or \tm{r} for observables even with dynamical \ac{KW} fermions in the sea. 
These arguments trivially carry over to perpendicular \ac{KW} fermions. 
\vskip1ex

Next, we employ either of the other two taste generators \tm{\vr_{1,2}}. 
We obtain two sets of restrictions,
\nequn{
\begin{aligned}
\vr_{2} ~\Rightarrow~|\mDKW| 
&~=~ |\mDKW|\big(~g^2,
+m_{0},-m_{3},+\mu_{0},-\mu_{3},
-r_0,-r_1,
+\xi_{f0,j},+\xi_{f1,D}
\big)
,~\\
\vr_{1} ~\Rightarrow~|\mDKW| 
&~=~ |\mDKW|\big(~g^2,
+m_{0},-m_{3},+\mu_{0},-\mu_{3},
+r_0,-r_1,
+\xi_{f0,j},+\xi_{f1,D}
\big)
,~
\end{aligned}
\label{eq:KW_det_vr1}
}%nequn
which are equivalent given Eq.~\eqref{eq:KW_det_vr3_upgrade}. 
Taking all constraints from the \tm{\{\vr\}} together, i.e. Eqs.~\eqref{eq:KW_det_vr3_upgrade}-~\eqref{eq:KW_det_vr1}, 
we have 
\nequn{
|\mDKW| 
= |\mDKW|\big(g^2,
~m_0,~\mu_{0},
~m_3^2,~\mu_{3}^2,~m_3\mu_{3},
~r_0^2,~r_1^2,~\xi_{f0,j},~\xi_{f1,D}
\big)
~.~
\label{eq:KW_det_sng0}
}%nequn 
If there is no \acl{TI} chemical potential \tm{\mu_{3}}, then the determinant \tm{|\mDKW|} is an even function of \acl{TI} mass \tm{m_3}, 
or if there is no \acl{TI} mass \tm{m_3}, then the determinant is an even function of the \acl{TI} chemical potential \tm{\mu_{3}}.
Then the two possible signs of \tm{r} and \tm{m_3} or \tm{\mu_{3}} can be chosen independently in the sea and in the valence sectors, and odd powers may be cancelled. 
Again, these arguments trivially carry over to perpendicular \ac{KW} fermions. 
\vskip1ex

The \ac{KW} operator is not invariant under time reflection. 
We may change the signs of some coefficients using \tm{\gaDf},
\nequn{
\gaDf ~\Rightarrow~|\mDKW| 
~=~ |\mDKW|\big(~g^2,
+m_{0},+m_{3},+\mu_{0},+\mu_{3},
-r_0,-r_1,
\xi_{f0,j},-\xi_{f1,D}
\big)
~.~\label{eq:KW_det_gD5}
}
Due to Eq.~\eqref{eq:KW_det_vr3_upgrade}, we already know that \tm{|\mDKW|} is independent of the sign of \tm{r}, 
and thus \tm{|\mDKW|} is an even function of \tm{\xi_{f1,D}}. 
The dual representation of Eq.~\eqref{eq:KW_dual} has the same eigenvalues and determinant, 
\nequn{
\gaD ~\Rightarrow~|\mDKW| 
~=~ |\mDKW|\big(~g^2,
+m_{0},+m_{3},+\mu_{0},+\mu_{3},
r_0,r_1,
-\xi_{f0,j},\xi_{f1,D}
\big)
~.~\label{eq:KW_det_gD}
}
Thus, \tm{|\mDKW|} is an even function of \tm{\xi_{f0,j}}, too.
As analogous arguments exist for perpendicular \ac{KW} fermions, we combine Eqs.~\eqref{eq:KW_det_gD5} and~\eqref{eq:KW_det_gD} with Eq.~\eqref{eq:KW_det_sng0} to obtain 
\nequn{
|\mDKW| 
= |\mDKW|\big(g^2,
~m_0,~\mu_{0},
~m_3^2,~\mu_{3}^2,~m_3\mu_{3},
~r_0^2,~r_1^2,~\xi_{f0,j}^2,~\xi_{f1,D}^2
\big)
~.~
\label{eq:KW_det_sng1}
}%nequn 
\vskip1ex

In the next step, we employ \tm{\gaf} as the involution. 
\tm{\gaf} Hermiticity is broken by real chemical potential, 
\nequn{
\gaf ~\Rightarrow~|\mDKW| 
~=~ |\mDKW|^\ast\big(~g^2,
+m_{0},+m_{3},-\mu_{0R},-\mu_{3R},+\mu_{0I},+\mu_{3I},
~r_0,~r_1,
~\xi_{f0,j},~\xi_{f1,D}
\big)
,~\label{eq:KW_det_g5}
}
or if we include the restrictions of Eqs.~\eqref{eq:KW_det_sng1}
\nequn{
|\mDKW| 
= |\mDKW|^\ast\big(g^2,
+m_{0},+m_{3}^2,-m_{3}\mu_{3R},+m_{3}\mu_{3I},
-\mu_{0R},+\mu_{0I},
+\mu_{3R}^2,+\mu_{3I}^2,-\mu_{3R}\mu_{3I},
r_0^2,r_1^2,
\xi_{f0,j}^2,\xi_{f1,D}^2
\big)
~.~
\label{eq:KW_det_sng2}
}%nequn
More simplification cannot be obtained from symmetry arguments alone. 
But we note that---as usual---some restrictions on the chemical potentials and masses may imply that \tm{|\mDKW|} is real. 
Namely, the determinant \tm{|\mDKW|} is real for purely imaginary chemical potential via \tm{\mu_{0R}=\mu_{3R}=0}, or for purely real isovector chemical potential \tm{\mu_{0}=\mu_{3I}=0} and vanishing isovector mass \tm{m_{3}=0}. 
The exact same arguments apply for perpendicular \ac{KW} fermions. 
\vskip1ex

All of this means that \tm{|\mDKW|} never generates contributions breaking either charge conjugation \tm{\mC}- \emph{or} time reflection \tm{\mT}-invariances, unless \emph{both} were already broken for the underlying gauge fields. 
This breaking of both symmetries cannot occur if the gauge fields were evolved by molecular dynamics using a \ac{KW} fermion force at vanishing or imaginary chemical potential, since the symmetry breaking is only induced by \acl{TI} terms. 
Yet a real part of \acl{TS} chemical potential, complex \acl{TI} chemical potential, or real \acl{TI} chemical potential with a \acl{TI} mass permit \acl{TS} forces breaking \emph{both} symmetries under charge conjugation \tm{\mC} \emph{and} time reflection \tm{\mT}. 
These are the origin of the sign problem, which is unrelated to the symmetry breaking in the original \ac{KW} operator in Eq.~\eqref{eq:D_KW}. 
One preserved symmetry for the gauge fields is enough to protect the determinant from terms breaking the other. 
In particular, the presence of a nonzero \tm{\Theta} term does not prevent applying these arguments, as it is \tm{\mC}-invariant. 
Despite any symmetry breaking and complexification induced by chemical potentials, odd powers in \tm{\aD} or \tm{r} remain prohibited on general grounds. 
\vskip1ex

\subsection{Pole structure due to mistuning the Karsten-Wilczek operator}\label{sec:KW_mistuning}

For a mistuned relevant coefficient \tm{c_{3}} the cancellation between the two extra terms at vanishing spatial momenta \tm{ap_j=0} fails. 
Thus, the temporal momenta of the propagator poles cannot match their tree-level values \tm{ap_D=0,\pi}. 
If that happened in a manner independent of the gauge coupling, reminiscent of the situation with \aclp{BCF} already at tree-level~\cite{Durr:2020yqa}, we could simply identify the physical tastes by expanding about the two actual poles---wherever they might be---but would not have a significant problem (besides the nuisances of giving up the equivalence of forward and backward \tm{D} direction and of algorithmic issues). 
Yet if the positions of the poles depend on the gauge coupling \tm{g^2} or the lattice spacing \tm{\aD}, then the nonrenormalized theory lacks a continuum limit. 
\vskip1ex

Let us consider in the following a fixed gauge coupling. 
Namely, we work at fixed yet sufficiently fine lattice spacing and assume that the relevant coefficient is mistuned by \tm{|\de c_{3}| \ll 1}. 
If the marginal coefficient in Eq.~\eqref{eq:S_KW_4} were mistuned by \tm{|\de c_{4}| \ll 1}, it would only mistune the fermion anisotropy \tm{\xi_{f1,D}} defined in Eq.~\eqref{eq:KW_xif1D}, which is irrelevant for the following discussion.  
For simplicity, we consider the (almost) dressed \ac{KW} propagator in the chiral limit \tm{m_0=m_3=0} and at vanishing chemical potential \tm{\mu_{0}=\mu_{3}=0}.
Its poles are determined by \tm{ap_j=0} and 
\nequn{
0 = \xi_{f1,D} \sin(\aD p_D) + r \de c_{3}
\quad\Leftrightarrow\quad \aD p_D = -\arcsin\left(\frac{r \de c_{3}}{\xi_{f1,D}}\right) 
~.
\label{eq:KW_mistuned}
}%nequn
These poles assume their tree-level positions for \tm{\de c_{3}=0} independent of \tm{\xi_{f1,D}}, which suggests that \tm{c_{3}} can and should be tuned before the fermion or boson speeds of light (\tm{\de c_{4}},~\tm{\de c_{{4g}}}). 
This sequence was used in all studies of \ac{KW} fermions so far. 
We define the two physical tastes at the two poles of the mistuned \ac{KW} propagator given in Eq.~\eqref{eq:KW_mistuned}, i.e. at 
\nequn{
\ka(r\de c_{3},\xi_{f1,D}) \equiv 
\arcsin\left(\frac{r \de c_{3}}{\xi_{f1,D}}\right)
,\quad
ap_D = 
\left\{\begin{aligned}
-\ka(r\de c_{3},\xi_{f1,D}) \\
\pi+\ka(r\de c_{3},\xi_{f1,D})
\end{aligned}\right.
,~\label{eq:KW_ka}
}%nequn
where \tm{\ka(r\de c_{3},\xi_{f1,D})} is the principal value of the arcsine function. 
\tm{\ka(r\de c_{3},\xi_{f1,D}) \ll 1}, since we have assumed \tm{\de c_{3} \ll 1}.
In the following we generally just write \tm{\ka} and suppress its arguments. 
If we had not restricted the consideration to the chiral limit, \tm{\ka} would be modified by quadratic corrections in the quark masses \tm{\aD m_{0}} and \tm{\aD m_{3}} from terms that naturally arise in at \tm{\mO(\aD)} in Symanzik \ac{EFT}~\cite{Weber:2015hib}. 
These are of no relevance for the discussion in the following, and they are very small both in the quenched approximation~\cite{Weber:2015oqf} and in full \ac{QCD}~\cite{Godzieba:2024uki}, too. 
\tm{\ka} is closely tied to the field redefinition suggested to absorb the relevant operator for \ac{KW} fermions~\cite{Bedaque:2008xs}. 
This unfortunately does not work due to the \acl{TN} structure of Eq.~\eqref{eq:D_KW_comp}. 
The two tastes would need different boundary conditions, but their propagation cannot be separated. 
This implies that the quark propagator cannot have zero energy solutions for parallel \ac{KW} fermions or zero \tm{j_{0}}-momentum solutions for perpendicular \ac{KW} fermions. 
\vskip1ex

In the next two sections, we look at two distinct consequences for meson correlation functions obtained with a mistuned \ac{KW} operator that can be used to nonperturbatively tune \tm{c_{3}} in numerical simulations. 
Both of these schemes have been used in previous studies without a first-principle explanation of how and why they work~\cite{Weber:2013tfa, Weber:2015hib, Weber:2015oqf, Vig:2024umj, Godzieba:2024uki}.
\vskip1ex

\subsection{Oscillations due to mistuning the Karsten-Wilczek operator}\label{sec:KW_oscillations}

To see the first consequences, we define the taste components in the mistuned theory. 
We assume vanishing chemical potential \tm{\mu_{0}=\mu_{3}=0}, but leave quark masses arbitrary. 
We start from the weak-coupling approach and modify Eq.~\eqref{eq:KW_proj_mom} in a mistuned momentum space representation
\nequn{
\begin{aligned}
\bar\ps(p) 
&= 
\bar\ps_{\uparrow }\Big(p+\frac{\ka}{\aD}e_{D}\Big) + \bar\ps_{\downarrow }\Big(p+\frac{\pi}{a_j}e_{j}-\frac{\ka}{\aD}e_{D}\Big)\gaD
,~\\
\ps(p) 
&= 
\ps_{\uparrow }\Big(p+\frac{\ka}{\aD}e_{D}\Big) + \gaD \ps_{\downarrow }\Big(p+\frac{\pi}{a_j}e_{j}-\frac{\ka}{\aD}e_{D}\Big)
,~
\end{aligned}
\label{eq:KW_proj_mom_mis}
}%nequn 
The momentum space representations \tm{\bar\ps_{f}(ap_j,ap_D)} or \tm{\ps_{f}(ap_j,ap_D)} are the same as in Eq.~\eqref{eq:KW_comp_mom}. 
The same taste generators \tm{\{\vr\}} defined in Eq.~\eqref{eq:KW_vr} still work, mapping the two tastes onto one another.\footnote{
This involves two subtleties. 
First, \tm{\vr_{2}} flips the sign of \tm{r}. 
Second, \tm{\vr_{1}} flips the sign of the second extra term, such that \tm{\de c_{3}} is mistuned with opposite sign.
Hence, in either case the sign of \tm{\ka} is changed by the nondiagonal taste generators.} 
Via an inverse Fourier transform we obtain instead of Eq.~\eqref{eq:KW_proj_mis} the mistuned coordinate space representation 
\nequn{
\bar\ps(n) 
=
\Big(\xi^{+\nD} \bar\ps_{\uparrow } P_{+}^{\vr}  + \xi^{-\nD} \bar\ps_{\downarrow } P_{+}^{\vr} \vr_{1} \Big)(n)
,\quad
\ps(n) 
=
\Big(P_{+}^{\vr} \ps_{\uparrow } \xi^{-\nD} + \vr_{1} P_{+}^{\vr} \ps_{\downarrow } \xi^{+\nD} \Big)(n)
~.~
\label{eq:KW_proj_mis}
}%nequn 
where we have included powers of the phase factor \tm{\xi=\ee^{\ri\ka}}. 
Anti-periodicity of \ac{KW} fermions, \tm{\bar\ps(n)=-\bar\ps(n +\ND {e}_{D})} and \tm{\ps(n)=-\ps(n +\ND {e}_{D})}, implies that the mistuned taste-components must incur \emph{\acl{TT} boundary conditions} in time, i.e.
\nequn{
\begin{aligned}
\bar\ps_{\uparrow}(n)
&= -\bar\ps_{\uparrow}(n \pm \ND {e}_{D}) \xi^{\pm \ND}
,~&
\bar\ps_{\downarrow}(n)
&= -\bar\ps_{\downarrow}(n \pm \ND {e}_{D}) \xi^{\mp\ND}
,~\\
\ps_{\uparrow}(n)
&= -\ps_{\uparrow}(n \pm \ND {e}_{D}) \xi^{\mp\ND}
,~&
\ps_{\downarrow}(n)
&= -\ps_{\downarrow}(n \pm \ND {e}_{D}) \xi^{\pm \ND}
~.~
\end{aligned}
\label{eq:KW_bc_mis}
}%nequn 
Aiming at the most simple correlation functions we contract the mistuned spinors \tm{\bar\ps(n)} and \tm{\ps(n)} into local pseudoscalar interpolating operators with all four possible taste structures. 
We ignore the scalar parity partners in the following. 
We employ Eq.~\eqref{eq:KW_proj_mis}, and convert each projector \tm{P^{\vr}_{\pm}} into \tm{P^{\vr}_{+}} by commuting its associated factors \tm{\vr_{1}} past it, see Eq.~\eqref{eq:KW_Pvr}. 
While factors \tm{\xi^{\pm \nD}} cancel exactly in the \acl{TD} bilinears, they persist in the \acl{TN} ones as \tm{\xi^{\pm 2\nD}}:
\nalign{
\Big(\bar\ps \gaf \ps\Big)(n) 
&= \Big( \bar\ps \big( \hspace{0.44em}\rho_3 \otimes \Gaf \big) \ps \Big)(n_{H}) 
 = \etf(n) \Big( \big( \bar\ps_{\uparrow } P_{+}^{\vr}\ps_{\uparrow }
 - \bar\ps_{\downarrow } P_{+}^{\vr} \ps_{\downarrow } \big) \Big) (n)
,~\label{eq:KW_bi_rho3_Ga5}\\
\Big(\bar\ps \gaD\etDf \ps\Big)(n) 
&= \Big( \bar\ps \big( \hspace{0.44em}\rho_1 \otimes \Gaf \big) \ps \Big)(n_{H}) 
 = \etf(n) \Big( \bar\ps_{\downarrow } P_{+}^{\vr} \vr_{1} \ps_{\downarrow } \xi^{-2\nD}
 + \bar\ps_{\downarrow } \vr_{1} P_{+}^{\vr}\ps_{\downarrow } \xi^{2 \nD} \Big)(n)
,~\label{eq:KW_bi_rho1_Ga5}
\\
\Big(\bar\ps \etf \ps\Big)(n) 
&= \Big( \bar\ps \big( \Id_{\rho} \otimes \Gaf \big) \ps \Big)(n_{H}) 
 = \etf(n) \Big( \big( \bar\ps_{\uparrow } P_{+}^{\vr}\ps_{\uparrow }
 + \bar\ps_{\downarrow } P_{+}^{\vr} \ps_{\downarrow } \big) \Big) (n)
,~\label{eq:KW_bi_Irho_Ga5}\\
\Big(\bar\ps \gaDf\etD \ps\Big)(n) 
&= \Big( \bar\ps \big( \hspace{0.44em}\rho_2 \otimes \Gaf \big) \ps \Big)(n_{H}) 
 = \ri\etf(n) \Big( \bar\ps_{\downarrow } P_{+}^{\vr} \vr_{1} \ps_{\downarrow } \xi^{-2\nD}
 - \bar\ps_{\downarrow } \vr_{1} P_{+}^{\vr}\ps_{\downarrow } \xi^{2 \nD} \Big)(n)
,~\label{eq:KW_bi_rho2_Ga5}
}%nalign 
\vskip1ex

The quark-connected correlation functions \tm{C_{X}^{\text{qc}} (n-m) = \big(\bar\ps X \ps\big)(m) \big(\bar\ps X \ps\big)(n)} are products of such operators. 
Expressing those via the original \ac{KW} spinor fields \tm{\bar\ps} and \tm{\ps}, which we contract to \ac{KW} propagators \tm{\mDKW\inv}, we obtain 
\nalign{
&C_{\gaf}^{\text{qc}} (n-m) \phantom{\,_{\etDf}} 
\equiv
C_{\big(\rho_3 \otimes \Gaf \big)} (n-m) ~ 
=
 -\Tr\Big( \big| \mDKW\inv(n-m) \big|^2 \Big) 
,~\label{eq:KW_corr_rho3_Ga5}\\
&C_{\gaD\etDf}^{\text{qc}} (n-m) 
\equiv
C_{\big(\rho_1 \otimes \Gaf \big)} (n-m) ~
=
 -\Real\Tr\Big( \vr_{2}(m) [\mDKW\inv(n-m)]\dag \vr_{2}(n) \mDKW\inv(n-m) \Big)
,~\label{eq:KW_corr_rho1_Ga5}\\
&C_{\etf}^{\text{qc}} (n-m) \phantom{\,_{\etDf}}
\equiv
C_{\big(\Id_{\rho} \otimes \Gaf \big)} (n-m) 
=
 -\Real\Tr\Big( \vr_{3}(m) [\mDKW\inv(n-m)]\dag \vr_{3}(n) \mDKW\inv(n-m) \Big)
,~\label{eq:KW_corr_Irho_Ga5}\\
&C_{\gaDf\etD}^{\text{qc}} (n-m) 
\equiv
C_{\big(\rho_2 \otimes \Gaf \big)} (n-m) ~ 
=
 -\Real\Tr\Big( \vr_{1}(m) [\mDKW\inv(n-m)]\dag \vr_{1}(n) \mDKW\inv(n-m) \Big)
~.~\label{eq:KW_corr_rho2_Ga5}
}%nalign 
Actual lattice correlators corresponding to these channels---although with mismatched signs and phases and with inappropriately smeared interpolating operators at the source---are shown in Fig.~\ref{fig:KW_corr_mes}. 
Equation~\eqref{eq:KW_corr_rho3_Ga5} is the only channel containing the one unbroken pseudo-\ac{NG} boson, while Eq.~\eqref{eq:KW_corr_rho1_Ga5} contains an only mildly broken pseudo-\ac{NG} boson that corresponds to the symmetric combination of the two nondiagonal pseudo-\ac{NG} bosons of the continuum theory. 
There is a subtlety, namely, the taste generators \tm{\vr_{i}} in the expressions in Eqs.~\eqref{eq:KW_corr_rho3_Ga5}{\markit{--\eqref{eq:KW_corr_rho2_Ga5}}} do not align naively with the \acl{ST} structure, since one factor of \tm{\gaf} at each end is absorbed into the Hermitian conjugation of one \ac{KW} propagator. 
Therefore, we refer to the correlators in the following via the \tm{\{\ga\} \otimes \{\et\} \otimes \{h\}} structure of their interpolating operators leaving out the trivial parts from the specification.  
\vskip1ex

Once these are reexpressed in terms of the taste components, the traces contain taste-component propagators defined via \tm{D_{fg}\inv(n-m)=\ps_{f}(n)\bar\ps_{g}(m)} and manifestly positive products of projection operators \tm{P^{\vr}_{+}(m) P^{\vr}_{+}(n)}. 
On the one hand, the \acl{TD} correlators are real and manifestly negative as we recognize from Eqs.~\eqref{eq:KW_corr_rho3_Ga5} or~\eqref{eq:KW_tastecorr_Irho_Ga5},\footnote{%
The oscillating part in the lower panel of Fig.\ref{fig:KW_corr_mes} is due to the poor choice of the interpolating operator lacking the phase factor \tm{\etf}. 
Moreover, the deviation at early time can be attributed to the parity partner, for whom the conclusion of a definite sign does not hold.%
}
\nalign{
C_{\gaf}^{\text{qc}} (n-m) 
&= 
 -\Tr\Big( P^{\vr}_{+}(m) P^{\vr}_{+}(n) \sum\limits_{f,g=\uparrow,\downarrow} 
 (-1)^{1+\de_{fg}}~\big| D_{fg}\inv(n-m) \big|^2 \Big)
,~\label{eq:KW_tastecorr_rho3_Ga5}\\
C_{\etf}^{\text{qc}} (n-m) 
&= 
 -\Tr\Big( P^{\vr}_{+}(m) P^{\vr}_{+}(n) \sum\limits_{f,g=\uparrow,\downarrow} 
 \big| D_{fg}\inv(n-m) \big|^2 \Big)
~.~\label{eq:KW_tastecorr_Irho_Ga5}
}%nalign 
On the other hand, the \acl{TN} ones are manifestly real, but do not have an unambiguous overall sign,  
\nalign{
C_{\gaD\etDf}^{\text{qc}} (n-m) 
&= 
 -\Tr\Big( P^{\vr}_{+}(m) P^{\vr}_{+}(n) \sum\limits_{f,g=\uparrow,\downarrow} 
 (1-\de_{fg})~\xi^{2[(-1)^{\de_{f\downarrow}}\nD+(-1)^{\de_{g\uparrow}}m_D]}
~\times~\nonumber\\
&\hskip2em~\times~\big( [D_{ff}\inv(n-m)]\dag D_{gg}\inv(n-m) 
 + [D_{fg}\inv(n-m)]\dag D_{gf}\inv(n-m) \big) \Big)
,~\label{eq:KW_tastecorr_rho2_Ga5}\\
C_{\gaDf\etD}^{\text{qc}} (n-m) 
&= 
 -\Tr\Big( P^{\vr}_{+}(m) P^{\vr}_{+}(n) \sum\limits_{f,g=\uparrow,\downarrow} 
  (1-\de_{fg})~\xi^{2[(-1)^{\de_{f\downarrow}}\nD+(-1)^{\de_{g\uparrow}}m_D]}
~\times~\nonumber\\
&\hskip2em~\times~\big( [D_{ff}\inv(n-m)]\dag D_{gg}\inv(n-m) 
 - [D_{fg}\inv(n-m)]\dag D_{gf}\inv(n-m) \big) \Big)
~.~\label{eq:KW_tastecorr_rho1_Ga5}
}%nalign 
These properties of reality and---for the \acl{TD} correlators---negativity hold for arbitrarily mistuned parameters. 
\vskip1ex

Next, we turn these correlation functions into point-point correlators with the source at \tm{m=0} and sum the spatial sink coordinates \tm{n_j} to obtain hadrons at rest.  
We associate the taste-hadron rest energies \tm{E_{i,\rho}} with those accessed by the taste-component propagators and justify this choice in Sec.~\ref{sec:KW_dispersion}. 
Namely, we have for the \acl{TD} channels 
\nalign{
C_{\gaf}^{\text{qc}} (\nD) 
&= 
 -\sum\limits_{i=0} |c_{i,\rho_{3}}|^2 \ee^{ -\aD E_{i,\rho_{3}} \nD }
,~\label{eq:KW_en_rho3_Ga5}\\
C_{\etf}^{\text{qc}} (\nD) 
&= 
 -\sum\limits_{i=0} |c_{i,\Id_{\rho}}|^2 \ee^{ -\aD E_{i,\Id_{\rho}} \nD }
,~\label{eq:KW_en_Irho_Ga5}
}%nalign
which yield the usual \tm{\cosh\big(\tfrac{\aD E_{i,\rho}}{2} (2\nD-\ND)\big)} form after accounting for (anti)-periodic  or \acl{TT} boundary conditions. 
Since the factors \tm{\xi^{\nD}} have already canceled in the interpolating operators at the source or the sink, the \acl{TD} hadrons must be and actually are unaffected by the \acl{TT} boundary conditions of Eq.~\eqref{eq:KW_bc_mis}. 
For the \acl{TN} channels the source at \tm{m=0} merges the two phase factors into \tm{\exp{\big(\pm\ri 2\ka(-1)^{\de_{f\downarrow}}\nD\big)}}, and we get 
\nalign{
C_{\gaDf\etD}^{\text{qc}} (\nD) 
&= 
 - \Real\big(\xi^{2\nD} \big) \sum\limits_{i=0} |c_{i,\rho_{2}}|^2 \ee^{-\aD E_{i,\rho_{2}} \nD }
,~\label{eq:KW_en_rho2_Ga5}\\
C_{\gaD\etDf}^{\text{qc}} (\nD) 
&= 
 - \Real\big(\xi^{2\nD} \big) \sum\limits_{i=0} |c_{i,\rho_{1}}|^2 \ee^{-\aD E_{i,\rho_{1}} \nD }
,~\label{eq:KW_en_rho1_Ga5}
}%nalign 
where the energies are real. 
At this stage we see precisely why the \acl{TT} boundary conditions of Eq.~\eqref{eq:KW_bc_mis} matter. 
The backward propagating state would be naively accompanied by \tm{\xi^{\pm 2(\ND-\nD)}}, but the \acl{TT} boundary phase factors \tm{\xi^{\mp 2\ND} } cancel the extra contribution. 
Therefore, we obtain the same oscillating factor  \tm{\Real\big(\xi^{2\nD} \big) = \cos \big(2\ka n_D \big)} for the forward or backward propagating mesons, and the correlators are of the form \tm{\cos(2\kappa \nD) \cosh\big(\tfrac{\aD E_{i,\rho}}{2} (2\nD-\ND)\big)}. 
The successful use of this form in fits was reported for full \ac{QCD} with \acl{SF} backgrounds~\cite{Godzieba:2024uki}, and had been tested much earlier in the quenched case~\cite{Weber:2015oqf}, too.
This shows the correctness of the \acl{TT} boundary conditions of Eq.~\eqref{eq:KW_bc_mis}. 
\vskip1ex

Before the \acl{ST} structure had been appreciated it had been suggested to consider instead a correlator ratio~\cite{Weber:2015oqf}, 
\nalign{
R(\nD) 
 = \frac{C_{\gaD\etDf}^{\text{qc}}}{C_{\gaf}^{\text{qc}}}(\nD)
&= \cos\big( 2\ka \nD \big) \frac
{\sum\limits_{i=0} |c_{i,\rho_{1}}|^2\ee^{-\aD E_{i,\rho_{1}} \nD }}
{\sum\limits_{j=0} |c_{j,\rho_{3}}|^2 \ee^{ -\aD E_{j,\rho_{3}} \nD }}
\to \cos\big( 2\ka \nD \big) \frac{
|c_{0,\rho_{1}}|^2}{|c_{0,\rho_{3}}|^2}
\ee^{-\aD [E_{0,\rho_{1}}-E_{0,\rho_{3}}] \nD }
,~\label{eq:KW_corr_ratio_mis}
}%nalign
which is dominated by the respective ground states at large times. 
The energy difference is expected to be small for fine lattices, i.e. \tm{\de E_{\rho_{1}} \equiv E_{0,\rho_{1}}-E_{0,\rho_{3}} = \mO(\aD^2,r^2, m_3^2)}. 
\tm{\de E_{\rho_{1}}} must be positive due to the pseudo-\ac{NG} boson nature of the ground state in Eq.~\eqref{eq:KW_corr_rho3_Ga5}. 
We derive the nature and origin of the taste-splittings in Sec.~\ref{sec:KW_splittings}. 
Thus, the ratio is almost a cosine due to \tm{\ka} with a modest contamination by a hyperbolic cosine, and can be subjected to a Fourier analysis. 
The two maxima in its frequency spectrum at \tm{\omega=\pm 2\ka} collapse to a single maximum once the relevant coefficient is tuned. 
Hence, tuning the oscillation frequency is a robust, nonperturbative scheme that permits finding \tm{\ka \to 0} from hadronic correlation functions, which had been realized a long time ago~\cite{Weber:2015hib}, albeit without providing a first-principles explanation for its origins. 
Equation~\eqref{eq:KW_corr_ratio_mis} had been a heuristic tool for precise tuning that eventually led to the discovery of its origins in the \acl{ST} structure. 
For an approximately tuned theory the frequency is smaller than the resolution of the discrete spectrum, i.e. \tm{2\ka \lesssim \sfrac{2}{\ND}}, and potentially of similar magnitude as the energy difference. 
Both issues could limit the precision achievable with this tuning strategy.  
However, a much better resolution is enabled by the tiling strategy~\cite{Godzieba:2024uki}, which permits enlarging \tm{\ND} by integer multiples for the purpose of resolving the frequency spectrum. 
Moreover, joint fits of decay and oscillation permits resolving both successfully. 
Thus, the mistuned oscillation frequencies can be determined with as much precision as required.  
\vskip1ex

An analogous result is expected for perpendicular \ac{KW} fermions, although it has not been worked out in detail yet. 
The corresponding spatial oscillation realizes a shift of quark and \acl{TN}l hadron \tm{j_{0}} momenta to values incompatible with the eigenmodes of the periodic lattice (averaging forward or backward propagation). 
In other words, \acl{TN} hadrons cannot be at rest unless \tm{\ka} is tuned. 
Finding a minimum of the \acl{TN} hadron energy excited by, e.g. the local fermion bilinear \tm{\bar{\ps} \ga_{j_{0}5}\etj \ps}, see Table~\ref{tab:KW_meson_per}, serves as the counterpart to merging the two oscillating modes into one. 
Instead of searching for the energy minimum at zero momentum one might determine \tm{\ka} from hadronic dispersion relations at multiple spatial momenta, similar to the established strategy for determining fermion anisotropy, e.g.,~see e.g.~\cite{Bazavov:2023ods}. 
Whether \acl{TN} hadrons with parallel or perpendicular \ac{KW} fermions are more convenient or efficient for tuning \tm{\ka} remains an open question to be addressed in future studies. 
\vskip1ex

\subsection{Modified dispersion relation due to mistuning the Karsten-Wilczek operator}\label{sec:KW_dispersion}

According to Eq.~\eqref{eq:KW_corr_rho3_Ga5} the correlator \tm{C_{\gaf}^{\text{qc}}(n-m)} is---as usual---just the modulus square of the \ac{KW} propagator, which we know explicitly in the weak-coupling regime after renormalization. 
We derive the quark dispersion relation from the \ac{KW} propagator in momentum space, Eq.~\eqref{eq:D_KW_prop}. 
The connection between hadron or quark dispersion relations is nontrivial in practice, in particular, since the former is physical and nonperturbative, whereas the latter is only a valid concept in a fixed gauge and---strictly speaking---in the weak-coupling regime. 
Yet we may expect that distinctive symmetry-breaking patterns are hereditary traits that feature in hadron dispersion relations, too. 
E.g., the improved isotropy in the quark dispersion relation for Brillouin Wilson fermions is inherited by the corresponding hadron dispersion relations~\cite{Durr:2012dw}. 
Therefore, we aim at understanding how mistuning of the relevant counterterm in Eq.~\eqref{eq:S_KW_3} and the resulting \tm{\ka} dependence affect quark dispersion relations for parallel or perpendicular \ac{KW} fermions. 
Then we compare the predicted qualitative patterns to observed \acl{TD} hadron rest energies. 
\vskip1ex

\subsubsection{Dispersion relation of parallel Karsten-Wilczek quarks}\label{sec:KW_par_dispersion}

For mesons at rest, quark and antiquark are at exactly opposite spatial momenta. 
Up to odd terms in the spatial momenta and up to charge conjugation, which flips the sign of the \tm{r} parameter (and, consequently, of \tm{\ka}), quark and antiquark have formally the same dispersion relation. 
If we neglect details of the (non-perturbative) binding and consider the weak-coupling regime, then the energy of the quark-antiquark state is the sum of the energies of the quark and antiquark, whose common dispersion relation we compute in the following. 
We assume vanishing chemical potential \tm{\mu_{0}=\mu_{3}=0} or isovector mass \tm{m_3=0}, but a nonzero \acl{TS} mass \tm{m_{0}}. 
An isovector mass \tm{m_3} would have no substantial impact for the following discussion and could be reinstituted by averaging \tm{m_0 \to m_0 \pm m_3}. 
We assume that all parameters have been renormalized with the exception of the mistuned \tm{c_{3}}. 
\vskip1ex

The mistuned parallel \ac{KW} propagator yields a quadratic dispersion relation, whose closed form solution is complex,
\nequn{
\sinh(\aD E) 
= 
\pm \om_{0}+\ri\big(r\om_{1}^2+\sin(\ka)\big)
,~
\left\{
\begin{aligned}
\om_{0}
&=
\aD E_0=\aD \frac{\sqrt{m_0^2 +\sum_{j=1}^{D-1}(\xi_{f0,j}\bar{p}_{j})^2}}{\xi_{f1,D}}
,~\\
\om_{1}^2
&=
\aD^2E_1^2=\aD^2\frac{\sum_{j=1}^{D-1}(\xi_{f0,j} \hat{p}_j)^2}{2\xi_{f1,D}}
,~
\end{aligned}\right.
\label{eq:KW_disp_mis}
}%nequn
where, up to fermion anisotropy \tm{\xi_{f,\mu}}, \tm{E_0} is just the energy of a naive fermion. 
\tm{\aD E_1^2} is the contribution due to the Laplacian, see Eq.~\eqref{eq:D_KW_mom}, and \tm{\ka} is defined in Eq.~\eqref{eq:KW_ka}. 
\tm{E} is a strictly even function of the spatial momenta. 
The real part \tm{\Real E} determines the exponential decay of the quark propagator, while the imaginary part \tm{\Imag E} drives \acl{TN} transitions between different tastes. 
We shall focus on the parametric \tm{\ka} dependence and divergent contributions. 
\vskip1ex

Let us consider fixed gauge coupling, and thus constant \tm{\ka}. 
For the slightly mistuned \ac{KW} operator, \tm{\ka \ll 1}, we may expand the energy, firstly\footnote{To the considered order, i.e. including \tm{\mO(\ka^2)}, the expansions in \tm{\ka} or in \tm{\ska} are the same.} in \tm{\ska}, and secondly in the (temporal) lattice spacing \tm{\aD},
\nalign{
\aD E 
&=
\pm \text{arsinh}
\Big( \om_{0} \pm \ri r\om_{1}^2 \Big)
+ \ri 
\frac{1}{\sqrt{1+\Big( \om_{0} \pm \ri r\om_{1}^2 \Big)^2}}
\ska
\pm 
\frac{ \Big( \om_{0} \pm \ri r\om_{1}^2 \Big)}{2\sqrt{1+\Big( \om_{0} \pm \ri r\om_{1}^2 \Big)^2}^3}
\sin^2(\ka)
+ \mO(\sin^3(\ka))
~\\
%%%%%%%%%%%%%%%%%%%%%%%%%%%%%%%%%%%%%%%%%%%%%%%%%%%%%%%%%%%
&=
\pm \aD E_{0}
\Big( 
\frac{2+\ka^2}{2} 
+ \aD^2 \big( E_{1}^2 r\ka 
- E_{0}^2\frac{2+9\ka^2}{12} \big)
\Big)
+ \ri r \Big( \frac{\markit\sin(\ka)}{r}
+ \aD^2 \big(
E_{1}^2
\frac{2+\ka^2}{2} 
- 
\frac{E_{0}^2}{2} \frac{\ka}{r}
\big)
\Big)
+ \mO(\aD^4,\ka^3)
~.
\label{eq:KW_disp_mis_series}
}%nalign
\tm{\Real(E)} corresponds to the energy of the quark and is independent of the sign of the \tm{r}-parameter or the sign of spatial quark momenta. 
Thus, \tm{\Real(E)} is indeed the same for quark or antiquark. 
We see in Eq.~\eqref{eq:KW_disp_mis_series} that the \ac{LO} contribution to \tm{\Real(E)} receives a positive, symmetric correction at \tm{\ka \neq 0} such that \tm{\Real(E)} would have a local minimum at \tm{\ka \to 0}. 
While higher-order corrections are suppressed by \tm{\aD^2}, these include odd powers in \tm{r\ka}, if the quark is not at rest, and these also permit explicitly negative contributions, too. 
The former implies that the stationary point \tm{\sfrac{\pad \Real(E)}{\pad\ka}=0} may be shifted away from \tm{\ka \to 0}, while the latter implies that it might not even be a local minimum in the first place. 
Thus, although a local minimum of \tm{\Real(E)} seems to be a good guess, it does not imply finding \tm{\ka \to 0}. 
\vskip1ex

\tm{\Imag(E)} is an odd function of the \tm{r} parameter, in part through \tm{\ka}, and an even function of spatial quark momenta. 
In the mistuned theory \tm{\Imag(E)} has a power-law divergence {\markit\tm{\ri\frac{\sin(\ka)}{\aD}}}, which would be removed by accounting for the physical tastes according to Eq.~\eqref{eq:KW_proj_mis} via \tm{\sinh(\aD E) \to \sinh(\aD E + \ri \ka) = \sinh(\aD E) \cos(\ka) + \ri\cosh(\aD E) \sin(\ka)}, where low-energy solutions with \tm{\Real(E)>0} permit choosing the positive root in \tm{\cosh(\aD E)=\pm\sqrt{1+\sinh^{2}(\aD E)}}. 
Due to its parametric form \tm{\Imag(E)} has opposite sign for quark or antiquark.
Thus, it is conceivable that it may cancel between both in many \acl{TD} hadron correlation functions, leaving those as real functions of \tm{\ka^2} as noted already in Sec.~\ref{sec:KW_oscillations}. 
\vskip1ex

As a result of a clever rewriting the dispersion relation remains quadratic in \tm{\sinh(\aD E)} and has closed form solutions
\nequn{
\sinh(\aD E) 
= 
\big( \pm\om_{0} +\ri r\om_{1}^2 +\ri \sin(\ka) \big)\cos(\ka)
\pm \ri \sin(\ka) 
\sqrt{1 + 
\big( \pm\om_{0} + \ri r\om_{1}^2 
+\ri \sin(\ka) \big)^2}
,~\label{eq:KW_proj_disp_mis}
}%nequn
where only the lower (minus) sign (in front of the root) yields the correct \tm{\ka \to 0} limit. 
In Eq.~\eqref{eq:KW_proj_disp_mis} or in its expansion the power-law divergence due to {\markit\tm{\ri \tfrac{\sin(\ka)}{\aD}}} is absent. 
Otherwise the expansion up to the considered orders is the same as Eq.~\eqref{eq:KW_disp_mis_series}. 
The higher order corrections to \tm{\Imag(E)}, which are suppressed by odd powers of \tm{\aD} against \tm{\Real(E)}, persist as usual. 
Thus, the same conclusions about its minimum still hold in this case, too. 
So we expect that \acl{TD} hadron rest energies with parallel \ac{KW} fermions are predominantly quadratic, concave, real functions of \tm{\ka}, as confirmed in~\cite{Weber:2015oqf}.
\vskip1ex

\subsubsection{Dispersion relation of perpendicular Karsten-Wilczek quarks}\label{sec:KW_perp_dispersion}

The situation is radically different for perpendicular \ac{KW} fermions. 
The mistuned perpendicular \ac{KW} propagator yields a quartic dispersion relation unless \tm{r^2=1},
\nequn{
(r^2-1)^2\sinh^4(\aD E) + 2\big((r^2-1)(A+B)-2r^2B\big)\sinh^2(\aD E) + (A+B)^2-4r^2B = 0
,~
}%nequn
where the contributions \tm{A=\om_0^2+r^2} and~\tm{B=\big(\om_1+r+r\om_2^2\big)^2} are a shorthand notation for 
\nequn{
\begin{aligned}
\om_0&\equiv\aD F_0=\aD \frac{\sqrt{m_0^2 +\sum_{j\neq j_{0}}^{D-1}(\xi_{f0,j}\bar{p}_{j})^2}}{\xi_{f0,D}}
,~\\
\om_1&\equiv\aD F_1=\aD\frac{(\xi_{f1,j_{0}}\bar{p}_{j_{0}})}{\xi_{f0,D}} + \ska
,~\\
\om_2&\equiv\aD^2 F_2^2=\aD^2\frac{\sum_{j\neq j_{0}}^{D-1}(\xi_{f0,j} \hat{p}_j)^2}{2\xi_{f0,D}}
,~
\end{aligned}
\label{eq:KW_per_omi}
}%nequn
where \tm{E_0 = \sqrt{F_0^2+(F_1-\ska)^2}} is just the energy of anisotropic naive fermions. 
The closed form solutions are real, 
\nequn{
\sinh(\aD E) 
= 
\pm \sqrt{ -\frac{(r^2+1)B-(r^2-1)A}{(r^2-1)^2} \Big(
1 \pm \sqrt{1+\frac{(r^2-1)^2}{(r^2+1)B-(r^2-1)A}\frac{(A+B)^2-4r^2B}{(r^2+1)B-(r^2-1)A}
}\Big)
}
,~
\label{eq:KW_disp_mis_per}
}%nequn
where only the lower (minus) sign (in front of the inner root) yields the correct \tm{r^2 \to 1} limit,
\nequn{
\sinh(\aD E) = \left.\pm \sqrt{
\frac{\big( \om_0^2+r^2 + ( \om_1+r+r\om_2^2 )^2 \big)^2-4( \om_1+r+r\om_2^2 )^2 }{4 ( \om_1+r+r\om_2^2 )^2}}
\right|_{r \to \pm 1}
~.
\label{eq:KW_disp_mis_per_r1}
}%nequn
Successive expansion of Eq.~\eqref{eq:KW_disp_mis_per} in \tm{\ka} and \tm{\aD} yields
\nalign{
\aD E 
&
%%%%%%%%%%%%%%%%%%%%%%%%%%%%%%%%%%%%%%%%%%%%%%%%%%%%%%%%%%%
 =\frac{\sqka}{2\aD E_0}\left(1-R_1^2\right)
 +\ska R_1 
 \left(\frac{4-3r\ka}{4}+\frac{r\ka}{4}\left(R_1^2-6 R_{2}^2\right) \right)
 ~\nonumber\\
&
%%%%%%%%%%%%%%%%%%%%%%%%%%%%%%%%%%%%%%%%%%%%%%%%%%%%%%%%%%%
 +\aD 
 E_0
 \Bigg( 
\frac{2-r\ka}{2}+\frac{9r^2-4}{16}\ka^2 
 -\frac{4r+2\ka-9r^2\ka}{8}\ka R_1^2 
 +\frac{4-3r\ka}{4}r\ka R_2^2 
\Bigg)
 ~\nonumber\\
&
%%%%%%%%%%%%%%%%%%%%%%%%%%%%%%%%%%%%%%%%%%%%%%%%%%%%%%%%%%%
-\aD^2 E_0^2 R_1 
\Bigg(
\left(\frac{r}{2}+\frac{4-9r^2}{8}\ka-\frac{54-75r^2}{32}r\ka^2\right)
-
\frac{6r+9\ka-25r^2\ka}{16}r\ka R_1^2 
-
\left(\frac{4-6r\ka-3\ka^2}{4}+\frac{45r^2\ka^2}{16}\right)r R_2^2
\Bigg)
 ~\nonumber\\
&
%%%%%%%%%%%%%%%%%%%%%%%%%%%%%%%%%%%%%%%%%%%%%%%%%%%%%%%%%%%
-\aD^3 E_0^3 \Bigg(
\frac{4-3r^2}{24}
-\frac{17 -15r^2}{48}r\ka
-\frac{144-720r^2+525r^4}{768}\ka^2
-\left(\frac{r+(3-5r^2)\ka}{3}r+\frac{(48-480r^2+525r^4)}{256}\ka^2\right)R_1^2
 ~\nonumber\\
  &\phantom{+\aD^3 E_0^3 \Bigg(}
  +\left(\frac{r}{2}+\tfrac{4-9r^2}{8}\ka-\frac{54-75r^2}{32}r\ka^2\right)r R_2^2
\Bigg)
+ \mO\left(\aD^4,\ka^3,\left(\frac{\bm{p}^2}{m_0^2}\right)^2\right)
,~\label{eq:KW_disp_mis_series_per}
}%nalign
where further terms due to higher powers of momenta have been omitted, i.e. higher powers of \tm{R_{1}\equiv\sfrac{F_1}{F_0}} or \tm{R_{2}\equiv\sfrac{F_2}{F_0}} that would be suppressed for lying eigenmodes. 
\begin{enumerate}
\item 
All terms that are even in powers of \tm{r} and of the lattice spacing \tm{\aD} are also even in terms of all spatial momenta. 
These terms are the same for quark or antiquark. 
\item 
There is a quadratic power-law divergence \tm{\tfrac{\sqka}{2\aD^2 E_0}\left(1-R_1^2\right)}, namely \tm{\mO\big(\aD^{-1}\big)} in the first line of Eq.~\eqref{eq:KW_disp_mis_series_per}. 
\item 
The continuum term, namely \tm{\mO\big(\aD\big)} in the second line of Eq.~\eqref{eq:KW_disp_mis_series_per}, contains odd powers in \tm{r\ka}.
The coefficient of the quadratic contribution in \tm{\ka} may be assumed to be positive; while it changes sign at \tm{r^2=\frac{4}{9}}, this is outside of the range of interest for actual simulations, i.e. \tm{r^2 \gtrsim 1}. 
For small values of \tm{\ka}, i.e. \tm{\mO(10^{-1})} or less, the linear term clearly dominates over the quadratic one.
Hence, there is no stationary point \tm{\sfrac{\pad \Real(E)}{\pad\ka}=0} near \tm{\ka \to 0}. 
\item 
We find terms that are odd in the spatial momentum in the \tm{j_{0}} direction, i.e. \tm{F_1} resp. \tm{R_1}, which are odd in powers of \tm{r} and of \tm{\aD}, too. 
Thus, these terms are the same for quark or antiquark. 
If parity invariance is enforced in the correlation function, e.g. by summing over all allowed fermion momenta, such odd terms cancel. 
\item 
The first among the odd terms, i.e. in the first line of Eq.~\eqref{eq:KW_disp_mis_series_per}, is a linear power-law divergence \tm{\tfrac{\ska}{\aD}R_1\left(\ldots\right)} (up to corrections \tm{\propto r\ka^2}) and thus vanishes for \tm{\ka \to 0}. 
The next odd term in the third line of Eq.~\eqref{eq:KW_disp_mis_series_per} includes a contribution \tm{-\aD E_0^2 R_1 \tfrac{r}{2}\big(1-2R_2^2\big)} that actually survives the \tm{\ka \to 0} limit. 
\item 
All power-law divergences are the same for quark or antiquark. Contrary to parallel \ac{KW} fermions the divergences cannot cancel between quark and antiquark contributions. 
\end{enumerate}

Next we account for the physical tastes at the actual propagator poles according to the analog of Eq.~\eqref{eq:KW_proj_mom_mis}  for perpendicular \ac{KW} fermions. 
Namely, focusing on the first pole\footnote{For the other pole at \tm{a_{j_{0}}p_{j_{0}} = \pi+\ka}, we get the same result.} at \tm{a_{j_{0}}p_{j_{0}} = -\ka} we replace \tm{\om_1} of Eq.~\eqref{eq:KW_per_omi} as 
\nequn{
\begin{aligned}
\om_1
\to 
\tilde{\om}_1
&=
\Big(\sin(a_{j_{0}}p_{j_{0}})\cos\left(\ka\right) - \cos(a_{j_{0}}p_{j_{0}})\sin\left(\ka\right) \Big)
+ \ska
\\
&=
\aD\frac{(\xi_{f0,j_{0}}\bar{p}_{j_{0}})}{\xi_{f0,D}}\cos\left(\ka\right) 
+ \Big(1-\cos(a_{j_{0}}p_{j_{0}})\Big)\ska
\approx
\om_1 \sqrt{1-\sqka} + \Big(1-\sqrt{1-\om_1^2}\Big)\ska
,~
\end{aligned}
\label{eq:KW_proj_per_omi}
}%nequn
where we have chosen the roots corresponding to small \tm{a_{j_{0}}p_{j_{0}}} and \tm{\ka} and neglected terms of order \tm{\mO(\ka^3)}. 
We obtain a modified \tm{B=\big(\tilde{\om}_1+r+r\om_2^2\big)^2}, which is plugged into Eq.~\eqref{eq:KW_disp_mis_per} and then consecutively expanded in \tm{\ka} and \tm{\aD} as
\nalign{
\aD E 
&
%%%%%%%%%%%%%%%%%%%%%%%%%%%%%%%%%%%%%%%%%%%%%%%%%%%%%%%%%%%
 =
 +\aD E_0
 \Big(
 \frac{2-R_1^2\ka^2}{2}
 \Big)
%%%%%%%%%%%%%%%%%%%%%%%%%%%%%%%%%%%%%%%%%%%%%%%%%%%%%%%%%%%
-\aD^2 E_0^2 R_1 
\Bigg(
\tfrac{2r-r\ka^2}{4}\left(1+2R_2^2\right) 
+\tfrac{2\ka-r\ka^2}{4} R_1^2
\Bigg)
 ~\nonumber\\
&
%%%%%%%%%%%%%%%%%%%%%%%%%%%%%%%%%%%%%%%%%%%%%%%%%%%%%%%%%%%
-\aD^3 E_0^3 \Bigg(
\tfrac{4-3r^2}{24}
-\tfrac{3r^2(2-3\ka^2)+4\ka(r+\ka)}{16}R_1^2
+\tfrac{r^2}{2}R_2^2
\Bigg)
+ \mO\left(\aD^4,\ka^3,\left(\tfrac{\bm{p}^2}{m_0^2}\right)^2\right)
,~\label{eq:KW_proj_disp_mis_series_per}
}%nalign
where further terms due to higher powers of momenta have been omitted. 
Equation~\eqref{eq:KW_proj_disp_mis_series_per} is dramatically better than Eq.~\eqref{eq:KW_disp_mis_series_per}. 
As for parallel \ac{KW} fermions, power-law divergences are gone. 
This explains why the \acl{TD} hadron rest energies are associated with the energies accessed by the taste-component propagators. 
The \ac{LO} term receives a \tm{\ka^2} correction for nonzero momentum \tm{p_{j_0}^2}. 
The higher-order terms contain an odd power in \tm{\ka} for nonzero momentum \tm{p_{j_0} \neq 0}, while terms simultaneously odd in \tm{R_1} and \tm{r} persist. 
So we expect that hadron energies with perpendicular \ac{KW} fermions are weakly curved, quadratic, convex, real functions of \tm{\ka}, whose stationary points are not expected to provide useful guesses of \tm{\ka \to 0}.
\vskip1ex

Let us summarize at this point the key features of quark dispersion relations, which differ significantly and parametrically for parallel or perpendicular \ac{KW} fermions. 
This should not come as a surprise, since these are two distinct lattice discretizations with different symmetry-breaking patterns and lattice artifacts. 
In the following we assume the same small magnitude \tm{|p|\sim\sfrac{\sqrt{\braket{\bm{p}^2}}}{(D-1)}} for contributions from any spatial momentum component.  
\begin{enumerate}
\item 
At \ac{LO} in \tm{\aD}, \tm{\Real(E)} is a quadratic function in \tm{\ka} for both parallel or perpendicular \ac{KW} fermions, concave for parallel and convex for perpendicular \ac{KW} fermions. 
For \tm{D > 2} the curvature of the latter is suppressed by \tm{R_1^2 \sim \sfrac{p^2}{(m_0^2+(D-2)p^2)} \le \frac{1}{D-2}}. 
Close to the continuum limit the quark energy difference for parallel or perpendicular \ac{KW} fermions has a genuine minimum near \tm{\ka \to 0}. 
\item 
Nonzero momentum modes contribute with odd powers in \tm{\ka} as \tm{\mO(\aD^2)} effects. 
These terms have different signs, i.e. \tm{-\aD^2 r\ka E_1^2} for parallel or \tm{+\aD^2 r\ka \sfrac{F_1^2}{4}} for perpendicular \ac{KW} fermions. 
Then \tm{- (\sfrac{3}{2})(\aD p)^2 r\ka} for parallel or \tm{+(\sfrac{1}{4})(\aD p)^2 r\ka} for perpendicular \ac{KW} fermions combine to 
\tm{-(\sfrac{7}{4})(\aD p)^2 r\ka} for the energy difference, whose stationary points are not at \tm{\ka \to 0} unless \tm{\mO(\aD^2)} terms and finite spatial momenta are in practice negligible. 
\end{enumerate}

\subsubsection{\tm{\ka} dependence of hadron masses}\label{sec:KW_ka_hadron}

\begin{figure}%
\includegraphics[width=0.38\txtw]{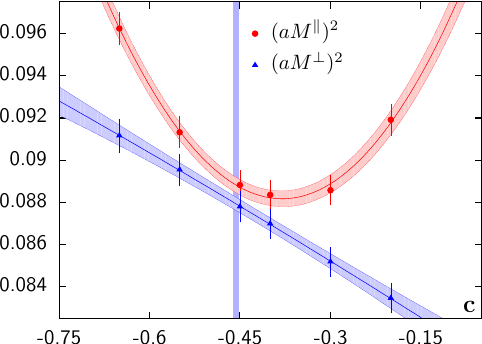}
\caption{%
Ground state rest energy of pseudoscalar correlation functions with parallel or perpendicular \ac{KW} fermions on a pure gauge background with Jacobi smearing at the source. 
Figure from~\cite{Weber:2015oqf}. For a detailed discussion see the text.% 
\label{fig:KW_aniso_energy}%
} 
\end{figure}%

We expect that ground state hadron rest energies of the \acl{TD} isovector pseudo-\ac{NG} boson correlator \tm{C_{\gaf}^{\text{qc}}(n-m)} defined in Eq.~\eqref{eq:KW_corr_rho3_Ga5} with parallel or perpendicular \ac{KW} fermions have features similar to the quark dispersion relations. 
Indeed, parametric \tm{\ka} dependence in line with the weak-coupling prediction was observed a long time ago~\cite{Weber:2015oqf}.
We reproduce a corresponding plot in Fig.~\ref{fig:KW_aniso_energy}, where the vertical band marks the minimal difference of squared energies at a ground state pseudo-\ac{NG} boson energy of about 630 MeV.  
In the range of interest the linear \tm{\ka} dependence is much stronger than the quadratic one for perpendicular \ac{KW} fermions. 
Although the first nonperturbative scheme ever suggested for \ac{KW} fermions was tuning the mass difference between parallel or perpendicular \ac{KW} fermions~\cite{Weber:2013tfa}, the derivation in this section shows that it is not an adequate or robust scheme. 
The energy minimum for parallel \ac{KW} fermions seems to be a less problematic guess than the one including perpendicular \ac{KW} fermions. 
Nonetheless and despite these solid arguments, the results obtained from tuning the energy difference were found to be consistent within their larger uncertainties with results from tuning the oscillation frequency~\cite{Weber:2015hib}. 
As we conclude that the lattice artifacts in the quark dispersion relations are diminished significantly in the tuned theory, 
we expect that this holds true on a qualitative level for hadron dispersion relations, too. 
\vskip1ex

\subsection{Karsten-Wilczek taste splittings}\label{sec:KW_splittings}

Many correlation functions are constrained to be even functions of \tm{\aD} and of the parameters \tm{r_0}, \tm{r_1}, \tm{m_3}, \tm{\xi_{f0,j}} and \tm{\xi_{f1,D}}. 
On the one hand, this is straightforward to show if the correlation functions have a finite, nonvanishing continuum limit, e.g. like Eqs.~\eqref{eq:KW_corr_rho3_Ga5} or \eqref{eq:KW_corr_rho1_Ga5}, which we prove in Sec.~\ref{sec:KW_parametric}. 
It follows a similar line of reasoning that we use to expose the parametric form of the \ac{KW} determinant, cf. Sec.~\ref{sec:KW_determinant}. 
We derive the leading \acl{TSB} cutoff effects and quark mass dependence of these two channels in Sec.~\ref{sec:KW_pNGB1}. 
On the other hand, if the correlation functions do not have a finite, nonvanishing continuum limit, namely like Eqs.~\eqref{eq:KW_corr_Irho_Ga5} or \eqref{eq:KW_corr_rho2_Ga5}, one specific part of the argument is not guaranteed for generic \tm{m_3}. 
A modified chain of arguments still applies at \tm{m_3=0}. 
Since the other pseudo-\ac{NG} bosons require a far more extensive discussion, we only summarize the main results in Sec.~\ref{sec:KW_other_pNGB_summary} and defer the derivation in its full glory to the Appendix~\ref{app:KW_other_pNGB_details}. 
While we assume parallel \ac{KW} fermions in the following, there is no apparent reason why these results should not carry over to perpendicular \ac{KW} fermions given the obvious replacements. 
\vskip1ex

\subsubsection{Parametric form of hadron correlators with Karsten-Wilczek fermions}\label{sec:KW_parametric}

In the following, we use the two correlation functions of Eqs.~\eqref{eq:KW_corr_rho3_Ga5} or \eqref{eq:KW_corr_rho1_Ga5} as examples. 
We first insert \tm{1=\vr_{3}^2} at both the source and the sink, commuting one \tm{\vr_{3}} at each past the \tm{\{\vr\}} that are remnants of the interpolating operators.
So possible signs cancel between source and sink.
Then we use each pair of \tm{\vr_{3}} to flip the signs of the respective \tm{r_0} parameters of the propagator \tm{\mDKW\inv(n-m)} or \tm{(\mDKW\inv)\dag(n-m)} between this pair. 
Thus, Eqs.~\eqref{eq:KW_corr_rho3_Ga5} -  \eqref{eq:KW_corr_rho1_Ga5} are independent of the sign of \tm{r_0}, i.e. they are even functions of \tm{r_0}. 
Existence of a continuum limit requires that, as one of these relevant coefficients is restricted to even powers, i.e. \tm{r_0}, the other relevant coefficient, i.e. \tm{r_1}, is restricted to even powers, too.
Thus, Eqs.~\eqref{eq:KW_corr_rho3_Ga5} or \eqref{eq:KW_corr_rho1_Ga5} are even functions of each \tm{r_{i}}, and thus of \tm{r}. 
Of course this argument cannot be applied without the existence of a finite, nonvanishing continuum limit, e.g. to Eqs.~\eqref{eq:KW_corr_Irho_Ga5} or~\eqref{eq:KW_corr_rho2_Ga5}. 
Next, we insert \tm{1=\vr_{2}^2} at both the source and the sink, commuting one \tm{\vr_{1}} at each past the \tm{\{\vr\}} that are remnants of the interpolating operators. 
Possible signs cancel again between source and sink. 
Then we use each pair of \tm{\vr_{2}} to flip the signs of the respective \tm{r} and \tm{m_3} parameters of both propagators. 
The former is irrelevant if a finite, nonvanishing continuum limit exists, since we have shown that the correlation function is an even function of \tm{r}.  
Thus, Eqs.~\eqref{eq:KW_corr_rho3_Ga5} or  \eqref{eq:KW_corr_rho1_Ga5} are even functions of \tm{m_3}, too.
Note that we could have used \tm{\vr_{1}} in place of \tm{\vr_{2}} and would have obtained the same information (after taking the information already known from \tm{\vr_{3}} into account). 
For correlation functions without a finite, nonvanishing continuum limit, we consider a restriction to \tm{m_3=0}. 
Then the argument based on \tm{\vr_{1,2}} constrains the correlation function to even powers of \tm{r}, and---in combination with even powers of \tm{r_0}---also even powers of \tm{r_1}. 
Thus, since odd powers in \tm{\aD} could only be due to odd powers in \tm{r}, there are no odd powers in \tm{\aD}. 
Finally, we use \tm{\gaD} or \tm{\gaDf} applying the same logic to show that Eqs.~\eqref{eq:KW_corr_rho3_Ga5} - \eqref{eq:KW_corr_rho1_Ga5} are even functions of \tm{\xi_{f0,j}} or \tm{\xi_{f1,D}}, respectively, completing the proof of our claim. 
A proof of the corresponding parameter dependence for perpendiuclar \ac{KW} fermions is analogous. 
\vskip1ex

In the following, we restrict ourselves to correlation functions that have a finite, nonvanishing continuum limit, e.g. Eqs.~\eqref{eq:KW_corr_rho3_Ga5} or \eqref{eq:KW_corr_rho1_Ga5}. 
Due to the parametric form of the correlator as a whole 
the real parts of (energy) levels or moduli square of overlap factors in these channels are constrained to be even functions of \tm{\aD} and of the parameters \tm{r_0}, \tm{r_1}, \tm{m_3}, \tm{\xi_{f0,j}} and \tm{\xi_{f1,D}}, too. 
This does not apply to imaginary parts of the (energy) levels of nonasymptotic states, e.g. quarks, or to the overlap factors themselves. 
To be consistent with the properties we proved, either of these has to be composed exclusively of either even or odd powers in \tm{m_3}, \tm{r}, and \tm{\aD}.
We know from the quark dispersion relation in the weak-coupling regime that the imaginary parts of quark energy levels include odd powers in \tm{r} and \tm{\aD}, see Sec.~\ref{sec:KW_dispersion}. 
Thus, even powers of \tm{m_3}, \tm{r}, and \tm{\aD} must be excluded from those imaginary parts. 
The overlap factors could be either even or odd functions of \tm{m_3}, \tm{r} and \tm{\aD}.
However, the existence of a finite, nonvanishing continuum limit of an overlap factor, e.g. in the form of a decay constant, requires that they have nonvanishing contributions that are even functions of \tm{m_3}, \tm{r} and \tm{\aD} in its real part. 
Since the correlation functions in Eqs.~\eqref{eq:KW_corr_rho3_Ga5} or \eqref{eq:KW_corr_rho1_Ga5} are real, as noted in Sec.~\ref{sec:KW_oscillations}, only even powers of any imaginary parts contribute. 
Thus, the imaginary parts of overlap factors may indeed be odd functions of \tm{m_3}, \tm{r} and \tm{\aD}. 
The same reasoning and constraints apply to any correlation function with a finite, nonvanishing continuum limit that has symmetric bilinear interpolating operators \tm{X} at source and sink, and thus has a well-defined \acl{ST} structure. 
Hence, we may write 
\nequn{
C_{X}^{\text{qc}} (\nD) 
= 
 -\sum\limits_{i=0} \big| c_{i,X}\big|^2(r^2,\aD^2,m_3^2)~\ee^{ -\aD \Real \big[ E_{i,X} \big](r^2,\aD^2,m_3^2) \nD }
~.~\label{eq:KW_en_finite}
}%nequn
In particular, this holds for the two-shift (\tm{\hDj}) taste-meson correlation functions with a finite continuum limit, whose local counterparts do not have a continuum limit due to the phase factors associated with source and sink, e.g. Eqs.~\eqref{eq:KW_corr_Irho_Ga5} or~\eqref{eq:KW_corr_rho2_Ga5}, see Appendix~\ref{app:KW_other_pNGB_details}. 
As this statement holds for the real parts of (energy) levels for each set of quantum numbers, which are independent of the details of interpolating operators (besides quantum numbers), it must also hold for the taste splittings between the corresponding (energy) levels in different channels, too. 
\vskip1ex

Next, we set up a generic, but fairly standard scheme for decomposing the inverse of a sum of operators, where we only assume invertibility.\footnote{Since this assumption does not hold in the chiral limit on topological backgrounds, we restrict our analysis to massive fermions or nontopological gauge backgrounds.} 
Equation~\eqref{eq:D_KW_comp} and Table~\ref{tab:KW_meson} allow us to understand symmetry properties of the \ac{KW} operator, propagator and functions thereof. 
We may take apart any operator \tm{X}---such as the \ac{KW} operator \tm{\mDKW}---that is a sum as 
\nequn{
X = \big( E+O \big) =
\begin{aligned}
\big( \Id + B \big)E & \quad\text{with}~B = 
OE\inv 
\end{aligned}
,~
\label{eq:op_decomp}
}%nequn
where we have assumed invertibility of 
\tm{E}. \tm{E} and \tm{O} 
might distinguish operators that are even or odd under some \tm{\{\vr\}} of Eq.~\eqref{eq:KW_vr}. 
The corresponding inverse operator, and thus the \ac{KW} propagator, is taken apart as well, 
\nequn{
X\inv = \big( E+O \big)\inv =
\begin{aligned}
E\inv \big( \Id + B \big)\inv 
&=
E\inv \big( \Id - B^2 \big)\inv \big( \Id - B \big)
~.~
\end{aligned}
\label{eq:prop_decomp}
}%nequn
Following the example of Eq.~\eqref{eq:KW_full}, we write the \ac{KW} operator \tm{\mDKW} as 
\nalign{
\mDKW(x,y) 
=& 
 \DN(x,y)
 + W_1(x,y) + W_0 \de(x,y) + M_{0}(x,y) + M_{3}(x,y) 
,~\label{eq:DKW_struct}\\
\text{with}&~
\begin{aligned}
W_1(x,y)
&=-\tfrac{\ri r_1}{\aD}\gaD \sum_{j=1}^{D-1} h_{j}^{c}(x,y)
&,\quad&
W_0
=\tfrac{\ri r_0}{\aD}\gaD
~.~
\end{aligned}
\nonumber
}%nalign
\tm{M_0} is proportional to the identity as defined in Eq.~\eqref{eq:KW_M0}, while \tm{M_{3}} could be from any among Eqs.~\eqref{eq:KW_M3_2s} - ~\eqref{eq:KW_M3_Ds}, 
and the coefficients \tm{r_{i}} are define in Eq.~\eqref{eq:KW_r01}. 
As we focus on the taste structure instead of the fermion anisotropy, we write 
\tm{\DN} in place of \tm{\sum_{j} \xi_{f0,j} \ga_{j} \nab_{j} + \xi_{f1,D} \gaD \nab_{D}}. 
So the full \ac{KW} propagator is 
\nequn{
\mDKW\inv(x,y) = 
\Big( \DN(x,y) + W_1(x,y) + W_0 \de(x,y) + M_{0}(x,y) + M_{3}(x,y) 
\Big)\inv
~.~
}%nequn
In the following we analyze the interplay of these five operators in correlation functions and perform a scaling analysis. 

\subsubsection{Pseudo-Nambu-Goldstone bosons in the \tm{\gaf} or \tm{\gaD\etDf} channels}\label{sec:KW_pNGB1}

The \tm{\big( \rho_3 \otimes \Gaf \big)} channel of Eq.~\eqref{eq:KW_corr_rho3_Ga5} (omitting the argument \tm{(n-m)}) is given in the notation of Eq.~\eqref{eq:DKW_struct} as 
\nalign{
C_{\gaf}^{\text{qc}}
&= -\Tr\Big( \mDKW\inv \mDKW\daginv \Big)
 = -\Tr\Big( \Big| \DN + W_1 + W_0 + M_{0} + M_{3} \Big|^2 \Big)\inv
~\nonumber\\
&=
 -\Tr\Big( 
 \Big| \DN + M_{0} \Big|^2 
 + \Big| M_{3} + W_1 + W_0 \Big|^2
 + \Big[ \big( M_{3} + W_1 + W_0 \big)\big( \DN + M_{0} \big)\dag +~\text{H.c.}\Big]
  \Big)\inv
,~
}%nalign 
where we have grouped terms by how they transform under \tm{\vr_{2}}. 
The last term contributing to the inverse is odd under \tm{\vr_{2}}. 
We may collect the contributions as 
\nequn{ 
K_2 = \DN + M_{0}
,\quad 
L_2 =  M_{3} + W_1 + W_0 
,~\label{eq:KW_KL2}
}%nequn 
and call upon Eq.~\eqref{eq:prop_decomp}, where \tm{ E_2 = |K_2|^2 + |L_2|^2 } and \tm{ O_2 = L_2 K_2\dag + K_2 L_2\dag }, which are both Hermitian. 
The second term in the last bracket of Eq.~\eqref{eq:prop_decomp}, i.e. \tm{\propto -B}, is odd in  \tm{L_2} and vanishes under the trace. 
It would eventually lead to odd terms in either \tm{r_0}, \tm{r_1}, or \tm{m_3}, while only manifestly even functions in \tm{r_0}, \tm{r_1}, \tm{m_3}, \tm{\xi_{f0,j}}, \tm{\xi_{f1,D}} and \tm{\aD} survive. 
The remaining term can be further simplified by noting that the operators have spectra that are either purely real, i.e. \tm{M_{0}} or \tm{M_{3}}, or purely imaginary, i.e.  \tm{\DN},~\tm{W_{0}}, or~\tm{W_{1}}. 
Thus, some products cancel in the real or imaginary parts or in the moduli, respectively, and the terms simplify as \tm{ E_2 = |\DN|^2 + M_0^2 + |M_{3}|^2 + |W_1 + W_0|^2 } and \tm{ O_2 = 2 M_{0} M_{3} + [ \DN, M_{3} ]_- -\big[ W_1 + W_0 , \DN \big]_+ }.
Finally, putting everything together, the correlator can be written as 
\nequn{
C_{\gaf}^{\text{qc}}
 =
 -\Tr\Big( \Big[
 \Big( \Id - \Big[ 
\frac{ 2 M_{0} M_{3} +[ \DN, M_{3} ]_- - \big[ W_1 + W_0 , \DN \big]_+ }{ E_2 } 
 \Big]^2 \Big)  
 E_2 \Big]\inv
 \Big)
~.~\label{eq:KW_corr_rho3_Ga5_decomp}
}%nequn
At this stage, we call upon a scaling analysis, namely, we consider the dominant scales for the operators, namely,  
\nequn{
M_0 \sim M_{3} \sim m_q
,\quad
\Imag(\DN) \sim \La
,\quad
\Imag(W_1+W_0) \sim r\aD \La^2
,\quad
\Imag(W_i) \sim (-1)^{i}\sfrac{r_{i}}{\aD}
~.~\label{eq:KW_scales}
}%nequn
We separate each operator into a nontrivial dimensionless operator times the dominant scale from Eq.~\eqref{eq:KW_scales} and omit any subleading contributions, 
i.e. 
\nequn{
\left\{\begin{aligned}
&\quad
M_0 = {c_{m}}{m_q}
,\quad
M_{3} = {c_{M}}{m_q}
,\quad
|\DN|^2 = {c_{|D|^2}}{\La^2}
,\quad
|W_1+W_0|^2 = {c_{|W_\Si|^2}}{(\aD\La^2)^2}
,\\
&\quad
\DN = {C_{D}}{\La}
,\quad
W_{0} = \frac{C_{W_0}}{\aD} 
,\quad
W_{1} = -\frac{C_{W_0}}{\aD} + {C_{W_\Si}}{\aD\La^2} 
,\\
&\quad
\big[ W_1 , \DN \big]_{\pm} = -C_{W_0 D}^{(\pm)}\frac{\La}{\aD} + C_{W_\Si D}^{(\pm)}{\aD \La^3} 
,\quad
\big[ W_{0} , \DN \big]_{\pm} = C_{W_0 D }^{(\pm)} \frac{\La}{\aD}
,\\
&\quad
\big[ \DN, M_{3}\big]_{-} = C_{D M}^{(-)}{\aD^2}{m_q}{\La^3}
,\quad
\big[ W_{1}, M_{3} \big]_{-} = C_{W_\Si M}^{(-)}{m_q\aD^3 \La^4} 
,\quad
\big[ W_{0} , M_{3} \big]_{-} = 0
~.~
\end{aligned}\right\}
\label{eq:KW_scales_operators}
}%nequn
\tm{c_{m}} is actually just a real scalar, while 
\tm{c_{M}}, \tm{c_{|D|^2}}, and \tm{c_{|W_\Si|^2}} are real scalars up to relative \tm{\mO(\aD^2\La^2)} corrections, too. 
These are related to renormalization constants of the two mass parameters, the wave function and the \tm{r} parameter. 
While \tm{C_{D}^2=-c_{|D|^2}} or \tm{C_{W_\Si}^2=-c_{|W_\Si|^2}} commute at their respective \aclp{LO}, too, 
all other \tm{C_{X}^{(\pm)}} have to be treated as noncommuting operators. 
The hierarchy for light quarks is \tm{m_q \ll \La \ll \sfrac{1}{\aD}}. 
Therefore, the contributions to the correlator scale as 
\nalign{
C_{\gaf}^{\text{qc}}
&=
 -\Tr\Big( \Big[
 \Big( \Id - \Big[ 
\frac{ 2 c_{m}c_{M} m_{q}^2 + C_{D M}^{(-)}{\aD^2}{m_q}{\La^3} - C_{W_\Si D}^{(+)}  \aD \La^3 }
{ c_{|D|^2}\La^2 + \big[ c_{m}^2+c_{M}^2 \big] m_{q}^2 + c_{|W_\Si|^2} \aD^2 \La^4 } 
 \Big]^2 \Big)  
 \left( c_{|D|^2}\La^2 + \big[ c_{m}^2+c_{M}^2 \big] m_{q}^2 
 + c_{|W_\Si|^2} \aD^2 \La^4 \right) \Big]\inv
 \Big)
~\nonumber\\
&=
 -\Tr\Big( \Big[
   c_{|D|^2}\La^2 
 + \big[ c_{m}^2 + c_{M}^2 \big] m_{q}^2 
 + \big\{ c_{|W_\Si|^2} - C_{W_\Si D}^{(+)} c_{|D|^2}\inv C_{W_\Si D}^{(+)} \big\} \aD^2 \La^4 
%  + \mO\left( \frac{m_q^4}{\La^2},~{m_q^2 \aD^2}{\La^2},~\aD^4\La^6 \right)
 \Big]\inv \Big)
 + \mO\left( \frac{m_q^4}{\La^6},~\frac{m_q^2 \aD^2}{\La^2},~\aD^4\La^2 \right)
,~\label{eq:KW_corr_rho3_Ga5_scaling}
}%nalign
where odd powers in \tm{C_{DM}^{(-)}{\aD^2}{m_q}{\La^3}} resp. \tm{m_3} or \tm{C_{W_\Si D}^{(+)}  \aD \La^3} resp. \tm{r} must cancel for the reasons laid out in Sec.~\ref{sec:KW_parametric}. 
The cancellation of power-law divergences occurs in \tm{W_{1}+W_{0}}, i.e. between first powers of the two relevant operators \tm{W_{i}}. 
Explicit symmetrization between positive or negative signs of the coefficients \tm{r_0}, \tm{r_1}, or \tm{m_3} is not needed. 
We realize from the underlying Dirac structure that each term contributing to the leading \ac{TSB} \tm{\mO(r^2\aD^2\La^2)} correction is negative by evaluating the anticommutators as 
\nequn{
\mB_{-} \equiv 
\big\{ c_{|W_\Si|^2}-C_{W_\Si D}^{(+)} c_{|D|^2}\inv C_{W_\Si D}^{(+)} \big\} \aD^2 \La^4
=
\big\{ C_{D} C_{W_\Si} C_{D}\inv C_{W_\Si} + C_{W_\Si} C_{D}\inv C_{W_\Si} C_{D} - c_{|W_\Si|^2} \big\} \aD^2 \La^4
< 0
~.~\label{eq:KW_LTB_rho3}
}%nequn
The \tm{\mO(m_q^2)} term looks as expected for the average of squared light-quark masses, \tm{\big[ c_{m}^2 + c_{M}^2 \big] m_{q}^2 \simeq \sfrac{\big[ m_\uparrow^2 + m_\downarrow^2 \big]}{2}}. 
For later use, we collect the terms contributing to the continuum limit in yet another abbreviation 
\nequn{
\mK_{+} = c_{|D|^2}\La^2 + \big[ c_{m}^2 + c_{M}^2 \big] m_{q}^2
,\quad
,~\label{eq:KW_mK+}
}%nequn
and simplify the correlator as 
\nequn{
C_{\gaf}^{\text{qc}}
 =
 -\Tr\Big( \Big[ 
 \mK_{+} + \mB_{-}
 \Big]\inv \Big)
 + \mO\left( \frac{m_q^4}{\La^6},~\frac{m_q^2 \aD^2}{\La^2},~\aD^4\La^2 \right)
~.~\label{eq:KW_corr_rho3_Ga5_KB}
}%nequn
\vskip1ex

Next, we consider the \tm{\big( \rho_1 \otimes \Gaf \big)} channel of Eq.~\eqref{eq:KW_corr_rho1_Ga5}. 
The two remaining factors \tm{\vr_{2}} are used to invert the signs of \tm{W_{0,1}} and \tm{M_{3}} in either one or the other of the quark propagators, \tm{\mDKW\inv} or \tm{(\mDKW\inv)\dag}. 
Thus, the full correlation function is the average of both versions, or the real part of either. 
While we use the same \tm{K_2} or~\tm{L_2} defined in Eq.~\eqref{eq:KW_KL2}, we have instead \tm{ E_2\pri = |K_2|^2 - |L_2|^2 } and \tm{ O_2\pri = L_2K_2\dag - K_2L_2\dag }.
For reasons analogous to the \tm{\big( \rho_3 \otimes \Gaf \big)} channel we may simplify these to \tm{ E_2\pri = |\DN|^2 + M_0^2 - |M_{3}|^2 - |W_1 + W_0|^2 } and  \tm{ O_2\pri = \big( 2M_{0}\big(W_1 + W_0\big) - [ \DN, M_{3} ]_+ - [W_1+W_0, \DN ]_- \big) }.
Finally, putting everything together, the correlator can be written as 
\nalign{
C_{\gaD\etDf}^{\text{qc}} 
&= -\Tr\Big( \mDKW\inv \vr_{2} \mDKW\daginv \vr_{2} \Big)
 =
 -\Real\Tr\Big( 
 \Big[ \DN + W_1 + W_0 + M_{0} + M_{3} \Big]
 \Big[ \DN - W_1 - W_0 + M_{0} - M_{3} \Big]\dag
  \Big)\inv 
~\nonumber\\
&=
 -\Real\Tr\Big( 
 \Big| \DN + M_{0} \Big|^2 
 - \Big| M_{3} + W_1 + W_0 \Big|^2
 + \Big[ \big( M_{3} + W_1 + W_0 \big)\big( \DN + M_{0} \big)\dag  -~\text{H.c.} \Big]
  \Big)\inv 
~\\
&=
 -\Tr\Big( \Big[
 \big( \Id - \Big[ 
 \frac{ 2M_{0}\big(W_1 + W_0\big) - [ \DN, M_{3} ]_+ - [W_1 + W_0, \DN]_- }{ E_2\pri }
 \Big]^2 \big)  
 E_2\pri \Big]\inv
 \Big)
,~\label{eq:KW_corr_rho1_Ga5_decomp}
}%nalign 
noting that the trace in Eq.~\eqref{eq:KW_corr_rho1_Ga5_decomp} is already manifestly real. 
We apply the same scaling analysis as in Eq.~\eqref{eq:KW_scales_operators} and obtain for the hierarchy of light quarks 
\nalign{
C_{\gaD\etDf}^{\text{qc}}
&=
 -\Tr\Big( \Big[
 \Big( \Id - \Big[ 
\frac{ 2c_{m} C_{W_\Si} m_{q}\aD \La^2 + 2c_{M} C_{D} m_{q} \La - C_{W_\Si D}^{(-)}\aD \La^3}
{ c_{|D|^2}\La^2 + \big[ c_{m}^2 - c_{M}^2 \big] m_{q}^2 - c_{|W_\Si|^2} \aD^2 \La^4 } 
 \Big]^2 \Big)  
~\nonumber\\
&\phantom{-\Tr\Big( \Big[
 \Big(}\times
 \left( c_{|D|^2}\La^2 + \big[ c_{m}^2 - c_{M}^2 \big] m_{q}^2 - c_{|W_\Si|^2} \aD^2 \La^4 \right) \Big]\inv
 \Big)
~\nonumber\\
&=
 -\Tr\Big( \Big[
   c_{|D|^2}\La^2 
 + \big[ c_{m}^2 + 3 c_{M}^2 \big] m_{q}^2 
 - \big\{ c_{|W_\Si|^2} + C_{W_\Si D}^{(-)} c_{|D|^2}\inv C_{W_\Si D}^{(-)} \big\} \aD^2 \La^4 
 \Big]\inv \Big)
 + \ldots
,~\label{eq:KW_corr_rho1_Ga5_scaling}
}%nalign
up to corrections at the same orders as in Eq.~\eqref{eq:KW_corr_rho3_Ga5_scaling}. 
Odd powers in \tm{c_{M} C_{D} m_{q} \La} resp. \tm{m_3} or \tm{C_{W_\Si D}^{(-)}  \aD \La^3} resp. \tm{r} must cancel for the reasons laid out in Sec.~\ref{sec:KW_parametric}. 
The same cancellation of power-law divergences occurs in \tm{W_{1}+W_{0}}, i.e. between first powers of the two relevant operators \tm{W_{i}}. 
Again, explicit symmetrization between positive or negative signs of the coefficients \tm{r_0}, \tm{r_1}, or \tm{m_3} is not needed. 
We see that the leading \ac{TSB} \tm{\mO(r^2\aD^2\La^2)} correction is less negative than \tm{\mB_{-}} of Eq.~\eqref{eq:KW_LTB_rho3} by evaluating the commutators as 
\nequn{
\mB_{+} \equiv 
-\big\{ c_{|W_\Si|^2} + C_{W_\Si D}^{(-)} c_{|D|^2}\inv C_{W_\Si D}^{(-)} \big\} \aD^2 \La^4
=
\big\{ C_{D} C_{W_\Si} C_{D}\inv C_{W_\Si} + C_{W_\Si} C_{D}\inv C_{W_\Si} C_{D} + c_{|W_\Si|^2} \big\} \aD^2 \La^4
,~\label{eq:KW_LTB_rho1}
}%nequn
since the first two terms agree with Eq.~\eqref{eq:KW_LTB_rho3}, while the last term is manifestly positive. 
We note at this point that the difference 
\nequn{
\delta\mB_{r^2} \equiv \mB_{+}-\mB_{-} = 2 c_{|W_\Si|^2} \aD^2 \La^{\markit 4}
~\label{eq:KW_LTB_delta}
}%nequn 
between the \tm{\big( \rho_1 \otimes \Gaf \big)}- or the \tm{\big( \rho_3 \otimes \Gaf \big)} channels constitutes \emph{a first type of \ac{TSB} cutoff effects at \tm{\mO(r^2\aD^2\La^2)}, which already exist at the tree level} (with trivial link variables). 
We have profound reason to expect that these cannot be ameliorated significantly by using smooth links obtained after any amount of link smoothing, e.g. gradient flow~\cite{Luscher:2009eq}, stout~\cite{Morningstar:2003gk} or hypercubic~\cite{Hasenfratz:2001hp} smearing. 
However, these can be ameliorated systematically by using an improved version of the \ac{KW} operator~\cite{Bakenecker:2024a, Godzieba:2024uki, Vig:2024umj}. 
There are some partial cancellations in the \acl{TI} mass contribution at \tm{\mO(m_q^2)}. 
The total coefficient is incompatible with the average of two squared light-quark masses, but differs only at the order \tm{m_3^2} as one expects from \ac{NLO} \ac{ChPT}~\cite{Gasser:1983yg}. 
For later use, we collect the terms contributing to the continuum limit as a new abbreviation 
\nequn{
\mK_- 
= c_{|D|^2}\La^2 + \big[ c_{m}^2 + 3 c_{M}^2 \big] m_{q}^2
= \mK_{+} + \de\mK_{m_3^2}
,\quad 
\de\mK_{m_3^2} \equiv 2 c_{M}^2 m_{q}^2
,~\label{eq:KW_mK-}
}%nequn
and simplify the correlator as 
\nequn{
C_{\gaD\etDf}^{\text{qc}}
 =
 -\Tr\Big( \Big[ 
 \mK_{+} + \de\mK_{m_3^2} + \mB_{-} + \de\mB_{r^2}
 \Big]\inv \Big)
 + \mO\left( \frac{m_q^4}{\La^6},~\frac{m_q^2 \aD^2}{\La^2},~\aD^4\La^2 \right)
~.~\label{eq:KW_corr_rho1_Ga5_KB}
}%nequn
\vskip1ex

\subsubsection{Summary of the other pseudo-Nambu-Goldstone bosons}\label{sec:KW_other_pNGB_summary}

Here we summarize the situation for the other two pseudo-\ac{NG} bosons, namely in the \tm{\big( \Id_{\rho} \otimes \Gaf \big)} or \tm{\big( \rho_2 \otimes \Gaf \big)} channels. 
The analysis for both channels needs substantially more effort. 
As the two channels work quite similarly, we begin with a general discussion. 
We recall the notation of Eq.~\eqref{eq:DKW_struct}. 
\vskip1ex

We begin with the correlators of these two channels \emph{with single-site local fermion bilinears}, i.e.~Eqs.~\eqref{eq:KW_corr_Irho_Ga5} or~\eqref{eq:KW_corr_rho2_Ga5}. 
Due to the two factors of either \tm{\vr_{1}} or \tm{\vr_{3}} being part of single-site local interpolating operators, an instance of \tm{W_{1} + M_{3}} or \tm{W_{0}}, respectively, flips its sign in either one or the other of the two quark propagators, \tm{\mDKW\inv} or \tm{(\mDKW\inv)\dag}. 
Thus, each correlation function is the average of two contributions with a sign flip in either \tm{\mDKW\inv} or \tm{(\mDKW\inv)\dag}. 
In the sign-flipped quark propagator the two power-law divergent operators \tm{W_{0,1}} fail to cancel at \tm{\mO(\sfrac{1}{\aD})}, i.e. the coupling to the two different, power-law divergent condensates fails to interfere destructively; see Sec.~\ref{sec:KW_doublet}. 
In contrast, the corresponding operators in the other, nondivergent quark propagator still contribute as \tm{W_{1} + W_{0}} at \tm{\mO(\aD\La^2)}. 
Since the full correlation function still has to be an even function of each parameter \tm{r_0} and \tm{r_1} (at least in the absence of \tm{m_3}), subtle cancellations among divergent terms could eliminate some divergent contributions while others may persist. 
% \vskip1ex
%
%
Both correlation functions indeed vanish in the continuum limit as their denominators contain a quadratic power-law divergence \tm{\mO(\sfrac{1}{\aD^2})}. 
We derive these results in Appendix~\ref{app:KW_pNGB2}. 
Such a power-law divergence conspires with higher order terms to yield  many extra contributions at order \tm{\mO(\aD^0)} that are even functions of \tm{r}, too. 
In terms of their \acl{ST} structure, such additional contributions include \acl{TS} operators that might contribute to the anomalous coupling in the \tm{\big( \Id_{\rho} \otimes \Gaf \big)} channel. 
Yet the structure at this order is far more complicated than Eqs.~\eqref{eq:KW_corr_rho3_Ga5_scaling} or~\eqref{eq:KW_corr_rho1_Ga5_scaling} might suggest, and there is no point working out these terms in full detail. 
\vskip1ex

\begin{figure}%
\begin{tikzpicture}
  \matrix (m) [matrix of math nodes,row sep=2em,column sep=5em,minimum width=6em]
  {
  \text{taste content} & \text{\acl{TD}} & \text{taste nondiagonal}\\[-1.5em]
  \hline
    \text{extended bilinears} & \Big( \Id_{\rho} \otimes \Gaf \Big) & \Big( \rho_{2} \otimes \Gaf \Big) \\
    \text{local bilinears} & \Big( \rho_{3}   \otimes \Gaf \Big) & \Big( \rho_{1} \otimes \Gaf \Big) \\
  };
  \path[->]
    (m-2-2) 
            edge [thick] node [below] {\tm{+\de\mK_{m_3^2} +\de\mB_{r^2}}} (m-2-3)
    (m-3-2) 
            edge [thick] node [left] {\tm{-\mB_{h} \mK_+\inv}} (m-2-2)
            edge [thick] node [above] {\tm{+\de\mK_{m_3^2} + \de\mB_{r^2}}} (m-3-3)
    (m-3-3) 
            edge [thick] node [right] {\tm{-\mB_{h} \mK_-\inv}} (m-2-3)
  ;
\end{tikzpicture}
\caption{%
Taste pattern of the pseudo-\ac{NG} bosons with \ac{KW} fermions. 
The unbroken pseudo-\ac{NG} boson has the local correlator 
of Eq.~\eqref{eq:KW_corr_rho3_Ga5_KB}. 
 The denominators of the other channels are modified according to the pattern given in the diagram. 
\label{fig:KW_taste_pattern}%
} 
\end{figure}
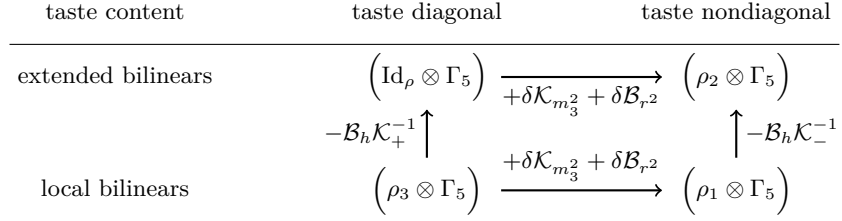%
The low-energy levels in these channels are still finite and meaningful, since we can access the same spectra in other correlation functions with extended fermion bilinears, cf.~Table~\ref{tab:KW_meson}, which are free from the divergences caused by the single-site local ones. 
Thus, the divergent behavior only causes vanishing overlap factors, while the physical energy levels are protected. 
We derive the \ac{LO} quark-mass dependence and \ac{LO} cutoff effects of these two channels with \emph{two-shift extended fermion bilinears} in Appendix~\ref{app:KW_pNGB3}.
There is respective agreement between the Eqs.~\eqref{eq:KW_corr_ext_Irho_Ga5_KB} or~\eqref{eq:KW_corr_ext_rho2_Ga5_KB} with Eqs.~\eqref{eq:KW_corr_rho3_Ga5_KB} or~\eqref{eq:KW_corr_rho1_Ga5_KB} up to the extra cutoff effects \tm{- \mB_{h} \mK_{\pm}\inv} and the overall factors \tm{c_{h}^2}, where \tm{\mB_{h}} and \tm{c_h} are defined in Eq.~\eqref{eq:KW_shift_scales}. 
Thus,the leading modificaton is an overall rescaling and a second type of \tm{\mO(\aD^2\La^2)} \ac{TSB} cutoff effect \tm{-\mB_{h} \mK_{\pm}\inv}. 
The details of \tm{\mB_{h}} depend on the interpolating operator, must be present for naive or \aclp{SF}, too, and depend on the \ac{TSB} cutoff effects of the \ac{KW} fermion propagators only in a subleading manner. 
In particular, \tm{\mB_{h}\mK_{\pm}\inv} could be made arbitrarily small with sufficiently smooth links, but not by using an improved \ac{KW} operator. 
This is a stark contrast to the first type of \tm{\mO(r^2\aD^2\La^2)} \ac{TSB} cutoff effects \tm{\delta\mB_{r^2}} between the \tm{\big( \rho_1 \otimes \Gaf \big)}- and the \tm{\big( \rho_3 \otimes \Gaf \big)} channels, i.e.~Eq.~\eqref{eq:KW_LTB_delta}, which are expected to be rather insensitive to smooth links, but much reduced by using an improved \ac{KW} operator. 
Moreover, \tm{\delta\mB_{r^2}} is obviously absent for naive or \aclp{SF}. 
This pattern of the two types of \ac{TSB} cutoff effects is unique to \ac{MDF} and sketched in Fig.~\ref{fig:KW_taste_pattern}.
The simplicity of the underlying arguments suggests that this pattern repeats within the taste multiplets of many different channels. 
\vskip1ex

We see in Fig.~\ref{fig:KW_taste_pattern} that there is---in contrast to a naive expectation from staggered hadrons with \aclp{SF}---not a naive \emph{1-2-1 taste multiplet} pattern. 
However, the physical charged pseudo-\ac{NG} bosons use linear combinations of two states with different \acl{ST} structure, namely, those excited by the single-site local fermion bilinears of the \tm{\big( \rho_{1} \otimes \Gaf \big)} channel or those excited by the extended fermion bilinears of the \tm{\big( \rho_{2} \otimes \Gaf \big)} channel. 
Thus, while their asymptotic behavior is controlled by the taste-symmetric \tm{\big( \rho_{1} \otimes \Gaf \big)} bilinears, there is a low-lying excited state due to the taste-antisymmetric \tm{\big( \rho_{2} \otimes \Gaf \big)} bilinears, which could be made degenerate at \tm{\mO(\aD^2\La^2)} with sufficiently smooth links. 
While the taste-symmetric \tm{\big( \rho_{1} \otimes \Gaf \big)} bilinears determine the ground states in hadronic correlation functions, the determinant obviously receives contributions from the antisymmetric \tm{\big( \rho_{2} \otimes \Gaf \big)} states, too. 
The quark-connected correlation functions of the taste-antisymmetric \tm{\big( \rho_{3} \otimes \Gaf \big)}- or  taste-symmetric \tm{\big( \Id_{\rho} \otimes \Gaf \big)}-bilinears could be made degenerate at \tm{\mO(\aD^2\La^2)} with sufficiently smooth links, too. 
Thus, these \acl{TD} channels are distinguished at the \ac{LO} only by their quark-disconnected contributions, which couple to the \acl{TS} or \acl{TI} masses, respectively, as expected from \ac{ChPT}~\cite{Gasser:1983yg}. 
Thus, there is indeed a \emph{1-2-1 taste multiplet} pattern of the physical states. 
The first splitting between the neutral or charged states is due to the \acl{TN} scalar (or pseudoscalar) operator, i.e. the second to rightmost operator structure in Eq.~\eqref{eq:D_KW_comp} and the corresponding condensate. 
The second splitting between the \acl{TS} or the \acl{TI} states is due to the \acl{TD} vector operator, i.e. the rightmost operator structure in Eq.~\eqref{eq:D_KW_comp} coupling to the \tm{U(1)} axial anomaly in the quark-disconnected contributions. 
\vskip1ex

The quark-connected correlation functions of these channels, shown in Fig.~\ref{fig:KW_corr_mes} for one lattice spacing and without a \acl{TI} mass term, show a modest 10\% difference between the ground state energies of, on the one hand, the 
\tm{\big( \rho_3 \otimes \Gaf \big)} channel with a smaller mass, or, on the other hand, the \tm{\big( \rho_1 \otimes \Gaf \big)}, \tm{\big( \rho_2 \otimes \Gaf \big)} or \tm{\big( \Id_{\rho} \otimes \Gaf \big)} channels with larger masses. 
This is in line with our predictions, despite that the interpolating operators were poorly chosen, see Sec.~\ref{sec:KW_meson}. 
\vskip1ex
% \clearpage
%%%%%%%%%%%%%%%%%%%%%%%%%%%%%%%%%%%%%%%%%%%%%%%%%%%%%%%%%%%%%%%%%%%%%%%%%%%%%%%%
% \input{\INCDIR/summary}
%% summary.tex
\acresetall

\section{Summary and conclusions}\label{sec:summary}

In this paper, we have derived the \emph{\acl{ST} representation of \acl{KW} fermions} from first principles with no additional assumptions, fully classified the properties of \ac{KW} fermion bilinear operators, and derived the most important implications of using of \ac{KW} fermions as dynamical or valence degrees of freedom in gauge theories. 
Our treatise covers parallel and perpendicular \ac{KW} fermions highlighting similarities and differences. 
\vskip1ex

In the first part we inferred the \acl{ST} representation from first principles and analytic symmetry arguments. 
For a long time it has been known~\cite{Weber:2016dgo} that the key observation to permit understanding the \acl{ST} structure is the \emph{mirror fermion symmetry}~\cite{Pernici:1994yj}, which must be interpreted as a \emph{tasted reflection} or \emph{tasted charge conjugation} symmetry. 
The corresponding \emph{representation of taste generators} \tm{\{\vr\}} is defined in Eq.~\eqref{eq:KW_vr} and leaves the naive Dirac operator invariant, but implies different \acl{ST} structures for the two relevant operators making up the \ac{KW} term. 
\emph{Taste projectors} \tm{P^{\vr}_{\pm}} defined in Eq.~\eqref{eq:KW_Pvr} extract the taste components from the \ac{KW} spinors. 
We have derived the \emph{\acl{ST} decomposition} of \ac{KW} spinors in position and momentum space, where Eq.~\eqref{eq:KW_proj} permits an interpretation of the \acl{ST} structure as a direct product, Eq.~\eqref{eq:KW_ST_int}.
This \emph{\acl{ST} representation} uses the same chiral representation for the spin gamma matrices of both tastes, \emph{ties each site-parity to the same chirality for both tastes}, and distributes the chiral components of each taste across the two boson sites making up one fermion hypersite. 
Hence, taste components require both boson sites to access their full spin content, or in momentum space from two maximally separated Fermi points, see Eqs.~\eqref{eq:KW_proj_mom} and~\eqref{eq:KW_comp_mom}. 
Due the structure of the taste generator \tm{\vr_{3}} the \acl{ST} components are subject to \emph{parity mixing}, Eq.~\eqref{eq:KW_parity_mixing}, and the \emph{symmetry under parity} simply swaps the association between chirality of the tastes and site parity. 
This ultimately leads to the \emph{\acl{ST} decomposition} of the \ac{KW} operator in Eq.~\eqref{eq:D_KW_comp}, which is consistent with the previous \acl{ST} identification~\cite{Kimura:2011ik}. 
While being largely compatible with our analysis, the origin and the physical implications of the \acl{ST} structure, or the appearance of fake \ac{TSB} terms had not been discussed~\cite{Kimura:2011ik}. 
In our formalism one of the two relevant operators is a scalar (or pseudoscalar term due to parity mixing) including a nontrivial phase factor, which does not have any continuum analog. 
It has to be pointed out that this nontrivial \acl{ST} representation and the remnant \acl{TD} isovector chiral symmetry \tm{\big( \rho_3 \otimes \Gaf \big)} have to be the underpinning of any Symanzik \ac{EFT} or chiral Lagrangian analysis. 
The symmetries have been accounted for in the discussion of the construction of the \tm{\mO(\aD)} operators of the Symanzik \ac{EFT}~\cite{Weber:2015hib}. 
However they have not been taken into account in the first attempt to construct a chiral Lagrangian~\cite{Shukre:2024tkw} that was based on the incorrect assertion of a \acl{ST} decomposition via time smearing~\cite{Tiburzi:2010bm}, which has not been refuted so far despite being obviously incompatible with the previous \acl{ST} identification~\cite{Kimura:2011ik}. 
\vskip1ex

In the middle part, we derive from the \acl{ST} representation on the quark level the \acl{ST} content of fermion bilinear operators, which excite parity partner states, too. 
On the one hand, we confirm in Sec.~\ref{sec:KW_meson} for the case of four tastes of pseudo-\ac{NG} bosons that the predictions in Table~\ref{tab:KW_meson} are consistent with existing lattice data~\cite{Weber:2015oqf} despite the deficiencies of the numerical simulation that have been realized only in hindsight. 
On the other hand, we discuss in Sec.~\ref{sec:KW_baryon} predictions for diquarks, Table~\ref{tab:KW_diquark}, or baryons, e.g. Eqs.~\eqref{eq:KW_proton_2L} or~\eqref{eq:KW_proton_0L}, which cannot be compared to actual lattice data, since there are none available yet. 
We also derive the pattern of spin \tm{1} partner states arising for interpolating operators with odd numbers of shifts in all spin \tm{0} channels with perpendicular \ac{KW} fermions, see Sec.~\ref{sec:KW_bilinears_per}, and show that the prediction, Table~\ref{tab:KW_meson_per}, is consistent with existing lattice data~\cite{Weber:2015oqf}. 
We understand from Table~\ref{tab:KW_meson} how to include further \acl{TS} or \acl{TI} bilinear operators in the \ac{KW} fermion action. 
This has been known for a long time for mass terms, see Sec.~\ref{sec:KW_mass}, and one of the proposed operators has been applied successfully for studying topology~\cite{Durr:2022mnz, Kishore:2025fxt}.
The derivation of the chemical potential terms in Sec.~\ref{sec:KW_chempot} is completely new. 
In particular, we demonstrate why previous approaches to interpret \ac{KW} fermions as fermions with a flavored chemical potential~\cite{Misumi:2012uu, Misumi:2012ky} cannot realize the theory that was originally intended. 
While \acl{TS} chemical potential can be implemented naively, \acl{TI} chemical potential involves various nontrivial subtleties and may require squire-movelike three-link terms. 
Having a robust understanding of \acl{TS} and \acl{TI} chemical potential is crucial for applying \ac{KW} fermions in \ac{QCD} thermodynamics at finite density with confidence. 
\vskip1ex

In the last part we address the use of \ac{KW} fermions in gauge theories. 
We clarify the \acl{ST} structure of the counterterms in Sec.~\ref{sec:KW_renormalization} and derive the parametric form and automatic \tm{\mO(a)} improvement of the \ac{KW} determinant in Sec.~\ref{sec:KW_determinant}. 
We move on to analyze the modification of quark propagators and hadronic correlation functions in the mistuned theory, where all nontrivial modifications can be expressed in terms of a single quantity \tm{\ka} defined in Eq.~\eqref{eq:KW_ka}. 
We derive in Sec.~\ref{sec:KW_mistuning} the appearance of oscillating modes in quark propagators at nontrivial frequencies \tm{\sfrac{\ka}{\aD}} that are not eigenfrequencies of the underlying space-time lattice. 
These oscillations are known~\cite{Weber:2015oqf} to feed through into \acl{TN} hadronic correlation functions. 
It has been known~\cite{Weber:2015hib} for a long time that the hadronic frequency \tm{\sfrac{2\ka}{\aD}} can be employed for reliably tuning the relevant counterterm, and \emph{remains reliable for tuning at arbitrary lattice spacing and without significant quark mass dependence}. 
We derive the fundamental, underlying mechanism in Sec.~\ref{sec:KW_oscillations} of this paper.
We derive in Sec.~\ref{sec:KW_dispersion} how this mistuning contributes to quark dispersion relations. 
Comparing quark dispersion relations of parallel or perpendicular \ac{KW} fermions we predict the qualitative \tm{\ka} dependence of hadronic energy levels, and confirm these predictions by comparing to existing lattice data~\cite{Weber:2015oqf}. 
Usually differences between temporal or spatial rest energies are a viable device for tuning the fermion anisotropy, see e.g.~\cite{Bazavov:2023ods}. 
However, for \ac{KW} fermions these differences are controlled by \tm{\ka} instead, whereas the influence of the anisotropy is a subleading effect. 
Although the \tm{\ka} dependence of the rest energies has been used in the first attempt to nonperturbatively tune the \ac{KW} operator~\cite{Weber:2013tfa}, \emph{our analysis shows that this approach to tuning is fundamentally flawed} and expected to be unreliable on coarser lattices. 
Lastly, we study an important subclass of hadronic correlation functions with \ac{KW} fermions (hadronic two-point functions with symmetric taste structure at source and sink) and identify properties analogous to those of the determinant, i.e. the parametric form and the automatic \tm{\mO(\aD)} improvement. 
We conduct a scaling analysis for all four tastes in the pseudoscalar sector (spin structure \tm{\Gaf}) and discern the leading quark mass dependence and leading \ac{TSB} cutoff effects. 
In particular, we find \emph{two qualitatively different types of \ac{TSB} cutoff effects}.
The first type of \ac{TSB} cutoff effects already exists at the tree level between two states whose interpolating operators differ by a taste generator \tm{\vr_{2}} and could be diminished for a given lattice spacing primarily through the Symanzik improvement program~\cite{Symanzik:1983dc, Symanzik:1983gh}. 
In its continuum extrapolation with only an \tm{\aD^2} term a small, negative remnant remains~\cite{Weber:2015oqf, Godzieba:2024uki}. 
The second type of \ac{TSB} cutoff effects is analogous to those of naive or \aclp{SF} and arises from the need to use extended interpolating operators (in order to have a finite continuum limit of the correlation function) and could be made arbitrarily small for sufficiently smooth links. 
These analytic predictions of the cutoff effects are qualitatively consistent with existent lattice data~\cite{Weber:2015oqf} and the quark mass dependence of \ac{NLO} \ac{ChPT}~\cite{Gasser:1983yg}. 
However, the pattern of cutoff effects we derived is incompatible with many operators at odd powers of the lattice spacing that were included in the first attempt to construct a chiral Lagrangian for \ac{KW} fermions~\cite{Shukre:2024tkw}. 
\vskip1ex

Existing lattice data broadly support the predicted \acl{ST} representation of \ac{KW} fermions, but are insufficiently precise to confirm or disprove our predictions. 
A key reason is that the numerical studies were restricted to a very small subset of observables with poorly chosen, nonoptimal interpolating operators, since the setups were not informed by an accurate understanding of the underlying \acl{ST} representation. 
Therefore, the \acl{ST} representation needs to be confirmed (or disproved) in a dedicated precision study with optimized interpolating operators, and must be extended to a wider range of hadronic channels. 
In particular, in order to employ \ac{KW} fermions for \ac{QCD} thermodynamics at finite density, which appears to date as one of the most evident use cases, the equation of state from a realisitic hadron gas model~\cite{Karsch:2003vd} including the \ac{TSB} cutoff effects of \ac{KW} fermions for all quantitatively relevant hadrons is a necessary baseline to compare to. 
Such lattice hadron resonance gas models are known to be very successful for different types of \aclp{SF}~\cite{Huovinen:2009yb}.
\vskip1ex

A very recent numerical study of \ac{TSB} with \ac{KW} fermions has been posted to the arXiv~\cite{Borsanyi:2025big}. 
It is no surprise that its results, in particular, the \ac{TSB} pattern of pseudo-\ac{NG} bosons comply with our predictions in Fig.~\ref{fig:KW_taste_pattern}, which were made before becoming aware of the upcoming numerical results. 
The actual \ac{TSB} cutoff effects in the squared energy of the \tm{\big( \rho_{2} \otimes \Gaf \big)} channel are to a good accuracy the sum of those in the \tm{\big( \Id_{\rho} \otimes \Gaf \big)} or \tm{\big( \rho_{1} \otimes \Gaf \big)} channels. 
Those between \acl{TN} and corresponding \acl{TD} channels strongly depend on improvement of the \ac{KW} operator, while those within the set of \acl{TD} or \acl{TN} channels are largely insensitive to it. 
The latter instead depend on smoothness of the links in the shifts of the interpolating operators. 
Thus, our analytic, assumption-free, nonperturbative predictions are validated consistently across all so far available numerical results, and we can declare that \emph{\acl{ST} representation of \aclp{KWF} is thoroughly understood from first principles}.

% \clearpage
%%%%%%%%%%%%%%%%%%%%%%%%%%%%%%%%%%%%%%%%%%%%%%%%%%%%%%%%%%%%%%%%%%%%%%%%%%%%%%%%
\section*{Acknowledgments}
The author thanks Michael Creutz, Stephan D{\"u}rr, Andreas Kronfeld, and Peter Petreczky for invaluable discussions over many years, which triggered some key insights leading up to this paper. 
The author's research has been funded by the Deutsche Forschungsgemeinschaft (DFG, German Research Foundation)---Projektnummer 417533893/GRK2575 ``Rethinking Quantum Field Theory''. 
The author acknowledges the support by the State of Hesse within the Research Cluster ELEMENTS (Project ID 500/10.006). 
% \clearpage
\section*{Data availability}
The data that support the findings of this article are not publicly available. 
The data are available from the authors upon reasonable request.
\appendix
%%%%%%%%%%%%%%%%%%%%%%%%%%%%%%%%%%%%%%%%%%%%%%%%%%%%%%%%%%%%%%%%%%%%%%%%%%%%%%%%
% \input{\INCDIR/kawi_other}
%% kawi_divergent.tex
% \acresetall

\section{The other pseudo-Nambu-Goldstone bosons}\label{app:KW_other_pNGB_details}

Here we discuss in detail the calculations for the other two pseudo-\ac{NG} bosons, namely in the \tm{\big( \Id_{\rho} \otimes \Gaf \big)} or \tm{\big( \rho_2 \otimes \Gaf \big)} channels. 
Namely, we calculate the divergence due to \emph{single-site local fermion bilinears} in Appendix~\ref{app:KW_pNGB2}, and the leading \ac{TSB} cutoff effects for \emph{two-shift extended fermion bilinears} in Appendix~\ref{app:KW_pNGB3}. 
The main results have been summarized already in Sec.~\ref{sec:KW_other_pNGB_summary}. 
\vskip1ex

\subsection{Pseudo-Nambu-Goldstone bosons with \tm{\gaDf\etD}- or \tm{\etf}-fermion bilinears}\label{app:KW_pNGB2}

We start with \tm{\big( \Id_{\rho} \otimes \Gaf \big)} with single-site local interpolating operators. 
The two remaining factors \tm{\vr_{3}} are used to invert the sign of \tm{W_{0}} in either one or the other of the propagators, \tm{\mDKW\inv} or \tm{(\mDKW\inv)\dag}. 
Hence, the full correlation function is the average of both versions. 
We have 
\nalign{
C_{\etf}^{\text{qc}} 
&=
 -\Tr\Big\{ 
 \mDKW\inv \vr_{3}
 \mDKW\daginv \vr_{3}
  \Big\} 
 =
 -\Real\Tr\Big\{ 
 \Big[ \DN + W_1 + W_0 + M_{0} + M_{3} \Big]
 \Big[ \DN + W_1 - W_0 + M_{0} + M_{3} \Big]\dag
  \Big\}\inv
~\label{eq:KW_corr_Irho_Ga5_step1}\\
&=
 -\frac{1}{2}\Tr\Big\{ 
 \Big[
 | K_{3+} |^2 - | L_{3} |^2
 - K_{3+}\dag L_{3} + L_{3}\dag K_{3+}
 \Big]\inv
 +
 \Big[
 | K_{3+} |^2 - | L_{3} |^2
 + K_{3+}\dag L_{3} - L_{3}\dag K_{3+}
 \Big]\inv
 \Big\}
~\label{eq:KW_corr_Irho_Ga5_step2}\\
&=
 -\frac{1}{2}\Tr\Big\{ 
  \Big[
 E_{3+}
 - O_{3}
 \Big]\inv
 +
 \Big[
 E_{3+}
 + O_{3}
 \Big]\inv
 \Big\} 
 =
 -\Tr\Big\{ 
 \big[E_{3+}-O_{3}E_{3+}\inv O_{3}\big]\inv
 \Big\}
,~\label{eq:KW_corr_Irho_Ga5_step3}
}%nalign 
where we have first defined \tm{K_{3\pm} \equiv
\DN \pm W_1 + M_{0} + M_{3}} and \tm{L_{3} \equiv W_{0}}, and later 
\nequn{
\begin{aligned}
E_{3\pm} 
&\equiv 
|K_{3\pm}|^2 - |L_{3}|^2 
= 
|\DN|^2 + |W_1|^2 -|W_{0}|^2 + |M_{0} + M_{3}|^2 \mp \big[ W_{1}, \DN \big]_{+}
,~\\
O_{3} 
&\equiv 
K_{3\pm}\dag L_{3} - L_{3}\dag K_{3\pm} 
= 
-\big[\DN, W_{0} \big]_{-} + 2(M_{0} + M_{3}) W_0
~.~
\end{aligned}
\label{eq:KW_EO3}
}%nequn 
In the second to last step, we have used that the operators have spectra that are either purely real, i.e., \tm{M_{0}} or \tm{M_{3}}, or purely imaginary, i.e.,  \tm{\DN},~\tm{W_{0}}, or~\tm{W_{1}}. 
\tm{E_{3\pm}} is a real function, and in the renormalized theory the quadratic power-law divergence cancels between 
\tm{|W_{1}|^2} and \tm{-|W_{0}|^2}. 
Yet, the last term \tm{\mp \big[ W_{1}, \DN \big]_{+}} is linear in \tm{r_{1}} and linearly power-law divergent. 
\tm{O_{3}} is anti-Hermitian and independent of \tm{r_{1}}, but linear in \tm{r_{0}} and linearly power-law divergent.
We have applied Eq.~\eqref{eq:prop_decomp} in the last step of Eq.~\eqref{eq:KW_corr_Irho_Ga5_step3}, where the second term in the last bracket of Eq.~\eqref{eq:prop_decomp}, i.e. \tm{\propto -B}, cancels in the sum under the trace.
While Eq.~\eqref{eq:KW_corr_Irho_Ga5_step3} is already manifestly even in \tm{r_0}, it is not yet even in \tm{r_1}, and in particular, contributions even or odd in \tm{r_1} are mixed in \tm{E_{3+}\inv}. 
Powers of those odd contributions could potentially cancel some power-law divergences from the two factors \tm{O_{3}}. 
We analyze the power-law divergence setting \tm{m_3=0} (resp. \tm{M_{3}=0}) to guarantee even powers of \tm{r} and \tm{\aD}. 
We define 
\nequn{
E_{3\pm} 
=
E_{31} \pm O_{31}
,\quad 
E_{31} 
\equiv
|\DN|^2 + |W_{1}|^2 -|W_{0}|^2 + M_{0}^2
,\quad
O_{31} 
\equiv
-\big[ W_{1}, \DN \big]_{+}
~.~\label{eq:KW_EO31}
}%nequn
In order to average \tm{\pm r_{1}} we split \tm{E_{3+}\inv} up via Eq.~\eqref{eq:prop_decomp} and reorganize terms via 
\nequn{
F_{31} \equiv \big[E_{31}-O_{31}E_{31}\inv O_{31}\big]
,\quad
E_L 
\equiv E_{31} - O_{3}F_{31}\inv O_{3}
,\quad
O_L 
\equiv O_{31} + O_{3}F_{31}\inv O_{31}E_{31}\inv O_{3}
,~\label{eq:KW_F31}
}%nequn
in order to invoke Eq.~\eqref{eq:prop_decomp} again and obtain 
\nalign{
C_{\etf}^{\text{qc}} 
&=
 -\frac{1}{2}\Tr\Big\{ 
 \Big[ E_{31}+O_{31}
 -O_{3}F_{31}\inv 
 \big[1 - O_{31}E_{31}\inv\big] O_{3} \Big]\inv
 +
 \Big[ E_{31}-O_{31}
 -O_{3}F_{31}\inv 
 \big[1 + O_{31}E_{31}\inv\big] O_{3} \Big]\inv
 \Big\} 
~\label{eq:KW_corr_Irho_Ga5_step4}\\
&=
 -\frac{1}{2}\Tr\Big\{ 
 \big[  E_L + O_{L} \big]\inv
 +
 \big[ E_L - O_L \big]\inv
 \Big\} 
 =
 -\Tr
 \Big\{ E_L
 -O_{L} E_L\inv O_{L} \Big\}\inv 
~\label{eq:KW_corr_Irho_Ga5_step5}\\
&=
 -\Tr\Big\{
 \big[ E_{31} - O_{3}F_{31}\inv O_{3} \big]
 - \big[ O_{31} + O_{3}F_{31}\inv O_{31}E_{31}\inv O_{3} \big] 
 \big[ E_{31} - O_{3}F_{31}\inv O_{3} \big]\inv 
 \big[ O_{31} + O_{3}F_{31}\inv O_{31}E_{31}\inv O_{3} \big] 
 \Big\}\inv 
~.~\label{eq:KW_corr_Irho_Ga5_step6}
}%nalign 
We see from Eqs.~\eqref{eq:KW_EO3},~\eqref{eq:KW_EO31},~and~\eqref{eq:KW_F31} that \tm{E_{31}} is finite, while \tm{O_{31}} or \tm{O_{3}} both linearly diverge as \tm{\sfrac{1}{\aD}}. 
\tm{F_{31}} is quadratically divergent as \tm{\sfrac{1}{\aD^2}}. 
After factoring out its divergence, we may reformulate \tm{F_{31}\inv} as a geometric series, 
\nalign{
F_{31}\inv
&=
\Big\{E_{31}-O_{31}E_{31}\inv O_{31}\Big\}\inv
=
\Big\{ \big[1 - \big(E_{31} O_{31}\inv\big)^2 \big] \big[-O_{31}E_{31}\inv O_{31}\big] \Big\}\inv
~\nonumber\\
&=
\big[-O_{31}E_{31}\inv O_{31}\big]\inv \big[1 - \big(E_{31} O_{31}\inv\big)^2 \big]\inv
 =
-O_{31}\inv \sum\limits_{l=0}^{\infty} \big(E_{31} O_{31}\inv\big)^{2l+1}
,~\label{eq:KW_corr_Irho_geom1}
}%nalign
such that all higher terms (\tm{l>0}) in the \tm{l}-sum produce contributions suppressed by \tm{\aD^{2l}}. 
We apply Eq.~\eqref{eq:KW_corr_Irho_geom1} to any \tm{F_{31}\inv} and obtain 
\nalign{
C_{\etf}^{\text{qc}} 
&=
 -\Tr\Big\{
 \Big( E_{31} + \underline{O_{3} O_{31}\inv} 
 \sum\limits_{l_0=0}^{\infty} \big(E_{31} O_{31}\inv\big)^{2l_0}
 E_{31} \underline{O_{31}\inv O_{3}} \Big)
 - O_{31} \Big( \Id - \underline{O_{31}\inv O_{3}}
 \sum\limits_{l_1=0}^{\infty} \big(O_{31}\inv E_{31}\big)^{2l_1}
 \underline{O_{31}\inv O_{3}} \Big) \times
 ~\nonumber\\
 &\phantom{-\Tr\Big\{}\times
 \Big( E_{31} + \underline{O_{3} O_{31}\inv} 
 \sum\limits_{l_2=0}^{\infty} \big(E_{31} O_{31}\inv\big)^{2l_2}
 E_{31} \underline{O_{31}\inv O_{3}} \Big)\inv 
 O_{31} \Big( \Id - \underline{O_{31}\inv O_{3}}
 \sum\limits_{l_3=0}^{\infty} \big(O_{31}\inv E_{31}\big)^{2l_3}
 \underline{O_{31}\inv O_{3}} \Big) 
 \Big\}\inv 
,~\label{eq:KW_corr_Irho_Ga5_step8}
}%nalign
where the underlined factors of \tm{\underline{G_R} = \underline{O_{3} O_{31}\inv}} or \tm{\underline{G_L} = \underline{O_{31}\inv O_{3}}} contribute at the leading order as \tm{\aD^0}. 
The inverse factor can be reformulated as yet another geometric series by extracting the \tm{l_2=0} term from the \tm{l_2}-sum, 
\nalign{
&\Big( E_{31} + \underline{G_R} \sum\limits_{l_2=0}^{\infty} \big(E_{31} O_{31}\inv\big)^{2l_2}
 E_{31} \underline{G_L} \Big)\inv 
=
\Big( \big[ E_{31} + \underline{G_R} E_{31} \underline{G_L} \big]
 + \underline{G_R} \sum\limits_{l_2=1}^{\infty} \big(E_{31} O_{31}\inv\big)^{2l_2}
 E_{31} \underline{G_L} \Big)\inv 
~\nonumber\\
&=
\sum\limits_{k=0}^{\infty} \Big( 
 -\big[ E_{31} + \underline{G_R} E_{31} \underline{G_L} \big]\inv 
 \underline{G_R} \sum\limits_{l_2=1}^{\infty} \big(E_{31} O_{31}\inv\big)^{2l_2} E_{31} \underline{G_L} 
 \Big)^{k} 
 \Big[ E_{31} + \underline{G_R} E_{31} \underline{G_L} \Big]\inv
,~\label{eq:KW_corr_Irho_geom2}
}%nalign
to arrive at the final result 
\nalign{
C_{\etf}^{\text{qc}} 
&=
 -\Tr\Big\{
 \Big( E_{31} + \underline{G_R} \sum\limits_{l_0=0}^{\infty} \big(E_{31} O_{31}\inv\big)^{2l_0}
 E_{31} \underline{G_L} \Big)
 - O_{31} \Big( \Id - \underline{G_L}
 \sum\limits_{l_1=0}^{\infty} \big(O_{31}\inv E_{31}\big)^{2l_1}
 \underline{G_L} \Big) \times
~\nonumber\\
 &\phantom{-\Tr\Big\{}\times
 \sum\limits_{k=0}^{\infty} \Big( 
 -\big[ E_{31} + \underline{G_R} E_{31} \underline{G_L} \big]\inv 
 \underline{G_R} \sum\limits_{l_2=1}^{\infty} \big(E_{31} O_{31}\inv\big)^{2l_2} E_{31} \underline{G_L}
 \Big)^{k} 
 \times
~\nonumber\\
&\phantom{-\Tr\Big\{}\times
 \Big[ E_{31} + \underline{G_R} E_{31} \underline{G_L} \Big]\inv
 O_{31} \Big( \Id - \underline{G_L}
 \sum\limits_{l_3=0}^{\infty} \big(O_{31}\inv E_{31}\big)^{2l_3}
 \underline{G_L} \Big) 
 \Big\}\inv 
,~\label{eq:KW_corr_Irho_Ga5_step9}
}%nalign
which permits the scaling analysis via Eq.~\eqref{eq:KW_scales}. 
Namely, we resolve \tm{E_{31}}, \tm{O_{31}}, and \tm{G_{R,L}} via Eqs.~\eqref{eq:KW_EO3} and~\eqref{eq:KW_EO31},  
\nalign{
E_{31} 
&\sim  \Big[ c_{|D|^2} -2 C_{W_{0}}C_{W_{\Si}} + \mO\left( \frac{m_q^2}{\La^2},\aD^2\La^2 \right) \Big] \La^2
,~\label{eq:KW_corr_Irho_scaling1}\\
O_{31} 
&\sim  \Big[ -C_{W_{0}D}^{(+)}
 + \mO\left( \aD^2\La^2 \right)
\Big] \frac{1}{\aD} 
,~\label{eq:KW_corr_Irho_scaling2}\\
\underline{G_R} 
= (\big[\DN, W_{0} \big]_{-})(\big[ W_{1}, \DN \big]_{+})\inv
 &\sim C_{W_{0}D}^{(-)} C_{W_0D}^{(+)\,-1} + \mO\left( \aD^2\La^2 \right)
,~\label{eq:KW_corr_Irho_scaling3}
\\
\underline{G_L} 
= (\big[ W_{1}, \DN \big]_{+})\inv(\big[\DN, W_{0} \big]_{-})
 &\sim C_{W_0D}^{(+)\,-1} C_{W_{0}D}^{(-)} + \mO\left( \aD^2\La^2 \right)
,~\label{eq:KW_corr_Irho_scaling4}
\\
\underline{G_L}
 \sum\limits_{l=0}^{\infty} \big(O_{31}\inv E_{31}\big)^{2l}
 \underline{G_L} 
&\sim 
 + \Big( C_{W_0D}^{(+)\,-1} C_{W_{0}D}^{(-)} \Big)^2 
 + \mO\left( \aD^2\La^2 \right)
,~\label{eq:KW_corr_Irho_scaling5}
\\
\Big( E_{31} + \underline{G_R} E_{31} \underline{G_{L}} \Big)
&\sim \Big[ c_{|D|^2} -2 C_{W_{0}}C_{W_{\Si}} 
+ C_{W_0 D}^{(-)} C_{W_0D}^{(+)\,-1} \big( c_{|D|^2} -2 C_{W_{0}}C_{W_{\Si}} \big) C_{W_0D}^{(+)\,-1} C_{W_0 D}^{(-)} 
\nonumber\\&
+ \mO\left( \frac{m_q^2}{\La^2},\aD^2\La^2 \right) \Big] \La^2
,~\label{eq:KW_corr_Irho_scaling6}\\
\Big( \underline{G_R} \sum\limits_{l=1}^{\infty} \big(E_{31} O_{31}\inv\big)^{2l} E_{31} \underline{G_L} \Big) 
&\sim \mO\left( {\aD^2 m_q^2 \La^2},\aD^2\La^4 \right)
~.~\label{eq:KW_corr_Irho_scaling7}
}%nalign
We point out that the leading contribution from the product \tm{C_{W_{0}}C_{W_{\Si}}} has a trivial Dirac structure, too. 
The \tm{k=0} term contributes to the divergence, since it scales as \tm{\aD^0}, while higher terms are suppressed by \tm{\aD^{2kl_2}}.  
Thus, we may ignore these higher terms in the following. 
On the one hand, we see that the subtrahend in the trace is a quadratic power-law divergence scaling as \tm{\mO(\aD^{-2})}, since Eq.~\eqref{eq:KW_corr_Irho_scaling5} is different from one at leading order. 
On the other hand, the minuend scales as \tm{\La^2}. 
Thus, the divergent contribution in the denominator is 
\nalign{
&
 \Big\{
 -\Big[ C_{W_{0}D}^{(+)} - C_{W_{0}D}^{(-)} C_{W_0D}^{(+)\,-1} C_{W_{0}D}^{(-)} \Big] 
 \Big[ \big( c_{|D|^2} -2 C_{W_{0}}C_{W_{\Si}} \big) + C_{W_0 D}^{(-)} C_{W_0D}^{(+)\,-1} 
 \nonumber\\&\times 
 \big( c_{|D|^2} -2 C_{W_{0}}C_{W_{\Si}} \big) C_{W_0D}^{(+)\,-1} C_{W_0 D}^{(-)} \Big]
 \Big[ C_{W_{0}D}^{(+)} - C_{W_{0}D}^{(-)} C_{W_0D}^{(+)\,-1} C_{W_{0}D}^{(-)} \Big] 
 \Big\}%\inv 
  \aD^{-2}
,~\label{eq:KW_corr_Irho_Ga5_divergence}
}%nalign
and the overall correlator vanishes in the continuum limit. 
\vskip1ex

We then turn our attention to the leading quark mass dependence and set \tm{r=0} (losing minimal doubling in the process). 
Then \tm{O_{3} = 0} and \tm{E_{3} = |\DN|^2 + |M_{0} + M_{3}|^2} and therefore \tm{ C_{\etf}^{\text{qc}} = -\Tr \big\{ \big[ |\DN|^2 + |M_{0} + M_{3}|^2 \big]\inv \big\} }. 
Due to the evenness in \tm{m_3}, which we showed on general grounds in Sec.~\ref{sec:KW_parametric}, the mixed term cannot contribute at leading order in \tm{m_3}. 
Thus, the leading \tm{m_q} scaling behavior is the same in Eq.~\eqref{eq:KW_corr_rho3_Ga5_scaling}, 
\nequn{
C_{\etf}^{\text{qc}}
 =
 -\Tr\Big\{ 
   c_{|D|^2}\La^2 
 + \big[ c_{m}^2 + c_{M}^2 \big] m_{q}^2 
%  + \mO\left( \frac{m_q^4}{\La^2},\aD^2\La^4 \right) 
  \Big\}\inv
 + \mO\left( \frac{m_q^4}{\La^6},\aD^2 \right) 
 =
 -\Tr~\mK_+\inv + \mO\left( \frac{m_q^4}{\La^6},\aD^2 \right)
,~\label{eq:KW_corr_Irho_Ga5_mq_scaling}
}%nequn
where we recognize the terms contributing to the continuum limit of the \tm{\big( \rho_3 \otimes \Gaf \big)} channel, Eq.~\eqref{eq:KW_mK+}. 
Thus, the \tm{\big( \Id_{\rho} \otimes \Gaf \big)} channel indeed contains \acl{TD} states. 
\vskip1ex

For the \tm{\big( {\rho}_2 \otimes \Gaf \big)} channel we proceed skipping most steps that would simply mirror the previous derivation. 
For \tm{m_3=0} (resp. \tm{M_{3}=0}) we obtain the final result directly from the one of the \tm{\big( \Id_{\rho} \otimes \Gaf \big)} channel, Eq.~\eqref{eq:KW_corr_Irho_Ga5_step9}, by swapping \tm{W_{1} \leftrightarrow W_{0}}, 
\nalign{
C_{\gaDf\etD}^{\text{qc}} 
&=
 -\Tr\Big\{
 \Big( E_{13} + \underline{G_R} \sum\limits_{l_0=0}^{\infty} \big(E_{13} O_{13}\inv\big)^{2l_0}
 E_{13} \overline{G_L} \Big)
 - O_{13} \Big( \Id - \overline{G_L}
 \sum\limits_{l_1=0}^{\infty} \big(O_{13}\inv E_{13}\big)^{2l_1}
 \overline{G_L} \Big) \times
~\nonumber\\
 &\phantom{-\Tr\Big\{}\times
 \sum\limits_{k=0}^{\infty} \Big( 
 -\big[ E_{13} + \overline{G_R} E_{13} \overline{G_L} \big]\inv 
 \overline{G_R} \sum\limits_{l_2=1}^{\infty} \big(E_{13} O_{13}\inv\big)^{2l_2} E_{13} \overline{G_L}
 \Big)^{k} 
 \times
~\nonumber\\
&\phantom{-\Tr\Big\{}\times
 \Big[ E_{13} + \overline{G_R} E_{13} \overline{G_L} \Big]\inv
 O_{13} \Big( \Id - \overline{G_L}
 \sum\limits_{l_3=0}^{\infty} \big(O_{13}\inv E_{13}\big)^{2l_3}
 \overline{G_L} \Big) 
 \Big\}\inv 
,~\label{eq:KW_corr_rho2_Ga5_final}
}%nalign
where the individual contributions are 
\nalign{
E_{13} 
\equiv
|\DN|^2 + |W_{0}|^2 -|W_{1}|^2 + |M_{0}|^2
,\quad
O_{13} 
\equiv
-\big[ W_{0}, \DN \big]_{+}
,\quad
O_{1} 
\equiv
-\big[ \DN, W_{1} \big]_{-}
~.~\label{eq:KW_EO13}
}%nalign
In analogy to the \tm{\big( \Id_{\rho} \otimes \Gaf \big)} channel the overlined factors \tm{\overline{G}_{R} = \overline{O_{1} O_{13}\inv}} or \tm{\overline{G}_{L} = \overline{O_{13}\inv O_{1}} } contribute at the leading order as \tm{\aD^{0}}. 
We move on to the scaling analysis via Eq.~\eqref{eq:KW_scales}, and obtain 
\nalign{
E_{13} 
&\sim  \Big[ c_{|D|^2} +2 C_{W_{0}}C_{W_{\Si}} + \mO\left( \frac{m_q^2}{\La^2},\aD^2\La^2 \right) \Big] \La^2
,~\label{eq:KW_corr_rho2_scaling1}\\
O_{13} 
&\sim  \Big[ +C_{W_{0}D}^{(+)}
 + \mO\left(\aD^2\La^2 \right)
\Big] \frac{1}{\aD} 
,~\label{eq:KW_corr_rho2_scaling2}\\
\overline{G_R} 
= (\big[\DN, W_{1} \big]_{-})(\big[ W_{0}, \DN \big]_{+})\inv
 &\sim C_{W_{0}D}^{(-)} C_{W_0D}^{(+)\,-1} + \mO\left( \aD^2\La^2 \right)
,~\label{eq:KW_corr_rho2_scaling3}
\\
\overline{G_L} 
= (\big[ W_{0}, \DN \big]_{+})\inv(\big[\DN, W_{1} \big]_{-})
 &\sim C_{W_0D}^{(+)\,-1} C_{W_{0}D}^{(-)} + \mO\left( \aD^2\La^2 \right)
,~\label{eq:KW_corr_rho2_scaling4}
\\
\overline{G_L}
 \sum\limits_{l=0}^{\infty} \big(O_{13}\inv E_{13}\big)^{2l}
 \overline{G_L} 
&\sim 
 + \Big( C_{W_0D}^{(+)\,-1} C_{W_{0}D}^{(-)} \Big)^2 
 + \mO\left( \aD^2\La^2 \right)
,~\label{eq:KW_corr_rho2_scaling5}
\\
\Big( E_{13} + \overline{G_R} E_{13} \overline{G_{L}} \Big)
&\sim \Big[ c_{|D|^2} +2 C_{W_{0}}C_{W_{\Si}} 
+ C_{W_0 D}^{(-)} C_{W_0D}^{(+)\,-1} \big( c_{|D|^2} +2 C_{W_{0}}C_{W_{\Si}} \big) C_{W_0D}^{(+)\,-1} C_{W_0 D}^{(-)} 
\nonumber\\&
+ \mO\left( \frac{m_q^2}{\La^2},\aD^2\La^2 \right) \Big] \La^2
,~\label{eq:KW_corr_rho2_scaling6}\\
\Big( \overline{G_R} \sum\limits_{l=1}^{\infty} \big(E_{13} O_{13}\inv\big)^{2l} E_{13} \overline{G_L} \Big) 
&\sim \mO\left( {\aD^2 m_q^2 \La^2},\aD^2\La^4 \right)
~.~\label{eq:KW_corr_rho2_scaling7}
}%nalign
Thus, the divergent contribution in the denominator of the \tm{\big( {\rho}_2 \otimes \Gaf \big)} channel is 
\nalign{
% C_{\gaDf\etD}^{\text{qc}} 
&
% \sim 
%  -\Tr
 \Big\{
 -\Big[ C_{W_{0}D}^{(+)} - C_{W_{0}D}^{(-)} C_{W_0D}^{(+)\,-1} C_{W_{0}D}^{(-)} \Big] 
 \Big[ \big( c_{|D|^2} +2 C_{W_{0}}C_{W_{\Si}} \big) + C_{W_0 D}^{(-)} C_{W_0D}^{(+)\,-1} 
 \nonumber\\&\times 
 \big( c_{|D|^2} +2 C_{W_{0}}C_{W_{\Si}} \big) C_{W_0D}^{(+)\,-1} C_{W_0 D}^{(-)} \Big]
 \Big[ C_{W_{0}D}^{(+)} - C_{W_{0}D}^{(-)} C_{W_0D}^{(+)\,-1} C_{W_{0}D}^{(-)} \Big]
%  \aD^{-2}
%  + \mO\left( m_q^2, \La^2\right)
 \Big\}%\inv 
 \aD^{-2}
%  + \mO\left( \aD^4 m_q^2, \aD^4 \La^2\right)
,~\label{eq:KW_corr_rho2_Ga5_divergence}
}%nalign
which differs from the divergence of the \tm{\big( \Id_{\rho} \otimes \Gaf \big)} channel only by the sign of the \tm{2 C_{W_{0}}C_{W_{\Si}}} terms, cf.~Eq.~\eqref{eq:KW_corr_Irho_Ga5_divergence}. 
\vskip1ex

We then turn our attention to the leading quark mass dependence and set \tm{r=0} (losing minimal doubling in the process). 
Then \tm{O_{1} = - \big[ M_{3}, \DN \big]_{+}} and \tm{E_{1} = |\DN|^2 + M_{0}^2 - M_{3}^2} and therefore 
\nequn{ 
C_{\gaDf\etD}^{\text{qc}} 
 =
 -\Tr\Big\{ 
 \big( c_{|D|^2}\La^2 + \big[ c_{m}^2 + 3 c_{M}^2 \big] \big)
  {m_q^2}
%  + \mO\left( \frac{m_q^4}{\La^2},\aD^2\La^4 \right)
 \Big\}\inv
 + \mO\left( \frac{m_q^4}{\La^6},\aD^2 \right)
,~\label{eq:KW_corr_rho2_Ga5_mq_scaling}
}%nequn 
which coincides with the leading quark mass dependence of the \tm{\big( \rho_1 \otimes \Gaf \big)} channel, cf.~Eq.~\eqref{eq:KW_corr_rho1_Ga5_scaling}. 
Namely, the \tm{\big( {\rho}_{2} \otimes \Gaf \big)} channel indeed contains \acl{TN} states. 
\vskip1ex

\subsection{Pseudo-Nambu-Goldstone bosons with \tm{\gaf \hDj}- or \tm{\gaD\etDf \hDj} fermion bilinears}\label{app:KW_pNGB3}

We start with \tm{\big( \Id_{\rho} \otimes \Gaf \big)} and construct the extended interpolating operators \tm{\bar{\ps} \gaf \hDj \ps} according to Table~\ref{tab:KW_meson}. 
At this stage, we do not have to specify the detailed definition of the shift \tm{\hDj}. 
The full correlation function is 
\nequn{
C_{\gaf \hDj}^{\text{qc}} 
 =
 -\Tr\Big\{ 
 \mDKW\inv \hDj
 \mDKW\daginv \hDj
  \Big\} 
 =
 -\Tr\Big\{ 
 \big[K_{2}+L_{2}\big]\inv \hDj \big[K_{2}+L_{2}\big]^{\dagger\,-1} \hDj
 \Big\}
,~\label{eq:KW_corr_ext_Irho_Ga5_step1}
}%nequn 
where the two operators \tm{K_{2}} or \tm{L_{2}} are defined in Eq.~\eqref{eq:KW_KL2}. 
We employ Eq.~\eqref{eq:prop_decomp} and have 
\nequn{
C_{\gaf \hDj}^{\text{qc}} 
 =
 -\Tr\Big\{ 
 K_{2}\inv \big[\Id \!-\!(L_{2} K_{2}\inv)^2\big]\inv \big[\Id -L_{2} K_{2}\inv \big] \hDj 
 \big(K_{2}\inv \big[\Id \!-\!(L_{2} K_{2}\inv)^2\big]\inv \big[\Id -L_{2} K_{2}\inv \big] \big)\dag
 \hDj
 \Big\}
~.~\label{eq:KW_corr_ext_Irho_Ga5_step2}
}%nequn 
Due to evenness in \tm{r_0}, \tm{r_1}, and~\tm{m_3}, see Sec.~\ref{sec:KW_parametric}, the terms odd in \tm{L_{2}K_{2}\inv}
combine with one another, i.e. 
\nalign{
C_{\gaf \hDj}^{\text{qc}} 
&=
 -\Tr\Big\{ 
 K_{2}\inv \big[\Id \!-\!(L_{2} K_{2}\inv)^2\big]\inv \hDj 
 \big(K_{2}\inv \big[\Id \!-\!(L_{2} K_{2}\inv)^2\big]\inv \big)\dag \hDj
~\nonumber\\
%  &\hspace{3em}+
%  K_{2}\inv \big[\Id \!-\!(L_{2} K_{2}\inv)^2\big]\inv L_{2} K_{2}\inv \hDj 
%  \big(K_{2}\inv \big[\Id \!-\!(L_{2} K_{2}\inv)^2\big]\inv L_{2} K_{2}\inv \big)\dag \hDj 
%  \Big\}
% ,~\label{eq:KW_corr_ext_Irho_Ga5_step3}\\
% &=
%  -\Tr\Big\{ 
%  \big[ K_{2} \!-\! L_{2} K_{2}\inv L_{2} \big]\inv \hDj 
%  \big[ K_{2} \!-\! L_{2} K_{2}\inv L_{2} \big]^{\dagger\,-1} \hDj
% ~\nonumber\\
 &\hspace{3em}
 +
 \big[\Id \!-\!(K_{2}\inv L_{2})^2\big]\inv K_{2}\inv L_{2} K_{2}\inv \hDj 
 \big(\big[\Id \!-\!(K_{2}\inv L_{2})^2\big]\inv K_{2}\inv L_{2} K_{2}\inv \big)\dag \hDj 
 \Big\}
,~\label{eq:KW_corr_ext_Irho_Ga5_step3}
}%nalign
where we have used \tm{K_{2}\inv f\inv(L_{2} K_{2}\inv) = f\inv(K_{2}\inv L_{2})  K_{2}\inv}. 
As required \tm{f(L_{2} K_{2}\inv) = \big[\Id -(L_{2} K_{2}\inv)^2\big]} is invertible and smooth (for nontopological modes and nonvanishing \tm{m_{0}}). 
Since the operators scale according to Eq.~\eqref{eq:KW_scales}, i.e. \tm{K_{2} = \mO(\La, m_q)},~\tm{L_{2} = \mO(\aD \La^2, m_q)},
both the inverses of \tm{\big[\Id -(L_{2} K_{2}\inv)^2\big]} or \tm{\big[\Id -(K_{2}\inv L_{2})^2\big]} can be interpreted as geometric series,  
whose higher summands are suppressed by powers of \tm{\aD \La} or \tm{\sfrac{m_q}{\La}}. 
Thus, the correlator simplifies to
\nalign{
C_{\gaf \hDj}^{\text{qc}} 
&=
 -\Tr\Big\{ 
 \sum\limits_{l_{0}=0}^{\infty} (K_{2}\inv L_{2})^{2l_{0}} K_{2}\inv \hDj 
 \big(\sum\limits_{l_{1}=0}^{\infty} (K_{2}\inv L_{2})^{2l_{1}} K_{2}\inv \big)\dag \hDj
~\nonumber\\
 &\hspace{3em}
 +
 \sum\limits_{l_{2}=0}^{\infty} (K_{2}\inv L_{2})^{2l_{2}} K_{2}\inv L_{2} K_{2}\inv \hDj 
 \big( 
 \sum\limits_{l_{3}=0}^{\infty} (K_{2}\inv L_{2})^{2l_{3}} K_{2}\inv L_{2} K_{2}\inv 
 \big)\dag \hDj 
 \Big\}
~.~\label{eq:KW_corr_ext_Irho_Ga5_step4}
}%nalign
For a scaling analysis via Eq.~\eqref{eq:KW_scales} we expand the shifts \tm{\hDj } with operator-dependent coefficients \tm{c_h},~\tm{C_h} as 
\nequn{
\hDj
= 
\Big[ c_{h} + C_{h}\aD^2\La^2 + \mO\left( \aD^4\La^4 \right) \Big]
=
c_{h} \Big[ \Id + \frac{\mB_{h}}{2} + \mO\left( \aD^4\La^4 \right) \Big]
,\quad
\mB_{h} \equiv 2\frac{C_{h}}{c_{h}} \aD^2\La^2
~\label{eq:KW_shift_scales}
}%nequn 
where \tm{c_{h} = c_{M}} is a legitimate choice for a certain combination of interpolating operator and \acl{TI} mass term. 
For other choices, odd terms in the lattice spacing could be canceled by choosing symmetrized shifts according to Eq.~\eqref{eq:cos_shift}, or an average of suitable combinations of different shifts at source and sink. 
We may assume that \tm{c_{h}} is a real scalar, i.e. a renormalization factor of order one, while \tm{C_{h}} is a generic, noncommuting operator, similar to the mixed terms in~Eq.~\eqref{eq:KW_scales_operators}. 
For smooth links, e.g. after a sufficient amount of gradient flow~\cite{Luscher:2009eq} or other types of link smoothing, e.g. stout~\cite{Morningstar:2003gk} or hypercubic~\cite{Hasenfratz:2001hp} smearing, \tm{c_{h} \simeq 1} and \tm{C_{h} \simeq 0}. 
Since \tm{K_{2} \sim \La}, we leave it unexpanded at first and obtain 
\nequn{
% K_{2}\inv 
% &\sim \Big[C_{D}\inv + c_m c_{|D|^2}\inv \frac{m_q}{\La} 
%  -2 c_m^2 c_{|D|^2}\inv C_{D}\inv \frac{m_q^2}{\La^2} 
%  + \mO\left( \frac{m_q^3}{\La^3}, \aD^2\La^2 \right) \Big] \La\inv
% ,~\\
K_{2}\inv L_{2}
\sim \Big[ K_{2}\inv \La \Big] \Big[ C_{W_{\Si}} \aD\La + c_{M} \frac{m_q}{\La} \Big] 
\Big[ \Id + \mO\left( \aD^2\La^2 \right) \Big] 
~.~
}%nequn
The leading contributions to the correlator are due to \tm{l_{0}=l_{1}=0} in the first line of Eq.~\eqref{eq:KW_corr_ext_Irho_Ga5_step4} combined with the leading term in Eq.~\eqref{eq:KW_shift_scales}. 
Thus, the correlator scales as 
\nalign{
C_{\gaf \hDj}^{\text{qc}} 
&=
 -\Tr\Big\{ 
 c_{h}^2 |K_{2}|^{-2} 
 \Big( 1 + \big[ (K_{2}\inv L_{2})^{2} + \text{h.c} \big] 
 + L_{2} |K_{2}|^{-2} (L_{2})\dag
 + \mB_{h} 
 + \mO\left( \frac{m_q^4}{\La^4}, {\aD^2 m_q^2},\aD^4\La^4 \right) 
 \Big)
 \Big\}
~.~\label{eq:KW_corr_ext_Irho_Ga5_step5}
}%nalign
Before we proceed, we note that \tm{|K_{2}|^{-2}} obviously commutes with \tm{K_{2}\inv} or its Hermitian conjugate, and also commutes at the leading order with \tm{C_{W_{\Si}}}, its Hermitian conjugate or with \tm{c_M}. 
Moreover, any contributions from \tm{(K_{2}\inv L_{2})^2} at even powers of the individual scales are Hermitian. 
Thus, we may reshuffle the \tm{\mO(r^2\aD^2\La^4)} terms due to two powers of \tm{C_{W_{\Si}}}. 
Furthermore, we may absorb the \tm{\mO(m_q^2)} term into the factor \tm{\mK_+\inv} defined in Eq.~\eqref{eq:KW_mK+}, 
\nalign{
C_{\gaf \hDj}^{\text{qc}} 
&=
 -\Tr\Big\{ 
 c_{h}^{2} |K_{2}|^{-2}
 \Big[ 
 1 + 
 \big\{K_{2}\dag C_{W_{\Si}} K_{2}\inv C_{W_{\Si}} + C_{W_{\Si}} K_{2}\inv C_{W_{\Si}}K_{2}\dag 
 + c_{|W_{\Si}|^2} \big\} |K_{2}|^{-2} \aD^2\La^4 
 + \mB_{h} 
 ~\nonumber\\
 &\hspace{3em}
 +(-2+1) c_{M}^2 |K_{2}|^{-2} {m_q^2}
 + \ldots \Big]
 \Big\}
~\nonumber\\
&=
 -\Tr\Big\{ 
 c_{h}^{2} \mK_+\inv
 \Big[ 
 1 \!-\! 
 \big\{K_{2} C_{W_{\Si}} K_{2}\inv C_{W_{\Si}} \!+\! C_{W_{\Si}} K_{2}\inv C_{W_{\Si}}K_{2} 
 \!-\! c_{|W_{\Si}|^2} \big\} |K_{2}|^{-2} \aD^2\La^4  
 \!+\! \mB_{h}
 \!+\! \ldots \Big]
 \Big\}
,~\label{eq:KW_corr_ext_Irho_Ga5_step6}
}%nalign
where the suppressed higher order terms are at the same orders as in Eq.~\eqref{eq:KW_corr_ext_Irho_Ga5_step5}. 
In the inner curly brackets the leading quark-mass dependence cancels, and we recognize the combination \tm{\mB_{-}} defined in Eq.~\eqref{eq:KW_LTB_rho3}. 
We pull both leading \ac{TSB} \tm{\mO(\aD^2)} effects into denominator and write 
\nequn{
C_{\gaf \hDj}^{\text{qc}} 
 =
 -\Tr\Big( 
 c_{h}^{2} \Big[ 
 \mK_+ + \mB_{-} - \mB_{h} \mK_+\inv
 \Big]\inv \Big) 
 + \mO\left( \frac{m_q^4}{\La^6},~\frac{m_q^2 \aD^2}{\La^2},~\aD^4\La^2 \right)
,~\label{eq:KW_corr_ext_Irho_Ga5_KB}
}%nequn
Upon comparing Eqs.~\eqref{eq:KW_corr_ext_Irho_Ga5_KB} and~\eqref{eq:KW_corr_rho3_Ga5_KB}, we find that the two correlators differ by an overall scalar factor \tm{c_{h}^2} of order one, and by extra \ac{TSB} cutoff effects due to a term \tm{-\mB_{h}\mK_{+}}. 
Both of these changes are entirely due to the shifts at source and sink, and we see that these extra \tm{\mO(\aD^2\La^2)} \ac{TSB} cutoff effects in the \tm{\big( \Id_{\rho} \otimes \Gaf \big)} channel are ameliorated for smooth links in their extended operators. 
\vskip1ex

We have included an extra factor \tm{\ri} to make the kernel operator Hermitian for the \tm{\big( {\rho}_2 \otimes \Gaf \big)} channel with extended interpolating operators \tm{\bar{\ps} \gaD\etDf \ri\hDj \ps}, see Table~\ref{tab:KW_meson}. 
We proceed skipping most steps that would simply mirror the previous derivation. 
The full correlation function is 
\nequn{
C_{\gaD\etDf \ri\hDj}^{\text{qc}} 
 =
 -\Tr\Big\{ 
 \mDKW\inv \vr_{2} \ri\hDj
 \mDKW\daginv \vr_{2} \ri\hDj
  \Big\} 
 =
 -\Tr\Big\{ 
 \big[K_{2}+L_{2}\big]\inv \hDj \big[K_{2}-L_{2}\big]^{\dagger\,-1} \hDj
 \Big\}
~.~\label{eq:KW_corr_ext_rho2_Ga5_step1}
}%nequn 
We follow the same steps as before and simplify the correlator to 
\nalign{
C_{\gaD\etDf \ri\hDj}^{\text{qc}} 
&=
 -\Tr\Big\{ 
 \sum\limits_{l_{0}=0}^{\infty} (K_{2}\inv L_{2})^{2l_{0}} K_{2}\inv \hDj 
 \big(\sum\limits_{l_{1}=0}^{\infty} (K_{2}\inv L_{2})^{2l_{1}} K_{2}\inv \big)\dag \hDj
~\nonumber\\
 &\hspace{3em}
 -
 \sum\limits_{l_{2}=0}^{\infty} (K_{2}\inv L_{2})^{2l_{2}} K_{2}\inv L_{2} K_{2}\inv \hDj 
 \big( 
 \sum\limits_{l_{3}=0}^{\infty} (K_{2}\inv L_{2})^{2l_{3}} K_{2}\inv L_{2} K_{2}\inv 
 \big)\dag \hDj 
 \Big\}
,~\label{eq:KW_corr_ext_rho2_Ga5_step4}
}%nalign
where only the sign of the second line has changed. 
The leading contributions to the correlator are due to \tm{l_{0}=l_{1}=0} in the first line of Eq.~\eqref{eq:KW_corr_ext_rho2_Ga5_step4} combined with the leading term in Eq.~\eqref{eq:KW_shift_scales}, and we may absorb the \tm{\mO(m_q^2)} term into the factor \tm{\mK_-\inv} defined in Eq.~\eqref{eq:KW_mK-}, 
\nalign{
C_{\gaD\etDf \ri\hDj}^{\text{qc}} 
&=
 -\Tr\Big\{ 
 c_{h}^{2} \mK_-\inv
 \Big[ 
 1 \!+\! 
 \big\{K_{2} C_{W_{\Si}} K_{2}\inv C_{W_{\Si}} \!+\! C_{W_{\Si}} K_{2}\inv C_{W_{\Si}}K_{2} 
 \!+\! c_{|W_{\Si}|^2} \big\} |K_{2}|^{-2} \aD^2\La^4  
 \!+\! 2\frac{C_{h}}{c_{h}} \aD^2\La^2 
 \!+\! \ldots \Big]
 \Big\}
,~\label{eq:KW_corr_ext_rho2_Ga5_step6}
}%nalign
where the suppressed higher order terms are at the same orders as in Eq.~\eqref{eq:KW_corr_ext_Irho_Ga5_step6}. 
In the curly brackets the leading quark-mass dependence cancels, and we recognize the combinations \tm{\mB_{+}} defined in Eq.~\eqref{eq:KW_LTB_rho1} and \tm{\mB_{h}} defined in Eq.~\eqref{eq:KW_shift_scales}. 
Thus, we pull all leading \ac{TSB} \tm{\mO(\aD^2)} effects into denominator and write 
\nequn{
C_{\gaD\etDf \hDj}^{\text{qc}} 
 =
 -\Tr\Big( 
 c_{h}^{2} \Big[ 
 \mK_{+} + \de\mK_{m_3^2} + \mB_{-} + \de\mB_{r^2} - \mB_{h} \mK_-\inv
 \Big]\inv \Big) 
 + \mO\left( \frac{m_q^4}{\La^6},~\frac{m_q^2 \aD^2}{\La^2},~\aD^4\La^2 \right)
,~\label{eq:KW_corr_ext_rho2_Ga5_KB}
}%nequn
Upon comparing Eqs.~\eqref{eq:KW_corr_ext_rho2_Ga5_KB} and~\eqref{eq:KW_corr_rho1_Ga5_KB}, we find a pattern analogous to the one seen already in the \tm{\big( \Id_{\rho} \otimes \Gaf \big)} channel. 
Namely, the two correlators differ by the same overall scalar factor \tm{c_{h}^2} of order one, and by the same extra cutoff effects due to a term \tm{- \mB_{h} \mK_-\inv}, both of which are ameliorated for smooth links in their extended operators. 
The difference between \tm{-\mB_{h} \mK_+\inv} or \tm{- \mB_{h} \mK_-\inv} is irrelevant at the order of the expansion. 
\vskip1ex
% \clearpage
%%%%%%%%%%%%%%%%%%%%%%%%%%%%%%%%%%%%%%%%%%%%%%%%%%%%%%%%%%%%%%%%%%%%%%%%%%%%%%%%
\bibliography{refs}
\end{document}